\documentclass{mythesis}

\usepackage{epsfig}
\usepackage{cite}
\usepackage{amsmath,amsfonts,amssymb}
\usepackage{mathbbol}
\usepackage{mathrsfs}
\usepackage{enumerate}
\usepackage{vmargin}
\usepackage{latexsym}

\numberwithin{equation}{section}

\newcommand{\al}{\ensuremath{\alpha}}
\newcommand{\bt}{\ensuremath{\beta}}
\newcommand{\ga}{\ensuremath{\gamma}}
\newcommand{\Ga}{\ensuremath{\Gamma}}

\newcommand{\e}{\ensuremath{\epsilon}}

\newcommand{\la}{\ensuremath{\lambda}}

\newcommand{\om}{\ensuremath{\omega}}

\newcommand{\g}{\ensuremath{\gamma}}

\font\mybb=msbm10 at 12pt
\def\bb#1{\hbox{\mybb#1}}

\newcommand{\A}{\ensuremath{{\cal A}}}
\newcommand{\B}{\ensuremath{{\cal B}}}

\renewcommand{\L}{\ensuremath{{\cal L}}}
\newcommand{\M}{\ensuremath{{\cal M}}}
\newcommand{\N}{\ensuremath{{\cal N}}}

\renewcommand{\S}{\ensuremath{{\cal S}}}

\def\PM{{\mathbb P}}
\def\CM{{\mathbb C}}

\newcommand{\ra}{\ensuremath{\rightarrow}}
\newcommand{\del}{\ensuremath{\partial}}

\newcommand{\half}{\ensuremath{\frac{1}{2}}}
\newcommand{\third}{\ensuremath{\frac{1}{3}}}
\newcommand{\quarter}{\ensuremath{\frac{1}{4}}}

\newcommand{\be}{\begin{equation}}
\newcommand{\ee}{\end{equation}}

\newcommand{\ba}{\begin{eqnarray}}
\newcommand{\ea}{\end{eqnarray}}

\newcommand{\gsim}{\raise.3ex\hbox{$>$\kern-.75em\lower1ex\hbox{$\sim$}}}
\newcommand{\lsim}{\raise.3ex\hbox{$<$\kern-.75em\lower1ex\hbox{$\sim$}}}

\newcommand{\nn}{\nonumber}
\newcommand{\wg}{\wedge}
\newcommand{\w}{\wedge}

\newcommand{\ir}[1]{\mathbf{#1}}
\newcommand{\irb}[1]{\overline{\ir{#1}}}
\newcommand{\irc}[1]{\ir{#1} + \irb{#1}}

\newcommand{\tc}[1]{\mathcal{W}_{#1}}

\newcommand{\Vol}{\mathcal{V}}

\newcommand{\cs}{\zeta}
\newcommand{\Omn}{\Omega}
\newcommand{\Omb}{\overline{\Omega}}

\newcommand{\im}{\mathrm{Im\;}}
\newcommand{\re}{\mathrm{Re\;}}

\newcommand{\one}{\mathbb{1}}

\newcommand{\bg}[1]{\mathring{#1}}

\newcommand{\ev}[1]{\left< #1 \right>}

\newcommand{\href}[1]{\underline{#1}}

\newcommand{\ep}{\epsilon}
\newcommand{\vp}{\varphi}

\newcommand{\Reals}{\mathbb{R}}

\begin{document}

\begin{center}
\resizebox{7.5cm}{!}{\includegraphics{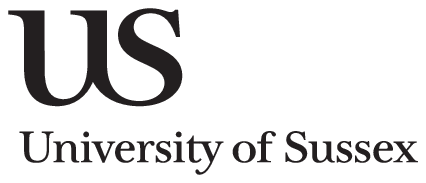}}

\vspace{5em}

\thispagestyle{empty}

\Huge{ASPECTS OF MODULI STABILISATION IN STRING AND M-THEORY\\[2em]}
\large{Eran Palti\\
\textit{Department of Physics and Astronomy\\[10em]
        Submitted for the degree of Doctor of Philosophy\\[2em] 
        July, 2006} }

\end{center}




\newpage
\begin{center}
{\Large ACKNOWLEDGEMENTS \\[2em] }
\end{center}
\noindent
I thank the tax paying citizens of the United Kingdom whose money has made my studies possible. 
\newline
\newline
I thank the University of Sussex for providing a comfortable and stimulating environment to work in, and Edmund Copeland, Mark Hindmarsh and Paul Saffin for guidance during 
my studies here. 
One of the most enjoyable features of research in physics are collaborations and so I thank Edmund Copeland, Thomas House, Andre Lukas, Andrei Micu, Paul Saffin and 
Jon Urrestilla whom I have had the privilege to work with. I also thank Beatriz deCarlos and David Bailin for stimulating discussions.
\newline
\newline
I thank my parents, Gilda and Carmel, for their encouragement throughout my education. Without their support, emotional and financial, this thesis 
would not have been possible. I therefore dedicate this thesis to them and to my brother, Itai, who together formed the platform from which I began 
this journey.
\newline
\newline
Finally, I thank my beautiful wife, Rosie, whose love and inspiration is beyond the scope of this acknowledgements section.

\newpage
\begin{center}
{\Large ABSTRACT \\[2em] }
\end{center}

In this thesis we study compactifications of type II string theories and M-theory to four dimensions.
We construct the four-dimensional $\N=2$ supergravities that arise from compactifications of type IIA string theory and M-theory 
on manifolds with $SU(3)$-structure. We then study their potential for moduli stabilisation and give explicit examples where all the moduli 
are stabilised. We also study the effective action for type IIB conifold transitions on Calabi-Yau manifolds. We find that, 
although there are small regions in phase space that lead to a completed transition, 
generically the moduli are classically trapped at the conifold point thereby halting the transition.

\tableofcontents

\starttext

%
\chapter{Introduction: To Eleven Dimensions}
\label{cha:intro}
%

The idea that the universe is governed by a set of rules and that we can discover what those rules are dates back to the time of the Ancient Greeks. This idea still forms the basis of physics today. However, the original proposition, which prevailed until the sixteenth century, was to deduce 
scientific rules by pure thought alone. This overestimated our reasoning ability and although the approach led to many successes ultimately it could only take us so far. The sixteenth century revolution of scientific thought argued that knowledge of nature should be gained through experiments.
These two approaches, rational deduction through thought and learning through experiments, 
have remained the driving forces behind physics to this day and it is the balance between the two that forms the motivation for this work.

By the end of the 1970s, the Standard Model (SM) and General Relativity (GR) could, between them, account for almost all the known observations. 
The problems that remained could be referred to as purely aesthetic. 
GR broke down at space-time singularities which meant it could never fully describe black holes or the Big Bang. 
The SM lacked explanation for the values of nineteen parameters in the theory which had to be put in by hand. 
Some of those parameters, such as the mass of the Higgs boson, had to be highly fine-tuned to match observations.
Apart from the problems faced by each of the theories separately the most theoretically troublesome issue was that 
the machinery behind the SM, that is Quantum Field Theory (QFT), could not be used to describe gravity and so the two theories remained separate. 

The beauty and unity of the laws of physics is not something easily ignored however. Indeed these two concepts were the driving force behind the discovery of the SM and GR themselves. So the physics community attempted to resolve these issues and from these motivations came ideas such as supersymmetry, string theory and extra dimensions that shape theoretical physics today.

The introduction to this work aims to review briefly the issues discussed above which culminated in M-theory. 
In section \ref{sec:smgr} we review the current standard model of particle physics 
and the current theory of gravity. 
Successive attempts at unification of the two theories are discussed in the following sections with unification of the 
symmetry groups discussed in section \ref{sec:susy}, quantum theory and gravity \ref{sec:stringtheory}, and the five string 
theories \ref{sec:mtheory}. 
Finally, section \ref{sec:back4} summarises the work presented here, which attempts to resolve the difficulties that arise when attempting to reconcile
the predicted eleven (ten) dimensions of M-theory (string theory) with the observed four.

%
\section{The Standard Model and gravity}
\label{sec:smgr}
%

There are four known forces in the universe: Gravity, the Weak force, the Strong force and the Electromagnetic force. The latter three, along with the 
known matter content, are described by the SM. 
The precise details of the SM are not  
discussed here but, in short, the matter content consists of three fermionic quark and lepton doublets and one scalar field called the Higgs \cite{Eidelman:2004wy}. 
The theory 
that describes the interactions between the matter and the forces is a QFT with a Lagrangian that has an internal symmetry of the 
group $SU(3)\times SU(2) \times U(1)$. The covariant derivatives on the fields that ensure this symmetry must contain spin-one fields that gauge this symmetry and these are the force carrier fields. More precisely the Strong force has eight massless carriers, the gluons, that are associated with the eight generators of the $SU(3)$. The Weak force has three massive carriers, the $W^{\pm}$ and $Z^0$, that are associated with an $SU(2)$ subgroup of the $SU(2) \times U(1)$. Finally the Electromagnetic force is mediated by the massless photon which is associated with a $U(1)$ subgroup of the $SU(2) \times U(1)$. A key concept of extensions to the SM is the unification of forces. 
An example of this process is 
the way in which the Electromagnetic force  
unifies what were once thought to be two separate forces, Electricity and Magnetism, into one force. 
More precisely, Electricity and Magnetism can be shown to be two properties of the same force. 
Within the SM there is evidence that all the forces can be unified into a Grand Unified Theory (GUT) where they are again differing properties of the same force. 
Such unifications are a more quantitative measure of what was 
previously referred to as the aesthetic quality of the theory. The more unified the theory the more aesthetically pleasing it is to the physicist. But there is more to unification than aesthetics: throughout history some of the greatest progress in physics has come from unifications. There remains one important unification that we have yet to address and that is of the SM forces with the force of Gravity. The pursuit of this unification takes us far (all the way to eleven dimensions), but first we summarise the current formulation of gravity. 

The current theory of gravity is purely classical and is formulated in terms of the metric on four-dimensional space-time, $g_{\mu\nu}$, where 
$\mu,\nu =0,1,2,3$. The metric describes the geometry of space-time and from it we can construct other useful geometrical quantities such as the Ricci tensor $R_{\mu\nu}$ and scalar $R$ which are defined in the appendix. 
GR gives the relation between the geometry of 
space-time and the energy present, which quantitatively is given by Einstein's equation
\be
R_{\mu\nu} - \half R g_{\mu\nu} = 8 \pi G T_{\mu\nu} \label{einseq} \mathrm{\;.}
\ee
$G$ is Newton's constant which can be used to define the Planck mass $M_p$
\be
M_p = \frac{1}{\sqrt{G}} = 1.22 \times 10^{19} \mathrm{GeV} \mathrm{\;,}
\ee
where we work in units where the speed of light $c$ is set to unity.
$T_{\mu\nu}$ is the energy-momentum tensor that describes the energy content.
There is also a Lagrangian formulation for GR where the action is given by
\be
S_{\mathrm{Gravity}} = \int_{\cal S}{\sqrt{-g}d^4x \left[ R + \L_{\mathrm{Matter}} \right]} \mathrm{\;,}
\ee
with $g$ standing for the determinant of the metric and $x$ are the space-time co-ordinates. $\L_{\mathrm{Matter}}$ is the Lagrangian density for all the matter content. In this formulation the Einstein equation (\ref{einseq}) is simply the equation of motion for the metric. 

An important part of the SM are the symmetries of the Lagrangian. The Lagrangian for gravity also has symmetries and these form the 
space-time 
symmetries of the Poincar\'e group. The Poincar\'e group is the sum of all space-time transformations that include translations, rotations and 
Lorentz boosts. Overall, this group has ten elements and has a representation in terms of vectors $P_{\mu}$ and 
symmetric $4 \times 4$ matrices $J_{\mu\nu}$ that satisfy the commutation relations
\ba
\left[ P_{\mu}, P_{\nu} \right] &=& 0 \mathrm{\;,} \nn \\
\left[ P_{\mu}, J_{\nu\rho} \right] &=& \eta_{\mu\nu} P_{\rho} - \eta_{\mu\rho} P_{\nu} \mathrm{\;,} \nn \\
\left[ J_{\mu\nu}, J_{\rho\sigma} \right] &=& - \eta_{\mu\rho} J_{\nu\sigma} - \eta_{\nu\sigma} J_{\mu\rho} + \eta_{\mu\sigma} J_{\nu\rho} + \eta_{\nu\rho} J_{\mu\sigma} \mathrm{\;,}
\ea
where the Minkowski metric $\eta=\mathrm{diag}(-1,1,1,1)$. As a first step in unifying GR with the SM we may attempt to unify the two symmetry 
groups and this is the topic of the next section. 

%
\section{Supersymmetry}
\label{sec:susy}
%

The Poincar\'e group is a symmetry group of external space-time symmetries. The SM's $SU(3)\times SU(2) \times U(1)$ are symmetries of internal degrees of freedom of
the fields. A first attempt at unification might be to unify the two types of symmetries. The only \cite{colman-mandula} possible extension to the Poincar\'e group 
by an internal symmetry is a graded Lie algebra of the form \cite{wess-bagger}
\ba
\left\{ Q^{i}_{A}, \bar{Q}_{Bj}  \right\} &=& 2 \sigma^{\mu}_{AB} \delta^i_j P_{\mu} \mathrm{\;,} \nn \\
\left\{ Q^{i}_{A}, Q_{B}^j  \right\} &=& - \hat{\epsilon}_{AB} Z^{ij} \mathrm{\;,} \nn \\
\left[ Q^i_A , J^{\mu\nu} \right] &=& -i \left(\sigma^{\mu\nu}\right)^B_A Q^i_B \mathrm{\;,} \nn \\
\left[ Q^i_A, T_{r} \right] &=& \left[ Q^i_A, P_{\mu} \right] = 0 \mathrm{\;,}
\ea
where $\sigma^{\mu\nu} = \quarter \left( \sigma^{\mu}\bar{\sigma}^{\nu} - \sigma^{\nu}\bar{\sigma}^{\mu}\right)$ with $\sigma^{\mu}$ being the Pauli spin matrices. The indices $A,B=1,2$ are Weyl spinor indices and $i,j=1,...,\N$ correspond to the number of sets of such generators that can be introduced with $Z$ labelled as the central charges of the theory. The generators $T_{r}$ denote the SM gauge groups.
In order to unify internal symmetries with the space-time symmetries we have introduced new generators and 
the natural question to ask is what symmetry is generated by the $Q^i_A$s? This new symmetry is called \textit{supersymmetry} and it is a symmetry that transforms between the bosonic and fermionic fields in the theory. To see this we introduce anti-commuting Grassman parameters 
$\xi^i_A$ which generate a supersymmetry transformations
\be
\delta_{\xi} \phi = i\left( \xi Q + \bar{\xi} \bar{Q} \right) \phi \mathrm{\;,}
\ee
where we have introduced a scalar field $\phi$ and have considered only $\N=1$ for simplicity. 
Then, in order for the supersymmetry algebra to close, $\phi$ should transform into another field, which should in turn transform under supersymmetry, eventually closing the algebra. The simplest possibility linear in $\xi$ is
\be
\delta_{\xi} \phi = a \xi \psi + b \bar{\xi} \bar{\psi} \mathrm{\;,}
\ee
where $a$ and $b$ are some constants and $\psi$ is a fermionic field. 
We see that a new field must be introduced which differs from the transformed field by spin-$\half$ and 
this new fermion is called the superpartner of $\phi$. To make the Lagrangian invariant under this symmetry it is sometimes useful to group all the superpartners of a field together into a supersymmetric invariant multiplet or a superfield. 
By writing the Lagrangian in terms of these superfields we ensure it is supersymmetric. 

We have seen that supersymmetry naturally emerges from unification. It also predicts that every particle must have a superpartner of equal mass and with a spin difference of a half. This prediction is clearly false since we see no such bosonic superpartners to the
SM fermions. We have arrived at a point that is a thread throughout this work: following unification we reach a theory that does not agree with 
the universe we observe. Nonetheless we continue with the hope that unification is a strong enough principle as guide whilst bearing in mind that 
at the end we should try to make contact with the observed universe. In this case the most obvious answer to the problem of the missing particles is that supersymmetry is a broken symmetry. Breaking supersymmetry induces a mass gap between the observed fermions and their missing superpartners so that if the energy scale of this mass gap, which is the energy scale of supersymmetry breaking, is larger than that which we are able to probe in particle accelerators, i.e. the TeV scale, the superpartners are just too massive to have been detected yet. The scenario of a TeV scale broken supersymmetry also offers possible solutions to current problems in cosmology and the SM. Most notably it explains why the mass of Higgs boson is of a 
TeV scale when without supersymmetry quantum loop corrections mean its mass should naturally be of the Planck scale. It also offers a candidate for the 
observed cosmological dark matter as the lightest supersymmetric particle and forms an important part of the 
unification of the forces in GUT theories.

Supersymmetry by itself is a global symmetry which means that the spinors $\xi^i_A$ do not vary over space-time. This is rather unnatural as there is no reason for such a global constraint, indeed the SM and Poincar\'e symmetries previously discussed are all local symmetries. Making supersymmetry local leads to a theory of \textit{supergravity} where the spin-$2$ graviton also has a spin-$\frac{3}{2}$ superpartner, the gravitino. Therefore, the simplest natural theory that unifies internal 
and space-time symmetries is a four-dimensional supergravity. However, in attempting the next step of unification, that is applying QFT to gravity, four-dimensional supergravity is not enough and an even more radical rethinking of the universe is required.  

%
\section{String theory}
\label{sec:stringtheory}
%

Quantum Electrodynamics (QED) is the quantum field theory that describes the Electromagnetic force. It has been successfully experimentally probed 
to a greater accuracy than any other theory in physics. There are also accurately tested QFTs for the Weak and Strong forces. 
Like the SM forces, a truly unified theory should also be a quantum field theory of gravity\footnote{The motivation of unification for a quantum theory of gravity has been emphasised here in keeping with the theme of the chapter. There is, 
however, a more fundamental reason for requiring such a theory and that is because gravity is coupled to matter which in turn we know is described by a quantum theory. Then making an observation of a quantum matter distribution collapses the wavefunction thereby altering its gravitational field instantaneously and, if the eigenstates are that of space-like separated matter, acausally.}.
Gravity, however, has a fundamental property different to the other forces that has made this construction so far unattainable. The SM forces are 
forces that act on a fixed space-time background, gravity is a force that is itself the geometry of space-time and so can not be defined on some 
fixed background. The key problem that arises from this is that causality becomes ill-defined since the metric, which should play the role of the quantum field, can be in some superposition of eigenstates where in one state points can be space-like related while being time-like related in another state. Abandoning causality would lead to even greater conceptual difficulties. This problem remains unsolved so far and is not discussed further in this work. 

The next best thing to a full quantum theory of gravity is a QFT of gravitational perturbations about a classical background, that is 
a quantum theory of gravitons. Taking a metric of the form
\be
g_{\mu\nu} = \eta_{\mu\nu} + h_{\mu\nu} \mathrm{\;,}
\ee
we can attempt to quantise the perturbation $h_{\mu\nu}$, that describes a graviton, about a classical Minkowski background metric $\eta_{\mu\nu}$. 
The problem with such a theory is that it is non-renormalisable. This can be attributed to the fact that the loop expansion parameter $G$ is 
dimensionful. Terms in the Lagrangian generated by successive loops must have an increasing number of dimensionful integrals to counter the 
dimension of $G$. Therefore an attempt to renormalise the theory by adding counter-terms requires an infinite number of different counter-terms 
corresponding to the infinite number of loops. This is by definition a non-renormalisable theory. 

The problem of non-renormalisability of a quantum theory of gravitons has a solution. The solution is that the fundamental constituents of the universe are not point 
particles but rather strings of a finite length. The theory describing a quantum string is a conformal QFT on the two-dimensional worldsheet of the string with physical 
states corresponding to harmonic excitations of the string. The mass of states $M$ on the string is given in units of the inverse string tension $\alpha'$ as
\be
\alpha' M^2 \sim  N - A \mathrm{\;,}
\ee
where $N$ is the number of quantised excitations on the string and $A$ is a (normal ordering) constant which takes the value of one for an open string and two for a closed 
string \cite{polchinski}. Let us define the creation operators for left-moving and right-moving harmonic excitations on the string as 
$\hat{\alpha}^{\mu}$ and $\hat{\tilde{\alpha}}^{\mu}$ respectively. The index $\mu$ corresponds to which space-time direction the excitation is in. Then a state 
$\epsilon$ of two excitations $N=2$ on the closed string is massless and is generated by the two operators acting on the vacuum state $|0>$ as
\be
\epsilon_{\mu\nu} \hat{\alpha}^{\mu} \hat{\tilde{\alpha}}^{\nu} |0> \mathrm{\;.}
\ee
The massless state $\epsilon$ is a general two-tensor and so can be decomposed into a traceless symmetric part, anti-symmetric part and a trace 
\be
h_{\mu\nu} \equiv \epsilon_{(\mu\nu)} \;\;,\;\;\; B_{\mu\nu} \equiv \epsilon_{[\mu\nu]} \;\;,\;\;\; \phi \equiv \epsilon_{\mu}^{\mu} \mathrm{\;.}
\ee
We now see that string theory necessarily has a massless spin-$2$ state $h_{\mu\nu}$ which can play the role of the graviton and is therefore a theory of gravity.
It also has a two-form $B_2$ which is called the Neveu-Schwarz Neveu-Schwarz (NS-NS) two-form, and a scalar $\phi$ called the dilaton. The vacuum expectation value of 
the dilaton also plays the role of the string self coupling with the self coupling $g_s$ given by
\be
g_s = e^{\ev{\phi}} \mathrm{\;.}
\ee
States on the string that are not massless have masses of order $\left(\alpha'\right)^{-\half}$. The value of the string tension is as yet unknown but given that it is a 
theory of gravity it is expected to be of order the Planck scale\footnote{There are however scenarios in string theory where the string scale is actually much lower and 
can be as low as the TeV scale.}. In general though, in terms of making contact with experiment, we are only interested in the lowest mass states of the theory and so 
can truncate the higher mass states, a truncation which is valid well below the string scale.

The property of string theory that solves the non-renormalisation problem is that there is no need for a cut-off or counter terms 
because all gravitational loop terms are actually finite. This key result is difficult to derive and so will not be derived here but rather the reader is referred literature \cite{polchinski}. There is however an intuitive argument as to why this is the case. The reasoning is that, unlike point particles, the coupling of strings 
to each other does not occur at a well defined point because of their length. This means that the gravitational interaction is 'smeared out' over a finite volume which 
resolves the divergences. The finiteness of loop amplitudes means that string theory can potentially be a theory that describes, in a fully consistent quantum field 
theoretic way, the SM forces and gravity together. Further, space-time supersymmetry, as discussed in section \ref{sec:susy} comes out naturally from a supersymmetric 
string, the superstring, which is described by a two-dimensional supersymmetric conformal QFT. Therefore, the low energy limit of superstring theory is a supergravity, 
the content of which can be determined from the massless spectrum of the string theory.

There are five known self-consistent superstring theories: type I, type IIA, type IIB, heterotic $SO(32)$ and heterotic $E_8 \times E_8$. They are all 
only anomaly-free in ten space-time dimensions. They all feature the 
NS states of $h$, $B$ and $\phi$, but also have further states that are particular to the type of theory. Type I is an $\N=1$ supersymmetric theory which has 
496 vectors with a gauge group $SO(32)$. Type IIA is $\N=2$ supersymmetric and has a vector $A_{\mu}$ and three-form $C_{\mu\nu\rho}$. Type IIB is again 
$\N=2$ supersymmetric but has a scalar $l$, a two-form $C_{\mu\nu}$ and a four-form $A_{\mu\nu\rho\sigma}$. The extra states in type II theories are called Ramond-Ramond (RR) 
states. Finally the heterotic string theory is $\N=1$ supersymmetric and has 496 vectors with gauge group $SO(32)$ or $E_8\times E_8$. In terms of the physical meaning 
of the extra states, the RR states in the type II theories are extra states on the closed string corresponding to world-sheet fermions with 
periodic boundary-conditions. The type I string arises from projecting out half these states using world-sheet parity, and the heterotic theories only have left-moving 
world-sheet fermions.

In the mid-nineties it was realised \cite{Polchinski:1995mt} that the spectrum of states in string theory is significantly richer than that of just the string. There are higher dimensional 
extended objects called branes on which open strings can end. The branes are charged under the states of the theory so that branes charged under the RR fields 
are called D-branes and branes charged under the common NS two-form are called NS-branes. The dimension of the brane is fixed by the degree of the form it couples to 
and we write a $p+1$ dimensional brane as a p-brane (the +1 difference accounts for the time direction).
Since the different string theories contain different states they also have different brane content. In particular D-branes only feature in type II string theories 
with type IIA containing D0, D2, D4, D6 and D8 branes, and type IIB having D1, D3, D5, D7 and D9 branes\footnote{The higher-dimensional branes couple to the Hodge duals of the field-strengths.}. 

String theory is the leading candidate for a unified quantum theory of the SM forces and gravity. As in the case of supersymmetry, there is a 
price to pay for the unification since superstring theory is only consistently defined in ten space-time dimensions. 
Again, this is a prediction that is in obvious contradiction 
with the observed four space-time dimensions. The resolution of this problem is discussed extensively in the upcoming sections.
There is another, slightly more aesthetic issue to address first. The idea of a quantum unified theory of all the interactions is really a hope for \textsl{the} 
theory of the universe. There are however five known string theories and the natural question is which one describes our universe? This is the topic of the next section.

%
\section{M-theory}
\label{sec:mtheory}
%

The final unification that is discussed in this chapter is that of the five string theories. 
It has been shown \cite{Witten:1995ex} that actually the five, ten-dimensional, string theories are all limits of a single, eleven-dimensional, 
theory that has been termed M-theory. This theory is not a theory of strings but rather a theory of two-dimensional extended objects, i.e. membranes.
The full version of the theory 
is as yet unknown but its states can be related in certain limits to those that exist in string theory. For example, taking one 
of the space dimensions to be in the shape of a circle, the theory is equivalent to type IIA string theory with the size of the circle playing the role 
of the string self coupling (the dilaton). Taking the circle to have a $Z_2$ symmetry leads to the heterotic $E_8 \times E_8$ string. 
There are then further dualities termed T, S and U dualities that connect all the string theories.

This unification has lead to the proposition that M-theory is indeed the correct theory of the universe. This idea is further backed by a seemingly unrelated fact. In section \ref{sec:susy} it was argued that supergravity is a natural first step towards unification. There are many different consistent supergravity theories that can be defined in four space-time dimensions and so this is not a strong enough principle to determine the theory. However, the number of possible supergravities decreases with increasing space-time dimensions culminating in the fact that eleven-dimensions is the largest number of dimensions for which a supergravity can be constructed \cite{Duff:1986hr} and there is a unique such theory which is precisely the low energy limit of M-theory. This supergravity contains the metric and a three-form $C_{\mu\nu\rho}$ which naturally couples to the membrane.

%
\section{Back to four dimensions}
\label{sec:back4}
%

The unification of the SM with gravity has lead us to a theory that has fascinating potential and is very aesthetically pleasing. M-theory has the look of a theory to describe all the known interactions. It only has one parameter, the membrane tension, and it predicts the number of space-time dimensions.
However, in looking for this theory we have strayed far from the observed universe, with the most obvious contradiction being that of the number of space-time dimensions. This thesis is concerned with the leading candidate for a possible resolution to this problem. The idea is that any extra unobserved spatial dimensions (seven in M-theory and six in string theory) are curled up (compact) with a radius much smaller than that which can be 
probed by experiments. Then, although the string/membrane propagates in a higher-dimensional universe, we observe an effective four-dimensional universe. 

The physics of ten-dimensional string theory and eleven-dimensional M-theory is fixed by the constraints of the 
theory as discussed in this introduction. The four-dimensional physics, however, depends on the extra dimensions 
or more precisely on the geometry of the manifold they specify. The task of recovering realistic models of the universe 
is therefore in large an exploration of the possible geometries spanned by the extra dimensions. In this thesis 
we study the effective four-dimensional physics that arises from particular choices of geometries. In particular 
we are concerned with the dynamics of the geometry that manifests itself in four-dimensional fields termed moduli. 
Determining the value of these moduli in the vacuum is an important first step towards recovering the standard model, and the observed cosmology, as an effective low energy four-dimensional theory. The motivation for this endeavour stems from the first paragraph of this introduction: theoretical aesthetics can only take us so far and it is only 
by making contact with the experimentally observed universe that we can be sure that we are on the right path to a 
true description of the universe rather than one of the many constructions that seem logically consistent to the human mind. 

This thesis is formed of five chapters. Following the introduction, chapter two reviews the idea of a compactification making the points discussed in the introduction more mathematically precise. 
The chapter also discusses some of the important problems that the resulting four-dimensional theories face highlighting the particular problem of moduli stabilisation. A brief review of the possible resolutions of this problem forms the end of the chapter and the introduction to the original work in this thesis. 
The three chapters that follow discuss moduli stabilisation in varying scenarios. Chapter \ref{cha:con} considers the case where the internal geometry is a Calabi-Yau manifold 
and studies the moduli dynamics near conifold points in the Calabi-Yau. 
Chapters \ref{cha:su3iia} and \ref{cha:su3mtheory} study flux compactifications of type IIA string theory and M-theory on 
six-dimensional and seven-dimensional manifolds with $SU(3)$-structure respectively.
The resulting four-dimensional theories are studied and in particular their potential for moduli stabilisation.

%
\chapter{Compactifications of String and M-theory}
\label{cha:compactifications}
%

In the introduction it was argued that a theory that unifies gravity with the standard model forces is naturally defined in eleven (or ten) space-time dimensions. A possible resolution of this contradiction with observations is that 
the extra dimensions are curled up with a radius smaller than can be probed with current experiments.
This idea is made more rigorous in the following sections which also assume knowledge of the mathematics 
reviewed in the appendix. Section \ref{sec:compactdim} discusses the idea of a compact spatial dimension.
Section \ref{sec:compactman} generalises this to six and seven compact dimensions and reviews the possible manifolds spanned by the dimensions. We then proceed to derive the effective four-dimensional action that results from the 
compactification which, for all the cases we consider, is an $\N=2$ supergravity.
In section \ref{sec:n=2sugra} we review the important features of general $\N=2$ supergravity in four dimensions. 
In section \ref{sec:cycompact} we go through a simple compactification to show how the four-dimensional effective action can be calculated and how it fulfils the conditions of $\N=2$ supergravity.
Section \ref{sec:vacdeg} discusses the main problem that the idea of compact dimensions leads to, that is the vacuum degeneracy problem, and the two approaches to resolving this problem that are explored in this thesis.

%
\section{Compact dimensions}
\label{sec:compactdim}
%

The idea of extra dimensions was introduced long before any considerations of string theory by Kaluza and Klein in an attempt to 
unify the two known forces at the time of electromagnetism and gravity \cite{kaluza,klein}. Although this is a much simpler example than the ones explored in this 
work it nevertheless includes the key issues and so forms a useful introduction to compactifications.
We begin with the five-dimensional Einstein-Hilbert action of pure gravity 
\be
S_{EH} = - \int_{\M_5}{d^5X \sqrt{-\hat{g}} \hat{R}} \mathrm{\;.}
\ee
We introduce here the notation where higher dimensional quantities are denoted by a $\hat{\;}$ which in this case are the five-dimensional metric 
$\hat{g}_{MN}$, with capital roman indices denoting higher dimensional co-ordinates $X^M$ where in this case $M,N = 0,1,2,3,4$. $\hat{R}$ denotes the five-dimensional Ricci scalar. We now consider a vacuum of this theory where the five-dimensional metric decomposes into a product of a 
four-dimensional part, which we take to be the space-time metric $g_{\mu\nu}$ where $\mu,\nu = 0,1,2,3$, and part along the fifth dimension $g_{55}$
\be
\left< \; \hat{g}_{MN}\right> dX^M dX^N = \left< g_{\mu\nu} \right> dx^{\mu} dx^{\nu} + \left< g_{55} \right> dy dy \mathrm{\;,}
\ee
where we have introduced the four space-time co-ordinates $x^{\mu}$ and the co-ordinate in the extra dimension $y$. The brackets $\left<...\right>$ denote 
vacuum expectation values of quantities. If the theory admits such a vacuum then we can perform a \textit{spontaneous compactification} of the theory. The idea being that instead of considering all the possible solutions of the 
theory we consider a particular ansatz which is a solution. This is important when we come to generalise this idea to string and M-theory in section \ref{sec:compactman} where the task of finding all the possible solutions 
is an extremely difficult one and again we resort to considering an ansatz. We then take this ansatz as the vacuum of the theory without considering the dynamics of how the theory reached that state.

To see if the theory admits a spontaneous compactification we solve the five-dimensional equations of motion which in this case is simply the Einstein equation
\be
\hat{R}_{MN} = 0 \mathrm{\;,}
\ee
which admits the solution 
\be
\left< g_{\mu\nu} \right> = \eta_{\mu\nu} \;\;\mathrm{\;,}\;\;\;\; \left< g_{55} \right> = 1 \mathrm{\;.}
\ee
We now consider fluctuations about the vacuum that are the fields of the theory. In this case we only have gravity and so all the fields arise from fluctuations of the five-dimensional metric which, following \cite{Duff:1986hr}, we write as 
\be
d\hat{s}^2 \equiv \hat{g}_{MN}dX^M dX^N = \phi^{-\third} \left[ \left( g_{\mu\nu} + \phi A_{\mu}A_{\nu} \right) dx^{\mu} dx^{\nu} 
+ 2 \phi A_{\mu} dx^{\mu} dy + \phi dy dy \right] \mathrm{\;.}
\ee
The effective fields in four dimensions are therefore the metric $g_{\mu\nu}$, a gauge field $A_{\mu}$ which is called the graviphoton, and a 
scalar $\phi$ which, anticipating future directions, we call the dilaton.
Taking the extra dimension as compact, and hence periodic, we preform a Fourier expansion of the dependence of the fields on it as 
\be
g_{\mu\nu} = \sum_{n=-\infty}^{+\infty} g^{(n)}_{\mu\nu}(x) e^{\frac{2\pi n i y}{l}} \;\mathrm{,}\;\;\;\;\;
A_{\mu} = \sum_{n=-\infty}^{+\infty} A^{(n)}_{\mu}(x) e^{\frac{2\pi n i y}{l}} \;\mathrm{,}\;\;\;\;\;
\phi = \sum_{n=-\infty}^{+\infty} \phi^{(n)}(x) e^{\frac{2\pi n i y}{l}} \mathrm{\;,}
\ee
where $l$ is the radius of the extra dimension. Consider the field $\phi$ which obeys the five-dimensional massless Laplace equation
\be
0 = \partial_M \partial^M \phi = \partial_{\mu}\partial^{\mu} \phi + \partial_y \partial_y \phi =  
\sum_{n=-\infty}^{+\infty} \left[ \partial_{\mu} \partial^{\mu} - \left( \frac{2\pi n}{l}\right)^2 \right]\phi^{(n)}(x) e^{\frac{2\pi n i y}{l}} \mathrm{\;.}
\ee
The four-dimensional modes $\phi^{(n)}$ have masses quantised in units of $2 \pi / l$. Taking the compactification radius to be very small, so that we do not observe it, we recover that all the excited modes are of a high mass. We can specify an effective theory keeping only the 
massless mode of $n=0$ and truncating the higher Kaluza-Klein (KK) modes. Finally, we integrate out the extra dimension to recover a low energy four-dimensional action which in this case reads
\be
S_{4D} = \int_{\cal S}{d^4x \sqrt{-g}\left[ R - \quarter \phi^{(0)} F^{(0)}_{\mu\nu}F^{(0)\mu\nu} - \frac{1}{6\left(\phi^{(0)}\right)^2}\partial_{\mu}\partial^{\mu}\phi^{(0)}\right]} \mathrm{\;,}
\ee
where $F^{(0)}_{\mu\nu}$ is the field strength of the zero mode gauge field $A^{(0)}$. 
We therefore recover four-dimensional gravity and electromagnetism as manifestations of five-dimensional gravity thereby unifying them.
Unfortunately we have also ended up with unwanted baggage in the form of a massless scalar mode of gravity, the dilaton. 
This is a typical by-product of compactifications and forms one of the major problems faced by such approaches. In fact 
the extra scalar mode was the reason for the rejection of the original KK proposition as it predicted 
unobserved long-range interactions. We return to this point in section \ref{sec:vacdeg}. 

The procedure of dimensional reduction of an action as reviewed above can be generalised to the case of six or seven extra dimensions. The full ten or eleven-dimensional 
fields can be decomposed in terms of eigenfunctions of the Laplacian into a tower of four-dimensional fields of which we only keep the lowest mass states. The internal manifold can then be integrated out to leave an effective four-dimensional action. The form of this action depends on the geometry of the manifold that the extra dimensions span. The problem of recovering the four-dimensional SM and gravity we observe from 
the higher dimensional supergravity actions (that are, in turn, the low energy limits of the string or M-theory actions) can therefore be cast into a geometric problem and this is the topic of the next section.

%
\section{Compactification manifolds}
\label{sec:compactman}
%

Applying the idea of compact dimensions to string and M-theory implies that instead of one compact dimension we should consider six and seven 
compact dimensions respectively. These dimensions span a compact manifold that we refer to as $\M$, with the remaining four space-time 
dimensions of the universe spanning the manifold $\S$. Since the time dimension lies in $\S$ we take the metric on $\M$ to be of Euclidean signature. 
Are there other constraints we can place of $\M$? In order to recover a well defined four-dimensional action we 
require the universe to be of the 
product form $\S \times \M$ or of the warped product form $\S \ltimes \M$ in which case the metric takes the form
\be
d\hat{s} = e^{2A(y)} g_{\mu\nu}(x) dx^{\mu}dx^{\nu} + g_{mn}(y) dy^{m} dy^{n} \mathrm{\;,}
\label{genmetdec}
\ee
where the internal co-ordinates $n,m$ range over the six or seven internal dimensions and $A(y)$ is a possible warp-factor that is still 
allowed for a consistent four-dimensional action. Within the spontaneous compactification scheme this puts a constraint on the geometry of $\M$ that 
it should form a solution of the full higher dimensional theory of the product type. This constraint is the only strict non-trivial constraint 
on the manifold, however there are other constraints that can be imposed to do with recovering a semi-realistic action in 
four dimensions. The most useful one of these is the requirement for the manifold to preserve some supersymmetry in four dimensions. Eventually we 
wish for supersymmetry to be broken but maintaining some minimal amount of supersymmetry serves as a good starting point form which we may later add effects that break supersymmetry in accordance with observations. 

In the following discussion the restriction for supersymmetry is made more rigorous through the introduction of the idea of $G$-structures. We then consider a compactification on the simplest, and best understood, example of a manifold that satisfies these constraints that is a Calabi-Yau manifold. The constraint 
of spontaneous compactification is shown to be satisfied for this example but a more general discussion of this constraint is left for chapters \ref{cha:su3iia} and \ref{cha:su3mtheory}.

%
\subsection{Constraints from supersymmetry: G-structures}
\label{sec:constsusy}
%

In section \ref{sec:susy} we showed that supersymmetry is associated with a Weyl spinor $\xi^i_A$. Consider the supersymmetry spinor in the higher dimensional theory, under the product ansatz it decomposes into a number of four-dimensional spinors in terms of a basis of globally well defined 
spinors on the internal manifold. Therefore the constraint of preserving some supersymmetry in four dimensions is a constraint on the number of 
independent globally well defined spinors on the internal manifold. 
A useful way to mathematically quantify this property is through the use of $G$-structures. In this section we 
review the $G$-structure formalism and its application to the particular cases relevant for this work. The discussion 
follows the work in \cite{Gauntlett:2002sc,Gauntlett:2003cy,Dall'Agata:2003ir,Behrndt:2005im}.

The frame bundle of a $d$-dimensional Riemannian manifold $\M$ is the bundle of all the orthonormal frames, or sets of basis vectors, on the manifold.
The manifold is said to have a $G$-structure, where $G$ is some group, if the structure group of the frame bundle is reduced from the most general $O(d)$ to some subgroup $G \subset O(d)$. 
An alternative way to define such manifolds is through the existence of globally defined tensors and spinors
that are invariants of transformations under $G$. 
By considering the set of frames where the tensors and spinors take the same form, i.e. are singlets of the structure 
group, we see that the structure group must indeed reduce to $G$ (or a subgroup of $G$). 
Hence a manifold can be said to have $G$-structure if it admits a set of $G$-invariant tensors and spinors. 
Since the number of globally defined spinors is directly related to the supersymmetry of the four-dimensional theory 
the $G$-structure of the internal manifold determines the amount of supersymmetry in four dimensions.
These ideas are quite abstract at this stage but hopefully a discussion of some examples will be illuminating.

%
\subsubsection{$SU(3)$-structure in six dimensions}
\label{sec:su3six}
%

To begin with, we consider compactifications of type II string theory only.  
Preserving the minimum amount of supersymmetry requires the minimum amount of globally defined spinors.
For six dimensions this is a single Weyl spinor $\eta_+$ and its complex conjugate $\eta_-$. 
In ten dimensions $\N=2$ supersymmetry is parameterised by two Majorana-Weyl spinors $\xi^i$ with $i=1,2$. 
The ten-dimensional $\Gamma_{11}$ projection matrix decomposes as
\be
\Gamma_{11} = \gamma_5 \otimes \gamma_7 \mathrm{\;,}
\ee
and so the ten-dimensional Weyl spinors decompose as
\ba
\xi^1 &\sim & \theta^1 \otimes \eta_+ \mathrm{\;,} \nn \\
\xi^2 &\sim & \theta^2 \otimes \eta_- \mathrm{\;,}
\label{4dsusypara}
\ea
where $\theta^i$ are four-dimensional Weyl spinors with positive chirality. Therefore the minimum amount of supersymmetry in four dimensions 
is $\N=2$.

Six-dimensional manifolds that have one globally defined spinor have $SU(3)$-structure. As outlined at the start of this section, the approach for determining the number of spinors is to study the number of singlets of the $Spin(d)$  group under decomposition of the structure group. For the case of $Spin(6)$ and $SU(3)$ we have that a Weyl spinor 
in six dimensions transforms as a $\ir{4}$ and decomposes to $SU(3)$ irreducible spinors as
\be
\ir{4} = \ir{1} + \ir{3} \mathrm{\;.}
\ee
Therefore we recover one singlet spinor as required. Hence we see that the minimal amount of supersymmetry for compactifications of type II string theory to four dimensions is preserved by six dimensional manifolds with $SU(3)$-structure and so these form a particularly interesting class of manifolds to consider.

To study compactifications on manifolds with $SU(3)$-structure it is more useful to deal with tensors than spinors. 
The spectrum of $G$-invariant tensors on the manifold can be constructed out of the possible spinor bilinears
\ba
J_{mn} & \equiv & -i \eta^{\dagger}_{+} \ga_{mn} \eta_{+} = i \eta^{\dagger}_{-} \ga_{mn} \eta_{-} \mathrm{\;,} \nn \\
\Omega_{mnp} & \equiv & \eta^{\dagger}_{-} \ga_{mnp} \eta_{+} \;\;,\;\; \Omb_{mnp} = -\eta^{\dagger}_{+} \ga_{mnp} \eta_{-} 
\mathrm{\;.}
\label{6dsu3def}
\ea
It is possible to show that the above tensors are the only objects that can be constructed out of the available spinors $\eta_{+}$ and $\eta_{-}$. 
If we took the manifold to have more globally defined spinors it would be possible to construct more tensors and so the structure group would reduce further. In this case the two objects $J$ and $\Omega$ \textit{define} an $SU(3)$-structure. 
To see the relation between the structure group and the invariant 
tensors we follow the same procedure as for spinors and decompose the two-forms, which are in the adjoint representation $\ir{15}$ of $SO(6)$, and the three-forms, which are in the $\ir{20}$ representation into 
$SU(3)$ representations as
\ba
\ir{15} &=& \ir{1} + \ir{8} + \irc{3} \mathrm{\;,} \\
\ir{20} &=& \ir{1} + \ir{1} + \irc{3} + \irc{6} \mathrm{\;.}
\label{formdec}
\ea
We recover one singlet two-form which we identify with $J$ and one complex singlet three-form which is identified 
with $\Omn$. 

We can obtain more information about the relations between $J$ and $\Omn$. 
Since there is no singlet in the decomposition of the five-forms we must 
have $J \wg \Omn = 0$ and since a six-form is itself a singlet we must have $\Omn \wedge \Omb \sim J \wg J \wg J$. Also since we can always define an almost complex structure on any even-dimensional manifold, which in this case is $J$, the total algebraic relations satisfied by the forms read\footnote{To show that $J$ is an almost complex structure we can use Fierz identities to evaluate explicitly $J_{m}^{\ p} J_{p}^{\ n}$ in its spinor bi-linear form \cite{Gurrieri:2003st}.} 
\ba
J_{m}^{\ p} J_{p}^{\ n} & = & - \delta_m^n \mathrm{\;,} \nn \\
J_{m}^{\ n} \Omega_{npq} & = & i \Omega_{mpq} \mathrm{\;,} \nn \\
\Omega \wg \overline{\Omega} & = & -\frac43 i J \wg J \wg J \mathrm{\;,} \nn \\
\Omega \wg J & = & 0  \mathrm{\;.} 
\label{6dsu3alg} 
\ea
We note that, in general, manifolds with $SU(3)$-structure are not necessarily K\"ahler or even complex, despite the existence of a globally defined almost complex structure $J$. More identities for $SU(3)$-structure manifolds 
are given in the appendix.

The $SU(3)$-structure also implies differential relations between the structure forms $J$ and $\Omn$. To derive these we first examine how the deformation away from $SU(3)$-holonomy is parameterised. $SU(3)$-holonomy means that the forms and spinors are constant under the Levi-Civita connection
\ba
\nabla^{LC} J = \nabla^{LC} \Omega = 0 \mathrm{\;,} \nn \\
\nabla^{LC} \eta = 0 \mathrm{\;.}
\ea
This need not be the case with $SU(3)$-structure, however it is always possible \cite{joyce} to find some connection $\nabla^{T}$ under which the structure forms and spinors are invariant
\be
\nabla^{T} \eta = 0 \mathrm{\;,}
\ee
and which differs from the Levi-Civita connection through torsion 
\be
\nabla^{T} = \nabla^{LC} - \kappa \mathrm{\;.}
\ee
$\kappa_{mnp}$ is the contorsion tensor on the manifold which relates to the torsion $T_{mnp}$ through $\kappa_{[mn]p} = T_{mnp}$ and is antisymmetric in its last two indices. Acting on the spinors with the Levi-Civita connection gives
\be
\nabla^{LC}_{m} \eta = \frac14 \kappa_{mnp} \ga^{np} \eta \label{contorsion} \; \mathrm{\;,}
\ee
which when applied to the spinor bilinears gives 
\ba
(d J)_{mnp} & = & 6 {\kappa_{[mn}}^r J_{p]r} \mathrm{\;,} \nn \\
(d \Omega)_{mnpq} & = &  12 {\kappa_{[mn}}^r \Omega_{pq]r} \mathrm{\;.}
\label{djtor}
\ea
The torsion on the manifold clearly plays an important role. It is possible to use the $G$-structure formalism 
to further classify the manifold in terms of its torsion. 
Since the contorsion has two antisymmetric indices and one other index, we have that
\be
\kappa \in \Lambda^1 \otimes \Lambda^2 \cong \Lambda^1 \otimes so(d) 
\cong \Lambda^1 \otimes (g \oplus g^{\perp}) \mathrm{\;,}
\ee
where $\Lambda^1$ and $\Lambda^2$ are the spaces of one and two-forms respectively, $g$ is the Lie algebra on $G$ and $g^{\perp}$ is its complement in $so(d)$. Given that the action of $g$ on the $G$-structure
must vanish by construction, we can decompose $\kappa$ according to
the irreducible representations of $G$ in $\Lambda^1 \otimes
g^{\perp}$. In the case of $SU(3) \subset SO(6)$, this gives
\be
\kappa \in \Lambda^1 \otimes su(3)^{\perp}
= (\irc{3}) \otimes (\ir{1} + \irc{3}) = (\irc{1}) + (\irc{8}) + (\irc{6}) + 2\times (\irc{3})  \mathrm{\;.}
\ee
We then associate each of these bracketed terms with a torsion class $\tc{}$ that can be used to decompose (\ref{djtor}) in terms of the torsion classes and the singlet spinors as
\ba
dJ & = & -\frac32 \im (\tc{1} \overline{\Omega}) 
+ \tc{4} \wg J + \tc{3}  \mathrm{\;,} \nn \\
d\Omega & = & \tc{1} J \wg J + \tc{2} \wg J 
+ \overline{\tc{5}} \wg \Omega \mathrm{\;.}  \label{torsionclasses}
\ea
In terms of the $SU(3)$ representations the torsion is parameterised by the singlet $\tc{1}$, the two vectors $\tc{4}$ and $\tc{5}$, the two-form $\tc{2}$ and the three-form $\tc{3}$ which are all complex. 
The non-vanishing torsion classes can be used to further classify the structure
manifold.  In particular manifolds where all the torsion classes vanish are called Calabi-Yau manifolds.
Manifolds where only $\im(\tc{1}) \neq 0$ are called nearly-K\"ahler and manifolds with $\re(\tc{1}) = \re(\tc{2}) = \tc{4} = \tc{5} = 0$ are called half-flat. 

To summarise we showed that requiring the minimal amount of supersymmetry to be preserved in four dimensions places strong restrictions on the possible manifold that the compact dimensions must span. In particular it must have $SU(3)$-structure. We could then further classify different types 
of $SU(3)$-structure manifolds through their non-vanishing torsion classes. The simplest example of such manifolds are the CY manifolds for which all the torsion classes vanish. These manifolds form the most studied class of compactifications that have been considered so far in the literature. In section \ref{sec:cycompact} we review 
 how the actual compactification proceeds for such manifolds. However we should keep in mind that these only form a small subset of all the possible $SU(3)$-structure manifolds and compactifications on general $SU(3)$-structure manifolds is the topic of chapter \ref{cha:su3iia}.
Before we proceed to the CY case there are two more important cases of $G$-structures that we should consider.

%
\subsubsection{$G_2$-structure in seven dimensions}
\label{sec:g2seven}
%

The manifolds relevant for M-theory compactifications are seven-dimensional and so we should consider $G$-structures in seven dimensions.
The minimal amount of globally defined spinors on seven-dimensional manifolds is a single Majorana spinor $\e$. 
Supersymmetry in eleven dimensions is parameterised in terms of a single Majorana spinor and so under decomposition 
we recover a single four-dimensional Majorana spinor which corresponds to $\N=1$ supersymmetry.
Manifolds with a single spinor have $G_2$-structure. To show this we follow the same arguments as in the previous section. A Majorana spinor is in the $\ir{8}$ representation of $Spin(7)$ which decomposes under $G_2$ as
\be
\ir{8} = \ir{1} + \ir{7} \mathbf{\;,}
\ee
giving one singlet spinor.

The unique spinor bilinear defines the $G_2$-structure form
\be
\vp_{mnp} = i \e^T \g_{mnp} \e \mathrm{\;.}
\label{g2def}
\ee
The space of three-forms $\Lambda^3$ decomposes as 
\be
\ir{35} = \ir{1} + \ir{7} + \ir{27} \mathrm{\;,}
\ee
giving one singlet three form which is $\vp$. 
The existence of the singlet three-form also implies there is a singlet four-form given by $\star \vp$. There are no neat algebraic relations analogues to (\ref{6dsu3alg}) that define a $G_2$-structure on 
a manifold although some useful identities are given in the appendix. Instead a manifold is said to have $G_2$-structure if it has a globally defined three-form that can be consistently mapped at each point to the three-form $\vp^0$ on $\Reals_7$ defined as\footnote{This definition also holds for the case of $SU(3)$-structure in six (or seven) dimensions where the explicit forms are given in the appendix.} \cite{joyce}
\be
\vp^0 \equiv dx^{136} + dx^{235} + dx^{145} - dx^{246} - dx^{127} - dx^{347} - dx^{567} \mathrm{\;,}
\ee
where $dx^{mnp} \equiv dx^{m} \wg dx^{n} \wg dx^{p}$ and $dx^{n}$ is the orthonormal basis on $\Reals_7$. 

As was the case for $SU(3)$-structure, $G_2$-structure manifolds can be further classified according to their 
torsion. Acting on $\vp$ with the Levi-Civita connection we find 
\be
\left( d\vp \right)_{mnpq} = 12 \kappa_{[mn}^{\;\;\;\;\;\;r} \vp_{r|pq]} \mathrm{\;.}
\ee
The contorsion decomposes under $G_2$ as
\be
\kappa \in \Lambda^1 \otimes g_2^{\perp}
= \ir{7} \otimes \ir{7} = \ir{1} + \ir{7} + \ir{14} + \ir{27} \mathrm{\;,}
\ee
which gives the differential relations
\ba
  \label{g2torsion}  
  \begin{aligned}
  d\vp = & \ \tc{1} \star \vp - \vp \wg \tc{2} + \tc{3} \mathrm{\;,} \\
  d\left( \star \vp \right) = & \ \frac43 \star \vp \wg \tc{2} + \tc{4} \mathrm{\;.}
  \end{aligned}
\ea
In terms of $G_2$ representations $\tc{1}$ is a singlet, $\tc{2}$ is a vector, $\tc{3}$ is a $\ir{27}$ and $\tc{4}$ transforms under the adjoint representation $\ir{14}$. Manifolds with only $\tc{1} \neq 0$ are called weak-$G_2$ and manifolds with all torsion classes vanishing have $G_2$-holonomy.

%
\subsubsection{$SU(3)$-structure in seven dimensions}
\label{sec:su3seven}
%

Seven-dimensional manifolds with $SU(3)$-structure have been less studied than their six-dimensional 
counterparts partly due to the fact that for the case of no torsion where the holonomy group of
the manifold is $SU(3)$ the seven-dimensional manifold is just a direct
product of a CY manifold and a circle \cite{Friedrich:1995dp}. Therefore studying M-theory on
such manifolds is equivalent to studying type IIA string theory on a
CY. Once some torsion classes are non-vanishing a non-trivial
fibration is generated thereby making such studies different to type IIA
compactifications.

An $SU(3)$-structure on a seven-dimensional manifold implies the existence of
two globally defined Majorana spinors $\ep_1$ and $\ep_2$
which are independent in that they satisfy $\ep_1^T \ep_2 = 0$. 
In terms of spinor representation we have the decomposition of the Majorana spinor as
\be
\ir{8} = \irc{1} + \irc{3} \mathrm{\;,}
\ee
giving the two singlet spinors. Having two independent spinors gives $\N=2$ supersymmetry in four dimensions.
In the following we find it more convenient to use two complex spinors $\eta_\pm$ defined as
\begin{equation}
  \label{xipm}
  \eta_{\pm} \equiv \frac1{\sqrt 2} \big(\epsilon^1\pm i\epsilon^2 \big) \mathrm{\;.}
\end{equation}
The $SU(3)$-invariant forms $\Omn$, $J$, $V$ are constructed as 
\begin{eqnarray}
  \label{OJVdef}
  \Omn_{mnp} & = & - \eta^{\dagger}_+\gamma_{mnp}\eta_- \mathrm{\;,} \;\; \Omb_{mnp} = \eta^{\dagger}_-\gamma_{mnp}\eta_+ \nn \mathrm{\;,} \\
  J_{mn} & = & i \eta^{\dagger}_+\gamma_{mn}\eta_+ = -i \eta^{\dagger}_-\gamma_{mn}\eta_- \mathrm{\;,}\\ 
  V_m & = & - \eta^{\dagger}_+\gamma_m\eta_+ = \eta^{\dagger}_-\gamma_m\eta_- \nn \mathrm{\;.}
\end{eqnarray}
In comparison with six-dimensional $SU(3)$-structures, in seven
dimensions there also exists a globally defined vector field $V$. It is
important to bear in mind that in general this vector is not a Killing
direction and thus the manifold does not have the form of a direct product between a six-dimensional manifold and a circle. In terms of tensor representations the $SU(3)$-structure is manifest through the decomposition of the one-forms, which are in the $\mathbf{7}$ representation of $SO(7)$, the two-forms in the $\mathbf{21}$ and the 
three-forms which are in the $\mathbf{35}$ representation, as
\ba
\ir{7} &=& \ir{1} + \irc{3} \mathrm{\;,}\nn \\
\ir{21} &=& \ir{1} + 2\times (\irc{3}) + \ir{8} \mathrm{\;,}\nn \\
\ir{35} &=& \ir{1} + \irc{1} + 2\times (\irc{3}) + \irc{6} + \ir{8} \mathrm{\;.}
\ea
Note that there are \textit{three} real singlet three-forms which correspond to $\Omn$ and $J\wedge V$.
Following the same procedure as for the six-dimensional case gives the set of relations for the forms 
\ba
J\wedge J\wedge J &=& \frac{3i}{4} \Omn \wedge \Omb \mathrm{\;,} \nn \\
\Omega\wedge J &=& V\lrcorner J = V \lrcorner \Omega = 0 \mathrm{\;,}\nn \\
V \lrcorner V &=& 1 \mathrm{\;,} \nn \\
J^m_{\;\;\;n}J^n_{\;\;\;p}&=&-\delta^m_p+V^m V_p \nn \mathrm{\;,} \\
J_m^{\;\;n}\Omega_{npq}&=&i\Omega_{mpq} \mathrm{\;.} 
\label{su3rel}
\ea
The differential relations of the forms can as usual be written using the contorsion and for the case of 
$SU(3)$-structure in seven dimensions read
\ba
\label{torel} 
\left( dV \right)_{mn} &=& 2 \kappa_{[mn]p}V^p \; , \\
\left( dJ \right)_{mnp} &=& 6 \kappa_{[mn}^{\;\;\;\;\;\;r} J_{r|p]}\; , \\
\left( d\Omega \right)_{mnpq} &=& 12 \kappa_{[mn}^{\;\;\;\;\;\;r}
\Omega_{r|pq]} \; . \nn 
\ea
The contorsion decomposes into $SU(3)$-modules as
\be
\kappa \in \Lambda^1 \otimes su(3)^{\perp}
= (\ir{1} + \irc{3})\otimes(\ir{1} + 2\times(\irc{3}) + \ir{8}) = 5\times \ir{1} + 4\times \ir{8} 
+ 2\times (\irc{6}) + 5\times (\irc{3}) \mathrm{\;.}
\ee
The differentials of the forms $\Omn$,$J$ and $V$ are given in terms of the torsion classes as 
\ba
dV &=& R J + \bar{W_1} \lrcorner \Omega + W_1 \lrcorner \bar{\Omega} + A_1 + V
\wedge V_1 \mathrm{\;,} \\
dJ &=& \frac{2i}{3}\left( c_1 \Omega - \bar{c_1} \bar{\Omega} \right) + J
\wedge V_2 + S_1 \nn \\ &~& \; + \;
V\wedge \left[ \frac{1}{3} \left( c_2 + \bar{c_2}\right) J +
  \bar{W_2}\lrcorner \Omega + W_2 \lrcorner \bar{\Omega} + A_2 \right] \mathrm{\;,}\\
d\Omega &=& c_1 J \wedge J + J \wedge T + \Omega \wedge V_3 + V \wedge \left[
  c_2 \Omega - 2 J \wedge W_2 + S_2 \right] \mathrm{\;,} \label{su3torsion} 
\ea 
where the torsion classes contain three singlet classes $R$ (real) and
$c_{1,2}$ (complex), five complex vectors $V_{1,2,3}$ and $W_{1,2}$,
three 2-forms $A_{1,2}$ (real) and $T$ (complex) and two complex 3-forms $S_{1,2}$.

Before concluding this section we make more precise the relation
between the $SU(3)$ and $G_2$-structures on a seven-dimensional
manifold. As $SU(3) \subset G_2$, an $SU(3)$-structure
automatically defines a $G_2$-structure on the manifold. In fact an $SU(3)$-structure 
on a seven-dimensional manifold implies the existence of two
independent $G_2$-structures whose intersection is precisely the $SU(3)$-structure. 
Concretely, using the spinor $\ep_1$ and $\ep_2$ defined above we
can construct the two $G_2$ forms $\vp^{\pm}$
\begin{equation}
  \label{phi+-}
  \begin{aligned}
    \left( \vp^{+} \right)_{mnp} \equiv & ~ 2i\epsilon_1 \gamma_{mnp} 
    \epsilon_1 \mathrm{\;,}  \\   
    \left( \vp^{-} \right)_{mnp} \equiv & ~ 2i\epsilon_2 \gamma_{mnp} 
    \epsilon_2 \mathrm{\;.}
  \end{aligned}
\end{equation}
The relation to the $SU(3)$-structure is now given by
\begin{equation}
  \label{phiOJV}
  \vp^{\pm} = \pm \Omega^{-} - J \wg V \mathrm{\;.}
\end{equation}
Throughout this work it is sometimes useful to use the $SU(3)$ forms and
sometimes the $G_2$ forms but we should keep in mind that the two formulations
are equivalent. 

%
\section{Four-dimensional $\N=2$ supergravity}
\label{sec:n=2sugra}
%

In section \ref{sec:constsusy} we showed how supersymmetry imposes constraints on the possible manifolds that the extra dimensions may span. For the cases of type II string theory and M-theory compactified on manifolds with $SU(3)$-structure, the resulting four-dimensional theory preserves $\N=2$ supersymmetry. The actions are 
therefore $\N=2$ supergravities and in this section we review the important features of this class of theories. 
The treatment is a purely four-dimensional one and the constraints on the theory come from requiring $\N=2$ 
supersymmetry. The features of the theory discussed here form constraints and guiding principles for the higher 
dimensional compactifications. 
We begin by reviewing ungauged $\N=2$ supergravity in four dimensions with only scalar and vector matter fields
\cite{deWit:1984px,D'Auria:1990fj,Andrianopoli:1996cm}. We then discuss the two possible extensions to this theory 
of a gauged supergravity \cite{deWit:1984px,D'Auria:1990fj,Andrianopoli:1996cm} and the addition of massive tensor 
multiplets \cite{Dall'Agata:2003yr,Sommovigo:2004vj,D'Auria:2004yi}. Finally we review the constraints placed on possible truncations to $\N=1$ supergravity \cite{Cecotti:1984rk,Cecotti:1985sf,Ferrara:1995xi,Taylor:1999ii,Andrianopoli:2001gm,Louis:2002vy,Gunara:2003td}.
We note that this chapter is a \textit{review} of the key concepts and so derivations are not presented here and 
the reader is referred to the literature for these. 

The gravitational sector of $\N=2$ supergravity is composed of a single $\N=2$ gravitational multiplet which contains 
the graviton $g_{\mu\nu}$, two spin-$\frac32$ Weyl gravitini $\psi^{1,2}$, and a vector $V^0$ called the graviphoton. 
There are two types of matter multiplets containing scalar fields that preserve $\N=2$ supersymmetry with the component fields in each multiplet transforming into each other 
under the supersymmetry transformation. There are vector multiplets that contain two real scalar fields, two fermions and one vector 
boson and there are hypermultiplets whose content is four real scalar fields and two fermions
\footnote{As we discuss in section \ref{sec:n=2maggauge}, there are also tensor multiplets that in the ungauged theory are dual to hypermultiplets. For simplicity we shall refer to these as hypermultiplets unless the distinction is important.}. 
One of the features of (ungauged) $\N=2$ supergravity is that the two sectors 
of scalar fields do not mix. More 
precisely the scalar fields in each sector can be modelled as non-linear sigma models that span two separate manifolds.
The scalar components of the vector 
multiplets are labelled as complex fields $z^a$ and the scalar components of the hypermultiplets as real fields $q^{\lambda}$ with index ranges $a=1,...,n_v$ and $\lambda=1,...,4n_h$, where $n_v$ and $n_h$ are the number of vector multiplets and hypermultiplets present. The full bosonic action reads \cite{Andrianopoli:1996cm}
\ba
S_{\mathrm{Kin}} &=& \int_{\cal S}{ \left[ \half R \star 1 - g_{a\bar{b}}(z,\bar{z})dz^a \wg \star d\bar{z}^{\bar{b}} - h_{\lambda\sigma}(q)dq^{\lambda} \wg \star dq^{\sigma} 
\right.} \nn \\
&+& \left.\half (\im{\M(z,\bar{z})})_{AB} F^A \wedge \star F^B + \half (\re{\M(z,\bar{z})})_{AB} F^A \wedge F^B \right]
 \mathrm{\;,} \label{n=2act}
\ea
where we have also written the kinetic term for the vector field components of the vector multiplets $V^a$ with field-strengths $F^a = dV^a$ and the graviphoton $F^0 = dV^0$ 
so that the index $A=0,...,n_v$. The matrix $\M$ is called the gauge kinetic matrix.
The metrics $g_{a\bar{b}}$ and $h_{\lambda\sigma}$ are on two separate manifolds $\M^{V}$ and $\M^{H}$ respectively, with the full scalar manifold being their direct product $\M=\M^V \times \M^H$. 
The hypermultiplets manifold is a \textit{quaternionic} (K\"ahler) manifold and the vector multiplets manifold is of the \textit{special K\"ahler} type. 
We proceed to highlight some of the important properties of each type of manifold. 

Consider a K\"ahler manifold $\M^V$ with co-ordinates $\left\{z^a,\bar{z}^a\right\}$ that has two types of vector bundles defined on it. The first is a complex line bundle ${\cal L}\ra {\M^V}$, and the second is a flat complex vector bundle of dimension $2n_v+2$, ${\cal SV} \ra \M^V$ with a symplectic structure group $Sp(2n_v+2,\Reals)$. 
The symplectic group has a representation in terms of the set of $2(n_v+1)\times 2(n_v+1)$ general linear matrices $\Lambda$ that satisfy
\be
\Lambda^T \left( \begin{array}{cc} 0 & \one \\ -\one & 0 \end{array} \right) \Lambda = \left( \begin{array}{cc} 0 & \one \\ -\one & 0 \end{array} \right) \mathrm{\;.} 
\ee
Consider a holomorphic section of the product bundle ${\cal H}\equiv{\cal SV}\otimes{\cal L}$ which takes the vector form
\be
\Omn^{sk}(z) = \left( \begin{array}{c} X^A(z) \\ F_A(z) \end{array} \right) \mathrm{\;,}
\ee
where the index $A=0,...,n_v$ and the vector entries are called the periods of $\Omn^{sk}$.
There is an unfortunate clash of notation of the periods $F_A(z)$ and the gauge field-strengths $F^A(x)$ in (\ref{n=2act}). The distinction between the two should be made by the context in which they appear and also by the 
position of the index.
The symplectic inner product on ${\cal H}$ is defined as 
\be
\ev{\Omn^{sk}|\Omb^{sk}} \equiv - \left(\Omn^{sk}\right)^T \left( \begin{array}{cc} 0 & \one \\ -\one & 0 \end{array} \right)   \Omb^{sk} = \bar{X}^A F_A - \bar{F}_A X^A \mathrm{\;.}
\ee
The period matrix $\N$ is defined as the symplectic matrix that transforms between the two halves of the holomorphic section
\be
F_A = \N_{AB} X^B \mathrm{\;.}
\ee
We can now define a special K\"ahler manifold as a K\"ahler manifold which admits the structure described above such that 
the K\"ahler potential is given by
\be
\label{skahler}
K^{sk} = - \mathrm{ln}\; i\ev{\Omn^{sk}|\Omb^{sk}} \mathrm{\;.}
\ee
Since the K\"ahler potential plays an important role it is worth going into a bit more detail as to its construction. A K\"ahler transformation is a transformation of the K\"ahler potential by some holomorphic function $f(z)$ of the type
\be
K^{sk} \ra K^{sk} + f(z) + \bar{f}(\bar{z}) \mathrm{\;.}
\ee
This type of transformation leaves the metric invariant and so is a symmetry of the action. Under this type of transformation the holomorphic section transforms as 
$\Omn^{sk} e^{-f}$ and so the set of periods $X^A$ can be regarded as homogeneous co-ordinates. Provided the Jacobian matrix 
\be
e^A_a(z) \equiv \partial_a \left( \frac{X^A}{X^0} \right) \mathrm{\;,}
\ee
is invertible we can fix the homogeneity by defining special co-ordinates $z^A \equiv \frac{X^A}{X^0} = (1,z^a)$. Then these co-ordinates are a good co-ordinate basis on the manifold so that 
the bottom half of the holomorphic section composed of the periods $F_A$ is actually a function of the top-half periods $F_A(X)$ and can be generated from a single 
holomorphic function of homogeneous degree two called the prepotential
\be
F_A(X) = \frac{\partial}{\partial X^A} {\cal F}(X) \mathrm{\;.}
\ee
The prepotential therefore encompasses the whole geometric structure and in particular the period matrix can be written as
\be
{\cal N}_{AB} = \bar{{\cal F}}_{AB} + \frac{2i(\im{{\cal F}})_{AC}X^C (\im{{\cal F}})_{BD}X^D}{(\im{{\cal F}})_{IJ}X^IX^J}
\mathrm{\;,} \label{n=2period}
\ee
where
\be
{\cal F}_{AB} = \partial_A \partial_B {\cal F} \mathrm{\;.}
\ee

The bundle ${\cal H}$ clearly plays an important role in the geometry of special K\"ahler manifolds and in particular 
its section $\Omn^{sk}$. It is important to understand how this section varies as a function of the co-ordinates 
of the manifold by which we mean how the covariant derivative acts on it. Since the symplectic bundle ${\cal SV}$ is flat, the only relevant connection is that of the line bundle ${\cal L}$ which is a one-form $\theta$ given by $\partial_{a}K^{sk} dz^a$. Now there exists a correspondence between line bundles and $U(1)$ bundles such that if 
a transition function on the complex line bundle is given by $e^{f(z)}$ the equivalent transition function on the 
$U(1)$ bundle is $e^{i\im(f)}$. In terms of the connection this means that the connection on the $U(1)$ bundle $Q$ is given by
\be
Q = \im(\theta) = -\frac{i}{2} \left( \partial_{a} K^{sk} dz^a - \partial_{\bar{a}} K^{sk} d\bar{z}^{\bar{a}} \right) \mathrm{\;.}
\ee
Now let $\tilde{\Omn}^{sk}$ be a section of the $U(1)$ bundle, then the covariant derivative acting on it is 
\be
\nabla \tilde{\Omn}^{sk} = \left( d + iQ \right) \tilde{\Omn}^{sk} \mathrm{\;,}
\ee
which in co-ordinates reads
\ba
\nabla_{a} \tilde{\Omn}^{sk} &=& \left( \partial_{a} + \half \partial_a K \right) \tilde{\Omn}^{sk} \mathrm{\;,} \label{u1secderiv} \\
\nabla_{\bar{a}} \tilde{\Omn}^{sk} &=& \left( \partial_{\bar{a}} - \half \partial_{\bar{a}} K \right) \tilde{\Omn}^{sk} \mathrm{\;.}
\ea
A covariantly holomorphic section of the $U(1)$ bundle is defined as $\nabla_{\bar{a}} \tilde{\Omn}^{sk} = 0$ and so under the map $\tilde{\Omn}^{sk} = e^{\half K^{sk}}\Omn^{sk}$ covariantly holomorphic sections of the $U(1)$ flow into 
holomorphic sections of ${\cal L}$. The covariant derivatives on the holomorphic section of ${\cal L}$ read
\ba
\nabla_{a}\Omn^{sk} &=& \partial_{a}\Omn^{sk} + \left(\partial_{a}K^{sk}\right)\Omn^{sk} \mathrm{\;,}\\
\nabla_{\bar{a}}\Omn^{sk} &=& \partial_{\bar{a}}\Omn^{sk} = 0 \mathrm{\;.}
\label{kahlercov}
\ea
This gives the covariant derivative, with respect to K\"ahler transformations, acting on $\Omn^{sk}$.

Supersymmetry implies that the vectors in the vector multiplet sector should follow the same geometry as the scalars and indeed this is the case. In fact this relation between the vectors and the scalars is the important constraint through which the whole geometric structure is derived. A particularly important concept is electric-magnetic duality 
which corresponds to the symplectic symmetry present in the scalar sector. We can define magnetic field-strengths 
$G_A$ in terms of the electric field-strengths $F^A$ as
\be
G_A \equiv \half \frac{\partial {\cal L}_{\mathrm{vec}} }{\partial F^A}  \mathrm{\;,}
\ee
where ${\cal L}_{vec}$ is defined as the vector part of the Lagrangian in (\ref{n=2act}). This gives explicitly
\be
G_A = (\re{\M})_{AB}F^B+(\im{\M})_{AB}\star F^B \mathrm{\;.}
\ee
It is easy to show that the equations of motion and Bianchi identity arising from ${\cal L}_{vec}$ for the gauge-fields 
are invariant under a symplectic rotation of the electric and magnetic field-strengths $F'=SF$ where $S$ is a symplectic matrix and the vector $F$ is defined as
\be
F \equiv \left( \begin{array}{c} F^A \\ G_A \end{array} \right) \mathrm{\;.}
\ee
Since such a rotation exchanges electric and magnetic field-strengths it corresponds to symplectic electric-magnetic 
duality.
The final important relation in the vector multiplet sector is that the period matrix $\N$ can be identified with the gauge kinetic matrix $\M$. 

The hypermultiplet sector spans a quaternionic manifold. Quaternionic manifolds are $4n$-dimensional, where $n$ is an integer, real manifolds that are endowed with a 
metric $h_{\lambda\sigma}$ and three complex structures $\left( J^x \right)^{\;\;\lambda}_{\sigma}$, $x=1,2,3$ that satisfy the quaternionic algebra
\be
\label{quatalg}
J^x J^y = -\delta^{xy} \one + \epsilon^{xyz} J^z \mathrm{\;,}
\ee
and with respect to which the metric is Hermitian. Using these it is possible to introduce a triplet of two-forms
\be
K^x \equiv \half K^x_{\lambda\sigma} dq^{\lambda} \wg dq^{\sigma} \equiv \half h_{\lambda\rho} \left( J^x \right)^{\;\;\rho}_{\sigma} dq^{\lambda} \wg dq^{\sigma} \mathrm{\;,}
\ee
that are the generalisation of the K\"ahler form on complex manifolds. These forms are then required to be covariantly closed with respect to some connection on the manifold
\be
\nabla K^x \equiv dK^x + \epsilon^{xyz} \om^y \wg K^z = 0 \mathrm{\;.}
\ee
If the connection $\om^y$ vanishes, i.e. the manifold is flat, the manifold is called hyperK\"ahler.

%
\subsection{Gauged $\N=2$ supergravity}
\label{sec:n=2gauge}
%

We have seen that $\N=2$ supersymmetry is a restrictive condition on the possible actions that can be written down. There are however ways that the action (\ref{n=2act}) can be extended whilst still preserving $\N=2$ supersymmetry. 
It is possible to gauge the matter fields, vector multiplets and hypermultiplets, with respect to the vectors in the vector multiplets and the 
graviphoton. The gauging is done with respect to the symmetries of the special K\"ahler and quaternionic sigma 
model manifolds. As long as the coupling of the scalars generated in the action are written in a way that preserves these symmetries the 
symmetries remain symmetries of the full action. 
To implement this gauging we first must find (holomorphic) Killing vectors of the 
manifolds, $k^{\lambda}_A(q)$ and $k^a_A(z)$, where the subscript index denotes the number of such Killing vectors and the superscript denotes 
the entry in the vector. Then it is possible to gauge the derivatives on the co-ordinates of the manifolds as
\ba
\nabla q^{\lambda} &=& dq^{\lambda} + g_{(A)} V^A k_A^{\lambda}(q) \mathrm{\;,} 
\label{gaugeq} \\
\nabla z^a &=& dz^a + g'_{(A)} V^A k_A^a(z) \label{gaugez}\mathrm{\;,}
\ea
where we have introduced coupling constants $g_{(A)}$ and $g'_{(A)}$ (with indices uncontracted) for each possible coupling. These 
can be absorbed into the definitions of the Killing vectors and so are generically dropped henceforth unless required for clarity. 
The interesting effect of such gaugings is that the new coupling induces 
shifts in the supersymmetry transformations of the spin-$\half$ fields of the theory which in turn means that a scalar potential 
is generated
\ba
V_{\mathrm{scalar}} &=&  \left( g_{a\bar{b}} k_A^a \bar{k}^{\bar{b}}_B + 4h_{\lambda\sigma} k^{\lambda}_A k^{\sigma}_B \right) e^K \bar{X}^A X^{B} 
+ g^{a\bar{b}} f_a^A \bar{f}_{\bar{b}}^B P^x_A P^x_B \nn \\
&-& 3 e^K \bar{X}^A X^B P^x_A P^x_B \mathrm{\;.}
\label{scalpotgaug}
\ea
We have introduced the covariant derivatives 
\ba
f_a^A &\equiv& \left( \partial_a + \half \partial_a K \right) e^{\half K} X^A \mathrm{\;,} \nn \\
\bar{f}_{\bar{a}}^A &\equiv& \left( \partial_{\bar{a}} - \half \partial_{\bar{a}} K \right) e^{\half K} \bar{X}^A \mathrm{\;,}
\ea
and also the quaternionic prepotentials $P^x_A$ (not to be confused with the prepotential ${\cal F}$ of the special K\"ahler manifold) that are prepotentials for the Killing vectors on the quaternionic manifold defined as 
\be
\label{momentummap}
2 k_A \lrcorner K^x = -\nabla P^x_A = -\left( dP^x_A + \epsilon^{xyz} \om^y P^z_A \right) \mathrm{\;.}
\ee 

For the purposes of comparing gauged $\N=2$ supergravity with compactifications of string and M-theory it is 
useful to consider the form of the mass matrix for the two gravitini. The kinetic terms and mass terms for the 
gravitini in the theory are given by \footnote{Note that we have different metric signature, and therefore $\gamma$ matrix, conventions to 
\cite{Andrianopoli:1996cm}.}
\be
S_{\mathrm{gravitini}} = \int_{\cal S}{d^4x\sqrt{-g}\; \left[ -\bar{\psi}^{\alpha}_{+\mu} \ga^{\mu\rho\nu} \nabla_\rho \psi^{\alpha}_{+\nu}  + 
S_{\alpha\beta} \bar{\psi}^{\alpha}_{+\mu } \gamma^{\mu\nu}\psi^{\beta}_{-\nu}
+ \mathrm{c.c.} \right] }
\label{n=2gravitini}
\ee
where $\alpha,\beta=1,2$ label the two gravitini and the $\pm$ subscript denotes the chirality.  
The action (\ref{n=2gravitini}) defines the gravitini mass matrix $S$. 
In terms of the geometric structure the mass matrix reads
\be
S_{\alpha\beta} = \frac{1}{2} \sigma^x_{\alpha\beta} P^x_A e^{\half K^{sk}}X^{A} \mathbf{\;,}
\ee
where $\sigma^x$ denote the Pauli matrices (\ref{pauli}) whose indices are raised and lowered by $\hat{\epsilon}_{\alpha\beta}$\footnote{The $\sigma$ matrices in the appendix are defined with one index up and 
one index down so that $\sigma^x_{\alpha\beta}$ is actually given by 
$\left( \begin{array}{cc} \delta^{x1} - i\delta^{x2}   & -\delta^{x3} \\ -\delta^{x3}& -\delta^{x1} - i\delta^{x2} \end{array} \right)$.}.  
We note here that the gravitini mass matrix determines all the relevant 
quantities of the gauged $\N=2$ supergravity with the exception of the gauged Killing vectors in the vector 
multiplet sector. This is analogous to the $\N=1$ case where the gravitino mass can be used to 
completely determine the scalar sector of the theory.

%
\subsection{$\N=2$ supergravity with massive tensor multiplets}
\label{sec:n=2maggauge}
%

In four dimensions a two-form is dual to a scalar. It is possible to dualise a sub-sector of the 
hypermultiplet scalars, say $q^I$ with $I=1,...,n_T<4n_H$, to anti-symmetric tensors $B_I$. 
The dualisation procedure leaves the vector multiplet sector unaffected. Dualisation of the ungauged theory is rather trivial and 
does not introduce any interesting features. However once the theory is gauged there appear non-trivial extensions 
to the usual gauged supergravity. The gauging process described in section \ref{sec:n=2gauge}, which we refer to as 
electric gauging for reasons that follow, remains unchanged apart from the constraint that the Killing vectors in the hypermultiplet sectors must correspond to isometries of the undualised submanifold of the full hypermultiplet quaternionic manifold that commute with the dualised co-ordinates. 
In order to carry out the dualisation, the dualised scalars must appear only as derivatives and so they must have the
translational isometry $q^I \ra q^I + \Lambda^I$ where $\Lambda^I$ is a constant. The dualised scalars can be gauged
 prior to the dualisation with respect to these isometries corresponding to constant Killing vectors $k^I_A = e^I_A$.
After the dualisation, there also appears the possibility of redefining the electric field-strengths with the new tensors as
\be
F^A \ra F^A + m^{AI}B_I \mathrm{\;.}
\label{ftob}
\ee
The key feature is that the full supersymmetry can still be maintained given an appropriate shift in the fermionic 
supersymmetry transformations and therefore a change to the scalar potential. Note that this deformation of the theory means that the tensors $B_I$ pick up an explicit mass term and so can not be dualised back to scalars. However they still remain part of the original multiplet which now becomes a tensor multiplet, so for example dualising $q^1$ gives
\be
\left\{q^1,q^2,q^3,q^4\right\} \ra \left\{B_1,q^2,q^3,q^4\right\} \mathrm{\;.}
\ee
It is interesting to note the coupling of the dualised tensors to the gauge fields. Substituting (\ref{ftob}) in the 
action (\ref{n=2act}), after the gauging (\ref{gaugeq}) with respect to the constant Killing vectors $e^I_A$ of the undualised hypermultiplets, gives the terms
\be
S_{B_I} \supset \int{ \half B_I \wedge \left( m^{IA} G_A - e^I_A F^A \right)} \mathrm{\;.}
\ee
The tensor couples to both electric and magnetic field-strengths. It is a 
dyon, a particle charged both electrically and magnetically. The mass parameters $m^{IA}$ correspond to the magnetic 
charges of the tensors and can be though of as constant magnetic Killing vectors in analogy with $e^I_A$. If we define 
the magnetic Killing vectors $\tilde{k}^{IA}$ as
\be
\tilde{k}^{IA} \equiv -\half m^{IA} \mathrm{\;,}
\ee
then we can assign them magnetic prepotentials $Q^{xA}$ through
\be
2 \tilde{k}^{IA} K^x_{IJ} \equiv -\nabla_J Q^{xA} \mathrm{\;.}
\ee
With these definitions the shifted gravitini mass matrix reads 
\be
S_{\alpha\beta} = \frac{1}{2} \sigma^x_{\alpha\beta} e^{\half K^{sk}}\left( g_{(A)} P^x_A X^{A} - Q^{xA} F_A \right) \mathbf{\;,}
\label{n=2magmass}
\ee
where we have reinstated the gauge coupling $g_{(A)}$ since it appears non-trivially.
The shifted scalar potential takes on a complicated form, which in none-the-less still fully determined by the 
gravitini mass matrix and the vector multiplets Killing vectors, and is given in \cite{Dall'Agata:2003yr}.

%
\subsection{Consistent truncations of $\N=2$ to $\N=1$}
\label{sec:n=2ton=1}
%

In chapters \ref{cha:su3iia} and \ref{cha:su3mtheory} we encounter partial supersymmetry breaking from $\N=2$ to $\N=1$ and so it is worthwhile  considering this phenomenon at this point. 
In the ${\cal N}=2$ theory the fields are grouped into vector multiplets and hypermultiplets.
Once supersymmetry is broken these multiplets split up into ${\cal N}=1$ multiplets. 
Since the number of supersymmetries is given by the number of massless gravitini, breaking one supersymmetry 
corresponds to one gravitino becoming massive.
The ${\cal N}=2$ gravitational multiplet therefore splits into a ${\cal
N}=1$ massless gravitational multiplet and a massive spin-$\frac{3}{2}$ multiplet
\be
 \left( g_{\mu\nu},\psi_1,\psi_2,V^0 \right) \ra 
  \mathrm{massless}\;\left( g_{\mu\nu},\psi_1 \right) \;\;+\;\; 
  \mathrm{massive}\;\left(\psi_2,V^0,V^1, \chi \right) \; .
\ee  
Here $V^1$ is a vector field which has to come from one of the vector
multiplets and $\chi$ is a spin-$\half$ fermion which
comes from a hypermultiplet.  Moreover, one also needs one Goldstone fermion and two
Goldstone bosons to be eaten by the gravitino and the two vector fields
respectively which become massive, and these additional Goldstone fields also
come from the hypermultiplet sector.
The $n_v$ ${\cal N}=2$ vector multiplets break into $\tilde{n}_v$ massless ${\cal N}=1$ vector multiplets and $n_c$
massless chiral multiplets (with the other fields forming massive
multiplets) such that the scalar components of the chiral multiplets
span a K\"ahler manifold which is a submanifold of the full vector multiplets special K\"ahler manifold. 
The $n_h$ ${\cal N}=2$ hypermultiplets break into $\tilde{n}_h$ massless ${\cal
N}=1$ chiral multiplets and $n_h - \tilde{n}_h$ massive chiral multiplets with
$\tilde{n}_h \leq \half n_h$. The scalar components of the massless chiral
multiplets span a K\"ahler submanifold of the original quaternionic manifold.
Identifying the massless fields that preserve the residual $\N=1$ supersymmetry is equivalent to 
finding the correct complex co-ordinates on the resulting submanifolds.
With the scalar fields of the $\N=2$ vector multiplets the situation is quite simple as they are
already complex coordinates on a (special) K\"ahler manifold. However, for the
hyper-scalars this is not the case, and it is in general non-trivial to find
the right combinations that represents the correct complex
coordinates. For simple cases, that we encounter in this paper, this can be
done and one can find explicitly the correct complex combinations which span
the $\N=1$ scalar K\"ahler manifold.

%
\subsubsection{$\N=1$ supergravity}
\label{sec:n=1sugra}
%

Following a consistent truncation of a $\N=2$ supergravity we reach a $\N=1$ supergravity. In this section we 
review briefly the scalar sector of $\N=1$ supergravities \cite{wess-bagger}. The gravitational multiplet contains the graviton and 
a single gravitino. There are also vector multiplets which contain a vector and two fermions. The only multiplet 
with scalar components is a chiral multiplet which contains two scalars and a fermion. The scalar sector is 
completely determined in terms of the superpotential $W$, which is a holomorphic function of the superfields, and the 
K\"ahler potential $K$, which is the K\"ahler potential for the scalar manifold spanned by the chiral multiplet scalars. 
If we consider a set of chiral superfields $\Phi^I$ the kinetic terms and scalar potential read
\be
{\cal L}_{\mathrm{N=1}} = - g_{I\bar{J}} \nabla_{\mu}\Phi^I \nabla^{\mu} \bar{\Phi}^{\bar{J}} -
e^K \left[ g^{I\overline{J}} D_I W D_{\overline{J}}  \overline{W} - 3\left| W \right|^2 \right]
 \; , \label{n=1genact}
\ee
where the covariant derivatives of the chiral superfields can be gauged with respect to the vector superfields and 
the kinetic metric is given by $g_{I\bar{J}}=\partial_I\partial_{\bar{J}} K$.
The K\"ahler covariant derivatives $D_I$ are given by
\be
D_IW = \partial_IW + \left( \partial_I K \right)W \mathrm{\;.}
\ee
The theory has a single gravitino whose mass is given by 
\be
M_{\frac{3}{2}} = e^{\half K} W \;. \label{n1gravmass} 
\ee
Therefore the gravitino mass completely determines the scalar sector of the theory.  Note that the only feature 
of the action that is not determined is the gauge group of the vector fields.

This concludes the summary of the essential features of general matter coupled $\N=2$ supergravity in four dimensions. 
Recall from section \ref{sec:compactman} that compactifications of type II string theories and M-theory on manifolds with 
$SU(3)$-structure lead to precisely such a theory. Throughout this work we use the constraints reviewed in this 
section to help construct and to check the resulting four-dimensional theories from such compactifications. In the next section we review how the resulting four-dimensional action can be derived for the simplest such compactification that is on a CY manifold.

%
\section{Calabi-Yau compactifications}
\label{sec:cycompact}
%

In this section we compactify type IIB string theory on a CY manifold and show how the resulting action matches the general ungauged $\N=2$ supergravity.
Although we consider the explicit type IIB example the generalities of the compactification procedure hold for all the string theories.
We do not concern ourselves with the fermionic sector of the actions for now. This sector can be deduced by supersymmetry from the bosonic sector and is non-chiral and so does not form a candidate for the standard model fermionic sector. The bosonic fields of the theory are the graviton $\hat{g}_{MN}$, the dilaton $\hat{\phi}$, the Neveu-Schwartz (NS) two-form $\hat{B}_2$, and the Ramond (R) scalar $\hat{l}$, two-form $\hat{C}_2$ and four-form $\hat{A}_4$.
The bosonic part of the action (in the String frame) reads \cite{polchinski}
\ba
S^{10}_{IIB}&=&
   \frac{1}{2K^2_{(10)}}\int_{\M_{10}}\left[e^{-2\hat{\phi}}\left(\hat{R}\star 1+4d\hat{\phi}\wedge\star d\hat{\phi}-\half \hat{H}_3 \wedge\star \hat{H}_3 \right)\right.\\\nonumber
                                &~&\left.\qquad\qquad\qquad
                                -d\hat{l}\wedge\star d\hat{l}-\hat F_{3}\wedge\star \hat F_{3}
                                -\half \hat F_{5}\wedge\star \hat F_{5}
                                - \hat{A}_{4}\wedge \hat{H}_3 \wedge d\hat{C}_{2}\right],
\label{10diibaction} 
\ea
where $K_{(10)}$ is the ten-dimensional Planck constant and the field strengths are defined as
\be
\hat{H}_{3}=d\hat{B}_{2}\;\mathrm{\;,}\;\hat F_{3}= d\hat{C}_{2}-\hat{l} \hat{H}_3\;\mathrm{\;,}\;\hat F_{5}=d\hat{A}_{4}-\hat{H}_3 \wedge \hat{C}_{2} \mathrm{\;.} 
\ee
The self-duality of the five-form $\hat{F}_5$ should be imposed at the equations of motion level since at the action level the corresponding kinetic term vanishes trivially. 
In analogy with section \ref{sec:compactdim} we consider a spontaneous compactification of this action to a vacuum with 
\ba
\ev{\hat{H}_3} &=& 0 \;\;\mathrm{\;,}\;\; \ev{\hat{F}_3} = 0 \;\;\mathrm{\;,}\;\; \ev{\hat{F}_5} = 0 \mathrm{\;,} \nn \\
\left< \; \hat{g}_{MN}\right> dX^M dX^N &=& \left< g_{\mu\nu} \right> dx^{\mu} dx^{\nu} + \left< g_{mn} \right> dy^m dy^n \mathrm{\;,}
\label{cysponcom}
\ea
with the index ranges $M,N=0,...,9$ and $m,n=1,...,6$.
The Einstein equation reads $\hat{R}_{MN}=0$ which is solved by 
\be
\ev{g_{\mu\nu}}=\eta_{\mu\nu} \;\;,\;\;\; \ev{g_{mn}} = \mathrm{any\;Ricci-flat\;Euclidean\;metric} \mathrm{\;.}
\ee
If we also impose the constraint from supersymmetry of $SU(3)$-structure we find the internal manifold must be CY. 

We now derive the effective four-dimensional action that results from a compactification of type IIB supergravity 
on CY manifolds first considered in \cite{Bodner:1989cg}. We follow the discussion outlined in \cite{Bohm:1999uk,Gurrieri:2003st}.
To derive the four-dimensional field spectrum we consider perturbations about the vacuum and expand them in terms of the harmonic forms on the CY. These lead to massless modes in four-dimensions as they are zero modes of the Laplacian and the higher mass KK states  are truncated as usual. The spectrum of harmonic forms on CY manifolds has been studied \cite{Candelas:1989bb} and is most neatly written in complex co-ordinates
which we define as $s^\alpha$ , $\alpha=1,2,3$ such that 
\be
s^1 = \frac{1}{\sqrt{2}} \left( y^1 + i y^2 \right) \;\mathrm{\;,}\;\; s^2 = \frac{1}{\sqrt{2}} \left( y^3 + i y^4 \right) \;\mathrm{\;,}\;\; s^3 = \frac{1}{\sqrt{2}} \left( y^5 + i y^6 \right) \mathrm{\;.}
\ee 
The non-vanishing Hodge numbers $h^{(p,q)}$, where $p$ labels the number of holomorphic indices and $q$ the number 
of anti-holomorphic, read 
\ba
h^{(0,0)} &=& h^{(3,0)} = h^{(0,3)} = h^{(3,3)} = 1 \mathrm{\;,}\nn \\
h^{(1,1)} &=& h^{(2,2)}\;\;\mathrm{,}\;\;\; h^{(2,1)} = h^{(1,2)} \mathrm{\;.}
\ea
The basis of forms then comprises of the $(1,1)$ forms $\om_{i}$, $i=1,...,h^{(1,1)}$ and their Hodge duals $\tilde{\om}^{i} \equiv \star \om_i$, which we choose to normalise so that 
\be
\int_{CY}{\om_i \wg \tilde{\om}^j } = \kappa \delta^j_i \mathrm{\;,}
\ee
where we have introduced $\kappa$ as some constant CY reference volume which we generally omit unless required for clarity (it can be replaced by dimensional analysis). 
The variable volume depends on the four-dimensional co-ordinates and is given by
\be
\Vol = \frac{1}{6} \int{J \wg J \wg J} \mathrm{\;.}
\ee
The spectrum of three-forms can be parameterised in two useful ways. The first is through the set of complex $(2,1)$ forms $\chi_a$, $a=1,...,h^{2,1}$ with the unique 
holomorphic $(3,0)$ form $\Omn^{cs}$ and their Hodge duals \footnote{The superscript on the $\Omega^{cs}$ denotes the difference from the $SU(3)$-structure $\Omega$ in section \ref{sec:su3six} and from the vector-multiplet $\Omn^{sk}$ in section \ref{sec:n=2sugra}. There are exact, but at times subtle, relations between the different $\Omn$s that are discussed in section \ref{sec:cyton=2} and in chapters \ref{cha:su3iia} and \ref{cha:su3mtheory} but at this point we denote them as different quantities.}.
The second basis is through the set of real three-forms $\alpha_A$, $A=0,...,h^{(2,1)}$ and $\beta^A$ that can be normalised as
\be
\int_{CY}{\alpha_A \wg \beta^B} = \kappa \delta^B_A \;\;\mathrm{\;,}\;\; \int_{CY}{\alpha_A \wg \alpha_B} = \int_{CY}{\beta^A \wg \beta^B} = 0 \mathrm{\;.}
\ee 
Using the harmonic basis forms it is simple to write down the massless bosonic field spectrum of the effective four-dimensional theory that arises from the decomposition of the ten-dimensional fields
\ba
\hat{l}(X) &=& l(x) \mathrm{\;,}\nn \\
\hat{B}_2(X) &=& B_2(x) + b^i(x) \om_i (y) \mathrm{\;,}\nn \\
\hat{C}_2(X) &=& C_2(x) + c^i(x) \omega_i(y) \mathrm{\;,}\nn \\
\hat{A}_4(X) &=& D_2^i(x) \wg \om_i(y) + \rho_i(x) \tilde{\om}^i(y) + V^A (x) \wg \alpha_A (y) - U_A (x)\wg \beta^A (y)
\mathrm{\;.}
\label{iibformexpan}
\ea
The resulting four-dimensional fields are referred to as axions.
There are also four-dimensional fields that arise from deformations of the metric on the CY that are the analogues of the dilaton in section \ref{sec:compactdim}. These fields are called moduli. 
The constraint on such deformations is that they maintain the CY condition
\be
R_{mn}(g_{mn} + \delta g_{mn}) = 0 \mathrm{\;,}
\label{maincy}
\ee
which leads to the Lichnerowicz equation 
\be
\nabla^l \nabla_l \delta g_{mn} - \left[ \nabla^l,\nabla_m \right] \delta g_{ln} - \left[ \nabla^l,\nabla_n \right] \delta g_{lm} = 0 \mathrm{\;,}
\ee
where $\nabla$ is the Levi-Civita connection. This is simply the Laplace equation for a symmetric two-tensor and so we should expand the metric deformations again in the basis of harmonic forms. The deformations are most neatly written using complex co-ordinates. Since for a K\"ahler manifold, 
the K\"ahler form $J_{\alpha\bar{\beta}} = i g_{\alpha\bar{\beta}}$, the $(1,1)$ variations of the metric are variations of the K\"ahler form and are 
therefore labelled K\"ahler moduli. They arise from the expansion in terms of the $(1,1)$ forms 
\be
\delta g_{\alpha\bar{\beta}} = -i v^i(x) \left( \om_i \right)_{\alpha\bar{\beta}} \mathrm{\;.}
\label{cykahler}
\ee
It is also possible to consider $(2,0)$ deformations of the metric. Since the metric on a K\"ahler manifold must be of a $(1,1)$ type these deformations correspond to 
deformations of the complex structure of the manifold and so are termed complex structure moduli. There are no $(2,0)$ forms on a CY with which to expand these 
deformations but rather they are expanded using the $(2,1)$ form basis as
\be
\delta g_{\alpha\beta} = \frac{i}{||\Omn^{cs}||^2} \bar{z}^a(x) \left( \bar{\chi}_a \right)_{\alpha\bar{\gamma}\bar{\delta}} \left(\Omn^{cs}\right)^{\bar{\gamma}\bar{\delta}}_{\;\;\;\;\beta} \mathrm{\;,}
\label{cycomplex}
\ee
where 
\be
||\Omega^{cs}||^2 = \frac{1}{3!}\left(\Omn^{cs}\right)_{mnp}\left(\Omb^{cs}\right)^{mnp} \mathrm{\;.}
\ee
We are now in a position to list the bosonic field content of the four-dimensional theory. 
It comprises of scalar fields $l$, $b^i$, $c^i$, $\rho^i$, $v^i$ and $z^a$. There are also the gauge fields $V^A$ and $U_A$ as well as the two-forms $B_2$, $C_2$ and $D^i_2$. 
The constraint of the self-duality of $\hat{F}_5$ eliminates the degrees of freedom $U_A$ and $D^i_2$ as these are Hodge dual to the fields $V^A$ and $\rho^i$ respectively.
The two-forms $B_2$ and $C_2$ are Hodge dual to scalars which we label $h_1$ and $h_2$ respectively.
The full field content is specified in table \ref{tab:N=2iibcontent} where the fields have been grouped as members of $\N=2$ multiplets.
\begin{table}
\begin{center} 
\begin{tabular}{||l|c||} \hline

 $g_{\mu\nu}, V^0$   & gravitational multiplet  \\ \hline
 $h_1, h_2, \hat{\phi}, l$  & universal hypermultiplet \\ \hline
 $z^a, V^a$ & $h^{2,1}$ vector multiplets \\ \hline
 $b^i, v^i, \rho^i, c^i $  & $h^{1,1}$ hypermultiplets \\ \hline

\end{tabular}
\caption{Table showing the ${\cal N}=2$ multiplets in compactifications of type IIB theory}
\label{tab:N=2iibcontent}
\end{center}
\end{table}   

To derive the effective four-dimensional action we substitute the expansions (\ref{iibformexpan}), (\ref{cykahler}) and (\ref{cycomplex}) into the action (\ref{10diibaction}) 
and integrate over the internal manifold. There are also three field rescalings that are performed during the calculation in order to bring the resulting action into a neat 
format. The first is a Weyl rescaling of the ten-dimensional metric by a dilaton factor to take us to the Einstein frame
\be
\label{weyl1}
\hat{g}_{MN} \ra \hat{g}_{MN} e^{\half \hat{\phi}} \mathrm{\;.}
\ee
Then follows a Weyl rescaling of the four-dimensional metric by a factor of the CY volume
\be
\label{weyl2}
g_{\mu\nu} \ra g_{\mu\nu} \Vol^{-1} \mathrm{\;,}
\ee
and finally there is a rescaling of the K\"ahler moduli by a dilaton factor
\be
v^i \ra v^i e^{\half\hat{\phi}} \mathrm{\;.}
\ee
We also define a new field which is the four-dimensional dilaton $\phi$ that differs from the ten-dimensional dilaton $\hat{\phi}$ by a factor of the CY volume
\be
\phi \equiv \hat{\phi} - \half \mathrm{ln} \Vol \mathrm{\;.}
\label{4ddildef}
\ee
Since the ten-dimensional action only contains kinetic terms the resulting four-dimensional effective action only contains kinetic terms for the fields and reads
\ba
\label{4diib}
S^{4D}_{IIB} &=&
       \frac{1}{K^2_{(4)}}\int_{\cal S}\left[ \half R\star 1  - g_{a\bar b}dz^a\wedge\star d\bar{z}^{\bar b} - h_{\lambda\sigma}dq^{\lambda}\wedge\star dq^{\sigma}\right.\\\nonumber
                &~&\qquad\qquad\left.+ \half (\im{\M})_{AB} F^A \wedge \star F^B + \half (\re{\M})_{AB} F^A \wedge F^B \right],
\ea
where we define the four-dimensional Planck constant
\be
K_{(4)}^2 = K^2_{(10)} \kappa^{-1}.
\ee
The hypermultiplets are denoted collectively by $q^u$ with $u=1,...,4 \times \left( h^{2,1} + 1 \right)$ with kinetic terms
\be
h_{\lambda\sigma}dq^{\lambda} \wg \star dq^{\sigma} =  - g_{i\bar{j}} dt^i \wg \star d\bar{t}^{\bar{j}} - d\phi \wg \star d\phi  
- \half \Vol e^{2\phi} dl \wg \star dl + K_{(h_1,h_2,c^i,\rho_i)} \mathrm{\;,}
\label{quatkin}
\ee
where we have introduced new complex fields
\be
\label{tidef}
t^i \equiv b^i - i v^i \mathrm{\;,}
\ee
and the last term of (\ref{quatkin}) denotes the kinetic terms for the subscripted fields that are complicated but are not important for the purposes of this work.
The gauge field-strengths are defined as 
\be
F^A \equiv dV^A \mathrm{\;,}
\ee
with the form of the gauge kinetic matrix $\M_{AB}$ given in terms of the basis forms in (\ref{smatrices}). The metrics on the moduli spaces are read off from the coefficients of the kinetic terms and are given by
\ba
g_{a\bar{b}} &=& - \frac{i}{\Vol ||\Omn^{cs}||^2 } \int_{CY}{ \chi_a \wg \bar{\chi}_{\bar{b}}  } \label{gabmet} \mathrm{\;,} \\
g_{i\bar{j}} &=& \frac{1}{4\Vol} \int_{CY}{ \om_i \wg \star \om_j } \label{gijmet} \mathrm{\;.}
\ea
This concludes the review of CY compactifications of type IIB supergravity. Compactifications of type IIA supergravity are very similar to the type IIB case with the 
only differences arising in the Ramond matter sector since the NS sector is common to both. In particular the geometrical moduli remain in exactly the same form. In fact 
there is an important symmetry between the two resulting four-dimensional actions called mirror symmetry which states that compactifying type IIA string theory on a 
CY is equivalent to compactifying type IIB on a mirror CY defined by interchanging the two hodge numbers $h^{(2,1)} \leftrightarrow h^{(1,1)}$. This symmetry is a manifestation of T-duality which is an exact symmetry of string theory which states that string theory compactified on a circle of radius $R$ is equivalent 
to the theory compactified on a circle with the inverse radius $\frac{1}{R}$. This is an interesting symmetry because it implies that string theory has a minimum scale that it can probe which is the self-dual scale $R=1$. Mirror symmetry then arises from T-dualising along three directions within the CY. 

%
\subsection{Calabi-Yau compactifications as an $\N=2$ supergravity}
\label{sec:cyton=2}
%

Having derived the key features of the effective four-dimensional action arising from a CY compactification it is informative to 
see how this action fits into the mould of general $\N=2$ supergravity. Since the action (\ref{4diib}) has no scalar potential it should 
correspond to an ungauged $\N=2$ supergravity. Comparing (\ref{4diib}) with (\ref{n=2act}) it is easy to see that the two actions are 
identical in form. The only thing to check is that the geometrical structure discussed in section \ref{sec:n=2sugra} is present. 
Consider the vector multiplet sector first. To show that the scalar components, that are the complex structure moduli $z^a$, 
span a special K\"ahler manifold we need 
to identify a holomorphic three-form to play the role of $\Omn^{sk}$ which is obviously $\Omn^{cs}$. Then using Kodaira's formula \cite{tian} for
the derivative of the holomorphic CY form
\be
\partial_a \Omega^{cs} =  k_a \Omn^{cs} + i\chi_a \mathrm{\;,}
\ee
where $k_a$ is some general function it is simple to show that metric $g_{a\bar{b}}$ is generated by a K\"ahler potential 
\be
K^{cs} = - \mathrm{ln} \;i\int_{CY}{\Omn^{cs} \wg \Omb^{cs}} = - \mathrm{ln\;} \Vol || \Omn^{cs} ||^2 \mathrm{\;.}
\label{cskahler}
\ee
The relation (\ref{cskahler}) gives the expression $k_a = -\partial_a K^{cs}$. An interesting observation that follows is that the metric (\ref{gabmet}) is given by 
\be
g_{ab} = \frac{i}{\Vol ||\Omn^{cs}||^2 } \int_{CY}{ D_a \Omega^{cs}\wg D_{\bar{b}}\Omb^{cs}  } \label{gabmetcov} \mathrm{\;,}
\ee
where the K\"ahler covariant derivatives $D_a$ are defined as
\be
D_a \equiv \partial_a + \left(\partial_a K^{cs}\right) \mathrm{\;.}
\ee
We have already noted that in $\N=2$ supergravity $\Omn^{sk}$ is homogeneous with rescalings of it corresponding to K\"ahler transformations. We see that the 
K\"ahler covariant derivatives ensure that the metric $g_{ab}$ is independent of such rescalings.
If we expand $\Omn^{cs}$ into its periods in terms of the three-form basis
\be
\label{omnexp}
\Omn^{cs} = X^A(z) \alpha_A - F_A(z) \beta^A \mathrm{\;,}
\ee
we recover the formula 
\be
\label{csk}
K^{cs} = - \mathrm{ln} \; i\left[ \bar{X}^A F_A - X^A \bar{F}_A \right] \mathrm{\;,}
\ee
which matches exactly the supergravity formula. It is worth noting that in special co-ordinates the periods are given by the complex structure moduli. 
Finally it is possible to show that the gauge-kinetic matrix in (\ref{4diib}) 
$\M$ also matches the formula for the period matrix (\ref{n=2period}) \cite{Gurrieri:2003st}.

The hypermultiplet section is more complicated due to the more complicated geometry of quaternionic manifolds. Nonetheless it is possible 
to show that the hypermultiplets do indeed span a quaternionic manifold \cite{Bodner:1989cg}. There is a further interesting feature to the 
hypermultiplet sector that arises because of mirror symmetry. Recall that mirror symmetry interchanges the hodge numbers $h^{(1,1)}$ and 
$h^{(2,1)}$. From the expansions (\ref{cykahler}) and (\ref{cycomplex}) this can also be viewed as an interchange of the complex structure moduli 
$z^a$ with the complex K\"ahler moduli and axions combinations $t^i$ defined in (\ref{tidef}). 
This means that, since in the mirror (IIA) picture the $t^i$ form the scalar 
components of the vector multiplets which should span a special K\"ahler manifold, the $t^i$ moduli should also span a special K\"ahler 
submanifold within the overall quaternionic manifold spanned by the full hypermultiplets. Indeed equation (\ref{quatkin}) shows that the 
$t^i$ moduli form a separate sub-manifold within the hypermultiplets with a metric given in (\ref{gijmet}). This metric arises from a K\"ahler potential 
\be
K^{km} = - \mathrm{ln} \; \frac{4}{3} \int_{CY}{J \wg J \wg J}= - \mathrm{log} \; 8\Vol  \mathrm{\;,}
\label{cykahlerpot}
\ee
which is again derived from a prepotential ${\cal F}^{km}$ which in this case can be evaluated explicitly 
\be
\label{cubicpre}
{\cal F}^{km} = - \frac{1}{3!} \frac{{\cal K}_{ijk} X^i X^j X^k}{X^0} \mathrm{\;,}
\ee
where we define the intersection numbers
\be
{\cal K}_{ijk} \equiv \int_{CY}{\om_i \wg \om_j \wg \om_k} \mathrm{\;,}
\ee
and we have introduced the homogeneous co-ordinates $X^I=(1,t^i)$ to keep the analogy with the special K\"ahler geometry.

In this section we reviewed some of the important aspects encountered when attempting to recover four-dimensional physics,
culminating in the effective four-dimensional action from CY compactifications of type IIB supergravity (\ref{4diib}).
The resulting action however does not resemble the standard model of physics and comparing it with observations leads to many difficulties. 
The following section reviews some of the major difficulties facing such an action. 

%
\section{The vacuum degeneracy problem}
\label{sec:vacdeg}
%

The effective four-dimensional action (\ref{4diib}) lacks the key features of the standard model or of its simplest supersymmetric extensions. 
The action preserves $\N=2$ supersymmetry which is non-chiral and so the fermions can not be the standard model fermions. 
Since the supergravity is ungauged neither are they charged under the gauge fields which are themselves 
only $U(1)$ abelian fields. These observations, and others similar in nature, are all features of the theory that are missing. It is possible to argue that 
this is not a major problem because string theory is very rich and the compactification only explores one sector of the theory and 
so it could be that the missing features are to be found in a different sector. This approach has been very successful in recent years. In type IIB string theory it has been possible to construct a chiral fermionic spectrum charged under non-abelian gauge fields as states corresponding to strings stretching between D3 and D7-branes \cite{Ibanez:2004iv}. In type IIA string theory states stretching between intersecting D6 branes can produce sectors that very strongly resemble the standard model \cite{Aldazabal:2000dg}. 
 
More serious problems arise from features of the theory that are present but contradict experimental observations. The original 1920s Kaluza-Klein theory was rejected 
because it predicted a massless scalar mode of gravity, the dilaton. This would lead to a new long range force, mediated by the dilaton, between all matter which 
in turn predicts changes to planet orbits that are ruled out by observations \cite{Will:2001mx}. The feature of massless scalar modes of gravity are common to all higher dimensional 
compactifications and string theory is no 
exception. For the case of CY compactifications we have $2\times h^{(2,1)}+h^{(1,1)}$ such scalar fields that are the complex structure and K\"ahler moduli. For 
a typical CY manifold these can reach the hundreds. The existence of massless moduli in string compactifications on CY manifolds rules them out as possible 
models of the universe. A possible way to get past this problem is modify the theory in such a way that these fields receive masses and this process is referred to as moduli stabilisation.

There is a further motivation for moduli stabilisation that stems from the fact that important quantities of the four-dimensional theory depend on the values that the moduli take 
in the vacuum. Since the moduli have no potential there is no particular value that is predicted and this problem is referred to as the vacuum degeneracy problem.
The features of the theory that depend on the moduli can be separated into conditions that are required for the consistency of the theory and quantities that 
can be measured experimentally. The first class includes the size of the extra dimensions and the value of the dilaton. The supergravity approximation of string theory 
is the low energy limit or equivalently the large scale limit. Therefore the size of the cycles on the internal manifold must be larger than the string scale. 
This translates into a condition on the moduli for their values to be large in units of the inverse string tension $\alpha'$. Another approximation is the use of perturbative 
string theory which means that the string self coupling should be weak. This in turn imposes a condition on the value of the dilaton. There is also a 
consistency condition that is imposed on any possible masses the moduli might obtain due to the truncation we performed of the higher mass states that arise from 
string excitations and from KK modes. In order for the truncation to have been a consistent one the moduli must have masses much smaller than the scale of these 
higher mass states. The second class of moduli dependent quantities that can be measured experimentally, and therefore are constrained, are more model dependent. These include 
the scale of supersymmetry breaking, the value of the gauge coupling function and the masses of any massive fields.  

The final important problem that is relevant to this work, and which is similar to the moduli problem discussed in the previous paragraph in terms of a lack of predictive 
power, is the topological degeneracy problem. The problem is that there are many CY manifolds of different topologies, and so different particle contents in 
four dimensions, and the compactification procedure does not differentiate between them. This means that we can not even predict the number of particles in the 
four-dimensional theory. The problem is made more serious by the fact that, as we show in chapter \ref{cha:con}, it is possible to move from a CY of one 
topology to one of a different topology. The problem of the choice 
of compactification manifolds becomes even more severe when we consider that CY manifolds are just a small subset of all the $SU(3)$-structure manifolds.

The problems highlighted in this section have so far been discussed in the context of string theory but most of them apply equally to the case of M-theory, in 
particular the moduli stabilisation problem. 
As we have shown the lack of a potential for the fields in the low energy four-dimensional effective theory of string and M-theory compactifications is the reason for the vacuum degeneracy problem. Any attempt to resolve these problems must include a process by which a potential is generated for the 
moduli and in this thesis we study two situations where such a potential is induced. The first potential is generated at particular points in the moduli space of CY 
manifolds called \textit{conifold points}. These points are also the points where the topology changing transition, known as a \textit{conifold transition}, from one CY to 
another occurs. A study of this potential can therefore shed light on both the moduli stabilisation problem and on the topological degeneracy problem. 
In chapter \ref{cha:con} we explore the dynamics associated with conifold transitions as a cosmology with an emphasis on whether the transition can be dynamically 
completed and on the fate of the moduli involved. We also look at the possibility of experimental signatures for such a transition through the formation of 
cosmic strings.
The second way to generate a potential is through the introduction of non-vanishing vacuum expectation values for the ten or eleven-dimensional field-strengths of the form fields, 
known as \textit{fluxes}. The presence of the fluxes changes the compactification procedure in a way that 
introduces a potential for the moduli. The fluxes also mean that in order to satisfy the conditions for spontaneous compactification the internal 
manifold is no longer a CY manifold but rather a more general $SU(3)$-structure manifold. In chapter \ref{cha:su3iia} we study compactifications of type IIA string theory 
on general manifolds with $SU(3)$-structure with fluxes present and in particular their potential for moduli stabilisation. Finally in chapter \ref{cha:su3mtheory} we 
perform a similar flux compactification but for the case of M-theory.

%
\chapter{Moduli Trapping and Conifold Transitions in Type IIB String Theory}
\label{cha:con}
%

As discussed in chapter \ref{cha:compactifications}, the problem of vacuum degeneracy in string theory compactifications on CY manifolds splits 
into two parts, namely the continuous degeneracy due to moduli fields and the discrete
one due to the large number of different topologies. It is well-known
that there are a number of different topology-changing processes
\cite{Candelas:1988di,Candelas:1989ug,Greene:1996cy} in string theory that connect the moduli spaces
associated with topologically different CYs. A detailed
understanding of these processes can be considered as a first
step towards resolving the topological degeneracy. In the vicinity of the transition point a potential is induced for the moduli thereby 
raising the possibility of a resolution to the degeneracy of the moduli fields.
In this chapter, we study a certain type of topology-changing process, the conifold
transition, as an explicit time-dependent phenomenon \cite{Lukas:2004du,Palti:2005kv}. 

A milder type of topology changing transition which arises in the
context of CY compactifications is referred to as a \textit{flop transition}.  
In such a transition, a two-cycle within the CY
space contracts to a point and re-expands as another two-cycle,
thereby leading to a topologically distinct CY space with
different intersection numbers. However, on either side of the
transition the Hodge numbers are the same and so the spectrum of
massless moduli is unchanged and only their interactions are
affected.  
Earlier work on black holes where
the internal manifold undergoes a flop as a function of radius is
given in \cite{Gaida:1998km}.
The study of five-dimensional cosmology associated with a flop of the
internal space in M-theory was initiated in \cite{Brandle:2002fa} and
followed up in \cite{Jarv:2003qx,Jarv:2003qy}. There it was
found that the tension of any M2-branes wrapping the collapsing
two-cycles caused the cycles to remain small, so the moduli forced the
CY to remain near the flop point. This ruled out a dynamical realisation of a flop transition.  

In the conifold transition, which we study here, a three-cycle on one
side of the transition collapses and then re-expands into a two-cycle,
thus changing the Hodge numbers of the CY. This results in a
different spectrum of massless moduli on either side of the
transition \cite{Greene:1996cy}.  
The transition proceeds through a space-time singularity termed the conifold point.
At least in the case where the CY manifolds are
quintics in $\CM\PM^4$ the generic singularity is a conifold
singularity and as such they form an important part of the geometric structure \cite{Greene:1995hu}.

Although conifold points naively give a singular low energy effective
theory it is a remarkable property of string theory that they can be
understood in a well defined manner\cite{Strominger:1995cz}. When
calculating the effective action for the low energy degrees of freedom
one integrates out the heavy modes, replacing their effect by altered
couplings between the light modes. It is only consistent to integrate
out the modes which are heavier than those present in the theory. 
As explained in \cite{Strominger:1995cz} it is precisely because some unseen light
modes are being integrated out that singularities appear in the
effective action for a conifold transition.  These extra modes
correspond to D3-branes wrapping the collapsing three-cycle, as the cycle
get smaller so these states become lighter.

This chapter begins with a review of conifold geometry and the string theory interpretation 
of the dynamics involved. In section \ref{sec:effectact} we construct the effective theory that 
describes such a transition as a gauged supergravity. We then proceed to study the dynamics of this 
theory. In section \ref{sec:conicosm} we 
consider an approximation where the fields are all homogeneous throughout the three-dimensional space.
We study the possible outcomes of a dynamical conifold transition in terms of the final state of the moduli as 
a function of the initial conditions. We find that it is possible to complete the transition given the appropriate 
initial conditions although for general initial conditions the moduli fields are trapped at the conifold point.
We follow this initial study with a study of the case where the moduli fields may be inhomogeneous in section 
\ref{sec:conicosmin}. We show that the inhomogeneities may help or hinder the moduli trapping mechanism depending 
on their amplitudes. We also study the possibility of cosmic strings forming after the transition and find 
that the cosmic strings, or any other structure, do not form. We summarise our findings in section \ref{sec:coniconc}.

%
\section{Review of conifold geometry}
\label{sec:conigeo}
%

\begin{figure}
\begin{center}
\epsfig{file=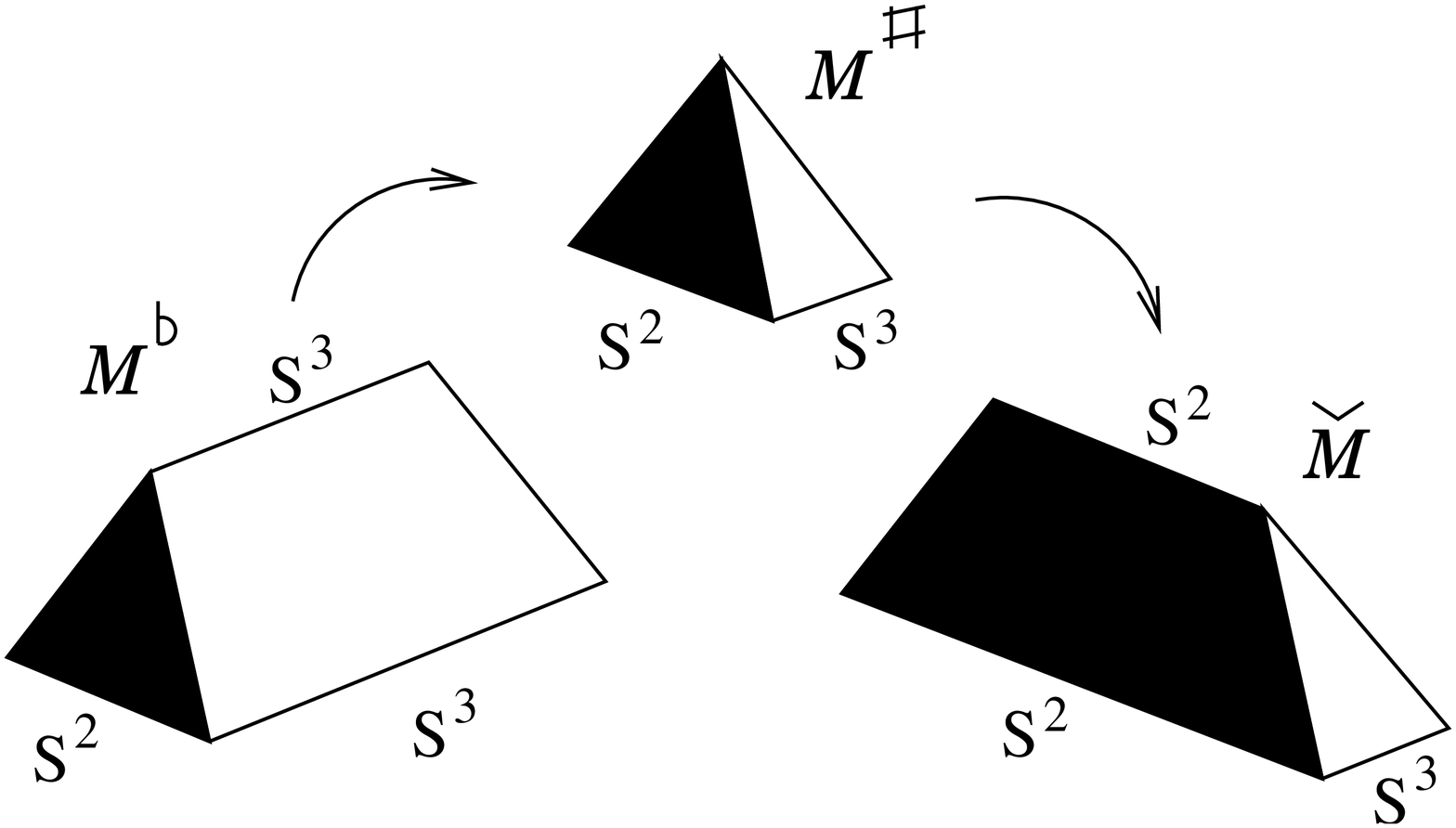,width=10cm}
\flushleft
\caption{
The deformed manifold ${\cal M}^\flat$ shrinks down to the conifold point ${\cal M}^\sharp$ and is resolved to $\check{\cal M}$. In terms of cycles a three-cycle 
turns into a two-cycle.
}
\label{fig:conifold}
\end{center}
\end{figure}

In this section we give a brief introduction to the geometry of conifold transitions, see 
\cite{Candelas:1989js,Hubsch:1992nu,Candelas:1988di,Candelas:1989ug,Green:1988bp,Green:1988wa,Candelas:1990rm,Greene:1996dh}
for a more complete discussion. 
It is possible to specify the geometry of CY manifolds through vanishing complex polynomials $P$. For example consider complex projective four-space $\mathbb{C}P^4$ with 
homogeneous co-ordinates $\left\{ z^1,z^2,z^3,z^4,z^5 \right\}$. The co-ordinates are homogeneous in the sense that points labelled by $\left\{ z^1,z^2,z^3,z^4,z^5 \right\}$ 
are identified with points $\left\{ \lambda z^1, \lambda z^2, \lambda z^3, \lambda z^4, \lambda z^5 \right\}$ where $\lambda$ is any complex number. Therefore a polynomial 
condition
\be
P = \sum_{i_1,i_2,i_3,i_4,i_5}{a_{i_1i_2i_3i_4i_5} z_1^{i_1}z_2^{i_2}z_3^{i_3}z_4^{i_4}z_5^{i_5}} = 0 \mathrm{\;,}
\label{cypoly}
\ee
where the $a$'s are complex numbers, defines a three-dimensional manifold since out of the five complex degrees of freedom one is fixed by the homogeneity and one is 
fixed by the polynomial constraint. The polynomial (\ref{cypoly}) in fact defines a CY manifold and the $a$'s correspond to complex structure moduli. The manifold 
associated with $P$ is not smooth for all values of the complex structure moduli. In particular the points where $P=0=dP$ are singular and the set of all values of 
the complex structure moduli for which this occurs is known as the discriminant locus. To see that this is singular at the points $dP=0$ recall that the vectors 
$\partial_i P$ fill the tangent space of the manifold and if they all vanish the tangent space has collapsed. In the vicinity of the singularity the manifold 
is called a node and takes the form 
\ba
\label{eqn:coniDef}
P=\sum_{A=1}^{A=4}(z^A)^2=0 \mathrm{\;,}
\ea
where $A=1,...,4$. We have fixed the homogeneous co-ordinates by choosing $\lambda$ so that $z_5=1$. 
This is the equation for a conifold singularity\cite{Candelas:1989js}. It is 
singular at $z^A=0$ as $P=0=dP$ at that point and it describes a 
conical shape because if $z^A$ lies on the conifold (\ref{eqn:coniDef}) then so does $\lambda z^A$.
To find the base of the cone we intersect it
with an $S^7$ of radius $r$ centred at the node, which is specified by the equation
\be
\sum_{A=1}^{A=4}|z^A|^2=r^2 \mathrm{\;.}
\ee
Writing $z^A=x^A+iy^A$ we find
\ba
x.x=y.y \mathrm{\;,}\qquad x.y=0\mathrm{\;,}\qquad x.x+y.y=r^2 \mathrm{\;.}
\ea
The equation $x.x=\half r^2$ defines an $S^3$ of radius $r/\sqrt{2}$, and $y.y=\half r^2$
combined with $x.y=0$ gives an $S^2$ fibred over the $S^3$. So, the base is
$S^3$ fibred by $S^2$. As all such fibrations are trivial the base of the
conifold is the product $S^3\times S^2$. The two distinct ways to make the
conifold regular correspond to blowing up either the $S^2$ to give the
(small) \textit{resolution} or by blowing up the $S^3$ to give the {\it deformation}. 
The conifold
transition then describes a CY going between these two regular manifolds.
We denote the conifold by ${\cal M}^\sharp$,
the deformed manifold by ${\cal M}^\flat$,
and the resolved manifold by $\check{\cal M}$. A nice picture of the transition
was presented in \cite{Candelas:1989js} and is given in Fig. \ref{fig:conifold}.
It shows the finite $S^3$ of the deformed conifold shrinking to zero and then
being replaced by an expanding $S^2$ of the resolved conifold. 

The conifold point is the location in the moduli space of CYs where the manifold acquires a node. 
In fact, we shall see that for the space to remain K\"ahler it must acquire a set of
nodes. Consider a CY containing $P$ such nodes
which have been deformed , thereby introducing $P$ three-cycles. Not all of these
need be homologically independent so we take there to be $Q$ homology relations
among them, giving $P-Q$ independent three-cycles. Now we pick the
standard homology basis for the independent three-cycles,
\ba
{\cal A}^A \cdot {\cal B}_B=\delta^A_B\mathrm{\;,} \qquad A,B=0,1,2,...h^{(2,1)} \mathrm{\;,}
\ea
where $\cdot$ denotes the intersection of two cycles.
This introduces the magnetic cycles ${\cal B}_A$, dual to the electric cycles
${\cal A}^A$. 
For this discussion we shall consider the case where the collapsing cycles are composed
solely of electric cycles in which case, because of the $Q$ homology relations,
each ${\cal B}_A$ intersects more than one collapsing cycle. Also, each vanishing cycle must
be involved in at least one homology relation if the manifold is to be Kahler,
as we now show.

\begin{figure}
\begin{center}
\epsfig{file=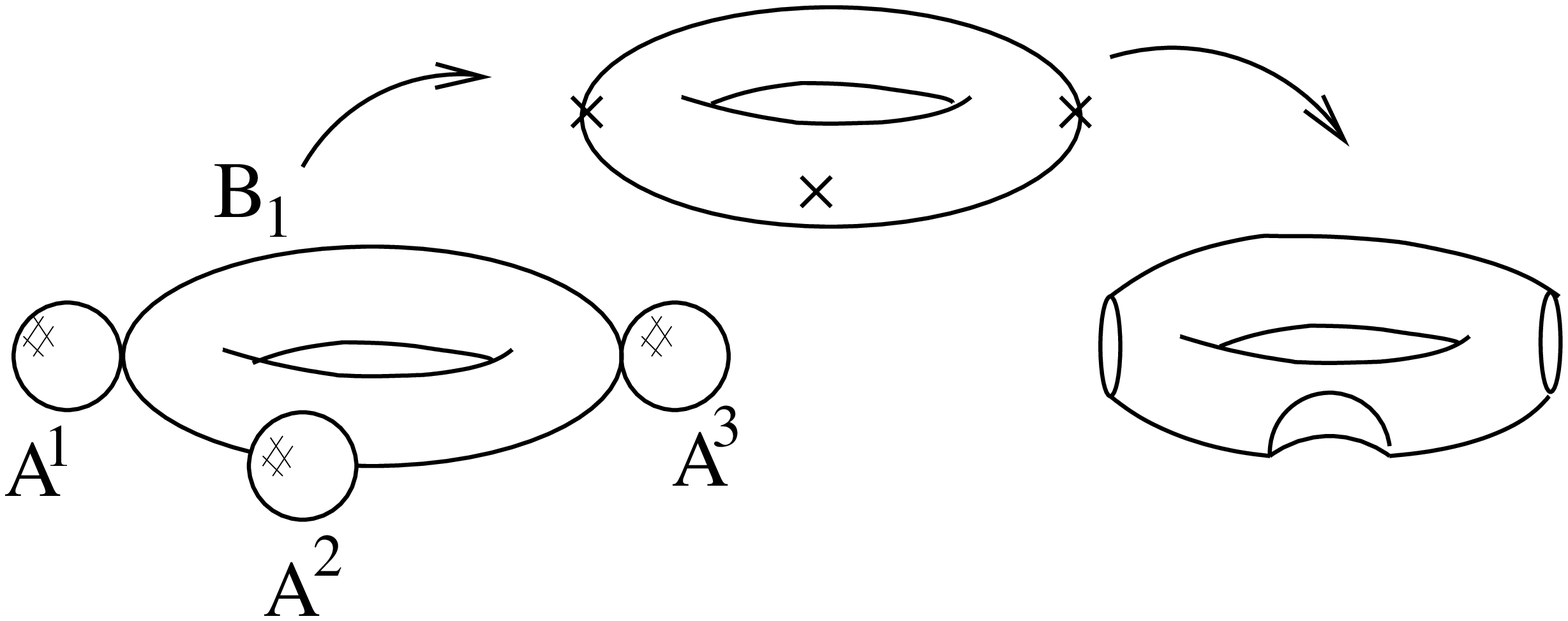,width=10cm}
\flushleft
\caption{ A magnetic three-cycle takes on the resolved two-cycles as its boundary thereby becoming a three-chain.}
\label{fig:cycle-chain}
\end{center}
\end{figure}

To see the effect of these homology relations consider the relationship
between ${\cal A}$ and ${\cal B}$ cycles. Fig. \ref{fig:conifold} shows a particular ${\cal A}$-cycle
three-sphere, ${\cal A}^1$, being blown down and then replaced by a two-sphere. The magnetic
dual of this three-cycle is constructed as follows. The shaded region in ${\cal M}^{\flat}$ is the ``cap''
$\bb{R}_{\ge 0}\times S^2$, which can be completed into a three-cycle when extended
away from the node. The picture shows that this cycle intersects
${\cal A}^1$ and can be chosen as its magnetic dual, ${\cal B}_1$. Also note that ${\cal B}_1$ remains
a three-cycle at the node, but when the node is resolved by a two-sphere then
${\cal B}_1$ takes on the $S^2$ as a component of its boundary and so becomes a three-chain.
Each ${\cal A}$-cycle that ${\cal B}_1$
intersects will provide an $S^2$ component to the boundary of ${\cal B}_1$ and as such
these two-spheres have a homology relation between them. Each of the magnetic
cycles provides a homology relation between the two-spheres and so we find $P-Q$
relations between the two-spheres of the resolved manifold $\check{\cal M}$.
We arrive at $P$ two-cycles with $P-Q$ homology relations giving $Q$ independent two-cycles.
The picture we are left with is illustrated in Fig. \ref{fig:cycle-chain} where
the magnetic cycle ${\cal B}^1$ touches the three-cycles ${\cal A}^1$, ${\cal A}^2$, ${\cal A}^3$ which shrink
to zero and then turn into boundary two-spheres thereby converting the cycle ${\cal B}^1$ into a chain.
If a ${\cal B}$-cycle had intersected only one ${\cal A}$-cycle then in the resolved manifold
this ${\cal B}$-chain would have a single $S^2$ boundary so we would have for the
K\"ahler form $J$,
\ba
\int_{S^2}J&>&0 \mathrm{\;,}\\\nonumber
\int_{\del B}J&>&0 \mathrm{\;,}\\\nonumber
\int_{B}dJ&>&0 \mathrm{\;.}
\ea
Which violates the K\"ahler condition, $dJ=0$, and so the resulting manifold can not be CY. This is why each vanishing cycle must be
involved in at least one homology relation. 

In summary, we have that $P-Q$ independent three-spheres collapse and then expand as
$Q$ independent two-spheres. This is a change in the topology of the CY and so is a change in the topology 
of space-time. The interesting thing is that string theory admits a consistent, non-singular, description 
of such a transition to which we now turn.

%
\section{A stringy description}
\label{sec:stringydes}
%

Singularities in physics usually manifest themselves as the breakdown of a theory. To understand how string theory resolves the conifold singularity we must 
first understand how the singularity manifests itself. The breakdown occurs in the sigma model of the complex structure moduli where the metric on the moduli 
space develops a curvature singularity. To show this, and for the rest of this section, we consider for simplicity the case of one degenerating cycle, i.e. a single 
conifold point. Consider the case of a CY with a single three cycle $\A$ and its dual $\B$ such that $\A . \B=1$. The cohomology associated with the cycles $\A$ and $\B$ is 
the two real three-forms $\alpha$ and $\beta$ respectively. Expanding the holomorphic three-form as in (\ref{omnexp}) we can define the usual periods in 
special co-ordinates
\be
z \equiv \int_{\A} \Omn^{cs} \;\;\mathrm{,}\;\; F(z) \equiv - \int_{\B} \Omn^{cs} \mathrm{\;.}
\ee
The conifold point where the cycle $\A$ vanishes corresponds to the moduli value $z=0$.
To know the form of the moduli space metric we need to know the form of the period $F(z)$. This is in general a difficult problem, however near the conifold point 
in moduli space we can determine an important property. To do this we consider the monodromy properties of the moduli space around the 
conifold point. There is a mathematical fact due to Lefshetz \cite{lefschetz} that states that if a cycle $\gamma$ is vanishing at the conifold point then another cycle $\delta$ 
undergoes monodromy 
\be
\delta \ra \delta + \left( \delta \cdot \gamma \right) \gamma \mathrm{\;,}
\ee
upon transport around this point in moduli space. 
Consider a parameterisation $z=re^{i\theta}$. 
Then near $r=0$ the period $F(z)$ must have the property that under a path in moduli space $\theta \ra \theta + 2\pi$
it transforms as $F(z) \ra F(z) + z$. Therefore $F(z)$ must take the form
\be
F(z) = \frac{1}{2\pi i} z \mathrm{ln} \; z + \mathrm{single\;valued\;terms} \mathrm{\;.}
\label{degf}
\ee
The first term in (\ref{degf}) is the cause of the metric singularity. Substituting (\ref{degf}) into the expression for the K\"ahler potential on the moduli 
space (\ref{csk}) we find that the metric takes the form \footnote{Note that we have assumed that there are other periods in the CY that are non-vanishing.}
\be
g_{z\bar{z}} \sim \mathrm{ln} \; \left( {z\bar{z}} \right) \mathrm{,}
\ee
which is a curvature singularity for $z=0$ at a finite distance \cite{Candelas:1990rm} and therefore a breakdown of the theory. 

Quantum string theory has an explanation for this singularity that also leads to a resolution \cite{Strominger:1995cz}. The idea is that there are states in string theory 
that correspond to D3-branes wrapping the degenerating three-cycle. The mass of these states is proportional to the volume of the wrapped cycle which is given 
approximately by $z$. In section \ref{sec:effectact} we calculate the mass exactly but for now the scale measure $z$ is good enough. Near the conifold point 
these states become very light and eventually massless at the conifold point itself. This means that it is inconsistent to not include them in the low-energy 
effective theory amounting to integrating out the states incorrectly. The breakdown of the theory can then be attributed to having integrated out the one-loop contribution 
that these states have to the quantum structure of the complex structure moduli space. To see this recall that in $\N=2$ supergravity the gauge kinetic matrix, that is the 
coupling between the different vector fields, is characterised by the second derivative of the prepotential on the vector multiplet moduli space
\be
\M_{AB} \sim {\cal F}_{AB} = \partial_{A} F_{B} \mathrm{\;.}
\label{gaugetoperiod}
\ee
The monodromy of the period, and therefore the singularity, manifests itself as a logarithm in the coupling of the gauge fields. Consider the effect 
that a hypermultiplet of fields, that we take to be the D3 states, charged under the gauge field associated with the degenerating cycle, i.e. the superpartner of $z$, 
have on the self coupling of that field $g$ through loop contributions. The $\beta$-function for the coupling is given by \cite{Klemm:1997gg}
\be
\beta(g) = \mu \frac{dg}{d\mu} = -\frac{g^3}{16\pi^2} \kappa \mathrm{\;,}
\label{betafun}
\ee 
where $\mu$ denotes the energy scale and
\be
\kappa = -\frac{2}{3} \sum_{i}{m_i T(R_i)_f} -\frac{1}{6} \sum_{i}{n_i T(R_i)_s} \mathrm{\;,}
\ee
where the sums are over the $m_i$ fermions and $n_i$ scalars in the loop that are in the representations $R_i$ such that the generators $T^a$ satisfy
\be
\mathrm{Tr}(T^a T^b) = T(R) \delta^{ab} \mathrm{\;.}
\ee
An $\N=2$ hypermultiplet has two fermions and four scalars which for a $U(1)$ field have $T(R)=\half$ thereby giving $\kappa = -1$.
We can solve the $\beta$-function equation (\ref{betafun}) to give the coupling at some energy scale $\tilde{\mu}$
\be
\M_{\tilde{\mu}} = \left(\frac{4\pi i}{g^2} \right)_{\tilde{\mu}} = - \frac{1}{2\pi i} \mathrm{ln} \; \tilde{\mu} \mathrm{\;.}
\ee
We now take this energy scale to be set by the energy scale that is the lower limit of the integrating out of the states. Recall that the states are a single hypermultiplet 
of mass scale $z$ thereby giving
\be
\M = - \frac{1}{2\pi i} \mathrm{ln} \; z \mathrm{\;,}
\ee
which from (\ref{gaugetoperiod}) gives
\be
F = \frac{1}{2 \pi i} z \mathrm{ln} \; z \mathrm{\;,}
\ee
as required.

We have shown that the breakdown of the theory can be directly attributed to incorrect integrating out of light states. This means that all we have to do 
in order to make the theory non-singular and perfectly consistent is include these states. The construction of such a theory is the subject of the next section.
Before we proceed with that it is interesting, although somewhat speculative, to consider the physical interpretation of the resolution of the conifold singularity by quantum 
string theory. A topology change of space-time requires the fabric of space-time to be somehow 'ripped' and 'glued' back together. It seems that string theory 
allows such a process without breaking down. A possible explanation for this might come from the fact that string theory can only probe scales down to the self-dual 
scale. What happens below this scale is as yet not understood, but with the hope that eventually string theory will be a full theory of gravity and so strings will 
form space-time, it is possible that its quantum nature on small scales allows this kind of behaviour. The necessary presence of the D3 branes which wrap the 
singularity points to some sort of 'shielding' mechanism of this process from the outside world. 

%
\section{The effective theory}
\label{sec:effectact}
%

In the previous section we showed how, by including D3-brane states, string theory allows a non-singular 
description of the conifold transition. In this section we construct an effective classical four-dimensional 
supergravity to describe this transition. Before we proceed it is important to understand the 
limitations of the supergravity description. The string theory picture of the transition is of a cycle 
reducing to the string scale before expanding out to a large size again. The supergravity description of 
string theory is only consistent at scales larger than the string scale and so must break down
during the transition near the point $z=0$. Therefore it can never give a complete description of the transition.
If we are interested in the question of whether the transition can be completed or not then the supergravity 
description is still a useful tool. By a completed transition we mean that the cycle has expanded out to a 
large size and so the supergravity description is again valid. The picture we should have then is of a supergravity description of 
dynamics towards and away from the string scale with the limitation that 
at the string scale we can not say much other than that the cycle is small, and not necessarily of vanishing 
volume as the supergravity description naively states. Another limitation is similar in nature and arises as 
a limitation of what the supergravity can tell us regarding the formation of a black hole from the kinetic energy 
of the brane states. The picture we have is of a brane wrapping a cycle that is shrinking in size and therefore the 
kinetic energy associated with the brane becomes concentrated in a shrinking region. If this region is smaller that the 
Schwartchild radius associated with the kinetic energy than a black hole will form. A calculation of whether this process occurs or 
not is beyond the scope of this thesis but the possibility of this phenomenon remains. There 
are yet more limitations to a classical description of the process as opposed to a quantum one. In fact it has 
been argued in \cite{Kofman:2004yc,Watson:2004aq} that quantum effects play an important part in understanding the dynamics of a conifold transition. We return to this point in
 section \ref{sec:conicosm} where we argue that these effects are mostly negligible with respect to the conclusions drawn. 
Having outlined the important shortcomings of the supergravity 
 description we proceed with the construction of the effective action whilst keeping in mind the conditions under 
 which it is valid.

The ten-dimensional actions that form our starting point are those of type IIB supergravity and of a D3-brane.
The type IIB action is reduced as outlined in section \ref{sec:cycompact} with the resulting four-dimensional action (\ref{4diib}).  
At this point it is interesting to explore the structure of the gauge fields of the theory. Recall that in the field 
expansion (\ref{iibformexpan}) there are two types of gauge fields in four dimensions $U_A$ and $V^A$, with respective field-strengths $G_A$ and $F^A$, which we stated can be eliminated using the self-duality of $F_5$. Imposing self duality $F_5=\star F_5$ on the field expansion gives
\ba
\label{eqn:magField}
G_A&=&(\re{\M})_{AB}F^B+(\im{\M})_{AB}\star F^B \mathrm{\;,}
\ea
where the action of the Hodge star on the basis forms is most neatly given in terms of matrices $S_1$, $S_2$, $S_3$, $S_4$ defined as functions of the gauge-kinetic matrix $\M$ \cite{Gurrieri:2003st}
\ba
\star\alpha_A&=&\left(S_1\right)_A^{\;\;B}\alpha_B+\left(S_2\right)_{AB}\beta^B \nn \mathrm{\;,} \\
\star\beta^A&=&\left(S_3\right)^{AB}\alpha_B+\left(S_4\right)^A_{\;\;B}\beta^B \mathrm{\;,} \\
\label{eqn:M}
S_1&=&-\left(S_4\right)^T=(\re{\M})(\im{\M})^{-1} \mathrm{\;,} \nn \\
S_2&=&-(\im{\M})-(\re{\M})(\im{\M})^{-1}(\re{\M}) \mathrm{\;,} \nonumber \\
S_3&=&(\im{\M})^{-1} \mathrm{\;.}
\label{smatrices}
\ea
Substituting the expression (\ref{eqn:magField}) into the action (\ref{4diib}) gives the gauge field terms
\be
S_{IIB}^{4D} \supset \half F^A \wedge G_A \mathrm{\;,}
\ee
which shows that the field strength $G_A$ is the magnetic dual of $F^A$. This understanding is important 
when describing the charges of the states arising from the D3-branes to which we now turn.

%
\subsection{The light states}
\label{sec:lightst}
%

The states arising from a D3-brane are a hypermultiplet. This has not been proved directly but can be inferred from the role they play in the transition as we now show.
First we proceed to calculate the mass and charge of these states by considering the action of a D3-brane \cite{polchinski}
\be
\label{eqn:10DD3}
S^{10}_{D3}= -\mu_3\int_{D3}d^4\xi e^{-\hat{\phi}}\sqrt{-\det[P(g^s_{\mu\nu})]}+\sqrt{2}\mu_3\int_{D3}\hat{A}_{(4)} \mathrm{\;,}
\ee
where the brane tension is related to the ten-dimensional Planck constant as
\be
\mu_3=\sqrt{\pi}/K_{(10)} \mathrm{\;.}
\ee
$P(g^s_{\mu\nu})$ denotes the pullback of the space-time metric onto the world-volume of the brane over which the integration is performed. We now consider 
the brane to be wrapped on some general three-cycle ${\cal C}$ which we can decompose in terms of the basis of three-cycles on the manifold $(\A^A,\B_A)$ as 
\be
{\cal C} = n_A \A^A+m^A \B_A,\qquad n_A,\;m^A\;\in \bb{Z} \mathrm{\;.}
\ee
The D3-brane wraps the cycle in such a way as to minimise its volume which in mathematical notation means it wraps a supersymmetric 
special Lagrangian three-cycle \cite{Joyce:2001nm,Becker:1995kb,Vafa:1995ta}.
Such cycles are calibrated by the form $\mathrm{Re}(e^{i\theta}\Omn^{cs})$, for some constant $\theta$, and saturate the following bound on their volume
\ba
\mathrm{Vol}({\cal C})\geq\frac{\sqrt{\Vol}\left|\int_{\cal C}{\Omn^{cs}}\right|}{\left| \int_{CY}{\Omn^{cs}\wedge\Omb^{cs}} \right|^{\half}}
      =\sqrt{\frac{\Vol}{\kappa}}\;e^{\half K^{cs}}\left|\int_{\cal C}\Omn^{cs}\right|.
\ea
Using the expansions for $\Omn^{cs}$ (\ref{omnexp}) and $\hat{A}_4$ (\ref{iibformexpan}) we can perform the spatial integration in (\ref{eqn:10DD3}) to find that the 
D3-brane action becomes, after the Weyl re-scalings of (\ref{weyl1}) and (\ref{weyl2}),
\be
\label{4DD3}
S^{4}_{D3} = -\frac{\sqrt{\pi}}{K_{(4)}}e^{\half K^{cs}}|n_A X^A - m^A F_A|\int d\tau
                                   +\frac{\sqrt{2\pi}}{K_{(4)}}n_A \int V^A
                                   -\frac{\sqrt{2\pi}}{K_{(4)}}m^A \int U_A \mathrm{\;.}
\ee
This is the action for a particle of mass 
\be
\label{branemass}
m = \frac{\sqrt{\pi}}{K_{(4)}}e^{\half K^{cs}}|n_A X^A-m^A F_A| \mathrm{\;,}
\ee
charged under the gauge fields $U_A$ and $V^A$, where we interpret $\tau$ as the proper time of the particle.
This relation between mass and charge is what is to be expected for an ${\cal N}=2$ extremal black hole
\cite{Ceresole:1995ca}. This black hole, or particle, is a \textit{dyon}, i.e. a particle charged both electrically and magnetically. 
Constructing theories with dyons is a difficult task and so from here on we take the brane to be wrapped on the purely electric cycles $\A^A$.
In the case of a single cycle the mass formula degenerates at the conifold point as discussed in section \ref{sec:stringydes}.

%
\subsection{The effective supergravity}
\label{sec:effectsugra}
%

In order to construct the effective supergravity description of the transition we use the constrained geometric structure imposed by the $\N=2$ supersymmetry reviewed 
in section \ref{sec:n=2sugra}. The new states are included as hypermultiplets and since they are charged under the gauge fields of the vector multiplets we 
expect a gauged supergravity. We address the construction of the supergravity in two parts. The first deals with the hypermultiplets sector of the theory 
and the second deals with the vector multiplet sector.

%
\subsubsection{The quaternionic geometry and the gauging}
\label{sec:quatgem}
%

The brane states should combine with the closed string hypermultiplets in table \ref{tab:N=2iibcontent} to span a quaternionic manifold. The gauging of the 
brane states has to be done with respect to the Killing vectors on this manifold and so a knowledge of the possible Killing vectors is required. 
This is a difficult task because we do not know the geometry of the manifold specified by the brane states nor that of the other hypermultiplets. A further 
complication arises due to the fact that a quaternionic manifold can not be a product of two quaternionic submanifolds \cite{Jarv:2003qx} and so there must 
be mixing between the brane hypermultiplets and the string hypermultiplets. The strategy we employ is a perturbative expansion in powers of the expectation value 
of the hypermultiplets in units of the Planck mass. 
It is possible to think of this expansion as an expansion in the effects of gravity since for global $\N=2$ supersymmetry we know that the hypermultiplets span a flat 
hyperK\"ahler manifold. To first order in this expansion we therefore expect that the brane states span a flat submanifold of the total quaternionic manifold. 
The constraint 
which provides a check on this statement comes from the mass of the states calculated in the previous section. Since the mass is given purely in terms of the 
complex structure moduli, which are vector multiplets, we know that certainly at that level there is no coupling to the hypermultiplet states. This allows us 
to, perturbatively, consistently truncate the string hypermultiplets and study the subsector of the theory formed by the brane states and the complex structure moduli.
We proceed to show more precisely that one recovers the correct mass term with this approach shortly. First it is worth noting that 
we do not expect the higher order terms to alter the qualitative behaviour of the dynamics.     
For example,the initial study of \cite{Brandle:2002fa} on flop transitions made the same approximation
which was amended in \cite{Jarv:2003qy}\cite{Jarv:2003qx} but lead to the same structure. More recently, dynamics of 
conifolds transitions in the context of M-theory were examined with both a first order approximation and using an explicit example of a full quaternionic 
manifold (a Wolf space) \cite{Mohaupt:2004pr} and the conclusion that the dynamics are unaffected was reached.

The hypermultiplets manifold of the brane states is spanned by real scalar fields $q^{au}$. The index $a=1,...,P$ ranges over the number of different 
hypermultiplets where each brane state wrapping a cycle corresponds to a separate hypermultiplet. The index $u=1,...,4$ runs over the component 
fields of the hypermultiplet. The flat geometry approximation then states that each hypermultiplet spans a submanifold with a flat metric $h_{uv}=\delta_{uv}$.
Since there are $Q$ homology relations among the $P$ wrapped cycles there are $P-Q$ independent wrapped cycles. Each independent cycle has 
associated with it a vector multiplet composed of a complex structure modulus $z^i$ and a vector field $V^i$ with the index $i$ running over the subset of 
all the three-cycles formed by the independent degenerating cycles $i=1,...,P-Q$.
The hypermultiplets may be charged under any of the electric vector gauge fields $V^i$ with charges $gQ_i^a$ where $Q_i^a$ are integers. 
To perform the gauging we must identify the Killing vectors 
on each hypermultiplet submanifold with respect to which we can gauge. As our metric is taken to be flat we have a choice of rotation or translation
Killing vectors. The constraint for the potential to vanish when the $q^{au}$ do singles out the rotation
Killing vectors as the relevant choice,
\be
\label{killvec}
k^{au}_i=Q^a_i t^u_{\;\;v}q^{av} \mathrm{\;,}
\ee
where $t$ is an anti-symmetric matrix which we take to be the same in each hypermultiplet, i.e. we gauge the same Killing vector within each hypermultiplet submanifold. 
It is easy to show that (\ref{killvec}) solves the Killing vector equation for flat space
\be
\partial_{bv} k^{au}_i \delta_{uw} + \partial_{bw}k^{au}_{i}\delta_{vu} = 0 \mathrm{\;.}
\ee
To derive the prepotentials from the Killing vectors we first introduce the 't Hooft symbols
\cite{'tHooft:1976fv}
\ba
\eta^x_{uv}&=&\bar\eta^x_{uv}=\epsilon^x_{\;uv} \;,\; \mathrm{if\;} u,v = 1,2,3 \mathrm{\;\;,}\\
\eta^x_{u4}&=&\bar\eta^x_{4u}=\delta^x_u \mathrm{\;,}
\ea
that are defined on the hyperK\"ahler submanifold for each hypermultiplet. Useful relations between the 't Hooft symbols can be found in the appendix.
We can parametrise the three complex structures on the manifold as
\be
\left(J^x\right)_{u}^{\;\;v} = - \bar{\eta}^x_{uw}\delta^{wv} \mathrm{\;.}
\ee
It is straight forward to check that this expression satisfies the quaternionic algebra (\ref{quatalg}).
The set of hyperK\"ahler forms is therefore given by
\ba
K^x_{uv}=-\bar\eta^x_{uv} \mathrm{\;.}
\ea
We wish to use (\ref{momentummap}) to derive the prepotentials associated with the Killing vector, for a flat metric this reads
\be
\label{flatkillpre}
2 k_i^{u} \bar{\eta}^x_{uv} = \partial_{v} P^x_i \mathrm{\;.}
\ee
The first thing to note is that (\ref{flatkillpre}) implies 
\be
\left[ \mathbf{t},\bar{\mathbf{\eta}}^x \right] = 0 \mathrm{\;.}
\ee
It is worth looking a little closer at the rotations we are considering. The rotations are in $SO(4)$ which is locally equivalent to $SO(3)\times SO(3)$ and the 
't Hooft symbol $\eta^x$ is a mapping between a vector in the first $SO(3)$ to self-dual two-tensors in $SO(4)$, with $\bar{\eta}^x$ mapping the other $SO(3)$ to 
anti self-dual tensors. Then we see that the rotations are rotations in one of the $SO(3)$s and we can expand them as
\ba
t_{uv}&=&n_x\eta^x_{uv} \mathrm{\;,}
\ea
where $\mathbf{n}$ is a unit vector. Explicitly we can take the matrix $\mathbf{t}$ to be of the form
\be
t_{12} = -t_{12} = 1 \;\;,\;\; t_{34} = -t_{43} = 1 \mathrm{\;,}
\ee
with all other components vanishing. 
Integrating equation (\ref{flatkillpre}) we arrive at the prepotentials
\be
\label{eqn:KillingPrePot}
P^x_i = \sum_{a}{Q^a_i q^{av}\left(\bar\eta^x_{uv}n_y\eta^{y,u}_{\;\;\;\;w}\right)q^{aw}} \mathrm{\;,}
\ee
where we have set the integration constants to zero as they would lead to a potential even in the absence of the light states.
The relevant quantities in the potential can be calculated to give
\ba
\label{eqn:Dterm}
V^{(D)}_{ij}&=&P^{x}_{i}P^{x}_{j} = \sum_{a,b} Q^{a}_{i}Q^{b}_{j} 
                           \left( 2 q^{av}q^{aw}q^{bv}q^{bw} 
                                 -  q^{av}q^{av}q^{bw}q^{bw} \right. \nn \\
                                 & & \;\;\;\;\;\;\;\;\;\;\;\;\;\;\;\;\;\;\;\;\;\;\;\;\;\;\;\;\;\;\;\; 
                                 \left. +\; 2 q^{av}q^{aw}q^{br}q^{bt}t_{wt}t_{vr} \right) \mathrm{\;,}\\
\label{eqn:mterm}
V^{(m)}_{ij} &=& h_{au,av} k^{au}_{i} k^{av}_{j} = \sum_{a} Q^{a}_{i} Q^{a}_{j} q^{aw} q_{aw} \mathrm{\;.}
\ea
The mass terms for the hypermultiplet states can be read off (\ref{scalpotgaug}) to give
\be
S_{\mathrm{mass}} = 4 g^2 e^{K^{cs}} X^i \bar{X}^j V^{(m)}_{ij} \mathrm{\;,}
\ee
which exactly matches the formula derived from the brane calculation (\ref{branemass}) for the case of electric cycles only with the identifications 
\be
g^2 = \frac{\pi}{4 K^2_{(4)}} \mathrm{\;,\;\;} Q^a_i = n^a_i \mathrm{\;,}
\ee 
where $n^a_i$ is the wrapping number for the cycle associated with the hypermultiplet index $a$.
This concludes our analysis of the hypermultiplet sector of the theory.

%
\subsubsection{The special K\"ahler geometry}
\label{sec:speckgem}
%

The vector multiplet sector of the theory can be completely determined in terms of the prepotential ${\cal F}$ of the special K\"ahler manifold spanned by 
the scalar components that are the complex structure moduli. The precise details of the prepotential depend on the CY in question and are difficult 
to calculate. There are some general facts that apply to all the prepotentials that can be used to determine important properties. The knowledge we have 
of the complex structure moduli prepotential on CY manifolds comes from mirror symmetry \cite{Candelas:1990rm}. 
The K\"ahler moduli are deformations associated with the volume of the cycles 
which means that their sigma model K\"ahler potential is given by the volume of the CY. This allowed us to write the cubic form (\ref{cubicpre}) for the prepotential. This 
prepotential is only accurate in the large volume, or large K\"ahler moduli, limit and receives corrections, in type IIA, from instanton effects 
that come into effect at small volumes. Applying mirror symmetry leads to a cubic prepotential for the complex structure moduli in type 
IIB which is valid for large complex structure and receives corrections at small complex structure. One major correction we have already encountered is the 
logarithmic term that arises from the monodromy properties. Having showed that this correction arises from incorrect integrating out of light states, by including 
the light states in the theory this correction is accounted for. The other corrections take the form of an analytic polynomial going up to cubic terms \cite{Candelas:1990rm}. 
As we are considering behaviour near the conifold point, which is at small complex structure, we can consider a perturbative expansion in the complex 
structure moduli. We therefore take a prepotential of the form
\be
{\cal F} = -\half i T_{IJ} X^I X^J \label{prepotentialans} \mathrm{\;,}
\ee
where we have introduced a constant \textit{coupling matrix} $T$ and the indices run over the degenerating cycles plus one $I,J=0,1,...,P-Q$.
In the expansion we dropped the cubic terms through the approximation and the linear term as a simplifying assumption. 
We do not expect this simplification to alter the dynamics of the transition qualitatively as it doesn't change the essential features of the action. 
Constant terms in the prepotential do not contribute to the action 
as it only depends on derivatives of the prepotential and so can be omitted. 
The prepotential (\ref{prepotentialans}) can also be thought of as specifying the (mirror) CY manifold through 
its intersection numbers. Of course it is a difficult task to prove that a CY exists with these intersection numbers. 

Having specified the prepotential we can use the constraints from supersymmetry discussed in section \ref{sec:n=2sugra} to write down the full resulting supergravity. 
We truncate the fields that do not feature in the transition which are all the matter hypermultiplets, the complex structure moduli that do not degenerate and their 
associated vector fields, and also the graviphoton $V^0$. In special co-ordinates $X^I=(1,z^i)$ the action reads
\ba
\label{completehyperaction}
S_{\mathrm{conifold}} = \nonumber \frac{1}{K^2_{4}} \int_{\cal S} \sqrt{-g}d^4x \left\{\;\; \half R_4\right.
                &-& \left(\frac{-T_{ij}}{\left<X|X\right>} + \frac{z^{k}\bar{z}^l T_{kj} T_{li}}{{\left<X|X\right>}^2} \right)
                   \partial_{\mu}z^{i}\partial^{\mu}\bar{z}^j \\ \nonumber
                &-& \nabla_{\mu}q^{au}\nabla^{\mu}q_{au} 
                - 2g^2 \left( \frac{z^i\bar{z}^j}{\left<X|X\right>} \right) 
                           V^{(m)}_{ij}  \\ \nonumber
                &-& g^2 \left( - \half (T^{-1})^{ij} - \frac{\bar{z}^i z^j}{\left<X|X\right>}
                          \right) V^{(D)}_{ij}  \\  
                &-& \left.\frac{1}{4} Im\left({\cal N}\right)_{ij}F^i_{\mu\nu} F^{j\mu\nu}\right\} \mathrm{\;,}
\ea
where
\be
\left<X|X\right> =  T_{IJ} X^I \bar{X}^J = T_{00} + T_{ij}z^i\bar{z}^j \mathrm{\;,}
\ee
and we have the gauge covariant derivatives
\be
\nabla_{\mu}q^{au} = \partial_{\mu}q^{au} + g Q_i^a t^{u}_{\;\;v}q^{av} V^i \mathrm{\;.}
\ee
One finds that in order to have positive kinetic terms for the scalar fields while still satisfying (\ref{skahler}) the coupling matrix 
$T_{IJ}$ must have signature (+,-,-,...)\cite{Ceresole:1995ca,Andrianopoli:1996cm}. 

To summarise, the action (\ref{completehyperaction}) describes the effective theory of $P-Q$ simultaneously degenerating 
independent cycles, with associated complex structure moduli $z^i$, $i=1,...,(P-Q)$, and vector fields $V^i$, where $P$ is the number of cycles and $Q$ is the number of 
homology relations between them. The cycles are wrapped by D3 branes that give rise to light states denoted by $q^{au}$ where here the double index notation denotes 
the number of cycles $a=1,...,P$ and hypermultiplet components $u=1,...,4$. The free parameters of the action are the matrix components of $T$. 
The action (\ref{completehyperaction}) is non-singular even at the conifold point and so we expect the complete transition to be described by it. So far we have 
concentrated on the shrinking three-cycles but that is only half the transition, there are also the expanding two-cycles on the other side of the transition 
the dynamics of which should also be included in this action. The complete model of the transition in terms of the effective low energy theory is described in the 
next section.

%
\subsection{Completing the transition}
\label{sec:completetran}
%

Consider how the conifold transition described in section \ref{sec:conigeo} should appear 
from the point of view of the low energy effective theory.  The relation between the CY topology and the matter spectrum is such that 
three-cycles correspond to complex structure moduli, and each independent three-cycle gives a vector multiplet, 
while the two-cycles correspond to K\"ahler moduli and each generates
a hypermultiplet. Recall that the conifold transition then sees $P$ three-cycles ${\cal C}^a$ degenerate with $Q$ homology relations and are 
resolved to $Q$ independent two-cycles. 
We therefore have $P-Q$ massless vector multiplets disappearing and $Q$ massless hypermultiplets appearing. 
To see the transition explicitly in the field theory it is helpful to simplify the scenario to the case where there is a single homology relation between the 
degenerating cycles $Q=1$, such that the sum of the cycles is homologically trivial
\be
\sum_{a=1}^{P}{{\cal C}^a} = 0 \mathrm{\;.}
\ee
In terms of the wrapping numbers or the charges with respect to the gauge fields this reads as
\be
\sum_{a=1}^{P} Q_{i}^{a} = 0 \mathrm{\;.}
\ee
Recall that the massless field spectrum far from the transition point, when the light states can be ignored $q^{au}=0$, consists of $z^i$ complex structure moduli. 
Near the transition point the moduli $z^i \ra 0$ and so the states $q^{au}$ become light and now both fields provide an effective mass for each other. 
The conifold point is where the moduli degenerate $z^i=0$ and at this point the mass term for the brane states vanishes. The term that remains in the potential 
is the one proportional to $V^{(D)}_{ij}$. Inspecting the formula (\ref{eqn:Dterm}) it can be seen that this potential has a flat direction along 
\be
\label{vacsol}
q^{au} = \pm q^{bu} \equiv \Lambda^u \;\; \forall \; a,b \mathrm{\;.}
\ee
The flat direction parameters $\Lambda^u$ parametrise a single hypermultiplet of fields which is then precisely the hypermultiplet we have gained by the 
single homology relation. So, as first discussed in \cite{Greene:1996dh,Greene:1995hu}, the brane states combine to form the new hypermultiplet. 
This is an interesting phenomenon because the hypermultiplet is interpreted 
as a string state while the brane states are extremal black hole states. This is an example of what is termed a string/black hole transition and hints at a deep 
connection between the two. The above argument can be generalised to the case of $Q$ homology relations by counting degrees of freedom. 

So far we have explained how we gain a hypermultiplet but we also should lose $P-1$ vector multiplets. It is easy to understand this as a Higgs transition since 
the hypermultiplets are charged under the gauge fields and subsequent to the transition they pick up a vev $\Lambda^u$. Explicitly 
the mass terms arise as
\be
\nabla_{\mu}q^{au}\nabla^{\mu}q_{au} \supset g^2 \sum_{a}{Q^a_i Q^a_j \Lambda^{v} \Lambda_{v} V^i V^j} \mathrm{\;.} 
\ee

Given the description of the transition in terms of a field theory picture we can investigate its dynamics by considering the dynamics of the action (\ref{completehyperaction}). 
In terms of the geometry the sizes of the cycles are given by $z^i$ and $\Lambda^u$ and so a completed transition would take the form of a configuration with 
$z^i=0$ and $\Lambda_u > 0$. We can investigate if this state can be reached from different initial conditions and this is the topic of the next section.

%
\section{Cosmology of conifold transitions: The homogeneous case}
\label{sec:conicosm}
%

In this section we study the dynamics of conifold transitions as a cosmology. We make the 
simplifying approximation that all the fields are homogeneous and isotropic throughout space-time, which means that 
there are no gauge fields induced through spatial currents. The more general case where the fields may be 
inhomogeneous is discussed in the next section. We take the space-time metric to be Friedmann-Robertson-Walker (FRW) with flat 
spatial sections
\be
ds^2=-dt^2+a(t)^2 d\mathbf{x}^2 \mathrm{\;,}
\ee
where $a(t)$ is the cosmological scale factor.
The equations of motion for the action (\ref{completehyperaction}) read
\ba
\ddot{q}_{av} + 3\left(\frac{\dot{a}}{a}\right)\dot{q}_{av}  + \half\partial_{av}V &=& 0 \mathrm{\;,}\\
\label{eqn:zeqn}
\ddot{z}^i + 3\left(\frac{\dot{a}}{a}\right)\dot{z}^i + \Gamma^i_{\;\;jk}\dot{z}^j\dot{z}^k 
+ g^{\bar j i} \partial_{\bar j} V &=&0 \mathrm{\;,} \\
2\left(\frac{\ddot{a}}{a}\right) + \left(\frac{\dot{a}}{a}\right)^2 &=& -g_{i\bar j}\dot{z}^i\dot{\bar{z}}^j 
-\dot{q}^{av}\dot{q}_{av} + V \mathrm{\;,}
\ea
where 
\be
V = 2g^2 \left( \frac{z^i\bar{z}^j}{<X|X>} \right)  V^{(m)}_{ij}
    + g^2 \left( - \half (T^{-1})^{ij} - \frac{\bar{z}^i z^j}{<X|X>} \right) V^{(D)}_{ij} \mathrm{\;.}
\ee
The connection on the complex structure moduli space is given by
\be
\Gamma^i_{\;\;jk}=g^{\bar l i}\partial_j g_{k\bar l} \mathrm{\;.}
\ee
We also recover the two constraint equations of charge and energy conservation
\ba
\label{eqn:chargeConstraint}
Q_i^a t^u_{\;\;v} q^{av} \partial_{\mu}q_{au}&=&0 \mathrm{\;,} \\
\label{eqn:FRWConstraint}
3\left(\frac{\dot{a}}{a}\right)^2 &=& g_{i\bar j}\dot{z}^i\dot{\bar{z}}^j + \dot{q}^{av}\dot{q}_{av} + V \mathrm{\;,}
\ea
which provide checks on the accuracy of the simulations. Note that (\ref{eqn:chargeConstraint}) is expressing the statement that there are no electric currents and so no induced gauge fields. 
Therefore in the simulations we have zero charge density, which corresponds to no net brane wrapping in the string theory picture i.e. same number of branes as anti-branes. We shall perform our simulations in units where
$g = \frac{\sqrt\pi}{2K_{(4)}} = 1$.

The difficulty in studying the equations of motion lies in determining the initial conditions. The initial 
conditions for the complex structure moduli correspond to their initial value in moduli space. At large 
complex structure, where we can ignore any brane wrapping effects, they are flat directions and so no 
particular value is singled out. In our scenario we consider the case where they are small $|z|<1$ as this is 
the situation where the conifold transition dynamics come into play. The initial value for the brane states is 
a difficult proposition since we have little knowledge of the mechanism by which such state are generated. The vev of 
the states $q^{au}$ can be thought of as a measure of the number of brane-antibrane pairs wrapping the three-cycle.
There are two main contenders for mechanisms that can generate these states. The first is a generalisation of the 
string gas scenario \cite{Alexander:2000xv,Battefeld:2005av} where the idea is that the very early universe realised a 'gas' of branes wrapping 
cycles. In terms of the transition, the initial conditions that this scenario leads to are non-trivial vevs for the 
$q^{au}$ states, but not to any prediction as to what this vev might be. The second mechanism by which such states 
can be created is through quantum particle production \cite{Kofman:2004yc,Watson:2004aq}. 
However, as we show following the simulations, we expect such quantum effects to be negligible in the scenario we consider. The ambiguities in the initial conditions 
suggest that a study of this system should attempt to classify the regions of initial conditions for which the different possible final states may arise and it is this approach that we adopt. The possible classes of final states for the system can be classified into three cases. Case I sees $\left(|z^i|>0,\Lambda^{u}=0\right)$ and corresponds to the conifold point never being reached. Case II is where $\left(|z^i|=0,\Lambda^{u}=0\right)$ and is the case where the CY is exactly at the conifold point. Case III has $\left(|z^i|=0,\Lambda^{u}>0\right)$ and 
corresponds to the conifold transition completing. Our study therefore comprises of determining if all three cases are realisable, and if so, the initial conditions they require. 

\begin{figure}
\begin{center}
\epsfig{file=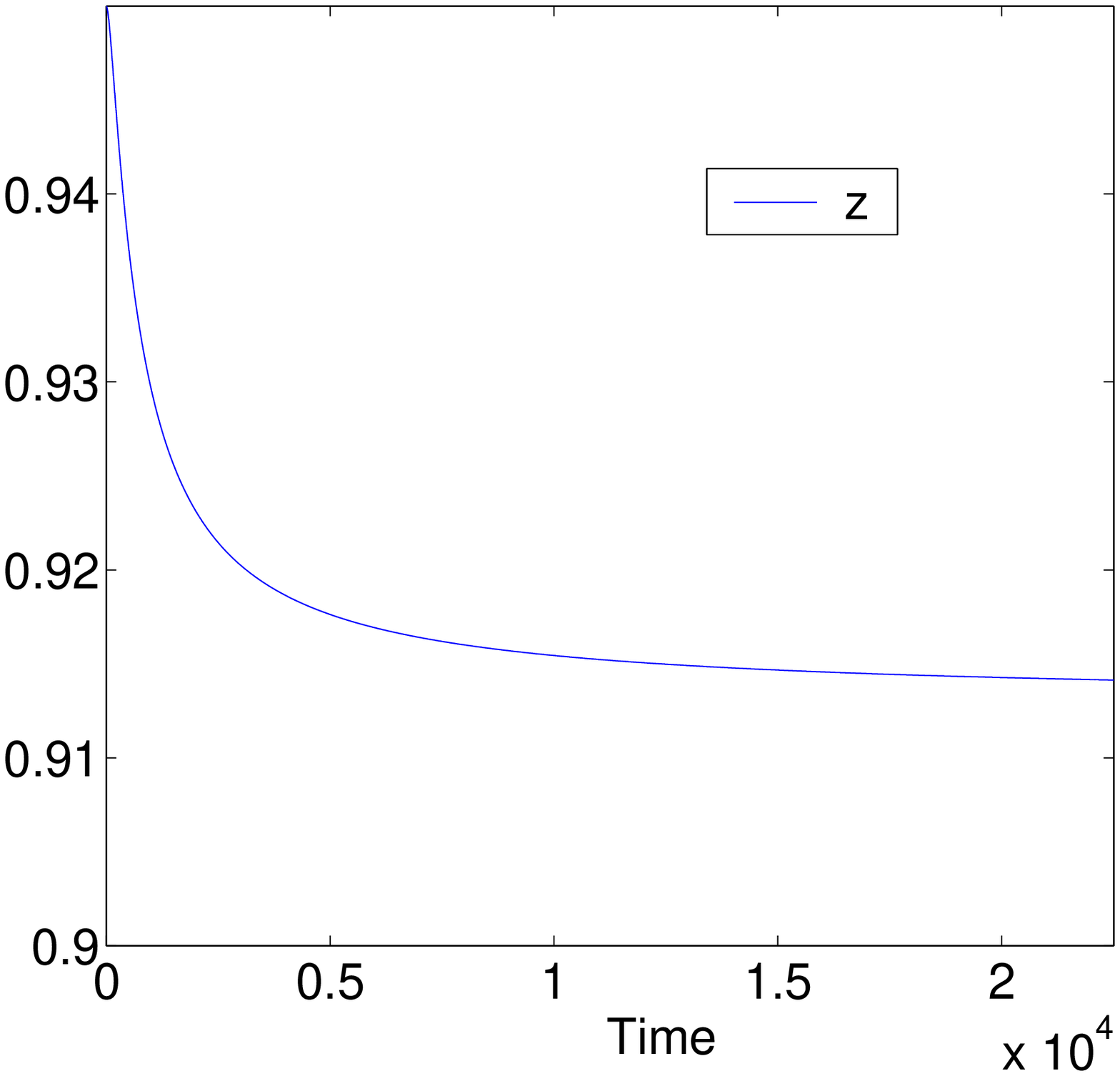,width=7.5cm}
\epsfig{file=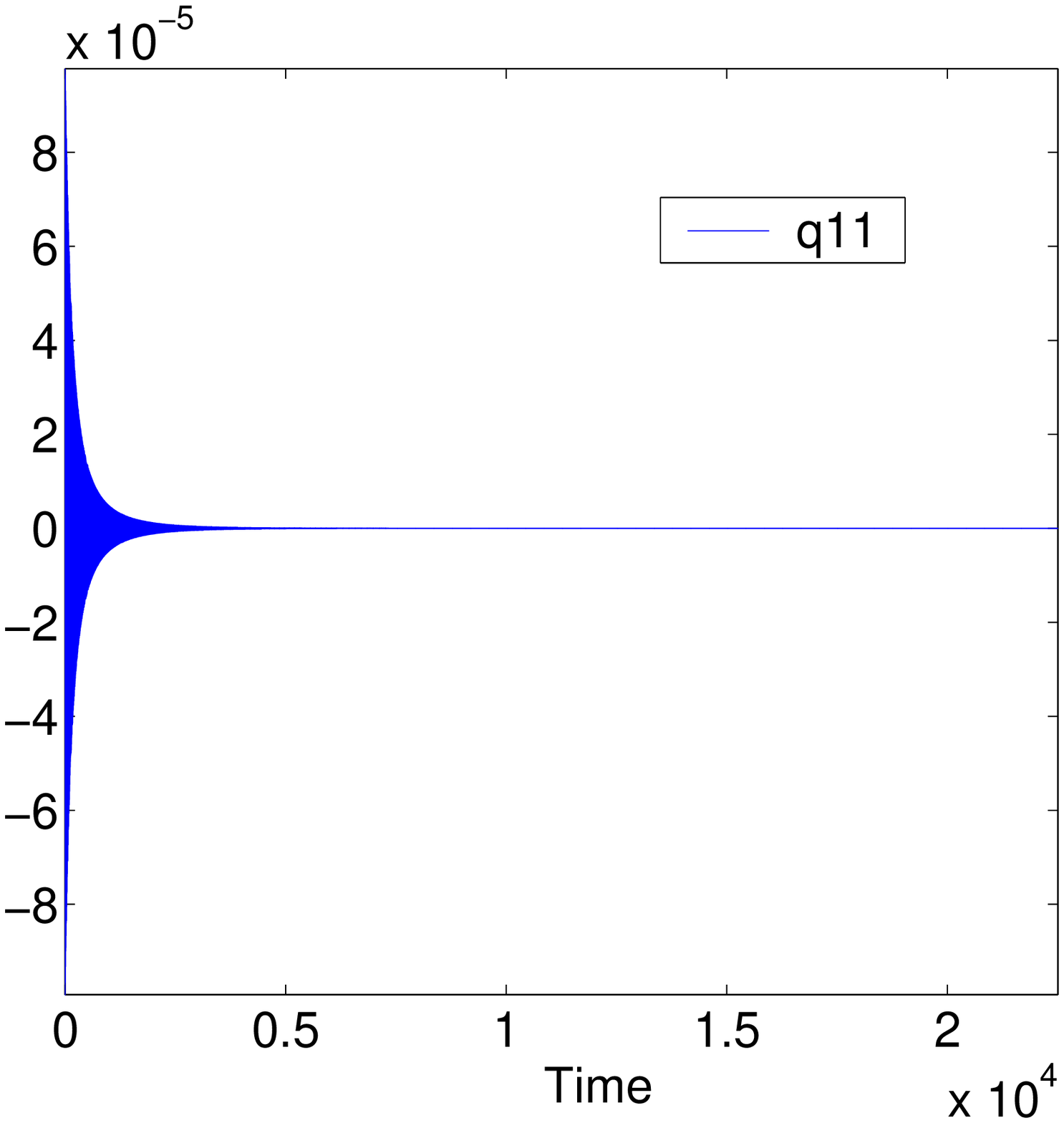,width=7.5cm} \\
 \begin{tabular}{||l|c|c|c|c|c|c|c|c||} \hline
   $z$ & $q^{11}$ & $q^{12}$ & $q^{13}$ & $q^{14}$ & $q^{21}$ & $q^{22}$ & $q^{23}$ & $q^{24}$ \\ \hline
   0.95 & 10$^{-4}$ & 1.5 $\times 10^{-4}$ & 2 $\times 10^{-4}$ & 2.5 $ \times10^{-4}$ & 2 $\times 10^{-4}$ & 3 $\times 10^{-4}$ & 4 $\times 10^{-4}$ & 5 
$\times 10^{-4}$ \\ \hline
 \end{tabular}
\flushleft
\caption{Figure showing the evolution of the fields $\mathrm{Re}(z)$, $q^{11}$  against time for case I where the conifold point is never reached. The initial values for 
all the fields are shown in the accompanying table.}
\label{fig:homI}
\end{center}
\end{figure}

To fix the theory that we explore we consider the case of two degenerating cycles with one homology relation 
between them such that their sum is homologically trivial. This gives a field content of one complex structure modulus 
$z^1 = z$ and two sets of hypermultiplets $q^{1u}$ and $q^{2u}$ with opposite charges $Q^1_1=+1$ and $Q^2_1=-1$. For 
this case we take a diagonal coupling matrix 
\be
T = \left( \begin{array}{cc} 1 &0  \\ 0 &-1  \end{array}  \right) \mathrm{\;.}
\ee
We find that all the three possible outcomes can be realised given the appropriate initial conditions. 
Figure \ref{fig:homI} shows the initial conditions and subsequent evolution of the moduli that leads to a realisation 
of case I. Only the first component of the first hypermultiplet is shown as all the other components followed 
 the same evolution. We see that initially both $z$ and $q^{11}$ are non zero though the defining character of the initial conditions is that $q^{11}\ll |z|$. Since 
the fields provide an effective mass term for each other they are both driven to zero. In this case $q^{11}$ reaches 
zero first and oscillates about it. The oscillations are damped by the Hubble expansion and so decay in amplitude 
quickly. The rate of decay of the amplitudes, crucially, is faster than the rate of decay of the $z$ towards zero which 
leads to a system that asymptotically tends towards the outcome of case I. Hubble friction therefore plays an important 
role in this scenario as without the damping the oscillations would continue to drive $z$ towards zero.
Therefore we have shown that if there are not enough brane states excited it is possible that the conifold point is never reached. 

\begin{figure}
\begin{center}
\epsfig{file=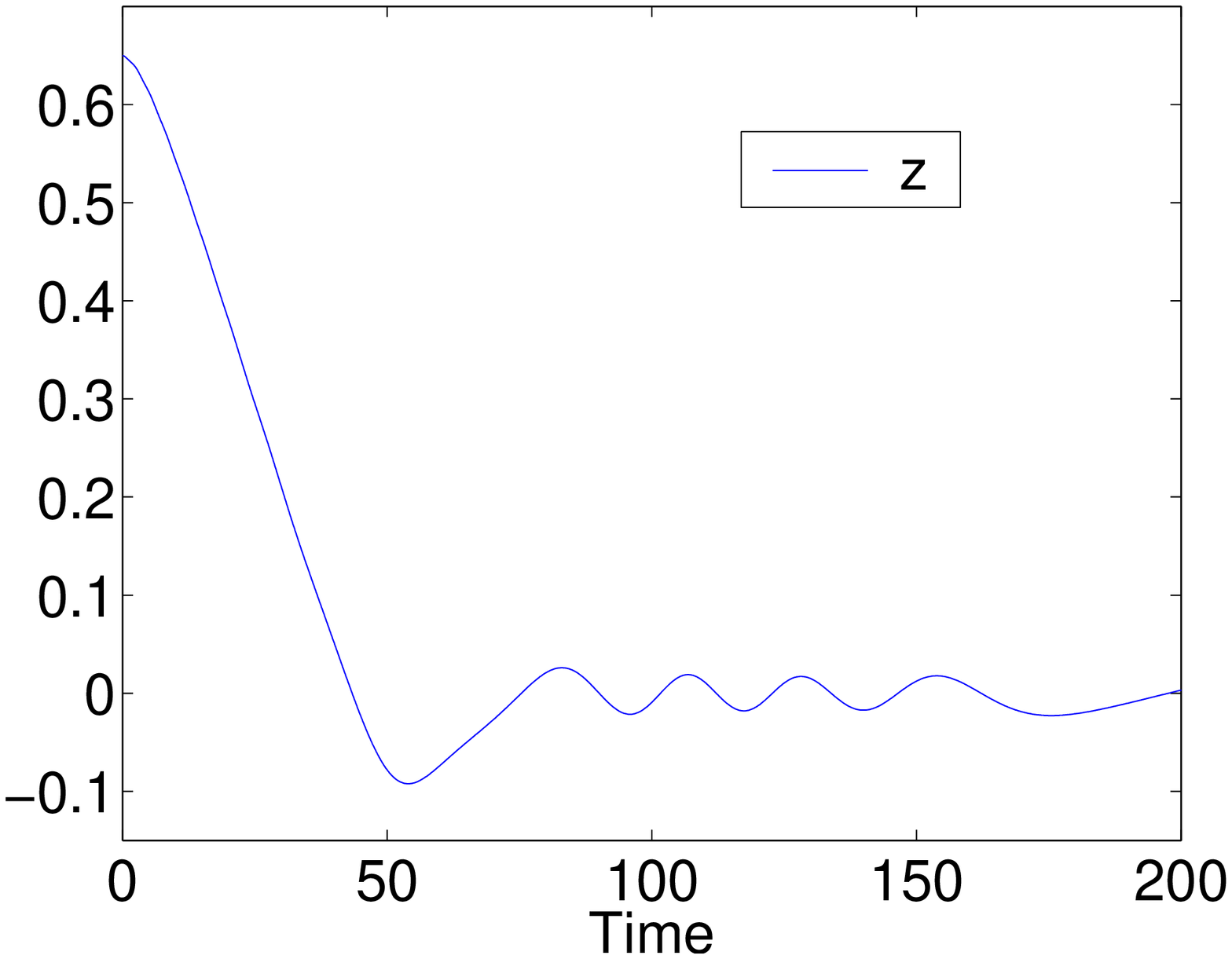,width=7.5cm}
\epsfig{file=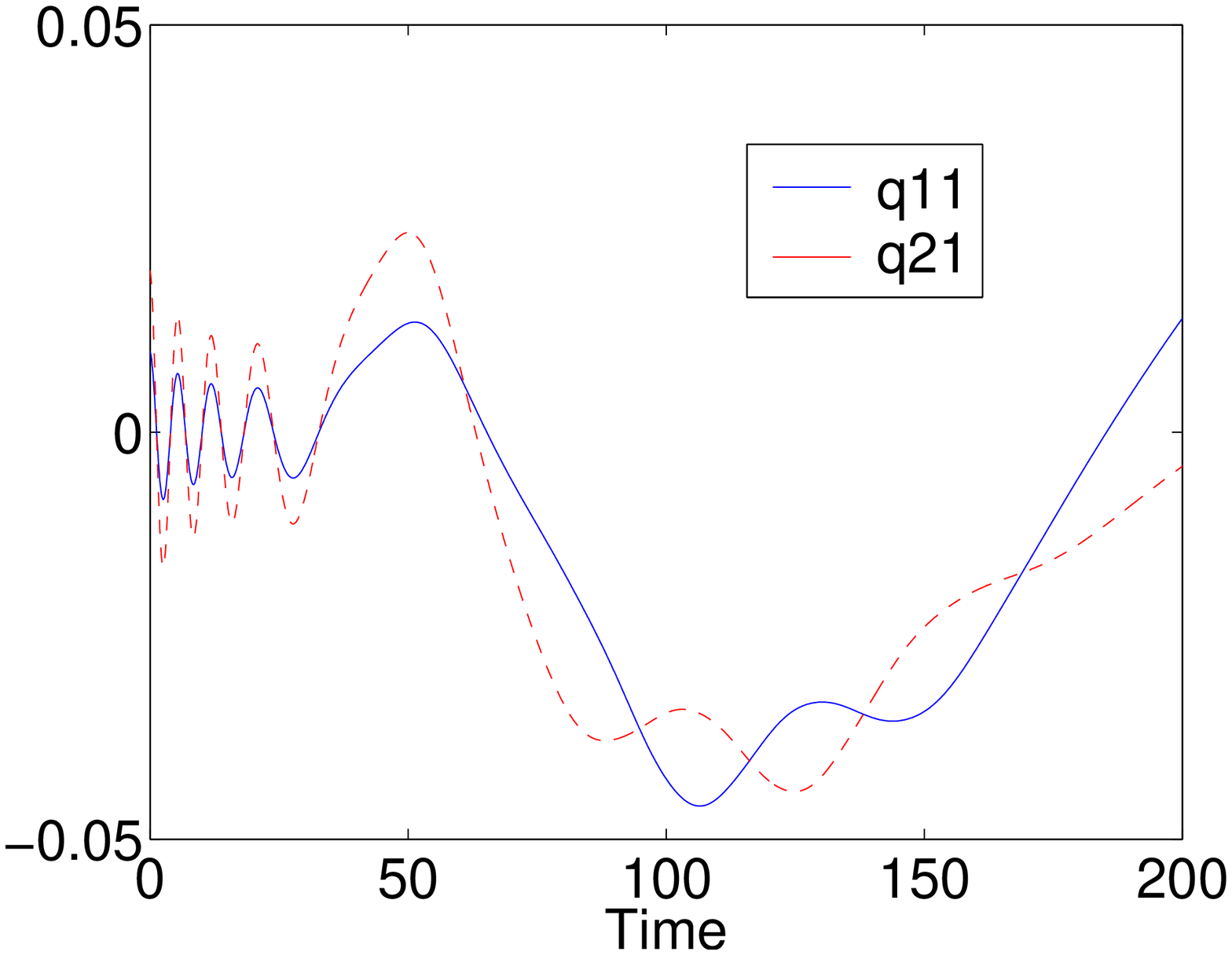,width=7.5cm} \\
 \begin{tabular}{||l|c|c|c|c|c|c|c|c||} \hline
   $z$ & $q^{11}$ & $q^{12}$ & $q^{13}$ & $q^{14}$ & $q^{21}$ & $q^{22}$ & $q^{23}$ & $q^{24}$ \\ \hline
   0.65 & 0.01 & 0.015 & 0.02 & 0.025 & 0.02 & 0.03 & 0.04 & 0.05 \\ \hline
 \end{tabular}
\flushleft
\caption{Figure showing the evolution of the fields $\mathrm{Re}(z)$, $q^{11}$ and $q^{21}$ against time for case II where the moduli are trapped at the conifold point. The initial values for all the fields are shown in the accompanying table.}
\label{fig:homII}
\end{center}
\end{figure}

Figure \ref{fig:homII} shows the realisation of case II. The initial conditions are of the type $|z| \sim q^{11}, q^{21}$. 
In this case we see the cycle being driven towards zero volume and then the system sets into oscillations about 
the conifold point. The oscillations are very lightly damped, as the energy is near vanishing, and seem chaotic in nature. This is a realisation of moduli trapping. The complex structure modulus is trapped at the conifold point with a mass given by the frequency of oscillations about the conifold point. 
This is a strong form of moduli trapping, the modulus $z$ is never a flat direction. If the system was more heavily 
damped, maybe by some outside energy densities, eventually the fields would settle at zero and the potential would 
vanish. This would be a weaker form of moduli trapping since the $z$ would strictly be a flat direction although it 
could be argued that because before the potential vanishes the point $|z|=0$ is an attractor, it is a favoured point in moduli space. In this scenario it is 
also possible for the $z$ to be trapped by quantum fluctuations in the fields $q^{au}$ giving them a vev and 
and creating a potential for $z$.

\begin{figure}
\begin{center}
\epsfig{file=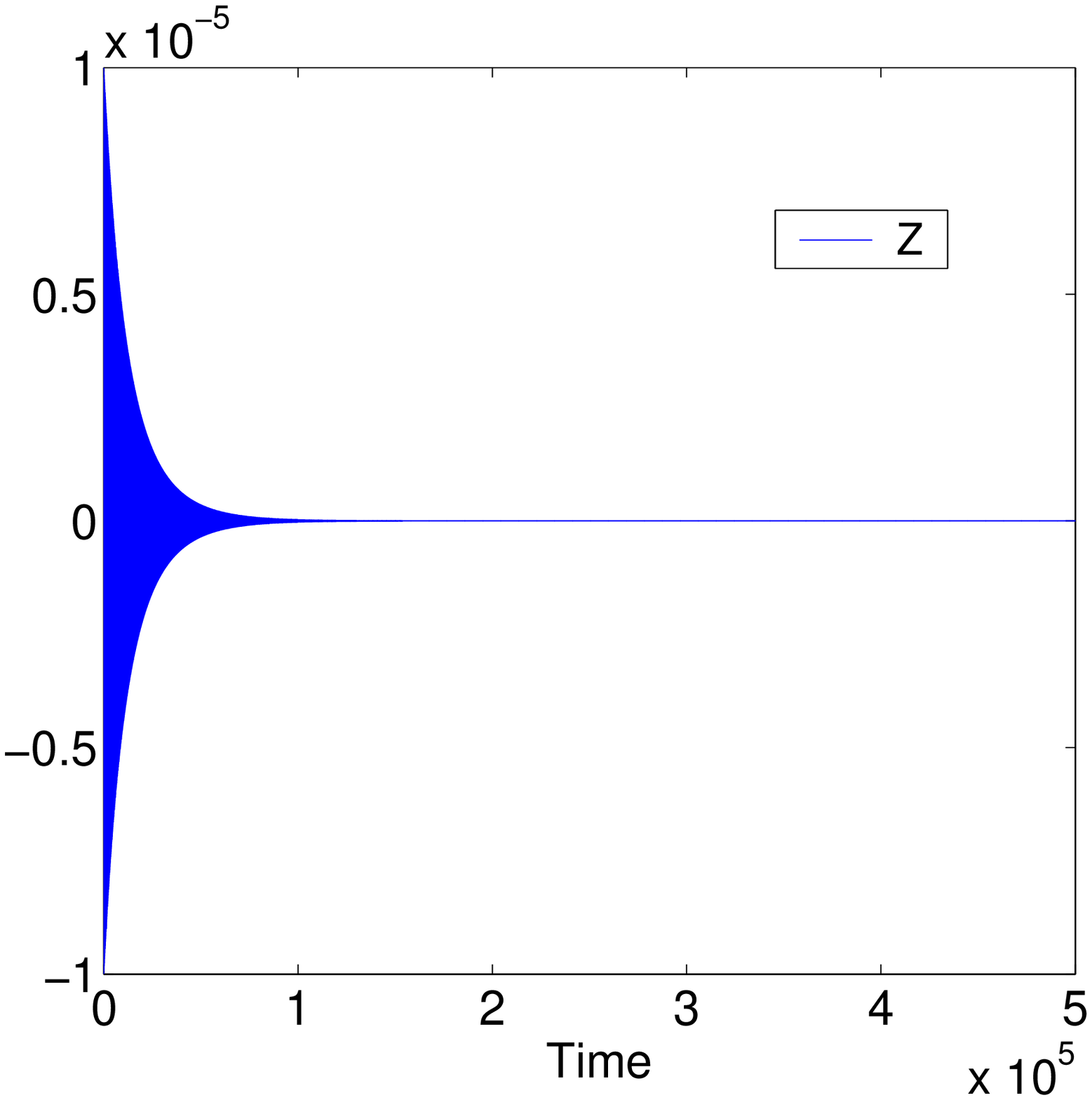,width=7.5cm}
\epsfig{file=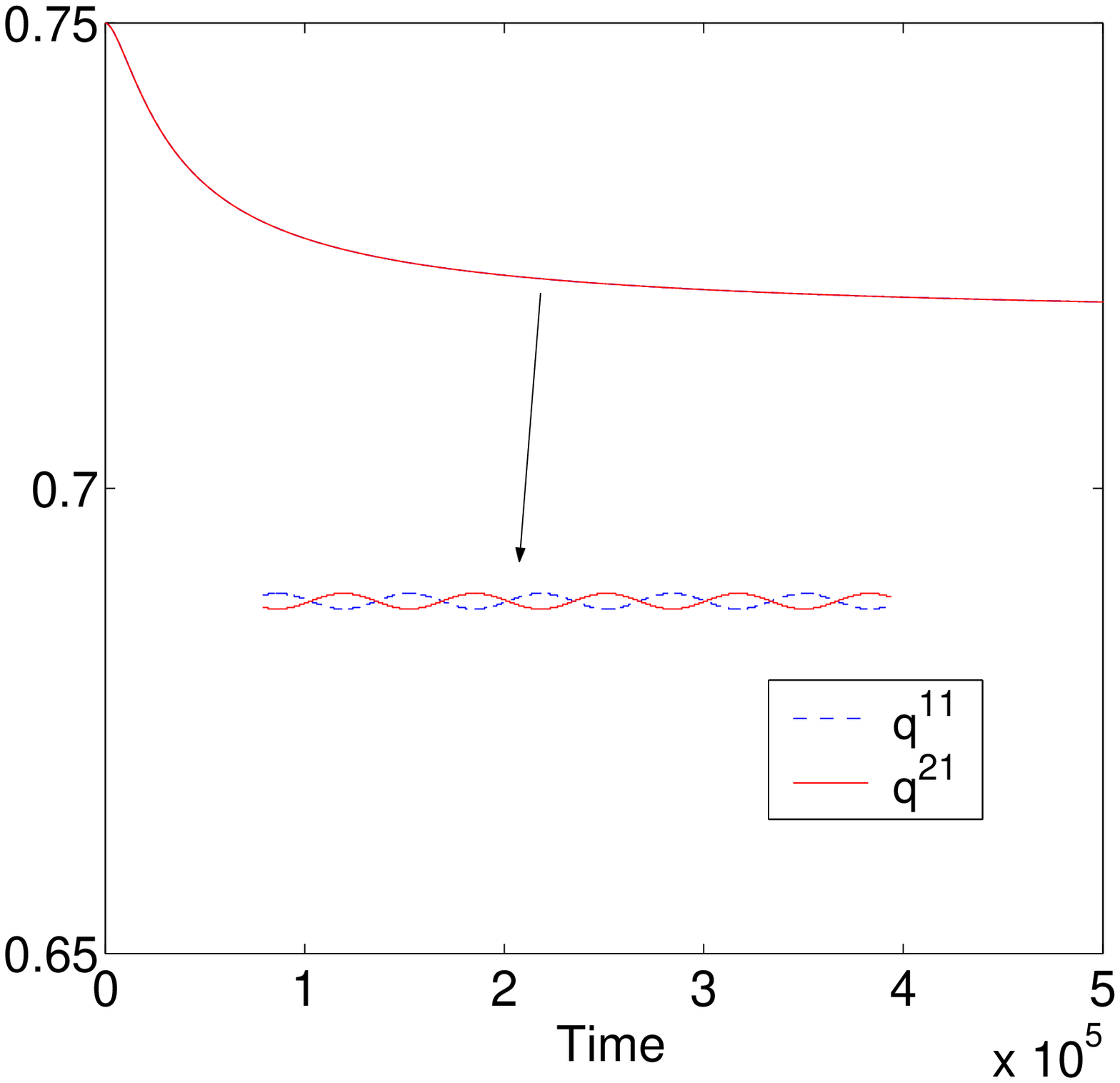,width=7.5cm} \\
 \begin{tabular}{||l|c|c|c|c|c|c|c|c||} \hline
   $z$ & $q^{11}$ & $q^{12}$ & $q^{13}$ & $q^{14}$ & $q^{21}$ & $q^{22}$ & $q^{23}$ & $q^{24}$ \\ \hline
   $10^{-5}$ & 0.75 & 0.75 & 0.75 & 0.75 & 0.75005 & 0.75005 & 0.75005 & 0.75005 \\ \hline
 \end{tabular}
\flushleft
\caption{Figure showing the evolution of the fields $\mathrm{Re}(z)$, $q^{11}$ and $q^{21}$ against time for case III where the conifold transition is completed. 
The initial values for all the fields are shown in the accompanying table.}
\label{fig:homIII}
\end{center}
\end{figure}

Figure \ref{fig:homIII} shows how a conifold transition may be completed. The initial conditions needed are $|z|\ll q^{11}$ and $q^{1u} \sim q^{2u}$.
The second condition can be parameterised in terms of a 'Higgsing' parameter
\be
\Delta \equiv \sum_{u} \left( |q^{1u}| - |q^{2u}| \right)^2 \mathrm{\;.}
\ee
The order of magnitude needed to complete a transition with $\Lambda^{u} \sim O(1)$ is $\Delta \sim O(10^{-9})$. The combination of this tight restriction and 
the small value of $z$ can be regarded as a very small region of parameter 
space. For such initial conditions we see that the Hubble friction leads to oscillations that decay sufficiently fast to allow $z$ to reach zero while $q^{11}$ has yet to 
reach zero. The plot also shows $q^{21}$ which follows $q^{11}$ thereby showing explicitly how the Higgs branch is realised. The asymptotic value that $q^{11}$ tends to is the size of the two-cycle on the other side of the transition $\Lambda^u$. It is important to highlight the role Hubble friction 
plays once more. It is only if there is enough friction that the oscillations of $z$ decay quickly enough and therefore we can draw the conclusion that in Minkowski space it is not possible to complete such a transition. Conversely with some background energy density, much larger values of $\Delta$ can lead to a completed transition.
In terms of moduli trapping this scenario strongly traps the complex structure moduli who gain a mass of order $\Lambda^u$, but leaves a remaining flat direction that is the value of $\Lambda^u$ or the size of the two-cycle.
We return to this flat direction in section \ref{sec:conicosmin} where we show that once inhomogeneities in the fields are included it too picks up non-trivial dynamics.

\begin{figure}
\begin{center}
\epsfig{file=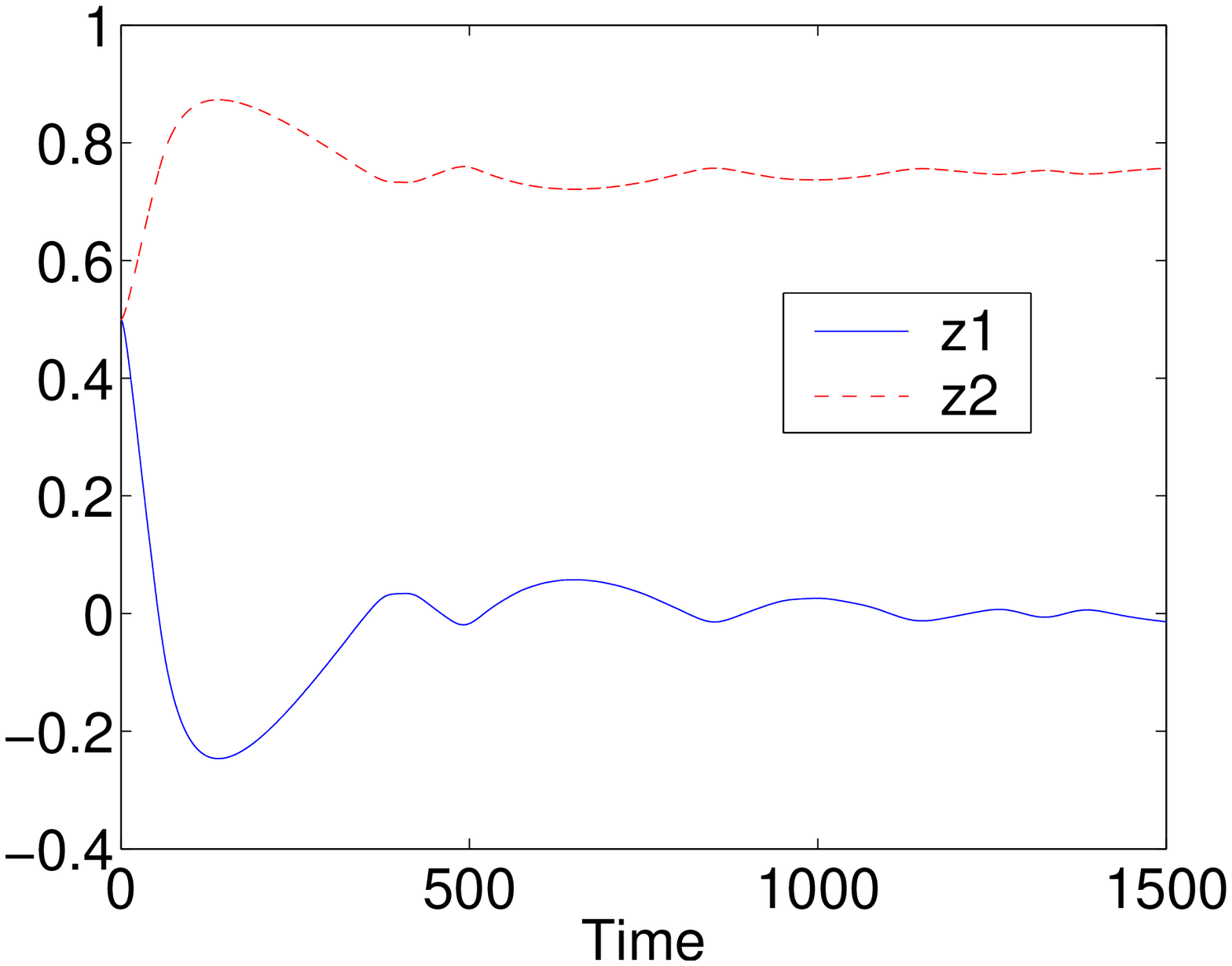,width=7.5cm}
\epsfig{file=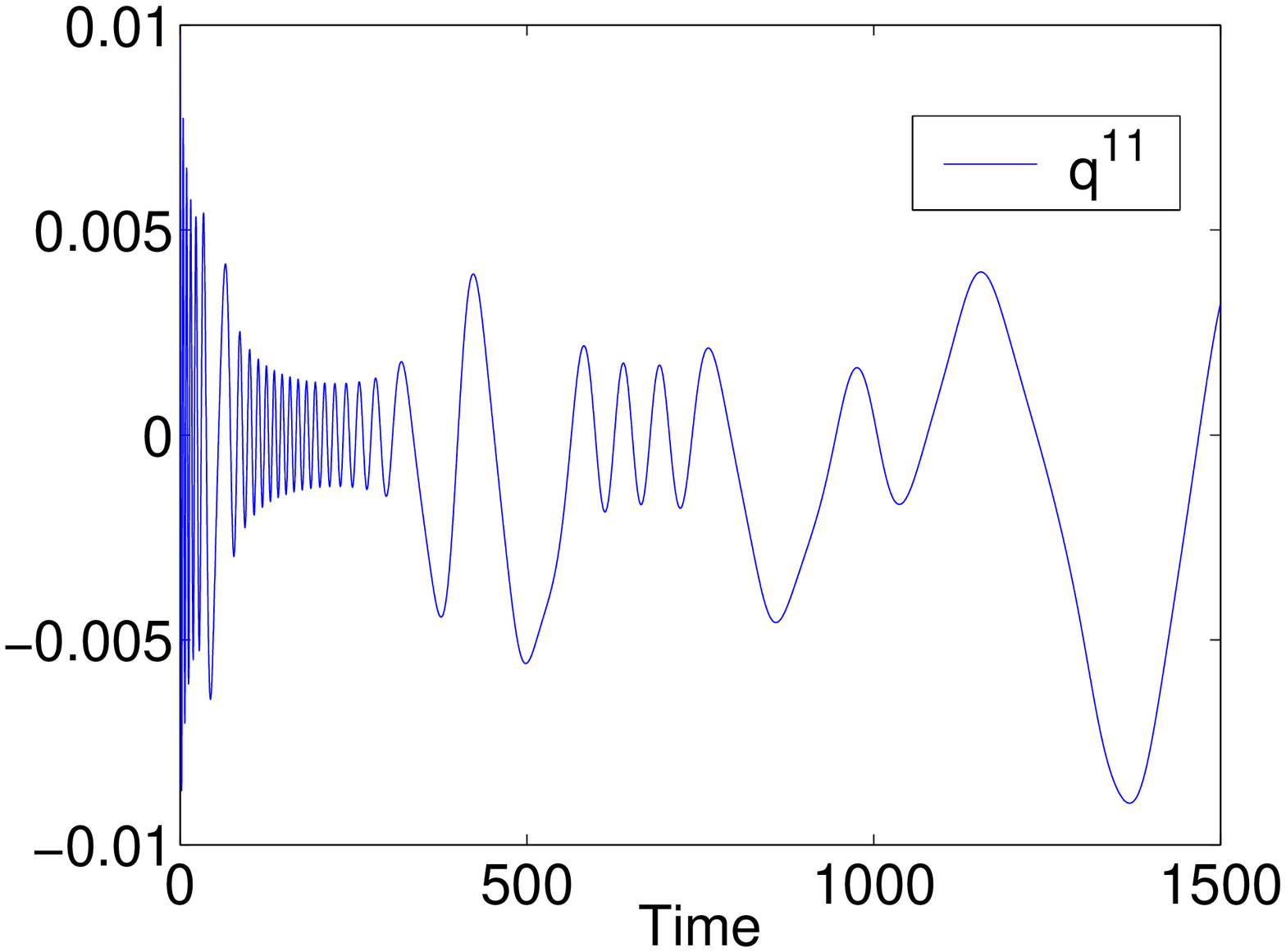,width=7.5cm} \\
\begin{tabular}{||l|c|c|c|c|c|c|c|c|c||} \hline
   $z^1$ & $z^2$ & $q^{11}$ & $q^{12}$ & $q^{13}$ & $q^{14}$ & $q^{21}$ & $q^{22}$ & $q^{23}$ & $q^{24}$ \\ \hline
   0.5 & 0.5 & 0.01 & 0.015 & 0.02 & 0.025 & 0.015 & 0.02 & 0.025 & 0.03 \\ \hline
 \end{tabular}
\flushleft
\caption{ Plots showing evolution of $q^{11}$, the wrapped modulus $\mathrm{Re}(z^1)$ and unwrapped modulus $\mathrm{Re}(z^2)$ with 
initial values given in the accompanying table. }
\label{fig:couple}
\end{center}
\end{figure}

The scenario of two degenerating cycles that has been explored can be extended to a larger number of degenerating cycles and 
we expect that the dynamics governing the wrapped cycles would be qualitatively the same. There is however the 
non-trivial extension of considering the dynamics of the cycles that are not participating in the transition. 
In the scenario we have considered there is no coupling between these cycles and the degenerating ones and so 
the two systems can be separated. It is possible to induce coupling between the cycles by considering a non-diagonal 
coupling matrix $T$ \footnote{This would correspond to non-trivial intersection numbers in the mirror picture.}. To see this consider the equation of motion for the complex structure moduli (\ref{eqn:zeqn}) with a 
general coupling matrix where we only impose $T_{i0}=0$
\ba
\label{eqn:unwrapped}
\ddot{z}^i + 3\left(\frac{\dot{a}}{a}\right)\dot{z}^i + \Gamma^i_{\;\;jk}\dot{z}^j\dot{z}^k
- \left( T^{-1} \right)^{ij}z^k\left( V^{(m)}_{jk} -  V^{(D)}_{jk}  \right) = 0 \mathrm{\;.}
\ea
As hypermultiplets are only charged under the gauge fields associated with the wrapped three-cycles,
one finds that those $z^i$, where the index $i$ corresponds to an unwrapped cycle,
are flat directions in the potential $g^{\bar j i} \del_{\bar j}V=0$. 
To see this, note from (\ref{eqn:Dterm}) and (\ref{eqn:mterm}) that $V^{(m)}_{jk}$, $V^{(D)}_{jk}$
vanish for those indices $j,k$ not associated to the charges. However it can also be seen that allowing terms 
in the (inverse) coupling matrix that couple the wrapped moduli to the unwrapped ones leads to a linear 
forcing term for those moduli or an effective quadratic potential. For example, consider an unwrapped modulus $z^2$ along with the original $z=z^1$, but now we move away from minimal coupling taking the coupling matrix to be of the form 
\be
T_{IJ}=\left( \begin{array}{ccc} 1 &0 &0 \\ 0 &-1  &-\half \\ 0 &-\half &-1  \end{array}  \right) \mathrm{\;.}
\ee
Fig. \ref{fig:couple} shows the evolution of both the complex structure moduli $z^1$, $z^2$,
along with the representative component for the hypermultiplets $q^{11}$. We see that the evolution
of the wrapped cycle corresponding to $z^1$ has created a potential for the unwrapped cycle
given by $z^2$. Unlike the case for $z^1$, the value around which $z^2$ eventually oscillates seems 
random and again highly dependent on initial condition. In this particular case it is repelled rather than 
attracted to the conifold point. It is therefore difficult to draw any general conclusions regarding the dynamics 
of such cycles but rather state that scenarios exist where they also participate in the transition.

The study of the moduli dynamics performed was purely classical. There are also quantum effects that play a role in the vicinity of the conifold.
More specifically it was pointed out in \cite{Kofman:2004yc,Watson:2004aq} that as $z$ oscillates about the conifold point it produces particles associated with the 
brane states that are light in the vicinity of the conifold point, these in turn drive $z$ further towards the conifold
point thereby trapping it. This is a slightly different scenario to the one discussed above in that 
instead of beginning at an initial configuration of stationary $z$ and some non-zero vev for $q^{au}$ the idea is to start with zero
vev for $q^{au}$ and a (fine tuned) initial velocity for $z$ in the direction of the conifold point. If both $z$ and $q^{au}$ are near the conifold
point quantum fluctuations create particle pairs of each type thereby keeping them at the conifold
point even though classically there are flat directions away from the conifold point. We should therefore consider if these effects modify our conclusions 
from the classical study. Case I is the case where $z$ never reaches the conifold point, this is, almost by definition, different to the scenario considered above
and we therefore do not expect any modifications. Case II is the case where both fields are oscillating about the conifold. At this point quantum effects of particle production do come into play but work both ways $q^{au} \leftrightarrow z$ thereby preserving the oscillations. This therefore does not change the strong moduli trapping 
scenario but when eventually the oscillations do decay, and we reach the weak moduli trapping scenario, it is quantum fluctuations about the conifold point that would act as the force trapping the moduli. Case III is the case where the transition is completed. Quantum particle production in this 
scenario is only in the direction $z \ra q^{au}$ since the particles associated with $z$ are massive and anyway $q^{au}$ is not oscillating. Therefore 
quantum effects facilitate the completion of the transition by increasing the expectation value of $q^{au}$ and providing added friction on the oscillations of $z$.
We conclude from this discussion that a classical analysis of the transition captures all the essential features of the dynamics with a possible small effect 
that helps complete the transition.

The scenarios explored in this section should be thought of as actual cosmologies. Unfortunately, cosmologically they 
are rather uninteresting, since none of the situations discussed produced cosmic acceleration. 
There are no predictions of possible observations that such a
transition could have occurred in the early universe. It is also unrealistic to consider fields that 
are homogeneous throughout space on a cosmological scale. In the next section we consider the more 
realistic case where the fields have spatial fluctuations and explore the effects that this has on the transition 
and on the possibility of generating a more interesting cosmology.

%
\section{Cosmology of conifold transitions: The inhomogeneous case}
\label{sec:conicosmin}
%

In this section we consider the generalisation of the cosmology studied in the previous section to the case where the fields have 
spatial inhomogeneities. The reason for considering such perturbations is that any realistic system is only correlated up to 
the correlation length of the system. There are a number of physical factors that may induce spatial perturbations in the fields such as
quantum fluctuations, finite temperature fluctuations or inhomogeneous effects in the brane wrapping mechanism. 
For the case where they arise from thermal fluctuations we can get an order of magnitude estimate of $\left< \delta \right>^2_T \sim T^2$. 
However, generically it is difficult to quantify the size and nature of the perturbations and so as an initial study we parameterise 
them and study their effects in terms of those parameters. Three-dimensional cosmological simulations incorporating these effects are performed 
in section \ref{sec:moduliinhom}. An important consequence of the inhomogeneities is that they induce spatial currents which in turn induce gauge fields. 
This means that we can explore the cosmology associated with the gauge field energy density and in 
particular look for stable solitonic solutions that could form a possible cosmological signature of such a transition.  

%
\subsection{Cosmic strings from conifold transitions}
\label{sec:conistrings}
%

The possibility of the formation of cosmic strings following a conifold transition was first raised in \cite{Greene:1996dh}. To see the motivation for this consider the case 
of a transition with two degenerating cycles and a homology relation between them. After the transition is completed there is a three-chain ${\cal C}_3$ connecting the 
two two-cycles $S^2_1$ and $S^2_2$ with the cycles forming its boundary. Now the total space in the CY compactification is of the form $\Reals^3 \times Y$ where $\Reals^3$ 
denotes the uncompactified space and $Y$ denotes the CY. Consider two points in $\Reals^3$, $x$ and $y$, separated by a line $I$. Then it is possible to form a three-cycle 
${\cal W}_3$ in the complete space by joining the four three-chains ${\cal U}^s$, $s=1,2,3,4$ constructed as
\ba
{\cal U}^1 &=& x \times {\cal C}_3 \mathrm{\;,} \nn \\
{\cal U}^2 &=& I \times S^2_1 \mathrm{\;,} \nn \\
{\cal U}^3 &=& y \times {\cal C}_3^* \mathrm{\;,} \nn \\
{\cal U}^4 &=& I^* \times S^2_2 \mathrm{\;,}
\label{stringgeo}
\ea
where $*$ denotes the same three-chain with the opposite orientation.
The three-cycle ${\cal W}_3$ looks like the line $I$ when projected onto the uncompactified space. Now the theory has states that correspond to D3-branes 
wrapping ${\cal W}_3$ which would look like strings in space-time. Recalling that the three-chain ${\cal C}_3$ corresponds to a magnetic three-cycle before the transition 
we see that a brane wrapping it should be identified with a monopole and so the string can be understood to be confining monopole anti-monopole pairs. 

This stringy picture has a realisation in the low energy effective field theory as cosmic string solutions that arise from the Higgsing of the gauge fields 
\cite{Greene:1996dh,Achucarro:1998er}. Consider parameterising the hypermultiplets as two complex scalar fields $h_{ai}$ with $i=1,2$
\ba
q^{a1} + iq^{a2} &=& h_{a1} = r_{a1}e^{i\theta_{a1}}   \mathrm{\;,} \nn \\
q^{a3} + iq^{a4} &=& h_{a2} = r_{a2}e^{i\theta_{a2}}   \mathrm{\;.}
\ea
The vacuum solution (\ref{vacsol}) is given by
\ba
r_{1i} &=& r_{2i} \mathrm{\;,} \nn \\
\theta_{11} - \theta_{12} &=& \theta_{21} - \theta_{22} \mathrm{\;.}
\ea
There is then a finite energy configuration of the fields given by the constraint that the total energy should vanish at infinity. More precisely working in 
cylindrical coordinates $(r,\phi,z)$ we look for a cylindrically symmetric static solution that has the property that as $r \ra \infty$ the total energy vanishes.
The solution takes the asymptotic form as $r \ra \infty$ 
\ba
h_{11} & \ra & c_1 e^{-i n \theta} \;\;,\;\; h_{12} \ra c_2 e^{-i n \theta} \mathrm{\;,} \nn \\ 
h_{21} & \ra & c_1 e^{i \Delta }e^{-i n \theta} \;\;,\;\; h_{22} \ra c_2 e^{i \Delta } e^{i n \theta} \mathrm{\;,} \nn \\
V^1_{\theta} & \ra & \frac{n}{r} \;\;,\;\; V^1_r \ra 0 \mathrm{\;.}
\ea
where $c_i$ are arbitrary complex numbers and $\Delta$ is real. This is the string solution and it has quantised flux $\int_{S^2}{F^1}=2\pi n$.

The existence of the solution should mean that inducing a gauge field density should lead to stable cosmic strings. However, as was shown in \cite{Achucarro:1998er} this 
need not always be the case. Consider the axio-symmetric solution 
\ba
h_{11} & = & g(r) e^{-i n \theta} \;\;,\;\; h_{12} = f(r) e^{-i n \theta} \mathrm{\;,} \nn \\ 
h_{21} & = & g(r) e^{i \Delta }e^{-i n \theta} \;\;,\;\; h_{22} = f(r) e^{i \Delta } e^{i n \theta} \mathrm{\;,} \nn \\
V^1_{\theta} &=& v(r) \mathrm{\;,}
\ea
where $v(r)$, $g(r)$ and $f(r)$ are arbitrary functions. This is the lowest energy state with respect to the potential and has vanishing potential energy. 
The energy per unit length of the string is then composed of the gradient energy of the fields and the magnetic flux energy.
Now for the family of expanding solutions 
\ba
h_{ai}\left(r,\theta\right) = \hat{h}_{ai}\left(\frac{r}{\lambda},\theta\right) \mathrm{\;,} \nn \\
V^1\left(r\right) = \frac{1}{\lambda}\hat{V}^1\left(\frac{r}{\lambda}\right) \mathrm{\;,}
\ea
the energy per unit length $E$ scales as
\be
E = E_G + \frac{E_F}{\lambda^2} \mathrm{\;,}
\ee
where $E_G$ denotes the gradient energy and $E_F$ denotes the magnetic flux energy. We can lower the energy by letting $\lambda \ra \infty$ which means 
that the core of the strings naturally wants to expand thereby losing its confining nature.

We have shown that because the potential energy of the strings vanishes they have an instability towards expanding their core. Such an instability 
deforms the string configuration of (\ref{stringgeo}) to more of a spherical 'blob' configuration with the length of $I$ being comparable to the length of ${\cal C}_3$.
However, the simulations of section \ref{sec:conicosm} showed that the potential energy does not vanish until the transition is completed and even then oscillations persist 
about the zero potential configurations. There therefore remains the possibility that some strings (or even blobs) do form and eventually decay. 
The study of this process forms part of this section.

%
\subsection{Cosmological simulations}
\label{sec:moduliinhom}
%

The scenario we consider in this study is the case of two degenerating cycles with one homology relation. 
The theory we work with is a gauged ${\cal N}=2$ super Yang-Mills with the action
\ba
S &=& \int_{\cal S}\sqrt{-g}d^4x \Bigg[ - \partial_{\mu} z \partial^{\mu} \bar{z}  \nonumber \\
 &-& \left( \partial^{\mu} q^{1u} + e A^{\mu} t^u_{\;v} q^{1v}
\right)\left( \partial_{\mu} q^{1u} + e A_{\mu} t^u_{\;w} q^{1w} \right) \nonumber \\
 &-& \left( \partial^{\mu} q^{2u} - e A^{\mu} t^u_{\;v} q^{2v}
\right)\left( \partial_{\mu} q^{2u} - e A_{\mu} t^u_{\;w} q^{2w} \right) \nonumber \\
 &-& \quarter F_{\mu\nu} F^{\mu\nu} \nonumber \\
 &-& |z|^2 \left( q^{1v}q^{1v} + q^{2v}q^{2v} \right) \nonumber \\
 &-& 2\left( \quarter q^{1v}q^{1w}q^{1v}q^{1w} + \quarter q^{2v}q^{2w}q^{2v}q^{2w} - q^{1v}q^{1w}q^{2v}q^{2w}  \right. \nonumber \\
 && \;\; \left. + \half q^{1v}q^{1v}q^{2w}q^{2w} - q^{1v}q^{1w}q^{2r}q^{2t}t_{wt}t_{vr} \right) \Bigg] \;, \label{action}
\ea
where
\be
F^{\mu\nu} = \partial^{\mu} A^{\mu} - \partial^{\nu} A^{\mu} \mathrm{\;,}
\ee
and we have renamed $A \equiv V^1$.
The charge $e$ is unity in our units but we leave it in so that we can study the case where the scalars decouple from the gauge fields by setting it to zero.
The action (\ref{action}) differs form the one studied in the homogeneous case in that it is not a supergravity. 
This is done because the simulations are much more involved which means we should consider the simplest theory that captures the relevant physical processes.
We now claim that
the action can be supplemented by Hubble friction in the equations of motion such that it captures all the essential features
of the physics involved. The above action
differs from the supergravity we considered in the homogeneous case in two ways. The first is that the Ricci scalar is missing. 
This term would lead to Hubble friction terms $ \sim 3H\dot{q}$ in the equations of motion, with $H$ given by the Friedman equation. 
We therefore include those Hubble friction terms in the numerical equations of motion.
The second is that the metric on the moduli space of $z$ is taken to be flat while in the
full supergravity it is not and would have to be calculated from the complex structure prepotential. However, in this thesis
we mainly consider the evolution of the light states $q^{au}$ rather than the evolution of $z$ and fix $z$ to be 
at the conifold point. In that case the metric on the space does not play a role. Furthermore the metric is well approximated as
flat near the conifold point with deviations from flatness leaving the behaviour of $q^{au}$ qualitatively unaltered \cite{Mohaupt:2004pr}, and so we may consider 
this action to be valid near the conifold point in the moduli space of $z$.

The simulations of the transition are full three-dimensional simulations using techniques from Hamiltonian lattice gauge theories \cite{Moriarty:1988fx}. 
The usual lattice link and plaquette operators are given by
\ba
& &U_i(x)=e^{-ielA_i(x)}\;,\\
& &Q_{ij}=U_j(x)U_i(x+x_j)U^\dagger_j(x+x_i)U^\dagger_i(x)\;,
\ea
respectively, where $l$ is the lattice spacing, the label $i$ takes the values
$1,2,3$ corresponding to the three spatial dimensions, and $A_i$ are the
gauge fields (the gauge choice $A_0=0$ has been made). 
By $x+x_i$, we denote the nearest lattice point in the $i$ direction from $x$. 
The plaquette operators are related to the gauge field strength  \cite{Moriarty:1988fx},
and the lattice link operator  is used to define
discrete covariant derivatives
\ba
D_i\phi^{1}(x)&=&\frac{1}{l}\left(U_i(x)\phi^{1}(x+x_i)-\phi^{1}(x)\right)
\;,\nonumber\\
D_i\phi^{2}(x)&=&\frac{1}{l}\left(U^{\dagger}_i(x)\phi^{2}(x+x_i)-\phi^{2}(x)\right)\;,
\ea
where $\phi^{1}$ corresponds to both $(q^{11}+iq^{12})$ and $(q^{13}+iq^{14})$, and $\phi^{2}$ to $(q^{21}+iq^{22})$ and $(q^{23}+iq^{24})$. 
Using the lattice link and plaquette operators, we transform  (\ref{action})
into a discretised Hamiltonian and derive the discretised equations of motion in the standard way. 
These equations were solved numerically in a cubic lattice using a staggered leapfrog
method. Several lattice spacings, time steps and cube sizes were used in order to check the code, and the results were fairly
 insensitive to these parameters. The actual plots shown in this work are for a $200^3$ cube with a ratio of time step to lattice spacing 
 $dt/l=0.2$. We also monitored Gauss's Law throughout the simulations to check the stability of the code.

There are two main ingredients in the simulations that are not very well constrained from the model: initial conditions and the damping
term. The approach to the initial conditions is the same as for the homogeneous case in that we consider the initial conditions 
that lead to the possible classes of final states. There is an added input parameter however which is the inhomogeneities in the fields.
The consequences of starting with inhomogeneities in the fields $q^{au}$ or their velocity was investigated and the most effective way of understanding the effect on the 
transition was by starting with zero velocity and inhomogeneities in the scalar fields $q^{au}$. 
Starting with zero fields but non-zero velocities resembles the case studied after a few time steps. 
Furthermore, for the study of formation of defects, we rely on previous works showing
that the evolution of related systems is fairly insensitive to the initial condition in the formation of defects \cite{Achucarro:1997cx,Urrestilla:2001dd}.
Therefore the initial condition chosen is given by a homogeneous value of the scalar fields $q^{au}$ that is perturbed by some inhomogeneities.
For the homogeneous case the Hubble damping was an important ingredient in the evolution of the system. 
We inherit that result, and include it in our simulation by adding a damping term 
proportional to the square root of the average energy density of the simulation.

Out of the possible cases of final states discussed in section \ref{sec:conicosm} we are mainly interested in the possibility of completing 
the transition. The case where $z$ never reaches the conifold point is rather trivial and the case where both $z$ and $q^{au}$ are
sitting at the conifold point does not contain any interesting dynamics.

From the homogeneous case we know that there are three important parameters in the initial conditions of the system
that determine whether the transition is completed. They are the initial vev for $z$, $\langle z_0 \rangle_x$, the initial vev for the
$q^{au}$, $\langle q^{au}_0\rangle_x$ (where the subscript $x$ denotes averaging over space) 
 and the initial value for the 'Higgsing' parameter $\Delta$ defined as
\be
\Delta \equiv \sum_{u} \left( \left| \langle q^{1u}_0\rangle_x \right| - \left| \langle q^{2u}_0\rangle_x \right| \right)^2 \label{bigdelta} \mathrm{\;.}
\ee  
In order to complete a transition we need a configuration of small or vanishing $z_0$, large $q^{au}_0$ and small $\Delta$. We do not
concern ourselves with the dynamics of $z$ as these are quite simple and remain unchanged under inhomogeneities. We therefore set
$z_0=0$ for the purpose of looking at the possibility of completing a transition whilst keeping in mind that an initial non-zero 
value for $z_0$ would make the transition less likely to complete. 

By introducing spatial inhomogeneities in the values of the fields we introduce a new important parameter $\tilde{\Delta}$ that measures
the effect the spatial inhomogeneities have on the Higgsing parameter. The initial configurations we
chose to simulate are given by a homogeneous vev for the $q^{au}$ given by $\langle q^{au}_0\rangle_x$, and superimposed on that, some
random inhomogeneities:
\be
q^{au}_0(x) = \langle q^{au}_0\rangle_x + \delta \; \hat{n}^{au}(x) \;,
\ee
where $\delta$ measures the size of the inhomogeneities and the unit vector, $\hat{n}^a$, randomly distributes 
the inhomogeneities among the hypermultiplet members. 
Defining
\be
\tilde{\Delta} \equiv \delta^2 \sum_{u} \left< \left(  n^{1u}(x)  - n^{2u}(x) \right)^2 \right>_x \;,
\ee
we can introduce a total 'Higgsing' parameter, ${\cal D}$, to indicate the effect of inhomogeneities
\be
{\cal D} \equiv \Delta + \tilde{\Delta} \;,
\ee
where $\Delta$ is calculated using the homogeneous part of $q^{au}$.
Due to the uncertainties in the origin of the perturbations, we encode the effects of inhomogeneities in the parameter $\tilde \Delta$, 
and study different ranges for its initial value.

\begin{figure}
\begin{center}
\epsfig{file=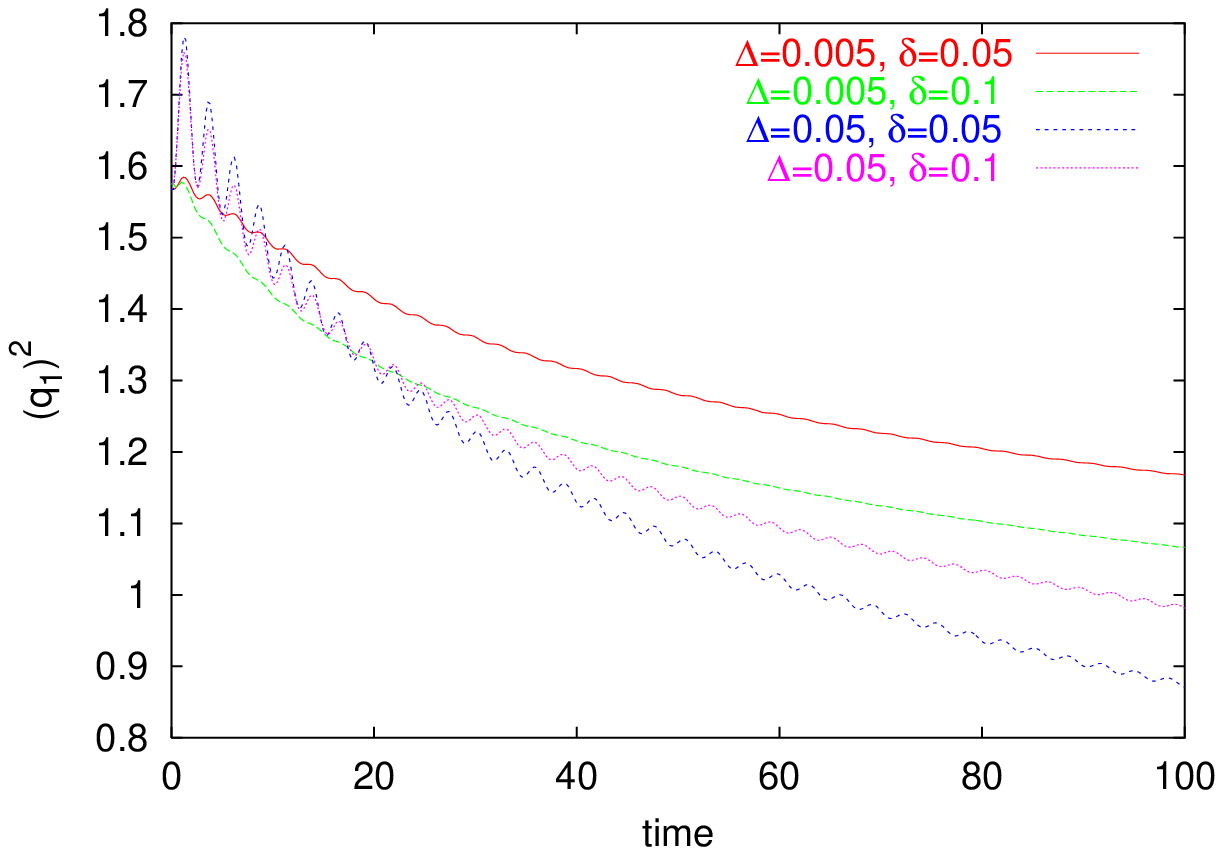,width=10cm}
\flushleft
\caption{
Figure showing the evolution in time of $\left( q^1 \right)^2$ for varying amplitudes of $\Delta$ and $\delta$.}
\label{fig:inhomdelta}
\end{center}
\end{figure}

Consider what kind of effects inhomogeneities have on the system. There is an increase in the 
Hubble friction due to the increase in the energy of the system through the contribution of gradient energies.  
Also currents induce gauge fields, and finally we see that $\tilde{\Delta}$ contributes positively to ${\cal D}$. 
The last two effects make it more difficult to complete the transition, as the gauge fields have an energy density which is 
minimised at $q^{au}=0$ and so drive the $q^{au}$ towards zero, and a larger ${\cal D}$ means it takes longer to reach the Higgs phase. 
The increase in damping however helps complete the transition as was discussed in section \ref{sec:conicosm}.
We therefore expect two different regimes to emerge where one effect dominates over the other. The regimes can be parameterised
as 
\ba
\mathrm{Case \;I\; :}&\;& \Delta \gg \tilde{\Delta} \mathrm{\;,}\\
\mathrm{Case \; II \; :}&\;& \Delta \ll \tilde{\Delta} \mathrm{\;.}
\ea
In case I larger perturbations help complete the transition, as increasing the fluctuations 
increases the gradient energy and so the Hubble damping, slowing down the fields and enabling them to settle
at a non zero value.
In case II the fluctuations are large to start with, increasing them further drives the $q^{au}$ towards the conifold point. 
We can see this behaviour in figure \ref{fig:inhomdelta}. The figures show the spatial average of the 
quantity $\left(q^1\right)^2$, defined as
\be
\left(q^1\right)^2 \equiv \frac{1}{V} \sum_u \sum_x \left( q^{1u} \right)^2 \mathrm{,}
\ee
against time for various sizes of $\delta$ and $\Delta$. The vev of $\left(q^2\right)^2$ followed the same type of evolution as $\left(q^1\right)^2$ with
both oscillating about each other. Figure \ref{fig:inhomdelta} shows how the possibility of completing the transition is manifested in the field theory. As occurred in the 
homogeneous case, the vev of 
$\left(q^{1}\right)^2$ tends towards a non-zero asymptotic value that corresponds to the size of the two-cycle on the other side. The magnitude of the asymptotic
value determines whether the transition is completed or the moduli are trapped.
The two lines with $\Delta \sim \delta$ ($\Delta=0.05$) correspond to case I and we see that increasing $\delta$ 
increases the asymptotic value for $\left(q^1\right)^2$ thereby helping complete the transition with a large two-cycle on the other side.
This behaviour can be expected to continue for the limit $\tilde{\Delta} \ra 0$ which is the homogeneous case explored in section \ref{sec:conicosm}. 
We saw in that case that the Higgsing parameter needed to complete a transition is many orders of magnitude smaller than what is needed here. 
The two lines with $\Delta \ll \delta $ ($\Delta=0.005$) correspond to case II and here we see that larger $\delta$ drives 
the asymptotic value further towards zero and so a small size for the cycle. 

\begin{figure}
\begin{center}
\epsfig{file=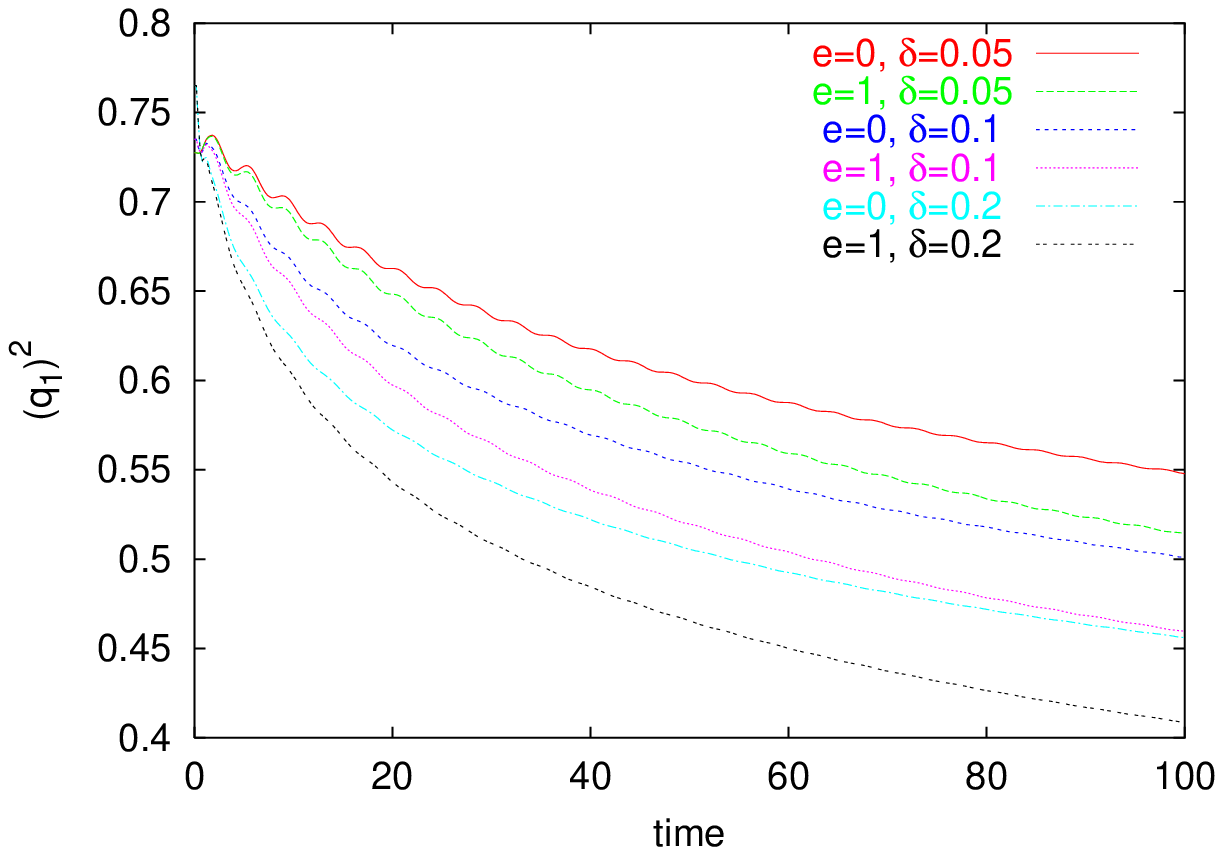,width=10cm}
\flushleft
\caption{
Figure showing the effects of coupling to gauge fields on the evolution of $\left( q^1 \right)^2$.}
\label{fig:inhomgauge}
\end{center}
\end{figure}

The effect of gauge fields is shown in Figure \ref{fig:inhomgauge}. We see plots for various perturbation sizes 
with the charge of the hypermultiplet fields, $e$, on and off.
We see that coupling the hypermultiplet to the gauge fields that are naturally induced always drives their vev towards zero thereby helping to trap them and 
hindering the completion of the transition. The plots shown are for the case where the initial conditions are of no gauge fields and so the gauge fields present are
the ones induced through the currents generated since the beginning of the simulation. There is of course the possibility of some initial gauge field density and this would amplify the effects shown in the simulations. 

Having shown that there are two regimes with quite different behaviour we might speculate on which is the more physical.
The first regime is  when $\Delta \gg \tilde{\Delta}$ (case I).
Physically this situation corresponds to the case where the spatial averages of the number of branes wrapping each of
the two cycles differ substantially and the spatial perturbations of the number of wrapping branes 
are small in comparison to this difference. It is difficult to think
of a scenario in string theory where such an initial condition could come about. The reason for this 
is that the two cycles are homologically related and so both must 
degenerate simultaneously. Given that they have an equal size the masses of states wrapping them are 
also equal and so it is unlikely that there will be much more
of one than the other. The second regime occurs when $\Delta \ll \tilde{\Delta}$ (case II). Physically this 
scenario corresponds to the case where the difference in the number of 
wrapping branes arises primarily due to spatial perturbations. This is a more likely scenario and in fact should be the case generically as it 
simply corresponds to the finite correlation length of the system. In this case the inhomogeneities 
help to trap the remaining moduli thereby stopping the transition from completing.

The final issue to address is the possible formation of structure in the magnetic field. In the simulations magnetic energy density 
was induced through the currents. Observing level surfaces of the magnetic energy density, we aimed at
looking for structures within the field. However, we were unable to find any shell of constant magnetic field. 
What we found was that the scalar field would form lumps while the gauge field 
flux appeared simply as a white noise background, not following the scalar field. 
This is likely to be a result of the damping in the system. Due to Hubble damping the magnetic energy density is rapidly decaying and also 
any movement of magnetic flux is slow. The formation of magnetic structures, by following the scalar field zero lines, relays upon the dynamical fact that 
the magnetic flux is driven quickly enough towards the scalar field structure. In all the simulations studied the magnetic field decayed too quickly 
for this to happen. This could be a result of not being able to find a configuration in which the induced
magnetic field lived long enough so as to be able to follow the scalar field. In any case, the simulations give very strong evidence that
there are simply no vestiges of the unstable vortices.

%
\section{Summary}
\label{sec:coniconc}
%

In this chapter we studied the dynamics of conifold transitions. The motivation for studying 
these transitions is the vacuum degeneracy problem. The topological degeneracy problem arises from the absence of any principle by which a particular CY 
manifold is singled out. An aspect of CYs that made this degeneracy more troubling is the fact that CYs of different topology could transform into each other 
through a conifold transition. We aimed to address this issue by studying whether it is possible for a conifold transition to dynamically complete.
The moduli degeneracy problem arises because the moduli fields in CY compactifications have no classical potential which leads to a lack of prediction power and to 
unobserved massless modes of gravity. We considered this issue by studying the dynamics of the moduli near the conifold point where a potential is 
generated. Finally we studied the possibility of a cosmological signature for a conifold transition through the formation of defects. 

Through an understanding of the geometry of conifold transitions and of the D3-brane states we were able to use the constrained structure of $\N=2$ supergravities 
to derive an effective action valid near the conifold point. We used this action to study the dynamics of the fields involved within a cosmological context.
We began by studying the simplified case of spatially homogeneous moduli fields and no gauge fields. We showed that the three possible outcomes of not reaching the 
conifold point, being held at the conifold point and completing the transition are all dynamically possible with Hubble friction playing a crucial role. 
We classified the initial conditions that lead to a realisation of each scenario. One conclusion drawn is that the completion of the transition required an initial 
condition where the brane states were strongly excited and argued that a possible mechanism for this is the early universe brane gas. 
This possibly sheds some light on the topological degeneracy problem in that as the brane anti-brane pairs in the universe annihilate eventually there will not be 
enough left to power such transitions. A further restriction on the initial conditions leading to a complete transition was that the initial values for the 
two hypermultiplets had to be very similar to a level that could be viewed as tuning.
The study also had implications for the moduli degeneracy problem. We showed that generically the complex structure moduli were trapped at the conifold point 
both strongly, where the moduli still have classical masses through oscillations, and weakly, where the classical mass vanishes but there is an 
increased likelihood of finding the moduli at the conifold point. The weak form of trapping could also be combined with quantum effects to form a stronger 
trapping mechanism. Following the classical trapping of the moduli at the conifold point quantum fluctuations in the fields induce an effective mass thereby 
keeping the moduli at the conifold point.

We have also seen how the structure of the CY, through its coupling matrix, can create a potential even for those cycles which are
not being wrapped. Depending on the form of this matrix then, it is possible for these other complex structure moduli to be involved in
the evolution.

Following the study of the homogeneous case we generalised to the case where the moduli fields have spatial inhomogeneities thereby inducing gauge fields.
By performing three dimensional simulations of the transition we showed that inhomogeneities in the fields can either help or 
hinder the completion of a transition. 
We found two regimes, parameterised by the relative magnitude of the inhomogeneities $\tilde\Delta$ and the 'Higgsing' parameter $\Delta$.
The first regime is when local fluctuations are small $\Delta \gg \tilde{\Delta}$ and
in this case the inhomogeneities help the transition complete. 
The second case is where the local fluctuations dominate, $\Delta \ll \tilde\Delta$ and for this case we found that increasing the inhomogeneity
tends to trap the moduli at the conifold point. We argued that the latter case is the more physically sensible thereby strengthening the conclusion drawn
from the homogeneous case that the vast majority of the initial conditions parameter space leads to trapped moduli and a failure to complete the transition.

A second issue in the inhomogeneous model is that of cosmic strings. Following a review of the motivation for looking for cosmic strings 
we studied the formation of structure subsequent to a transition.
Although there was structure formed by the scalar fields we did not find any trace of structure formation in the magnetic field.
This gives strong evidence that there is no structure formed by the magnetic field. 

%
\chapter{Compactification of Type IIA String Theory on Manifolds with $SU(3)$-Structure}
\label{cha:su3iia}
%

This chapter addresses two important features of string theory compactifications to four dimensions that were discussed in chapter \ref{cha:compactifications}. 
The first is the issue of the type of manifolds that preserve the minimum amount of supersymmetry in four dimensions. These were shown to be the general 
class of manifolds with $SU(3)$-structure. Section \ref{sec:cycompact} outlined a compactification of string theory on a particular subset of these manifolds that are 
the CY manifolds. This chapter extends this procedure to a general $SU(3)$-structure manifold for the case of type IIA string theory \cite{House:2005yc}. 
Similar work was done in \cite{Derendinger:2004jn,Villadoro:2005cu,Camara:2005dc,Grana:2005ny,Acharya:2006ne} and also for the case of the Heterotic string  \cite{Curio:2000dw,Becker:2002sx,Curio:2003ur,Gurrieri:2004dt,Micu:2004tz,deCarlos:2005kh,Manousselis:2005xa,Anguelova:2006qf} 
and type IIB  \cite{Gurrieri:2002iw,Behrndt:2005bv}.
These type of compactifications are motivated by more than just their generality and to understand this we discuss the second issue addressed in this chapter that is moduli stabilisation. 

The absence of a potential for the four-dimensional scalar fields in CY compactifications leads to many problems as reviewed is section \ref{sec:vacdeg}.
One way to obtain a potential for some of the moduli in CY compactifications was studied in chapter \ref{cha:con}. Another way is through the 
introduction of non-vanishing field-strengths in the vacuum for the form fields known as fluxes. The theory we consider in this chapter is type IIA string theory which 
has the form fields $\hat{A}_1$, $\hat{B}_2$ and $\hat{C}_3$. A flux for say $\hat{B}_2$ corresponds to
\be
\ev{\hat{F}_3} \equiv \ev{d\hat{B}_2} \neq 0 \mathrm{\;.}
\ee
Consider the CY case where we have a basis of three-cycles $\alpha_A$ and $\beta^A$ defined on the manifold in terms of which we can decompose the flux 
\be
\ev{\hat{F}_3} = e^A \alpha_A + m_A \beta^A \mathrm{\;,}
\ee
where the constants $e^A$ and $m_A$ are termed electric and magnetic flux parameters respectively. We note here that they are Dirac quantised in units of $\alpha '$ 
\cite{polchinski}.
For non-zero flux parameters a scalar potential is induced in four dimensions. 
To illustrate this we consider a particular term in the type IIA action (\ref{iia10daction}) and for simplicity turn on electric fluxes only 
\ba
S^{10}_{IIA} \supset \int_{\M_{10}}{\left[-\quarter e^{\hat{\phi}} \hat{F}_3 \wedge \star \hat{F}_3 \right]} 
&\supset & \int_{\M_{10}}{-\quarter e^{\hat{\phi}} e^A e^B \left( \alpha_A \wedge \star \alpha_B \right)} \nn \\
&=& \int_{\cal S}{-\quarter e^{\hat{\phi}} e^A e^B \left(S_2\right)_{AB}} \mathrm{\;,}
\ea
where the matrix $S_2$ is defined in (\ref{smatrices}). Since the matrix $S_2$ is a function of the complex structure moduli they develop a potential 
(as does the dilaton in this case). The K\"ahler moduli however do not feature in this potential and so remain as flat directions. 
The requirement for a flux compactification that successfully stabilises the moduli is that all the moduli fields feature in the potential and that this potential is one 
where a minimum exists, i.e. not of a run-away behaviour. If this is the case then we have no massless scalar modes of gravity and have gained a prediction 
for the values of the moduli in the vacuum. 

The approach taken to find potentials with minima for all the moduli is to turn on all the possible fluxes in the theory and calculate the resulting scalar potential. 
Before discussing the case of type IIA string theory it is worth reviewing the case of type IIB. In type IIB turning on all the 
fluxes induces a potential for the complex structure moduli $z^a$, the dilaton $\phi$ and the scalar $l$ \footnote{There is also a dependence on the K\"ahler moduli $v^i$ but 
it is a 'trivial' one in the sense that it appears as an inverse of the overall CY volume which is runaway to large volume.}\cite{Louis:2002ny}. 
The reason that a potential is only 
induced for the complex structure moduli can be seen through the fact that in type IIB string theory only three-form fluxes exist which give a potential 
only to the fields corresponding to three-cycles. The potential does have a minimum for the complex structure moduli, the scalar $l$, and the dilaton and so they are 
stabilised \cite{Giddings:2001yu}. 
It is also possible to stabilise the remaining fields $b^i$, $v^i$, $\rho^i$, $c^i$, $h_1$ and $h_2$ by first performing an orientifold projection which projects out 
all the fields apart from $\rho^i$ and $v^i$ \cite{Grimm:2004uq} and then inducing a potential with a minimum for those remaining fields 
through non-perturbative effects such as gaugino condensation or instanton corrections \cite{kklt,Balasubramanian:2005zx}. 
This procedure was the first case of a string compactification where all the moduli were stabilised.

We now turn to the case of type IIA string theory. We have the fields $\hat{A}_1$, $\hat{B}_2$ and $\hat{C}_3$ which means that it is possible to turn on 
both two-form, four-form and three-form fluxes and so we expect a non-trivial potential to be induced for all the fields. There is, however, a subtlety 
that makes turning on both RR and NS fluxes a difficult proposition. 
To understand this consider integrating the flux $\hat{F}_3$ over a three-cycle ${\cal A}^1$ which is the dual to the three-form $\alpha^1$
\be
e^1 = \int_{{\cal A}^1}{\hat{F}_3} = \int_{{\cal A}^1}{d\bg{B}_2} \mathbf{\;,}
\label{flux}
\ee
where $\bg{B}_2$ denotes the background value for the field $\hat{B}_2$ which gives rise to the flux. By Stokes' theorem the last term in (\ref{flux}) implies that, since 
the cycle ${\A}^1$ has no boundary, $\bg{B}_2$ can not be globally well defined. Generally integrals 
involving 'naked' $\bg{B}_2$, i.e. without a covering derivative, can not be performed in practice because of the patch dependence. The problem that arises in type IIA flux compactifications is that if both three-form and two-form 
(or four-form) fluxes are turned on precisely such naked terms appear. This problem  means that such compactifications are a difficult proposition and has meant 
type IIA flux compactifications have been less popular than their type IIB counterparts. A breakthrough was made in \cite{Kachru:2004jr} 
where a solution was proposed to the problem where the action is modified so that naked terms do not appear. 
The modified action however has no covariant uplift to M-theory. 
With a modified action it is possible to turn on both types of fluxes in IIA and indeed a potential is induced for 
all the fields \cite{Kachru:2004jr}. 

In summary fluxes induce a potential for the moduli fields in four dimensions. 
We have also argued that CY compactifications only form 
a small subset of the more general $SU(3)$-structure compactifications. The connection between flux compactifications and $SU(3)$-structure compactifications lies in the 
spontaneous compactification constraint. Recall that in order to perform a spontaneous compactification the internal manifold should be a solution of the 
ten-dimensional equations of motion. However when flux is present a CY manifold is no longer a 
solution\footnote{In type IIB string theory the solution to the equations of 
motion is a manifold that is \textit{conformally} CY.} \cite{Behrndt:2004km,Behrndt:2004mj,Lust:2004ig}. This can be viewed as the energy density 
of the flux back-reacting on the geometry of the internal manifold so that it is no longer Ricci-flat. Therefore the flux back-reaction implies that the compactification 
manifold should not be a CY but rather a more general manifold of $SU(3)$-structure. The particular type of flux present determines which torsion classes are 
non-vanishing and so which type of $SU(3)$-structure manifold forms the appropriate solution.

This chapter addresses flux compactifications of type IIA string theory on general manifolds with $SU(3)$-structure and their implications for moduli stabilisation.
We begin by reviewing the most general solution of type IIA string theory with fluxes that preserves the minimum amount of supersymmetry thereby quantifying the flux back-reaction on the geometry.
In section \ref{sec:miiasugra} we derive the effective $\N=2$ four-dimensional action following a compactification on a general $SU(3)$-structure manifold. In section 
\ref{sec:iian=1theory} we restrict the study to a sub-class of $SU(3)$-structure manifolds and show that for those cases the theory exhibits spontaneous partial supersymmetry 
breaking to $\N=1$. We go on to derive the effective $\N=1$ theory and discuss its potential for moduli stabilisation.
In section \ref{sec:su3u1u1} we consider an example of an $SU(3)$-structure manifold that shows how the more general results of the previous sections 
are explicitly realised. We summarise in section \ref{sec:iiasummary}.

%
\section{Spontaneous compactification of massive type IIA supergravity}
\label{sec:iiafluxback}
%

The ten-dimensional action that we consider is that of massive type IIA supergravity \cite{Romans:1985tz} which reads, up to two fermion terms,
\ba
S^{10}_{IIA} &=& 
\int_{\M_{10}}\left( \nn
 \half \hat{R} \star 1  
 - \quarter d\hat{\phi} \wedge \star d\hat{\phi} 
 - \quarter e^{-\hat{\phi}} \hat{F}_3 \wedge \star \hat{F}_3
 - \quarter e^{\half \hat{\phi}} \hat{F}_4 \wedge \star \hat{F}_4 \right.\\ \nn
 &-& m^2 e^{\frac{3}{2}\hat{\phi}} \hat{B}_2 \wedge \star \hat{B}_2
 - m^2 e^{\frac{5}{2}\hat{\phi}} \star 1 \\ \nn
 &-& \left. \quarter d\hat{C}_3 \wedge d\hat{C}_3 \wedge \hat{B}_2
 - \frac{1}{6} m d\hat{C}_3 \wedge \hat{B}_2 \wedge \hat{B}_2 \wedge \hat{B}_2 
 - \frac{1}{20} m^2 \hat{B}_2 \wedge \hat{B}_2 \wedge \hat{B}_2 \wedge \hat{B}_2 \wedge \hat{B}_2
   \right) \\ \nn
 &+& \left. \int_{\M_{10}} \sqrt{-\hat{g}} d^{10} X \right[
 - \hat{\overline{\Psi}}_M \Ga^{MNP} D_N \hat{\Psi}_P 
 - \half \hat{\overline{\la}} \Ga^M D_M \hat{\la}
 - \half (d \hat{\phi})_N \hat{\overline{\la}} \Ga^M \Ga^N \hat{\Psi}_M \\ \nn
 &-&\frac{1}{96} e^{\frac{1}{4}\hat{\phi}} (\hat{F}_4)_{PRST} \left( 
 \hat{\overline{\Psi}}^M \Ga_{[M}\Ga^{PRST}\Ga_{N]}\hat{\Psi}^N + 
 \frac{1}{2}\hat{\overline{\la}}\Ga^{M}\Ga^{PRST}\hat{\Psi}_M 
 + \frac{3}{8} \hat{\overline{\la}} \Ga^{PRST} \hat{\la} 
 \right) \\ \nn
 &+&\frac{1}{24} e^{-\frac{1}{2}\hat{\phi}} (\hat{F}_3)_{PRS} \left(
 \hat{\overline{\Psi}}^M \Ga_{[M}\Ga^{PRS}\Ga_{N]}\Ga_{11}\hat{\Psi}^N + 
 \hat{\overline{\la}}\Ga^{M}\Ga^{PRS}\Ga_{11}\hat{\Psi}_M \right) \\ \nn
 &+& \quarter m e^{\frac{3}{4}\hat{\phi}} \hat{B}_{PR} \left(
 \hat{\overline{\Psi}}^M \Ga_{[M}\Ga^{PR}\Ga_{N]}\Ga_{11}\hat{\Psi}^N 
 + \frac{3}{4}\hat{\overline{\la}}\Ga^{M}\Ga^{PR}\Ga_{11}\hat{\Psi}_M + 
 \frac{5}{8}\hat{\overline{\la}}\Ga^{PR}\Ga_{11}\hat{\la} \right) \\ 
 &-& \left. \half m e^{\frac{5}{4}\hat{\phi}}\hat{\overline{\Psi}}_M \Ga^{MN} \hat{\Psi}_N
 - \frac{5}{4} m e^{\frac{5}{4}\hat{\phi}}\hat{\overline{\la}} \Ga^{M} \hat{\Psi}_M
 + \frac{21}{16} m e^{\frac{5}{4}\hat{\phi}}\hat{\overline{\la}}\hat{\la} \right]
\label{iia10daction} \; ,
\ea
where
\ba
\hat{F}_4 & = & d\hat{C}_3 + m \hat{B}_2 \wedge \hat{B}_2 \mathrm{\;,} \\
\hat{F}_3 & = & d\hat{B}_2 \; . \label{fieldstrenghts}
\ea
The indices $M,N \ldots$ run from 0 to 9, and the ten-dimensional coordinates are
$X^M$. In the NS-NS sector the action contains the bosonic fields $\hat{\phi}$, $\hat{B}_2$ and $\hat{g}$, which
are the ten-dimensional dilaton, a \textit{massive} two-form and the metric,
together with the fermionic fields $\hat{\Psi}$ and $\hat{\la}$, which are
the gravitino and dilatino. The RR sector contains the three-form $\hat{C}_3$. 
We have fixed our units by setting 
\be
\kappa^2_{10} = \half \left( 2\pi\right)^7 \left( \alpha'\right)^4 = 1 \;.
\ee

This action differs from the 'massless' type IIA supergravity that is the low energy limit of type IIA string theory in two important ways. The first is the 
presence of a mass term for the NS two-form $\hat{B}_2$ and the second is the absence of the RR one-form $\hat{A}_1$. The explanation for this is that 
the original action has a gauge symmetry under $\hat{B}_2 \ra \hat{B}_2 + d\hat{A}_1$. $\hat{A}_1$ is therefore a Stucklberg field which can be gauged away breaking 
the gauge invariance and leaving the two-form $\hat{B}_2$ with a mass in a Higgs-type mechanism. 
The action (\ref{iia10daction}) can also be interpreted as the low energy limit of type IIA string theory in the background of D8-brane flux. A D8-brane couples to 
a ten-form flux which is dual to a constant $m$. Therefore in the string embedding of the supergravity the parameter $m$ is also quantised in units of $\alpha '$.

We now turn to the spontaneous compactification solution. Because fluxes are now present the vacuum configuration is not as simple as the CY case (\ref{cysponcom}). 
The most general solution that preserves $\N=1$ supersymmetry was found in \cite{Behrndt:2003ih, Behrndt:2004km, Behrndt:2004mj,Lust:2004ig} and takes the form
\ba
\hat{g}_{MN}(X) dX^M dX^N &=& g_{\mu\nu}(x) dx^{\mu}dx^{\nu} + g_{mn}(y) dy^m dy^n \mathrm{\;,} \label{vacmetricdecomp} \\
 m \hat{B}_2 & = & \frac{1}{18} f e^{- \half \hat{\phi}} J + m \breve{B}_2 \mathrm{\;,}\nn \\
 \hat{F}_3 & = & \frac45 m e^{\frac74 \hat{\phi}} \Omn^+ \nn \mathrm{\;,}\\
 \hat{F}_4 & = & f \star 1 + \frac35 m e^{\hat{\phi}} J \wg J \mathrm{\;.}
\label{lustfields} 
\ea
All the quantities correspond to their vacuum expectation values.
The internal manifold is an $SU(3)$-structure manifold with metric $g_{mn}$ and the usual forms $\Omn$ and $J$, with $\Omn^{+}$ denoting the real part of $\Omn$ and 
$\Omn^{-}$ denoting the imaginary part. The parameter $f$ is the purely external part of the vacuum expectation value of $\hat{F}_4$
\be
\left(\hat{F}_4\right)_{\mu\nu\rho\sigma} = f\epsilon_{\mu\nu\rho\sigma} \mathrm{\;.}
\label{iiafflux}
\ee
The two form $\breve{B}_2$ is the traceless part of $\hat{B}_2$ (the $\ir{8}$) and so satisfies $\breve{B}_2 \wedge J \wedge J =0$.
The internal manifold is constrained to have all the torsion classes (\ref{torsionclasses}) vanishing in the vacuum  apart from 
\ba
\tc{1} & = & - i \frac49 f e^{\quarter \hat{\phi}} \mathrm{\;,}\nn \\
\tc{2} & = & - 2 i m e^{\frac34 \hat{\phi}} \breve{B}_2 \mathrm{\;.}
\label{lusttcs} 
\ea
Within the classification of section \ref{sec:constsusy} the manifold is a subset of half-flat manifolds with $\tc{3}=0$. 
A property of the solution which we use in section \ref{sec:iiapartialsusybreaking} is that the gravitino mass takes the form\footnote{This can be deduced from the fact that in a supersymmetric vacuum the scalar potential is given by $-3\left| M_{\frac32} \right|^2$ which can be compared 
with the quantity $W$ in \cite{Lust:2004ig}.}
\be
  M_{\frac32} = - \frac{1}{10} m e^{\frac54 \hat{\phi}} + \frac{i}{12} f e^{\quarter \hat{\phi}} 
 \;. \label{lustuseful}
\ee
The solution (\ref{lustfields}) constrains the fluxes that are allowed to be present. To see this we can decompose the internal fluxes in terms of $SU(3)$-structure modules
using (\ref{formdec}).
The flux modules for $\hat{F}_3$ are a complex singlet $H^{(1)}$, a complex vector $H^{(3)}$ and a $(\irc{6})$ which 
we denote $H^{(6)}$. The internal part of the four-form flux $\hat{F}_4$ decomposes into a real singlet $G^{(1)}$, a complex vector $G^{(3)}$ and an $\ir{8}$ which we denote $G^{(8)}$.
Since $\Omn^+$ and $J \wg J$ are singlets, we see that the solution only has $H^{(0)}$ and $G^{(0)}$ present and so the fluxes are not the most general type possible.

The solution (\ref{lustfields}) can form the basis for a spontaneous compactification and indeed we use the information on the torsion classes (\ref{lusttcs}) in section \ref{sec:iian=1theory} where we look for $\N=1$ supersymmetric vacua. However, the conclusions that we can draw from the solution are more general. 
Consider a vacuum with fluxes which preserves no supersymmetry. 
This vacuum is not included in the solution and so the compactification manifold need not take the form (\ref{lusttcs})
but can take some other form. However, we would still expect the manifold to be of a general $SU(3)$-structure with torsion since the $\N=1$ solution has taught us that 
fluxes induce torsion on the manifold. The same reasoning can be applied to all the different possible vacuum solutions and also to the action away from the vacuum. 
The conclusion is that flux compactifications of type IIA string theory should be considered on a general $SU(3)$-structure manifold.

%
\section{The four-dimensional action}
\label{sec:miiasugra}
%

In this section we dimensionally reduce the action (\ref{iia10daction}) to four dimensions on a product 
manifold $\M_{10} = {\cal S} \times \M_6$ using a metric ansatz
\be
\hat{g}_{MN}(X) dX^M dX^N = g_{\mu\nu}(x) dx^{\mu}dx^{\nu} + g_{mn}(x,y) dy^m dy^n \mathrm{\;,} \label{metricdecomp} 
\ee
where we take the internal manifold to be a general $SU(3)$-structure manifold. The metric ansatz now includes the perturbations of the metric as opposed 
to (\ref{vacmetricdecomp}) which is the vacuum expectation value. The metric ansatz (\ref{metricdecomp}) is not the most general type as we have not included 
a possible warp-factor as in (\ref{genmetdec}). 
Warping may be induced by flux in the same way that torsion is. The ansatz we consider therefore requires that the flux present 
does not induce warping. An example of fluxes that do not induce warping are $H^{(0)}$, $G^{(0)}$ and $f$, since we know they correspond to an unwarped solution. 
Some flux modules may induce warping and the main contenders are the vector flux modules since they could take the form $H^{(3)}_m \sim \partial_m A(y)$. It is therefore tempting to conclude, although is not yet proved, that in the absence of vector flux modules there is no warping induced and the ansatz 
(\ref{metricdecomp}) is valid. A final comment regarding warping is that even if warping is induced it may be possible to consistently neglect it as is the case in 
type IIB for large CY volume.

As was the case for CY compactifications the resulting effective four-dimensional action should be an $\N=2$ supergravity. However since fluxes are present we expect 
a potential to be induced and therefore the supergravity should be a gauged supergravity. In the upcoming sections we keep the analogy with $\N=2$ supergravity 
as explicit as possible. In this compactification we are primarily concerned with the scalar sector of the theory, however as shown in section \ref{sec:n=2sugra}, 
once the geometry of the scalar sector is specified it is possible to deduce also the gauge field sector and then also the fermionic sector by supersymmetry.

The derivation of the action follows two steps. We first derive the relevant terms in the action without assuming anything other than $SU(3)$-structure. 
In sections \ref{sec:iiaricciscalar} and \ref{sec:iiaaxions} we derive the kinetic terms of the action. Section \ref{sec:iiaflux} discusses the possible flux that 
can be present in the theory and in section \ref{sec:iiamass} we derive the four-dimensional 
gravitini mass matrix from which it is possible to deduce the scalar potential. 
The next step is to assume there exists a basis of forms on the manifold in which we can expand the ten-dimensional fields into four-dimensional 
components. The justifications for assuming such a basis and a summary of its properties are given in section \ref{sec:iiakkbasis}. 
Finally we write down the complete action and show its $\N=2$ structure in section \ref{sec:iian=2sugra}.

%
\subsection{The Ricci scalar}
\label{sec:iiaricciscalar}
%

In section \ref{sec:cycompact} we outlined how the kinetic terms for the geometrical moduli are derived from the Ricci scalar for CY manifolds. In summary, 
there were two types of metric variations, the $(1,1)$ variations which we termed K\"ahler moduli and the $(2,0)$ variations that are the complex structure moduli. 
The variations were shown to be harmonic and so could be decomposed in terms of the harmonic bases on the CY. Substitution of these decompositions into the 
Ricci scalar lead to the kinetic terms for the moduli which were subsequently shown to match the geometric form expected from $\N=2$ supergravity. 

The approach we take to finding the geometric moduli for a general $SU(3)$-structure manifold is different. The reason is that we have much less information 
about the geometry of the internal manifold. In general it need not be K\"ahler or even complex. It does not have to be Ricci-flat and so the metric variations 
need not be harmonic, further we do not even know what the harmonic forms on the manifold are or if any exist at all. The only information we have is the existence of 
the $SU(3)$-structure forms and the fact that the resulting four-dimensional action should be an $\N=2$ supergravity. The first step in the method we adopt is to relate 
the possible internal metric deformations to the $SU(3)$-structure on the manifold and this is the topic of this section. 

Having $SU(3)$-structure on a manifold is a stronger condition than having a metric.  
In fact the forms $J$ and $\Omega$ induce a metric on the space via the relation
\ba
g_{mn} & = & s^{-1/8} s_{mn} \mathrm{\;,} \nn \\
s_{mn} & = & - \frac{1}{64} ( \Omega_{mpq} \overline{\Omega}_{nrs} + \Omega_{npq} \overline{\Omega}_{mrs} ) J_{tu} \hat{\epsilon}^{pqrstu} \label{indmet} \; ,
\ea
where $s$ is the determinant of $s_{mn}$. This relation was derived by writing down the combination of structure forms with the correct symmetry properties and then 
using the $SU(3)$-structure identities in the appendix to evaluate the right-hand-side of (\ref{indmet}). Taking variations of (\ref{indmet}) we reach
\ba
 \delta g_{mn} & = & 
  - \frac18 {(\delta \Omn)_{(m}}^{pq} \Omb_{n)pq} - \frac18 {(\delta \Omb)_{(m}}^{pq} \Omn_{n)pq}  - (\delta J)_{t(m} {J^t}_{n)} \nn \\ & & 
  + \left[ \frac{1}{64} (\delta \Omn) \lrcorner \Omb   + \frac{1}{64} (\delta \Omb) \lrcorner \Omn   - \frac18 (\delta J) \lrcorner J \right] g_{mn}
 \label{indvar} \;.
\ea
Equation (\ref{indvar}) expresses variations of the metric in terms of variations of the structure forms. The possible variations are not arbitrary but rather 
we require that the varied structure forms themselves define an $SU(3)$-structure. This is part of the assumptions that form the idea of compactifying on manifolds 
with $SU(3)$-structure and is analogous to the condition (\ref{maincy}) where we required variations of the metric to keep the manifold CY.
The variations of the metric can be classified into variations induced through variations of $J$ and variations induced through $\Omn$.
In analogy with the CY case we call the variations induced through $J$ K\"ahler moduli and the variations induced through $\Omn$ complex structure moduli. It is important 
to remember that $J$ is not necessarily the K\"ahler form nor is there a complex structure defined on the manifold. Note that we have not introduced complex co-ordinates.

The separation of the metric variations into K\"ahler and complex structure moduli is the first step towards classifying the relevant low energy fields. This separation 
should done be done carefully however and in particular we should make sure that the two types of variations are distinct. The $SU(3)$-structure relations (\ref{6dsu3alg}) 
show that in fact this is not the case since a rescaling of $\Omn$ is equivalent to an appropriate rescaling of $J$. In order not to count the same degrees of freedom twice 
we introduce a new three-form $\Omn^{cs}$ which we define as
\begin{equation}
  \label{omegacs}
    e^{\half K^{cs}}\Omega^{cs} \equiv \frac{1}{\sqrt{8\Vol}}\Omega \; , \\ 
\end{equation}
where we have also introduced the K\"ahler potential $K^{cs}$ defined as
\begin{equation}
  \label{defocs}  
  K^{cs} \equiv - \mathrm{ln}\; \left( || \Omega^{cs} ||^2 \Vol
  \right) = -\mathrm{ln}\;i<\Omega^{cs}|\bar\Omega^{cs}> = -\mathrm{ln}\;i\int_{\M_{6}}{\Omega^{cs} \wedge \bar \Omega^{cs}} \;,
\end{equation}
and the volume of the internal manifold $\Vol$ is given by integrating the unique volume form 
\be
  \label{iiavol}
  \Vol \equiv \int_{\M_6}{\sqrt{g_6} d^6y } = \frac16 \int_{\M_6}{J \wedge J \wedge J} \; .
\ee
The definition (\ref{omegacs}) means that under a rescaling of $\Omn$ the left-hand-side remains invariant and so these rescalings are now parameterised only in $J$ 
and are not present in $\Omn^{cs}$. The extra factor of $e^{\half K^{cs}}$ is introduced in anticipation of the $\N=2$ geometric structure that the complex structure moduli 
span. At this point it appears somewhat arbitrary but it turns out to be the correct K\"ahler potential on the complex structure moduli space.

As the metric is determined uniquely in terms of the structure forms, all the metric variations can be treated as variations of the structure forms. The converse however is not true as it is possible that different structure forms give rise to the same or equivalent metrics. Therefore, when
expressing the metric variations in terms of changes in the structure forms we must take care not to include the spurious variations as well. In particular the 
expression (\ref{indmet}) shows that the metric is invariant under an arbitrary phase rotation of $\Omn$. We can understand this phase freedom by noting that under a
rescaling $\Omn^{cs} \ra \Omn^{cs}e^{-f(z)}$ where $f(z)$ is a function holomorphic in the complex structure moduli we find 
\be
e^{\half K^{cs}}\Omega^{cs} \ra e^{\half K^{cs}}\Omega^{cs} e^{i \im{f}} \mathrm{\;,}
\ee
which through (\ref{omegacs}) precisely corresponds to a phase rotation of $\Omn$. 
We have come across this structure in section \ref{sec:n=2sugra} where we calculated the covariant derivative 
on $\Omn^{sk}$ by identifying a correspondence with a section of a $U(1)$ bundle, which in this case is played by $\Omn$.
The spurious variations therefore correspond to the K\"ahler transformations which are gauged by the K\"ahler 
covariant derivative as given in (\ref{u1secderiv}).

Having identified the appropriate metric variations we can go on to calculate how such variations appear in the action and replace them with variations of the structure forms. The full calculation is given in the appendix and here just outline how it proceeds. Consider the metric ansatz (\ref{metricdecomp}) with four-dimensional metric 
fluctuations
\be
\hat{g}_{MN}(X) dX^M dX^N = g_{\mu\nu}(x) dx^{\mu}dx^{\nu} + \left[ g^0_{mn}(y) + \delta g_{mn}(x,y) \right]dy^m dy^n \mathrm{\;.} 
\label{iiametfluc}
\ee 
Substituting (\ref{iiametfluc}) into the ten-dimensional Ricci scalar and only keeping terms to order $\delta^2$ we find
\ba
R^{(4)}_{EH} &=&\int_{\M_{10}}{d^{10}x\sqrt{-\hat{g}} \left[ \half \hat{R}  - \quarter \partial_M \hat{\phi} \partial^M \hat{\phi} \right]} = \\
& &\int_{\cal S}{ d^4x\sqrt{-g_4} \left[  \half R - \partial_{\mu} \phi \partial^{\mu} \phi + 
\frac{1}{8\Vol} \int_{\M_6}{ d^6y \sqrt{g_6}\; \left( 4 e^{2\phi}R_6 - g^{mp} g^{nq} \partial_{\mu} g_{mn} \partial^{\mu} g_{pq} \right)} \right] }\label{s4ehd} \nn \; .
\ea
where we define the four-dimensional dilaton as in the CY case (\ref{4ddildef}). 
We have also performed the Weyl rescalings 
\ba
g_{\mu\nu} &\ra & \Vol^{-1} g_{\mu\nu} \nn \;, \\
g_{mn} &\ra & e^{-\half \hat{\phi} }g_{mn} \;.
\label{iiaweyl}
\ea
We now use the expressions for the metric variations (\ref{indvar}) to arrive at the terms
\ba
R^{(4)}_{EH} = \int_{\cal S}{\sqrt{-g}d^4x \; \bigg[ \half R}  &-& \partial_{\mu} \phi \partial^{\mu} \phi  \\ 
  &-& e^{K_{cs}} \int_{\M_{6}}{d^6y \sqrt{g_6}\; D_{\mu} \Omega^{cs} \lrcorner D^{\mu} \bar{\Omega}^{cs}} \nn \\
  &-&\frac{1}{4\Vol}\int_{\M_{6}}{ d^6y \sqrt{g_6}\; \partial_{\mu}J \lrcorner  \partial^{\mu}J} \bigg] \;, \nn  
\label{iia4deh}
\ea
where we dropped the $R_6$ term in (\ref{s4ehd}) to leave the kinetic terms. The action (\ref{iia4deh}) contains the kinetic terms for the dilaton, the complex structure moduli and the K\"ahler moduli. 
$D_{\mu}$ is the K\"ahler covariant derivative (\ref{kahlercov}).
The important result is that the kinetic terms for each sector have decoupled from the others. In section \ref{sec:iian=2sugra} we show that these terms have the structure expected from $\N=2$ supergravity.

%
\subsection{The form fields}
\label{sec:iiaaxions}
%

The reduction of the kinetic terms for the form fields in the action is a much simpler task than the Ricci scalar. The relevant terms in the ten-dimensional action read
\be
S_{BC} = \int_{\M_{10}}{\left[  - \quarter e^{-\hat{\phi}} d\hat{B}_2 \wedge \star d\hat{B}_2
 - \quarter e^{\half \hat{\phi}} d\hat{C}_3 \wedge \star d\hat{C}_3     \right]} \mathrm{\;.}
\ee
These terms give after the appropriate Weyl rescalings 
\be
S_{BC} = \int_{\cal S}{\sqrt{-g}d^4x \; \bigg[ -\frac{1}{4} e^{2\phi} \int_{\M_{6}}{d^6y\sqrt{g_6}\; \partial_{\mu} \hat{C}_3 \lrcorner \partial^{\mu} \hat{C}_3} -
    \frac{1}{4\Vol}  \int_{\M_{6}}{d^6y\sqrt{g_6}\; \partial_{\mu} \hat{B}_2 \lrcorner \partial^{\mu} \hat{B}_2 } \bigg]} \;.
\label{bckinetic}
\ee
There is not much to say except that already at this stage the K\"ahler moduli and the $\hat{B}_2$ fields pair up into the 
complex combination $T = \hat{B}_2 - iJ$.

%
\subsection{Flux}
\label{sec:iiaflux}
%

In this section we summarise the fluxes that could be present in the theory. The flux parameter $m$ already appears in the ten-dimensional action and we have argued that 
it can be interpreted as D8-brane flux. It takes integer values quantised in units of $\alpha'$ \cite{DeWolfe:2005uu}
\be
m = \frac{m_0}{2\sqrt{2\pi\alpha'}} \mathrm{\;,} \; m_0 \in \mathbf{Z} \mathrm{\;.}
\ee
From the form fields the internal background fluxes that preserve Lorentz invariance are 
\footnote{Note that any two-form RR flux in the massless formulation of IIA string theory can be absorbed into the vev of $b$ in the massive formulation.} 
\ba
H_3 + \tilde{H}_3 &\equiv & d\bg{B} + \ev{db} \mathrm{\;,} \\
G_4 + \tilde{G}_4 &\equiv & d\bg{C} + \ev{dc} \mathrm{\;.}
\ea
We have decomposed the fluxes into a exact and a non-exact parts. The non-exact parts, $H_3$ and $G_4$, are the usual fluxes, as discussed at the beginning of this chapter, 
that arise from the forms $\bg{B}$ and $\bg{C}$ which are not globally well defined. These fluxes are similar to $m$ in that they come from branes and are 
quantised. There are also exact parts of the fluxes $\tilde{H}_3$ and $\tilde{G}_4$ which arise from the vacuum expectation values of the scalar fields coming 
from $\hat{B}_2$ and $\hat{C}_3$ which we denote collectively as $b$ and $c$ to distinguish them from the parts that are not globally well defined. See (\ref{bdecompose}) and (\ref{cdecompose}) for the explicit decomposition. 
This second type of 
flux can not arise in CY compactifications since the basis of forms in which we decompose $\hat{B}_2$ and $\hat{C}_3$ is closed. For a general $SU(3)$-structure manifold 
this need not be the case and so we must allow for such a possibility.

We also have the purely external flux (\ref{iiafflux}) which is termed 'Freud-Rubin' flux in analogy with a similar flux that occurs in M-theory. The Freud-Rubin flux, $f$, is not the true free parameter but also depends on the internal value of the form fields. To see this and to determine the true free 
parameter of the theory we must dualise the purely external part of the three-form $\hat{C}_3$ which we write as $C(x)$. 
Reducing the relevant terms in (\ref{iia10daction}) gives the four-dimensional action for $C(x)$
\be
S^{(4)}_C = \int_{\cal S}{ \left[ -\quarter \Vol e^{\half \hat{\phi}} 
(dC + mB\wg B) \wg \star (dC + mB \wg B) + \half A dC \right] }\; ,
\label{4dc}
\ee 
where
\ba 
A \equiv &-& \int_{\M_6}{ \left[ d\bg{C}\wg\bg{B} + b\wg d\bg{C} + dc\wg\bg{B} + dc \wg b
  + \frac{1}{3}m\bg{B}\wg\bg{B}\wg\bg{B} + m\bg{B}\wg\bg{B}\wg b \right.} \nn \\
  & & \;\;\;\;\;\;\;\;\;\;\; \left. +\; m\bg{B}\wg b \wg b + \frac{1}{3} m b \wg b \wg b \right] \; . \label{Adef}
\ea
The field $B$ is the purely external part of $\hat{B}_2$.
To dualise $C$ we eliminate it using its equation of motion from (\ref{4dc}) which gives 
\be
 \star (dC + m B \wg B) 
  = \Vol^{-1} e^{-\half\hat{\phi}}\left( A + \la \right) 
  = - f
 \label{f} \; ,
\ee
where $\lambda$ is an integration constant which is now the true free parameter of the theory\footnote{Note that the 
two terms $\int_{\M_6}{d\bg{C}\wg\bg{B}}$ and $\frac{1}{3}m\int_{\M_6}{\bg{B}\wg\bg{B}\wg\bg{B}}$ can be absorbed into $\lambda$ but we choose to keep them separate so that $\bg{B}$ and $\bg{C}$ remain on the same footing as $b$ and $c$.}.

%
\subsection{The gravitini mass matrix}
\label{sec:iiamass}
%

As mentioned before, the effect of the fluxes is to gauge the $\N=2$ supergravity theory and induce a potential for the scalar
fields. These effects can be best studied in the gravitini mass matrix to which we now turn.  In an $\N=1$ supersymmetric theory, the gravitino mass
is given by the K\"ahler potential and superpotential (\ref{n1gravmass}), while in an $\N=2$ theory we have a mass matrix which is constructed out of the Killing
prepotentials of the hypermultiplet sector (\ref{n=2magmass}). The mass matrix therefore determines the prepotentials and in turn most of the scalar potential. 
The exact procedure for calculating these quantities is discussed in section \ref{sec:iian=2sugra}.

The gravitini mass matrix appears in the supersymmetry transformations of the
four-dimensional gravitini 
\be
\delta \psi_{\alpha\mu} = \nabla_{\mu}\theta_{\alpha} + i\gamma_{\mu} S_{\alpha\beta}\theta^{\beta} \;, 
\ee
where $\theta^{\alpha}$ are the supersymmetry parameters given in (\ref{4dsusypara}) and $S_{\alpha\beta}$ is the mass matrix. 
Therefore its value in the vacuum gives information about the amount of supersymmetry which is preserved. 
In particular an unbroken supersymmetry requires a vanishing \textit{physical mass} for the gravitino associated with it.
The emphasis of physical mass arises because in anti-de Sitter (AdS) backgrounds physically
massless particles can have non-zero mass parameters in the Lagrangian
\cite{Breitenlohner:1982jf, Breitenlohner:1982bm, Gunara:2003td}. 
If we consider the mass parameter of the gravitino in an $\N=1$ theory, $M_{\frac{3}{2}}$, then the physical 
mass in AdS is given by 
\begin{equation}
  \label{mphys32}
  M_{phys} = M_{\frac{3}{2}} - l \; ,
\end{equation}
where $l$ is the AdS inverse radius and is defined as
\begin{equation}
  R = -12 l^2 \; ,
\end{equation}
with $R$ the corresponding Ricci scalar. In an $\N=2$ theory it is the eigenvalues of the mass matrix that are required to be physically massless.
This is the case here and so although the masses $S_{11}$ and $S_{22}$ in
\eqref{m1m2d} are non-zero for non vanishing fluxes one of them may
still be physically massless.
Non-vanishing physical mass eigenvalues of the gravitini mass matrix in the vacuum imply partial or complete spontaneous supersymmetry breaking. 
In the case of partial supersymmetry breaking of
an $\N=2$ theory, the superpotential and D-terms of the resulting $\N=1$ theory are completely determined by the $\N=2$ mass matrix.
In a compactification from a higher-dimensional theory there are several ways to determine the gravitini mass matrix in the four-dimensional theory.
If we have explicit knowledge of the four-dimensional degrees of freedom we can derive the complete bosonic action and from the potential and gaugings
derive the $\N=2$ Killing prepotentials. Alternatively we can perform a computation in the fermionic sector and directly derive the
gravitino mass matrix. The advantage of the latter method is that we can obtain a generic formula for the mass matrix in terms of integrals over the internal
manifold without explicit knowledge of the four-dimensional fields. This is essential to keep things as general as possible.
Once these fields are identified in some expansion of the higher-dimensional fields one
can obtain an explicit formula for the mass matrix which should also be identical to the one obtained from a purely bosonic computation. 

In the following we determine the gravitino mass matrix by directly identifying all the possible contributions to the gravitino mass from ten
dimensions. For this we first identify the four-dimensional gravitini. For conventions and notations regarding spinors in various dimensions 
the reader is referred to the appendix.
As discussed in section \ref{sec:constsusy} the internal manifold with $SU(3)$-structure supports a single globally defined,
positive-chirality Weyl spinor $\eta_+$ and its complex conjugate $\eta_-$, which has negative chirality. 
To decompose the ten-dimensional gravitino, which for $\N=2$ supersymmetry is parameterised by a single Majorana spinor, we consider two four-dimensional 
Majorana gravitini $\psi_{\mu}^{\al}$ with $\al=1,2$ which give the $\N=2$ supersymmetry in four dimensions. 
Since in both ten and four dimensions Majorana spinors are real we can construct the ten-dimensional gravitino by 
taking the two independent real combinations of $\eta_+$ and $\eta_-$ giving the decomposition
\be
\hat{\Psi}_M = a \psi_{M}^{1} \otimes ( \eta_+ + \eta_- ) + i b \psi_{M}^{2} \otimes ( \eta_+ - \eta_- ) 
\label{gengravitino} \;.
\ee
where $a$ and $b$ are real parameters which have yet to be determined. 
The four-dimensional gravitini are the components where $M=\mu$. There are also four-dimensional spin-$\half$ fields $\psi^{1,2}_{m}$.
In the reduction, in order not have cross terms between the gravitini and the spin-$\half$ fields the gravitini need to be redefined with some combination of the
spin-$\half$ fields. This does not affect the mass of the gravitini however, and so is not performed here.
It is more conventional to work with four-dimensional Weyl gravitini and so we further decompose the four dimensional Majorana gravitini into Weyl gravitini
\be
\psi_{\mu}^{\alpha} = \half \left( \psi_{+\mu}^{\alpha} + \psi_{-\mu}^{\alpha} \right)\;,
\ee
where $\alpha, \beta = 1,2$ and the chiral components of four-dimensional
gravitini satisfy 
\be
\gamma_5\psi^\alpha_{\pm\mu} = \pm\psi^\alpha_{\pm\mu} \; .
\ee
We can constrain the ansatz further by requiring that it should yield canonical kinetic terms when reduced, which for the case of $\N=2$ supergravity in four 
dimensions are given in (\ref{n=2gravitini}).
The kinetic term for the ten-dimensional gravitino reads
\be 
S^{10}_{\mathrm{k.t.}} = \int_{\M_{10}}{d^{10}X \sqrt{-\hat{g}} \; \left[\; - \hat{\overline{\Psi}}_M \Ga^{MNP} D_N \hat{\Psi}_P \; \right]} \;. \label{10dkineticterm}
\ee
Substituting \eqref{gengravitino} into \eqref{10dkineticterm} and performing the Weyl rescalings we arrive at the result that the four-dimensional 
gravitini kinetic terms are of the form (\ref{n=2gravitini}) if the ansatz (\ref{gengravitino}) takes the form
\be
\hat{\Psi}_M = \frac{1}{2\sqrt2} \Vol^{-1/4} \left[ \left( \psi^1_{+M} + \psi_{-M}^1 \right)\otimes \left(\eta_+ + \eta_- \right)
 -i \left( \psi^2_{+M} + \psi_{-M}^2 \right) \otimes \left(\eta_+ - \eta_- \right) \right] \;. \label{iiagravitinoansatz}
\ee

The terms in the ten-dimensional action (\ref{iia10daction}) that contribute to the gravitino masses in four dimensions are
\ba
S^{10}_{\mathrm{mass}} = \left. \int_{\M_{10}} d^{10}X\sqrt{-\hat{g}}\; \right[
 &-& \hat{\overline{\Psi}}_\mu \Ga^{\mu n \nu} D_n \hat{\Psi}_\nu \nn \\
 &-&\frac{1}{96} e^{\frac{1}{4}\hat{\phi}} (\hat{F}_4)_{prst}  
 \hat{\overline{\Psi}}^\mu \Ga_{[\mu}\Ga^{prst}\Ga_{\nu]}\hat{\Psi}^\nu \nn \\ 
 &-&\frac{1}{96} e^{\frac{1}{4}\hat{\phi}} (\hat{F}_4)_{\rho\sigma\delta\epsilon}  
 \hat{\overline{\Psi}}^\mu \Ga_{[\mu}\Ga^{\rho\sigma\delta\epsilon}\Ga_{\nu]}\hat{\Psi}^\nu \nn \\ 
 &+&\frac{1}{24} e^{-\frac{1}{2}\hat{\phi}} (\hat{F}_3)_{prs} 
 \hat{\overline{\Psi}}^\mu \Ga_{[\mu}\Ga^{prs}\Ga_{\nu]}\Ga_{11}\hat{\Psi}^\nu \nn \\  
  &+& \quarter m e^{\frac{3}{4}\hat{\phi}} \hat{B}_{pr} 
 \hat{\overline{\Psi}}^\mu \Ga_{[\mu}\Ga^{pr}\Ga_{\nu]}\Ga_{11}\hat{\Psi}^\nu \nn \\
 &-& \left. \half m e^{\frac{5}{4}\hat{\phi}}\hat{\overline{\Psi}}_\mu \Ga^{\mu\nu} \hat{\Psi}_\nu \right]
\; . \label{givegravitinomasses}
\ea
The derivation of the four-dimensional mass matrix proceeds by substituting the ansatz (\ref{iiagravitinoansatz}) into the terms and using the 
definition of the structure forms in terms of the internal spinors (\ref{6dsu3def}) to write the result as integrals over the structure forms.
The full calculation is given in the appendix and here we quote the result.
After performing the Weyl rescalings (\ref{iiaweyl}) the mass
matrix $S$, defined in (\ref{n=2gravitini}), reads \footnote{This expression corrects a factor of $-1$ in the equivalent expression in \cite{House:2005yc}.}
\ba
 S_{11} & = & -\frac{e^{2\phi}}{16\sqrt{\Vol}} \Bigg[ i\la 
  +\int_{\M_6}{\left( idT \wg U -\frac{i}{3}m T \wg T \wg T - iG_4 \wg T \right)} \nn \\
  & & \;\;+\int_{\M_6}{\left( iH_3 \wg U -im \bg{B} \wg T \wg T -im \bg{B} \wg \bg{B}\wg T \right.} \nn \\ 
  & & \;\;- \left. \frac{i}{3}m\bg{B} \wg \bg{B} \wg \bg{B} -id\bg{C} \wg \bg{B} -idc \wg \bg{B} \right)  \Bigg]\;,
  \nn \\
 S_{22} & = & S_{11}|_{U \rightarrow \overline{U}} \;,\nn \\
 S_{12} & = & \frac{e^{2\phi}}{16\sqrt{\Vol}} 
  \int_{\M_6}{ \left[ dT \wg \left( e^{-\hat{\phi}} \Omn^+ \right)  + H_3 \wg \left( e^{-\hat{\phi}} \Omn^+ \right) \right] } \;,\nn \\
 T & \equiv & b-iJ \;, \nn \\
 U & \equiv & c - i e^{-\hat{\phi}}\Omn^-
  = c - i \sqrt{8} \Vol^{-\half} ||\Omn^{cs}||^{-1} e^{-\phi}\Omn^{cs-} 
\;. \label{m1m2d}
\ea
We note here that in a generic vacuum the off diagonal components of the
mass matrix are non-vanishing and therefore the gravitini as defined in
equation \eqref{gengravitino} are not mass eigenstates. The masses of the two
gravitini are then given by the eigenvalues of the mass matrix evaluated in
the vacuum. If these masses are equal and the two gravitini are physically
massless then the full $\N=2$ supersymmetry is preserved in the vacuum. 
However this is not the case in general and then one encounters partial (when one gravitino is
physically massless) or total spontaneous supersymmetry breaking. We shall
come back to this issue in section \ref{sec:iian=1theory}.

%
\subsection{The Kaluza-Klein basis}
\label{sec:iiakkbasis}
%

So far in this chapter we have derived the important quantities of the effective four-dimensional supergravity without assuming anything regarding 
the spectrum of forms that exist on the internal manifold. This has meant that we were unable to evaluate any of the internal integrals that feature in the four-dimensional action. 
To proceed with the compactification we must specify a basis of forms in which we expand the ten-dimensional fields to arrive at the four-dimensional spectrum. For the case of the CY compactification this basis was formed by the set of harmonic two-forms and three-forms on the manifold, which implied massless modes in four-dimensions. Forms that were not harmonic corresponded to the next level KK massive modes which were truncated leading to an effective theory below the KK scale.
The generalisation of this procedure is such that we should consider the basis of forms that are the lowest mass states of the Laplacian and truncate the higher mass modes. The difference lies in the fact that the lowest mass modes need not necessarily be massless, or in terms of the cohomology, the form basis need not be harmonic. There also exists the possibility that some of the forms are still harmonic and the rest are not harmonic but still of a lower mass than the next KK level.
The construction, or even proof of existence, of such a lowest mass basis for general $SU(3)$-structure manifolds has yet to be developed. 
However we proceed to define such a basis and then argue, using supersymmetry and examples of manifolds, that such a basis is generic to six-dimensional $SU(3)$-structure manifolds. 

The lowest mass basis\footnote{We henceforth use the notation of referring to the mass of the internal forms by which we mean the mass of the four-dimensional modes that 
come from the expansion of the ten-dimensional fields in those forms.} is composed of a finite set of two-forms $\left\{ \om_i \right\}$ where the index range of $i$ is 
as yet unknown. The definition of the Laplacian (\ref{lapdef}) shows that their four-form Hodge duals $\left\{ \tilde{\om}^i \right\}$ have the same mass and so are also part of the basis. We also include a finite symplectic set of three-forms and their Hodge duals $\left\{ \alpha_A,\beta^A \right\}$. 
There is also the unique six-form that is the volume form. However we retain no one-forms or their five-form duals. 
The forms therefore satisfy the following basis relations
\ba
  \label{defot}
  \int_{\M_6}{\om_i \wg \tilde{\om}^j} &=& \delta^{j}_{i}  \;, \\
  \int_{\M_6}{\alpha_{A} \wg \beta^{B}} &=& \delta^{B}_{A} \; , \label{defab}\\ 
  \int_{\M_6}{\alpha_{A} \wg \alpha_{B}} &=&  \int_{\M_6}{\beta^{A} \wg \beta^{B}}  = 0 \;,\\
  \om_{i} \wg \alpha_{A} &=& \om_{i} \wg \beta^{A} =0 \;.
\ea
The truncation to this basis of forms also means that their possible differential relations are limited and the most general construction takes the form 
\cite{Tomasiello:2005bp,Grana:2005ny}
\ba
 d \om_i & = & E_{iA}\beta^A - F_i^A\alpha_A \mathrm{\;,}\\
 d \al^A & = & E_{iA} \widetilde{\om}^i \mathrm{\;,}\\ 
 d \bt_A & = & F_i^A \widetilde{\om}^i \mathrm{\;,}\\
 d \widetilde{\om}^i & = & 0  \;. \label{formsdiffgen}
\ea
The matrices $E_{iA}$ and $F_i^A$ are symplectic matrices that are for now arbitrary. 

Having defined the lowest mass basis we turn to its justification. The first motivation arises from the expectation of an $\N=2$ supergravity in four dimensions.
This means that the field spectrum should come in a similar form to that of CY compactifications. In particular the fact that we do not expect any spin-$\frac32$ multiplets 
was shown \cite{Grana:2005ny} 
to imply that there should be no one-forms. The argument relies on the observation that the four-dimensional part of the ten-dimensional gravitino 
decomposes under $SU(3)$ into two types of spin-$\frac32$ fields
\be
\Psi_{\mu} \ra \ir{1}_{\frac32} + \ir{3}_{\frac32} \mathrm{\;.}
\ee
The singlet gives the four-dimensional gravitini and the triplet spinors give spin-$\frac32$ fields that sit in their own multiplet. Therefore in order to avoid these 
we require that no triplets, which are one-forms, be present in the basis. By Hodge duality this rules out the presence of five-forms as well.

The second type of motivation are examples of $SU(3)$-structure manifolds that are not CY. In particular it was shown in \cite{Gurrieri:2002wz} 
that the mirror manifolds to CY with NS flux 
are half-flat manifolds. Their basis forms could then be derived through mirror symmetry and were shown to have the expected structure. Explicit examples of such manifolds 
are twisted-tori which were studied in \cite{Camara:2005dc}. In section \ref{sec:su3u1u1} we give an example of a $SU(3)$-structure manifold where again it is possible to 
calculate the lowest mass forms explicitly.

Finally, a comment is in order regarding the index ranges of the forms. In the CY case the forms were harmonic and so their number was specified by the topological Hodge 
numbers. For the case of general $SU(3)$-structure it was shown in \cite{Tomasiello:2005bp} 
that basis forms do not carry topological information. This means that we do not expect 
the index range to be given by some topological quantity like the Hodge numbers. In the CY case the forms could be thought of as the forms that calibrate independent 
special Lagrangian submanifolds. This means that they were the forms that, when integrated over the cycle, formed the lowest bound on its volume. 
It is possible to think of them as the forms that minimise the brane action of a brane wrapping those cycles since the brane tension is proportional to the volume 
of the cycle. When flux is present the brane action is no longer minimised by a minimum volume cycle but rather by a balance between the increase in energy of the flux 
for small cycles and increase in tension for large cycles. The forms that form the lowest bound on the energy now are called generalised calibrating 
forms \cite{Gutowski:1999tu}. 
Since it is the flux back-reaction which induces the departure from CY manifolds it is tempting to conclude that the basis forms are now the forms that 
generalised calibrate independent sub-manifolds. This has yet to be shown however and so for now the form indices ranges should be taken as arbitrary.

%
\subsection{The action as a gauged $\N=2$ supergravity}
\label{sec:iian=2sugra}
%

The field content in four dimensions is given by the KK expansion of the ten-dimensional fields
\ba
J(X) &=& v^i(x)\om_i(y) \label{jexpansion} \mathrm{\;,}\\
\Omn^{cs}(X) &=& X^A(x) \al_A(y) - F_A(x) \beta^A(y) \mathrm{\;,} \label{omndecompose}\\
\hat{B}_2(X) &=& B(x) + \bg{B}(y) +  b(X)\mathrm{\;,} \label{bigbdecompose}\\
\hat{C}_3(X) &=& C(x) + \bg{C}(y) + A^i(x) \wg \om_i(y) + c(X) \label{bigcdecompose} \mathrm{\;,} 
\ea
where the scalar field components of the matter fields are given by
\ba
b(X) &\equiv & b^i(x)\om_i(y) \;, \label{bdecompose} \\
c(X) &\equiv & \xi^A(x)\al_A(y) - \widetilde{\xi}_A(x)\beta^A(y) \;. \label{cdecompose}
\ea
The periods of $\Omn^{cs}$ are homogeneous functions of the complex structure moduli $z^a(x)$.
The three-form field $C(x)$ carries no degrees of freedom and is dual to the constant $\lambda$.
The rest of the fields form $\N=2$ multiplets as shown in table \ref{iiaN=2multiplets}.
\begin{table}
\begin{center}
 \begin{tabular}{||l|c||} \hline

 $g_{\mu\nu}, A^0$   & gravitational multiplet  \\ \hline
 $\xi^0, \widetilde{\xi}_0, \phi, B$  & tensor multiplet \\ \hline
 $b^i, v^i, A^i$ & vector multiplets \\ \hline
 $\xi^a, \widetilde{\xi}_a, z^a$  & hypermultiplets \\ \hline

\end{tabular}
\caption{Table showing the ${\cal N}=2$ multiplets arising from type IIA theory on manifolds with $SU(3)$-structure.}
\label{iiaN=2multiplets}
\end{center}
\end{table}   
Note the presence of a tensor multiplet which can not be dualised to a hypermultiplet since the field $B$ is massive.
Its mass term is directly inherited from mass term of the ten-dimensional field $\hat{B}_2$.
It is sometimes useful to distinguish the four-dimensional fields that come from the ten-dimensional form-fields from the metric fields. We refer to 
$\left\{b^i,\xi^A,\tilde{\xi}_A,B\right\}$ as axions and to $\left\{ v^i,z^a\right\}$ as moduli. The axionic label is inherited from CY compactifications and is slightly misleading since if the basis form from which the four-dimensional field arises is not a cycle the field does not have axionic symmetries. 

The complex structure moduli kinetic term reads (\ref{iia4deh}) 
\ba
R^{(4)}_{EH} &\supset & \int_{\cal S}{\sqrt{-g_4}d^4x \; \left\{ -e^{K_{cs}} \int_{\M_{6}}{d^6y \sqrt{g_6}\; D_{\mu} \Omega^{cs} \lrcorner D^{\mu} \bar{\Omega}^{cs}} \right\} } \nn\\
&=&\int_{\cal S}{\sqrt{-g_4}d^4x \; \left\{ - e^{K_{cs}} \left[ \int_{\M_{6}}{d^6y \sqrt{g_6}\; D_{a} \Omega^{cs} \lrcorner D_{\bar{b}} \bar{\Omega}^{cs}} \right] \partial_{\mu}z^{a}\partial^{\mu}\bar{z}^{\bar{b}} \right\} }\nn \\
&=&\int_{\cal S}{\sqrt{-g_4}d^4x \; \left\{ -\partial_{a}\partial_{\bar{b}} \left[ -\mathrm{ln}\; i\int_{\M_{6}}{\Omn^{cs} \wedge \Omb^{cs}} \right]\partial_{\mu}z^{a}\partial^{\mu}\bar{z}^{\bar{b}} \right\} }\mathrm{\;.}
\ea
The complex structure moduli manifold is therefore a special K\"ahler manifold with the K\"ahler potential (\ref{defocs}).
The K\"ahler moduli $v^i$ pair up with the axions $b^i$ to form the complex fields $t^i=b^i-iv^i$. Their 
kinetic terms read 
\ba
R^{(4)}_{EH} &\supset &\int_{\cal S}{\sqrt{-g_4}d^4x \; \left\{ -\frac{1}{4\Vol} \int_{\M_{6}}{ d^6y \sqrt{g_6}\; \partial_{\mu}T \lrcorner  \partial^{\mu}\bar{T}} \right\} }\nn \\
&=&\int_{\cal S}{\sqrt{-g_4}d^4x \; \left\{ -\left[ \frac{1}{4\Vol}\int_{\M_{6}}{\omega_i \wedge \star \omega_j} \right] \partial_{\mu}t^i \partial^{\mu}\bar{t}^{\bar{j}} \right\} }\nn \\
&=&\int_{\cal S}{\sqrt{-g_4}d^4x \; \left\{ -\partial_{i}\partial_{\bar{j}} \left[ - \mathrm{ln} \; 8\Vol \right] \partial_{\mu}t^i \partial^{\mu}\bar{t}^{\bar{j}} \right\} } \mathrm{\;.}
\label{jkahler}
\ea
Therefore the K\"ahler moduli sector also spans a special K\"ahler manifold with K\"ahler potential as in (\ref{cykahlerpot}).

Having shown that the moduli fields follow the geometry expected from $\N=2$ supergravity we turn to the potential. 
The two expressions for the mass matrix (\ref{m1m2d}) and (\ref{n=2magmass}) can be compared by going to special co-ordinates where 
the periods $X^A(t^i)=(1,t^i)$. The prepotential of the vector multiplets special K\"ahler manifolds can be deduced from 
the form of the K\"ahler potential (\ref{jkahler}) and takes the form (\ref{cubicpre}). From this we can calculate the form of the 
period $F_0$ which reads
\be
F_0 \equiv \partial_0 {\cal F} = \frac{1}{3!} {\cal K}_{ijk} t^i t^j t^k \mathrm{\;.}
\ee 
We are now in a position to determine the quaternionic electric and magnetic prepotentials. To do this 
we restrict to the case where the ill defined field $\bg{B}$ is vanishing
\be
\bg{B}=0 \;.
\ee
This only leaves integrals that can be evaluated explicitly. The non-vanishing prepotentials for that case read 
\ba
P_0^2 &=& \frac{e^{2\phi}}{\sqrt{8}} \lambda \mathrm{\;,}\nn \\
P_i^1 &=& -\frac{e^{2\phi}}{\sqrt{8}} \int_{\M_6}{\omega_i \wedge dU^-} \mathrm{\;,}\nn \\
P_i^2 &=& -\frac{e^{2\phi}}{\sqrt{8}} \int_{\M_6}{\omega_i \wedge \left( dU^+ + G_4 \right)} \mathrm{\;,}\nn \\
P_i^3 &=& \frac{e^{2\phi}}{\sqrt{8}} \int_{\M_6}{ \omega_i \wedge d\left( e^{-\hat{\phi}} \Omn^+ \right)} \mathrm{\;,}\nn \\
Q^{20} &=& \frac{e^{2\phi}}{\sqrt{8}}2m \mathrm{\;,}
\ea
where $U^+$ and $U^-$ denote the real and imaginary parts of $U$. 
It can be explicitly seen from the kinetic terms (\ref{iia4deh}) and (\ref{bckinetic}) that there is no coupling 
between the vector multiplet scalars and the gauge fields and so the vector multiplets Killing vectors 
always vanish. We have therefore completely specified the four-dimensional $\N=2$ theory resulting from 
compactifications of type IIA string theory on manifolds with $SU(3)$-structure where the non-exact part of the 
NS flux $H_3$ is vanishing.

%
\section{The $\N=1$ theory}
\label{sec:iian=1theory}
%

Compactifying on manifolds with $SU(3)$-structure leads to an $\N=2$ supergravity in four dimensions. The vacuum 
of the theory however need not preserve the full supersymmetry. In this section we show that, depending on the type of manifold and the fluxes present, 
the full supersymmetry can be spontaneously broken to either $\N=1$ or no supersymmetry. We go on to examine a particular case of manifold where the 
supersymmetry is broken to $\N=1$. For that case we derive the effective $\N=1$ theory about the $\N=1$ preserving vacuum and argue that generically all 
the moduli fields are stabilised for such manifolds. We go on to consider an explicit example of such a manifold and recover the expected properties.

%
\subsection{Spontaneous partial supersymmetry breaking}
\label{sec:iiapartialsusybreaking}
%

The gravitini mass matrix is a useful object for studying the supersymmetry of a theory. The number of supersymmetries 
that the theory preserves is given by the number of massless gravitini. 
This follows directly from the supersymmetric variations of the gravitini which must vanish to preserve supersymmetry.
We can use this constraint to study how much supersymmetry is preserved by the class of $SU(3)$-structure manifolds we are compactifying on. Consider the case where we have the following vanishing torsion classes
\be
\re(\tc{1}) = \re(\tc{2}) = \tc{3} = \tc{4} = \tc{5} = 0 \;.
\label{hftc}
\ee
These type of manifolds are a subset of half-flat manifolds where 
also $\tc{3}=0$ and the solution in section (\ref{sec:iiafluxback}) showed that they form the most general $\N=1$ vacuum. However it is important to note that 
this only tells us about the vacuum, in general away from the vacuum more torsion classes may be non-vanishing. Our choice of manifold constrains these torsion classes to 
also vanish away from the vacuum and so the manifold is not the most general one that could preserve $\N=1$ supersymmetry in the vacuum. Nonetheless they still form 
a wide range of possible manifolds with nearly-K\"ahler manifolds being an important subset. 
The non-exact NS flux $\bg{B}$ is turned off as it is inconsistent with the induced torsion classes. This can be verified by noticing that, with the present torsion classes, the NS flux in the solution (\ref{lustfields}) is exact.

The $S_{12}$ terms in the mass matrix (\ref{m1m2d}) vanish for all manifolds that satisfy (\ref{hftc}) and the mass matrix diagonalises. The gravitini are then 
mass eigenstates and we can evaluate the mass gap $\Delta M^2$ between the two gravitini which is given by
\ba
 \Delta M^2 & = & |S_{11}|^2 - |S_{22}|^2 \nn \\
   & = & \frac{e^{4\phi}}{64\Vol} \Bigg\{  
    \int_{\M_6}{ \left( dJ \wg c + \frac{m}{3} J \wg J \wg J - G_4 \wg J - m b \wg b \wg J \right)} 
    \int_{\M_6}{ db \wg \left( e^{-\hat{\phi}}  \Omn^- \right)}
 \nn \\ & & -
     \left[ \lambda + \int_{\M_6}{ \left(  db \wg c - \frac{m}{3} b \wg b \wg b + m b \wg J \wg J - G_4 \wg b  \right)} \right] \nn \\
    & & \times \int_{\M_6}{ dJ \wg \left( e^{-\hat{\phi}}  \Omn^- \right)} \Bigg\} \;, \label{massgap}
\ea
where all the quantities are evaluated in the vacuum.
For general fluxes, the masses of the gravitini are non-degenerate. 
This implies that we no longer have ${\cal N}=2$ supersymmetry. 
Indeed such a mass gap corresponds to partial supersymmetry breaking with ${\cal N}=2 \rightarrow {\cal N}=1$ for a
physically massless lighter gravitino or full supersymmetry breaking with ${\cal N}=2 \rightarrow {\cal N}=0$ for a physically massive
lighter gravitino. 
We have already encountered a flux configuration that gives a solution preserving $\N=1$ supersymmetry in section \ref{sec:iiafluxback}. We can see how this is 
realised in terms of gravitino masses and partial supersymmetry breaking by studying the mass matrix for the solution. 

We can check that one of our gravitini is indeed
physically massless by substituting the solution (\ref{lustfields}) 
into our mass matrix \eqref{m1m2d} and checking
that one of the gravitini has a mass corresponding to the gravitino
mass found in the solution. Putting the solution (\ref{lustfields}) into the gravitino mass matrix
and performing the rescalings (\ref{iiaweyl}), we find firstly that $S_{12} = 0$. This means
that $\psi^{1,2}$ are both mass eigenstates, with eigenvalues given by\footnote{In order to compare this to the expression (\ref{lustuseful}), we have evaluated the mass in the 
conventions of \cite{Lust:2004ig}.}
\ba
 S_{11} & = & \frac{1}{10} m e^{\frac54 \hat{\phi}} - \frac{i}{12} f e^{\quarter \hat{\phi}}  \;,\nn \\
 S_{22} & = & - 3 S_{11} \; .
\ea
Comparison with (\ref{lustuseful}) gives that $| S_{11} | = | M_{\frac{3}{2}} |$.
We therefore see that for the background solution we have considered, a mass gap opens up for
the two gravitini such that the $\psi^1$ is physically massless and
$\psi^2$ is physically massive.  With a slight abuse of terminology
we shall therefore refer to the lower mass gravitino as massless and
the higher mass one as massive.

Before concluding this section we mention some subtle issues
related to the spontaneous $\N=2 \; \ra \N=1$ breaking. It has been shown
\cite{Cecotti:1984rk,Cecotti:1985sf,Ferrara:1995xi,Louis:2002vy,Gunara:2003td}
that in Minkowski space spontaneous partial supersymmetry breaking can only
occur if the symplectic basis in the vector-multiplet sector is such that no
prepotential exists. However these results do not apply to the cases we
discuss in this paper for the following reasons. First of all, the no-go result
above has been obtained for purely electric gaugings of the $\N=2$
supergravity. Here we have magnetic gaugings as well and
going to purely electric gaugings requires performing some electric-magnetic
rotation which, in special cases, can take us to a symplectic basis where no
prepotential exists. The second argument is that we encounter the
phenomenon of spontaneous partial supersymmetry breaking in AdS space and in
such a case it is not clear how to extend the no-go arguments of
\cite{Cecotti:1984rk}.

%
\subsection{The superpotential and K\"ahler potential}
\label{sec:iiasuperfields}
%

For the cases where there is a mass gap between the two gravitini it is possible to consider an effective theory below the mass scale of the higher mass gravitino. This effective theory may be an $\N=1$ supergravity, if the other gravitino is massless, or not supersymmetric if it is massive. 
The former case is easier to analyse because of the supersymmetry and we have already discussed a scenario where the partial supersymmetry breaking is realised. 
We therefore consider this case and proceed to derive the effective $\N=1$ theory about the vacuum. 

We begin by calculating the $\N=1$ superpotential and
K\"ahler potential. In the effective ${\cal N}=1$ theory the remaining gravitino mass is given by (\ref{n1gravmass}).
It is only this K\"ahler-invariant combination of $W$ and $K$ that
has any physical significance, although it is still  natural to decompose 
\eqref{n1gravmass} as
\ba
 e^{\half K} & = & \frac{e^{2\phi}}{\sqrt{8\Vol}} 
   \label{kahlerpotential} \;,\\
 W & = & \frac{-i}{2\sqrt{8}} \Bigg[ \la 
     + \int_{\M_6}\left( -\frac{1}{3} m T \wg T \wg T - G_4\wg T + dT \wg U \right)
 \Bigg] \label{superpotential} \; .
\ea
Note that, as discussed in section \ref{sec:iiapartialsusybreaking}, 
to retain consistency with the torsion classes we have set $\bg{B}=0$.
This gives a general form for the superpotential and K\"ahler
potential coming from the $\N =1$ effective action following
spontaneous breaking of the $\N =2$. The theory may also have D-terms
corresponding to the off-diagonal elements of the $\N =2$ gravitini
mass matrix, $S_{12}$ in \eqref{m1m2d}, which vanish for type of manifolds we are considering (\ref{hftc}).
Note that we have taken $\psi^1$ to be the lower mass gravitino, the case where $\psi^2$ has the lower mass corresponds to complex conjugation of $U$.

To determine the theory fully we have to identify the correct degrees of freedom to truncate. This amounts to identifying the massive
$\N=1$ superfields that arise from the $\N=2$ multiplets. Useful constraints are imposed by the requirements for a consistent truncation of $\N=2$ to $\N=1$ 
discussed in section \ref{sec:n=2sugra}. These constraints are however not enough to determine the truncation completely for the case of a general $SU(3)$-structure manifold. However we now argue that for the class of manifolds 
we are considering (\ref{hftc}), this is possible.

For the choice of manifold (\ref{hftc}) the basis form differential relations (\ref{formsdiffgen}) reduce to 
\ba
d \om_i & = & E_{i} \beta_0 \;,\nn \\
d \alpha_0 & = & E_{i} \widetilde{\om}^i \;,\nn \\
d \widetilde{\om}^i & = & 0 = d \beta^A = d \alpha_{A\neq 0} 
\; , \label{formsdiff}
\ea
for $E_i \equiv  E_{0i}$. Applying \eqref{formsdiff} to \eqref{torsionclasses} we arrive at
\ba
 dJ & = & E_{i} v^i \beta^0  
  = -\frac{3}{2} \im\left( \tc{1} \right) \re \left( \Omb \right)  
  \label{nocs1} \;,\\
 d\Omn & = & Z^0 E_i \widetilde{\om}^i
  = i \im \left( \tc{1} \right) J \wg J 
  + i \im \left( \tc{2} \right) \wg J
 \label{nocs2} \; .
\ea
Equation \eqref{nocs1} is the motivation that lies behind this choice of manifold. We see that 
$\re \left( \Omb \right)$ only has one component, which is $\beta^0$. This means that 
$\Omn^{cs}$ only has non-vanishing periods $X^0$ and $F_0$. The homogeneity of $\Omn^{cs}$ means that the two periods are not physical and so we reach the conclusion 
that for the type of manifolds we are considering $\Omn^{cs}$ carries no degrees of freedom or, in terms of the field content, there are no complex structure moduli.
There are therefore no hypermultiplets present in the theory and just a single tensor multiplet. The truncation of the $\N=2$ hypermultiplet sector to 
$\N=1$ superfields is trivial and this is precisely the difficult sector to truncate. With this simplification we are able to perform the truncation.

From the constraints in section \ref{sec:n=2ton=1} we know that in
order to have partial supersymmetry breaking we need at least two massive vectors $V^1$ and $V^0$ to form the massive spin-$\frac32$ multiplet. 
The vectors should become massive by eating two Goldstone bosons from the hypermultiplet or tensor multiplet sector. In the scenario at hand this is realised by one massive 
vector $V^1$ arising as the dual of the massive two-form $B$, and the other $V^0$ becoming massive by eating one of the scalar components of the tensor multiplet.
In the model at hand we have two possible Pecci-Quinn shift symmetries which can be gauged in this way. 
They correspond to the two scalar fields which arise from the expansion of
the three-form $c$ in the basis of three-forms $\left\{\alpha_0, \beta^0\right\}$ given in (\ref{cdecompose}). In
order to gauge one of these two directions, or a combination thereof, we
need that the corresponding combination of the forms $\alpha_0$ and $\beta^0$
is exact. Without loss of generality we will assume that $\beta^0$ is exact.
Consistency with \eqref{defab} implies then that
$\alpha_0$ is not closed. We therefore see that the scalar field which comes
from the expansion in the form $\beta^0$, which is $\tilde \xi_0$, is a
Goldstone boson and is eaten by one (or a combination) of the vector
fields which come from the expansion of $\hat{C}_3$. 
Therefore we learn that the fields which survive
the truncation in the $\N=1$ theory are the dilaton and the second scalar
field from the expansion of $c$ which is $\xi^0$. 
Together they form the scalar components of an $\N=1$ chiral superfield.
The final thing which we need to do is to identify the correct complex combination of these
two fields which defines the correct coordinate on the corresponding K\"ahler
submanifold. Knowing that the $\N=2$ gravitini mass matrix entries become the
superpotential in the $\N=1$ theory, which has to be holomorphic in the chiral
fields, we are essentially led to the unique possibility
\begin{equation}
  \label{superu}
  U^{0} \equiv \xi^{0} - ie^{-\phi}\left( \frac{-4iX^0}{F_0}
  \right)^{\half} \;.
\end{equation}
The quantity $-4iX^0/F_0$ is a positive real number as in
the particular choice of symplectic basis we have made ($\beta_0$ is exact)
$X^0$ is purely imaginary.

To check that this is indeed the correct superfield we should make sure we recover the moduli space metric from the K\"ahler potential in 
the gravitino mass. Inserting (\ref{bigcdecompose}) into (\ref{iia10daction}) we get the
kinetic term
\be
 S_{kin}^{U} =   \int_{\cal S} \sqrt{-g} d^{4}x \left[ 
  - \left( \frac{F_0}{-4iX^0} \right)e^{2\phi} \partial_\mu 
  \left(\xi^0 - i e^{-\phi} \left( \frac{-4iX^0}{F_0}\right)^\half \right) 
  \partial^\mu \left( \xi^0 + i e^{-\phi} 
  \left( \frac{-4iX^0}{F_0}\right)^\half \right)  \right]
  \; .
\ee
We see that taking the second derivatives,
\be
 -\partial_{U^0}\partial_{\bar{U}^0} \ln\left[ \frac{e^{4\phi}}{8\Vol} \right] 
 = \left( \frac{F_0}{-4iX^0} \right)e^{2\phi}  \; ,
\ee
and so \eqref{kahlerpotential} is indeed the correct K\"ahler potential
and \eqref{superu} is the correct superfield. Determining the superfields arising from the $\N=2$ vector 
multiplets is a much easier task as they are just the natural pairing found in (\ref{bigt})
\be
T^i \equiv b^i - iv^i \label{supert},
\ee
where the index $i$ now runs over the lower mass fields.

Having identified the $\N=1$ superfields we can write the superpotential for those fields
\be
W = \frac{-i}{2\sqrt{8}}\left( \la - \frac{1}{3} m {\cal K}_{ijk} T^i T^j T^k - E_{i}T^iU^0 -  e_iT^i  \right)\;, 
 \label{simplesuperpot} 
\ee
where we have decomposed the RR flux 
\be
G_4 = e_i \tilde{\om}^i \;.
\ee
Note that this flux becomes exact if 
\be
e_i = Q E_{i} \;,
\ee 
for any constant $Q$. In that case it simply corresponds to a rescaling of the superfield $U^0$ and can be dropped. 
The K\"ahler potential follows simply from (\ref{kahlerpotential}). The superpotential is 
particularly interesting since all the fields feature in it non-trivially. This raises the possibility that all the moduli fields are stabilised and indeed 
this is generally the case. To show this, and to make explicit the whole construction discussed in this chapter we consider an example manifold. 

%
\section{The coset $SU(3)/U(1)\times U(1)$}
\label{sec:su3u1u1}
%

Having derived in section \ref{sec:iiasuperfields} the form of the ${\cal N}=1$ effective theory on a general manifold with torsion
classes (\ref{hftc}), in this section we study an explicit example of such a manifold. 
Denoting the internal manifold by ${\cal Y}$ we consider the coset space
\be
{\cal Y} = \frac{SU(3)}{U(1)\times U(1)}\;. 
\label{coset}
\ee
Cosets are particularly useful as examples of structure manifolds because the spectrum of forms that respect the coset symmetries is
highly constrained. This allows us to calculate the structure forms, the expansion basis and their differential relations.
The particular case ${SU(3)}/{U(1)\times U(1)}$ was first considered in \cite{Mueller-Hoissen:1987cq},
and there are more details about cosets in general and about this coset in the appendix. As well as forming an example of a manifold that leads to an $\N=1$ theory 
the coset makes the whole construction of manifolds with $SU(3)$-structure explicit and so serves to clarify the ideas presented in this chapter and in 
section \ref{sec:constsusy}.

The derivation of the form spectrum on the  coset is given in the appendix and here we summarise the results. 
The metric on the coset is derived by calculating the most general symmetric two-tensor that respects the coset symmetries which for this case is given by
\be
g = \left( \begin{array}{cccccc} a  & 0  & 0  & 0 & 0  & 0   \\
    0  & a  & 0  & 0 & 0 & 0   \\
    0  & 0 & b  & 0 & 0 & 0  \\
    0  & 0  & 0  & b & 0 & 0   \\
    0  & 0  & 0  & 0  & c & 0   \\
    0  & 0  & 0  & 0 & 0  & c  \\
    \end{array}  \right) \; ,
\ee
where all the parameters are real. The parameters of the metric are the geometric moduli and we see that we have three real moduli fields. Note that the volume of the 
coset is given by
\be
\Vol = abc \;.
\label{cosetvolume}
\ee

There exists a basis two-forms on the coset $\left\{\om_1,\om_2,\om_3 \right\}$ and their Hodge duals. 
There are also two independent three-forms $\left\{ \alpha_0,\beta^0\right\}$. The form bases satisfies the 
algebraic relations (\ref{defot}) and have the only non-vanishing intersection number
\be
{\cal K}_{123} = 1\;.
\ee
To find the structure forms $J$ and $\Omn$ we impose the $SU(3)$-structure conditions (\ref{6dsu3alg}) 
on the most general two-form and three-form that respect the coset symmetries.
This uniquely determines their structure in terms of the metric parameters and the basis forms
\ba
 J & = & a \omega_1 + b \omega_2 + c \omega_3 \;,\nn \\
 \Omn & = & \sqrt{abc} \left( i \alpha_0 - 4\beta^0 \right) \;. 
\label{jandomegadecompose}
\ea
In terms of the moduli classification we see that the coset has three K\"ahler moduli and no complex structure moduli.

The coset also restricts how the differential operator acts on forms and using this we can determine the differential relations of the basis forms which read
\ba
d \om_i & = & 4 \beta_0 \;, \nn \\
d \alpha_0 & = & 4 \left( \widetilde{\om}^1 + \widetilde{\om}^2 + \widetilde{\om}^3 \right) \;,\nn \\
d \widetilde{\om}^i & = & 0 = d \beta^A = d \alpha_{A\neq 0} 
\;. \label{cosetformsdiff}
\ea
We see that these are of the form (\ref{formsdiff}) where the free parameters $E_i$ have been fixed by the geometry. With these relations the structure forms satisfy 
the differential relations (\ref{torsionclasses}) with torsion classes
\ba
 \tc{1} & = & \frac{2i}{3}
  \frac{a + b + c}{\sqrt{abc}} \;,\nn \\
 \tc{2} & = & -\frac{4i}{3}
  \frac{1}{\sqrt{abc}} \left[  a(2a - b - c) \om_1 +  b(2b - a - c) \om_2 +  c(2c - a - b) \om_3 \right]\;. 
\label{cosetsu3tcs}
\ea
The expression for the torsion classes \ref{cosetsu3tcs} shows the explicit dependence of the torsion on the moduli. An important feature is that for particular values 
of the moduli some torsion classes vanish which for general values are not zero. This makes explicit the point made in section \ref{sec:iiasuperfields} that 
the torsion classes in the vacuum are a subset of the torsion classes of the action.

%
\subsection{$\N=1$ supersymmetric minima}
\label{sec:iiasusyminima}
%

The structure of the coset exactly matches the requirements of section \ref{sec:iiasuperfields} and so the results derived in the section apply for this case.
We can therefore go on to derive the effective $\N=1$ theory on the coset. We begin by identifying the fields of the theory. 
Comparing \eqref{jexpansion} with \eqref{jandomegadecompose} we are able to relate the K\"ahler moduli to the metric parameters
\be
v^1 = a, \ v^2 = b, \ v^3 = c \;.
\ee
There are no geometric moduli associated with complex structure deformations.  
In the effective theory we therefore have three superfields $T^1, T^2, T^3$ from the K\"ahler sector and the
superfield $U^0$ coming from the tensor multiplet. Using the
decomposition of $\Omn^{cs}$ in \eqref{omndecompose}, together with
\eqref{omnsd} and \eqref{jandomegadecompose}, gives $F_0 = -4iZ^0$,
and so the superfields are
\ba
 T^i & = & b^i - i v^i \;,\nn \\
 U^0 & = & \xi^0 - i e^{-\phi} \; .
\ea
In terms of the superfields the superpotential (\ref{simplesuperpot}) and K\"ahler potential (\ref{kahlerpotential}) are given by
\ba
 W & = & \frac{-i}{2\sqrt{8}} \left[ 
   \lambda - 2m T^1 T^2 T^3 - 4\left( T^1 + T^2 + T^3 \right) U^0 \right. \nn \\
   & & \;\;\;\;\;\;\;\;\;\;\;\; \left. - e_1 T^1 - e_2 T^2 -e_3 T^3
  \right]\;, \label{cosetsuperpot}\\
 K & = & -4\ln \left[i\half\left( U^0 - \bar{U^0} \right)\right] 
   - \ln \left[-i
   \left( T^1 - \bar{T}^1\right)
   \left( T^2 - \bar{T}^2\right)
   \left( T^3 - \bar{T}^3\right)
  \right]\;.
\label{cosetkahlerpot} 
\ea
This completely specifies the ${\cal N}=1$ low energy effective theory. 

We are now in a position to consider moduli stabilisation. This amounts to finding minima of the scalar potential with respect to all the fields.
Finding all the minima of the scalar potential is a difficult task, however we can look for particular minima for which the analysis is simplified. 
It is a well known result that supersymmetric minima are stable vacua which are determined by the solutions to the F-term equations for the superpotential
(\ref{cosetsuperpot}), which read
\ba
 D_{T^1} W & = & -2mT^2 T^3 - 4U^0 - e_1 - \frac{W}{T^1 - \bar{T}^1} = 0 \nn \;,\\ 
 D_{T^2} W & = & -2mT^1 T^3 - 4U^0 - e_2 - \frac{W}{T^2 - \bar{T}^2} = 0 \nn \;,\\
 D_{T^3} W & = & -2mT^1 T^2 - 4U^0 - e_3 - \frac{W}{T^3 - \bar{T}^3} = 0 \nn \;,\\ 
 D_{U^0} W & = & -4 \left( T^1 + T^2 + T^3 \right) - \frac{4W}{U^0 - \bar{U^0}} = 0  
\label{feqs} \; ,
\ea
where the K\"ahler covariant derivative is given by $D_T=\partial_T + (\partial_T K)$. We arrive at four independent complex equations which serve to 
fix the four complex scalar fields present. To show that this is the case we solve them explicitly for a simplifying choice of parameters. We choose 
\be
e_1 = e_2 = e_3 \;,
\ee
which means that the flux $G_4$ becomes exact and so can be absorbed into the definition of $U^0$. In effect this is equivalent to setting $e_i=0$. 
We can then look for a solution where the K\"ahler superfields are equal $T^1=T^2=T^3 \equiv T$. 
In this case the equations simplify to the form
\ba
  U^0 & = & \frac{1}{24T\overline{T}} \left[ -T ( \lambda - 2m\overline{T}^3 ) 
            + 3\overline{T} ( \lambda - 2mT^3 ) \right] \;, \\ 
     0 & = &  6mT^2\overline{T}^3  
           - \lambda T\overline{T} +2m\overline{T}T^4 
      + 3\lambda T^2 - 2\lambda \overline{T}^2 +4mT^3\overline{T}^2 
      - 12mT\overline{T}^4  \label{tequation} \; .
\ea
The unique solution for $m>0$ has $\lambda<0$ and the vacuum expectation values for the
superfield components are
\ba
 \ev{b^1} = \ev{b^2} = \ev{b^3} & = & 
   - \frac{5^{\frac23}}{20} \left( \frac{-\lambda}{m} \right)^{\frac13}\nn \;,\\
 \ev{v^1} = \ev{v^2} = \ev{v^3} & = & 
   \frac{\sqrt{3}5^{\frac16}}{4}\left( \frac{-\lambda}{m} \right)^{\frac13} \;,\nn \\
 \ev{\xi^0} & = & 
   \frac{5^{\frac13}}{20}\left(m\lambda^2\right)^{\frac13} \nn \;,\\
 \ev{e^{-\phi}} & = & 
   \frac{\sqrt{3}5^{\frac56}}{20}\left( m\lambda^2\right)^{\frac13}
\label{solution} \; .
\ea  
This forms an explicit example of a minimum where all the moduli are stabilised thereby solving the difficult vacuum degeneracy problems outlined in 
section \ref{sec:vacdeg}.  The vacuum forms the first and, so far, 
only example of a purely perturbative vacuum in string theory where all the fields (moduli and axions) are stabilised. Non-perturbative effects are not 
expected to play a role here since all the fields appear at tree-level in the superpotential.
It does not require the use of orientifolds and the supersymmetry breaking occurs spontaneously rather than through an orientifold projection.  
The absence of non-perturbative effects means that the four-dimensional vacuum can be uplifted to the full ten-dimensional solution and it can be checked 
that the vacuum (\ref{solution}) satisfies the ten-dimensional constraints (\ref{lustfields}). This is a non-trivial result and forms an independent check on the procedure 
of supersymmetry breaking and, more importantly, on the basis of forms used in the compactification.

Substituting the vacuum (\ref{solution}) into the expression for the torsion class $\tc{2}$ (\ref{cosetsu3tcs}) we find that in the vacuum $\tc{2}$ vanishes. 
This makes explicit the difference between the torsion classes in the vacuum and in the action. 
The vanishing of $\tc{2}$ can be directly attributed to the simplification we made in turning off the non-exact flux $e_i$. This in turn allowed a 
solution where the $T^i$ superfields took the same values. In the presence of $e_i$ flux the superfields would take differing values in the vacuum and the 
torsion class $\tc{2}$ would be non-zero. 

We now turn to some brief phenomenological properties of the vacuum (\ref{solution}). The first thing to check is the consistency of the solution in terms of 
large volume and weak coupling. Substituting the vacuum solution into the expressions for the volume (\ref{cosetvolume}) and the string coupling, 
$g_s=e^{\hat{\phi}}$, gives that 
we can go to arbitrary large volume and weak-coupling by taking $|\lambda|>>|m|$.
Although the vacuum is consistent it lacks a number of key phenomenological features. In terms of the matter fields the fact that there are only two three-cycles means 
that it is not possible to embed a realistic intersecting $D6$-branes model. Also all the axionic fields are massive which rules them out as 
candidates for the QCD axion. 

\begin{figure}
\begin{center}
\epsfig{file=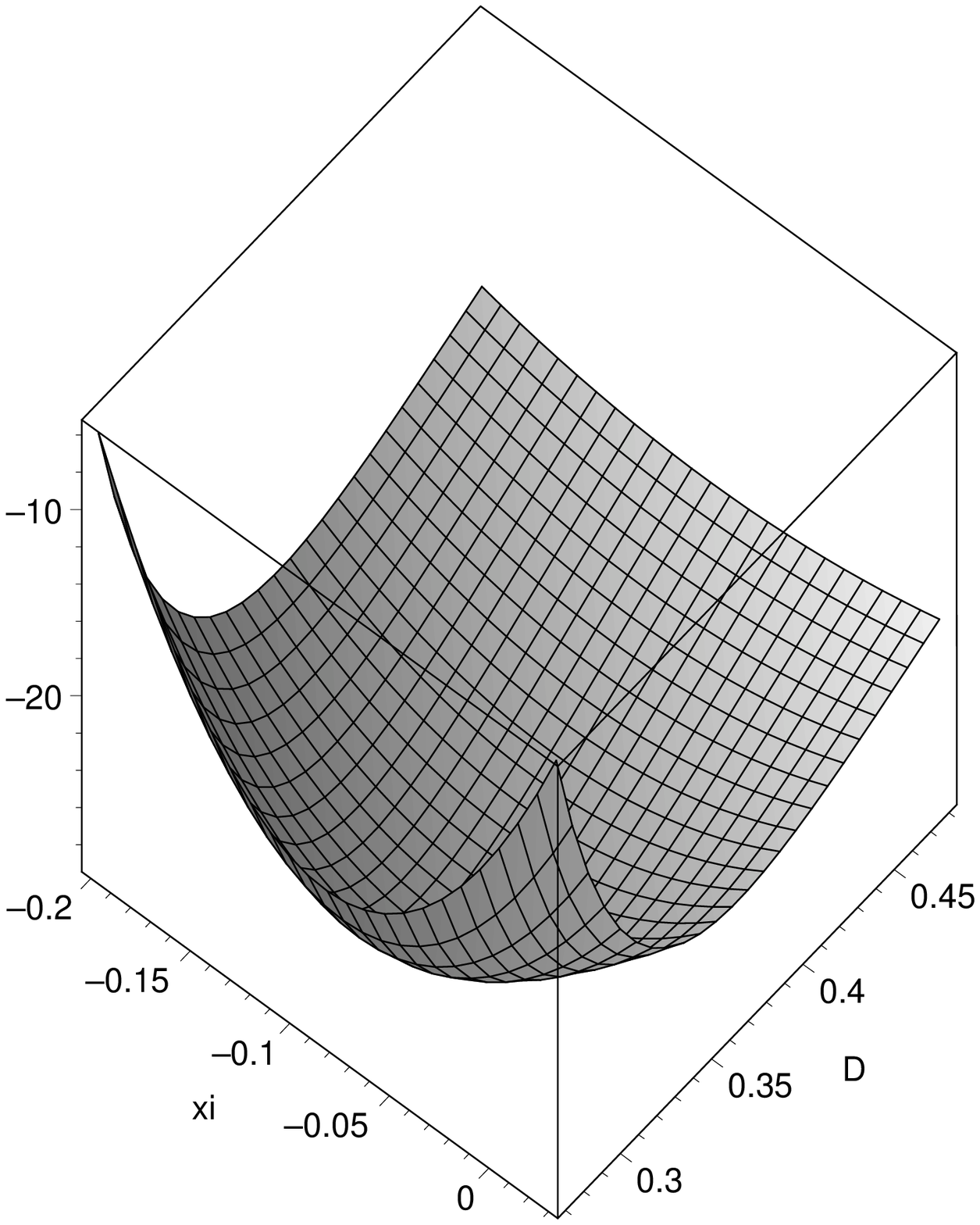,width=7.5cm,height=7.5cm}
\flushleft
\caption{Plot showing the scalar potential for the directions 
 $\xi^0$ and $e^{-\phi}$ (denoted as $D$). }
\label{fig:dilatonpot}
\end{center}
\end{figure}

The most serious problem is that the vacuum is anti deSitter (AdS) which can be seen from the value of the scalar potential in the vacuum 
\be
 \ev{V} = -3 e^K |W|^2 = \frac{- 7.25}{\left( -m \lambda^5 \right)^{\frac13}} 
 \label{adsback} \;.
\ee 
Although the limit $|\lambda|\ra\infty$ leads to Minkowski, the cosmological constant is generally negative. This of course is a problem that plagues most 
of the vacua found in supergravities. One possible resolution to this problem is the introduction of new effects into the potential such as anti-$D6$-branes which 
could uplift the potential to a positive or zero value. Such a mechanism has yet to be constructed in type IIA string theory but does exist in the type IIB case\cite{kklt}.
The possibility of such an uplift motivates studying the local form of the potential near the minimum. The fact that the vacuum is a
supersymmetric AdS vacuum means that it is stable even if it is a saddle point \cite{Breitenlohner:1982jf,Breitenlohner:1982bm}. However following a possible uplift 
to Minkowski or deSitter space saddle points become unstable vacua. 
We can construct a Hermitian block matrix from the second derivatives of the
potential with respect to the superfields evaluated at the solution
\ba
 H & \equiv &  \left( \begin{array}{cc} V_{I\overline{J}} & V_{IJ} \\ 
   V_{\overline{I}\overline{J}} & V_{J\overline{I}} \end{array} \right) \;,\\ 
   V_{I\overline{J}} &=& e^K K^{L\overline{M}}\partial_{L}\left( D_I W \right) 
   \partial_{\overline{M}} \left( D_{\overline{J}} \overline{W} \right) 
   - 2e^K K_{I\overline{J}}\left| W \right|^2 \;, \\
 V_{IJ} &=& - \overline{W} e^K \partial_{I} \left( D_J W \right)  \; .
 \label{secondderivs}
\ea
Then for the solution to be a local minimum in all the directions
associated with the components of the superfields the matrix $H$ must
be positive definite. Inserting the solution \eqref{solution} into
\eqref{secondderivs} we find that out of the eight real eigenvalues
only six are positive. This means that there are two real directions
for which the potential is at a maximum. 
The vacuum is therefore unstable under a direct uplift to Minkowski or deSitter space. It is possible however that since an uplift requires new terms in the potential 
the local structure may be modified by these new terms so that it is stable. 
The tachyonic directions are in the K\"ahler superfields sector as can be seen by directly plotting 
the potential in the axio-dilaton direction as in figure \ref{fig:dilatonpot}.
Therefore it is that sector that would have to be modified to solve the uplift problem. 

%
\section{Summary}
\label{sec:iiasummary}
%

In this chapter we studied compactifications of type IIA string theory on manifolds with $SU(3)$-structure. The presence of fluxes was shown to  
generate a potential for the moduli fields in four dimensions. It was then argued, using ten-dimensional solutions, that the flux energy density back-reacts on the 
geometry of the manifold inducing torsion so that the manifold is no longer CY. This formed the motivation for considering compactifications on manifolds with general 
$SU(3)$-structure. We proceeded to derive the four-dimensional action and its interpretation as a gauged $\N=2$ supergravity with electric and magnetic gauging. 
Through the form of the four-dimensional gravitini mass matrix we were able to study the phenomenon of spontaneous partial supersymmetry breaking and, by 
restricting the present torsion classes, derive the resulting effective $\N=1$ supergravity. We showed that this supergravity has a 
stable vacuum where all the moduli are stabilised in a perturbative manner thereby solving the vacuum degeneracy problem.

%
\chapter{Compactification of M-theory on Manifolds with SU(3)-Structure}
\label{cha:su3mtheory}
%

In this chapter we study compactifications of M-theory on seven-dimensional manifolds with $SU(3)$-structure \cite{Micu:2006ey}. Similar work can be found in 
\cite{Curio:2000dw,Curio:2003ur,Anguelova:2006qf,newcvetic}.
Unlike the type IIA case studied in chapter 
\ref{cha:su3iia} these manifolds are not the type that preserve the minimum amount of supersymmetry in four-dimensions, these are the manifolds with $G_2$-structure, for 
which compactifications have been studied in 
\cite{Lambert:2005sh,House:2004pm,Dall'Agata:2005fm,D'Auria:2005rv}. 
However, as we saw in chapter \ref{cha:su3iia}, the supersymmetry of the action need not be fully preserved by the vacuum.
Eleven-dimensional solutions that explore the structure of the vacuum have
been studied for the cases of $SU(2)$, $SU(3)$ and $G_2$-structure in
\cite{Kaste:2003zd,Dall'Agata:2003ir,Behrndt:2004mx,Lukas:2004ip,Behrndt:2004bh,Gauntlett:2004hs,Franzen:2005ve,Behrndt:2005im}.
An interesting point to come out of these studies is that compactifications on
manifolds with $SU(3)$-structure have a much richer vacuum spectrum than
manifolds with $G_2$-structure. Indeed there are solutions that preserve only
$\N=1$ supersymmetry in the vacuum putting them on an equal phenomenological
grounding with $G_2$ compactifications in that respect. There are however many
phenomenologically appealing features that are not present in the $G_2$
compactifications such as warped anti-deSitter solutions and solutions with
non-vanishing internal flux. Therefore manifolds with $SU(3)$-structure form an important set of compactifications to consider.

As was the case in type IIA, the presence of fluxes induces torsion on the manifold and therefore the manifolds do not preserve $SU(3)$-holonomy but are
of a more general $SU(3)$-structure. This is important since, as discussed in section \ref{sec:constsusy}, compactifications of M-theory on manifolds with $SU(3)$-holonomy 
are equivalent to compactifications of type IIA string theory on CY manifolds. In this chapter we show that because the $SU(3)$-structure naturally picks out
a vector on the internal manifold these compactifications can be cast into a form that is similar to 
the type IIA compactifications on $SU(3)$-structure manifolds considered in chapter \ref{cha:su3iia}.
Indeed the structure of this chapter is very similar in nature to chapter \ref{cha:su3iia} and many of the arguments used there apply for this case. 
For this reason the issues that arise in this chapter which have a parallel in the IIA case are not discussed extensively  and the reader is referred to 
chapter \ref{cha:su3iia} for more detail.

We begin this chapter with a reduction of eleven-dimensional
supergravity on a general manifold with $SU(3)$-structure deriving the 
resulting $\N=2$ theory in four dimensions.
The four-dimensional gravitini mass matrix is then used to explore the
amount of supersymmetry preserved by various manifolds. 
We begin by
looking at vacua that preserve $\N=2$ supersymmetry in section
\ref{sec:pren=2}. We derive the most general $\N=2$ solution and
use it as a check on the mass matrix. We then show how this solution can
be used to find explicit vacua of an example manifold.  In section
\ref{sec:pren=1} we move on to the more phenomenologically interesting
$\N=1$ vacua and show that some classes of $SU(3)$-structure manifolds induce spontaneous partial
supersymmetry breaking that leads to an $\N=1$ effective theory. We derive this theory and go through an explicit example of 
moduli stabilisation. 

%
\section{Reduction of eleven-dimensional supergravity}
\label{sec:11dred}
%

The theory that we consider is the low energy limit of M-theory that is
eleven-dimensional supergravity.  The action of the theory is given by \cite{Duff:1986hr}
\ba
S_{11}&=&\frac{1}{\kappa_{11}^2}\int_{\M_{11}}\sqrt{-g_{11}}d^{11}X\left[\half \hat{R} 
        - \half\bar\Psi_{M}\hat\Gamma^{MNP}\hat D_{N}\Psi_{P}
        -\quarter\frac{1}{4!}\hat{F}_{MNPQ}\hat{F}^{MNPQ} \right. \label{mtheoryaction} \\\nonumber
        &~& \hspace{2cm} + \half\frac{1}{(12)^4}\epsilon^{LMNPQRSTUVW}\hat{F}_{LMNP}\hat{F}_{QRST}\hat{C}_{UVW} \\\nonumber
        &~& \hspace{2cm} - \left. \frac{3}{4(12)^2}( \bar\Psi_{M}\hat\Gamma^{MNPQRS}\Psi_{N}
                           +12\bar\Psi^{P}\hat\Gamma^{QR}\Psi^{S})
                                                F_{PQRS}\right] \; . \\ \nonumber
\ea
The field spectrum of the theory contains the eleven-dimensional graviton
$\hat{g}_{MN}$, the three-form $\hat{C}_{MNP}$ and the gravitino,
$\hat{\Psi}_{P}$.  The indices run over eleven dimensions
$M,N,..=0,1,...,10$. For $\gamma$ matrix and $\epsilon$ tensor conventions see
the Appendix.  $\kappa_{11}$ denotes the eleven-dimensional Planck
constant which we shall set to unity henceforth thereby fixing our units.

In this section we consider this theory on a space which is a direct
product $\M_{11} = {\cal S} \times \M_7$ with the metric ansatz
\be
ds_{11}^2 = g_{\mu \nu}(x) dx^\mu dx^\nu + g_{mn}(x,y) dy^m dy^n \label{metric},
\ee
where $x$ denotes co-ordinates in four dimensions and $y$ are the co-ordinates
on the internal compact manifold. This ansatz
is not the most general ansatz possible for the metric as we have not included
a possible warp factor. 
As discussed in chapter \ref{cha:su3iia} there are many
compactifications that can consistently neglect such a warp factor because
either a warp factor is not induced by the flux or it can be perturbatively
ignored if the internal volume is large enough. 
For now we proceed with an 
unwarped ansatz bearing in mind that this is only consistent for certain
compactifications.

The four-dimensional effective theory is an $\N=2$ gauged supergravity.
In the upcoming sections we derive the
quantities necessary to specify this theory. The kinetic terms for the low
energy fields are derived from the Ricci scalar and the kinetic term for
the three-form. The prepotentials are derived from the
four-dimensional gravitini mass matrix and the vector multiplet Killing vectors are calculated 
through explicit dimensional reduction.

%
\subsection{The Ricci scalar}
\label{sec:11dricciscalar}
%

The reduction of the Ricci scalar follows the same methodology as in chapter \ref{cha:su3iia}. We begin by writing the metric variations 
in terms of the structure forms and then decompose the Ricci scalar appropriately.
In fact we now show that the presence of the vector $V$ allows us to cast the metric deformations into a form very similar to the IIA case 
considered in chapter \ref{cha:su3iia}. However, it is important to remember that a decomposition of the seven-dimensional manifold into a six-dimensional 
sub-manifold and a circle is never assumed and is not generally the case. The analogy with IIA is done at the 'four-dimensional' level while all internal forms 
have a general dependence on all the seven internal co-ordinates.

The $SU(3)$-structure induces a metric on the manifold that we
can write in terms of the invariant forms as
\begin{equation}
  \begin{aligned}
    g_{ab} \equiv & \; |s|^{-\frac{1}{9}} s_{ab} \;, \\
    s_{ab} \equiv & \; \frac{1}{16} \left[\frac{1}{4} \left(\Omega_{amn}
        \bar\Omega_{bpq} + \bar\Omega_{amn} \Omega_{bpq} \right) 
      + \frac{1}{3} V_a V_b J_{mn} J_{pq} \right] J_{rs} V_t \;\;
    \hat\epsilon^{mnpqrst} \; .
  \end{aligned}
\end{equation}
This expression for the metric can be checked by performing the contractions on the right-hand-side using the appropriate $SU(3)$-structure identities.
Varying the formula above we can write the metric deformations as
\ba
 \delta g_{ab} &=& \frac{1}{8} \delta \Omega_{(a}^{\;\;\;mn} \bar{\Omega}_{b)mn} + \frac{1}{8} \Omega_{(a}^{\;\;\;mn} 
\delta \bar{\Omega}_{b)mn} 
 + 2V_{(a}\delta V_{b)} + V_a V_b \left(J\lrcorner \delta J \right) + J_{(a}^{\;\;\;m} \delta J_{b)m} \nonumber \\
 &\;& \; + V^m V_{(a}J^{n}_{\;\;b)} \delta J_{mn} 
 - \frac{1}{3} \left( \frac{1}{4} \delta \Omega \lrcorner \bar{\Omega} + \frac{1}{4} \Omega \lrcorner \delta \bar{\Omega} 
+ J \lrcorner \delta J \right) g_{ab} \; . \label{su3metvar}
\ea
This is very similar to CY compactifications where
the metric variations were expressed in terms of K\"ahler class and complex
structure deformations. Keeping the terminology we refer to the scalar
fields associated with $\delta J$ and $\delta \Omega$ as K\"ahler moduli and
complex structure moduli respectively. Furthermore we denote the scalar
associated to $\delta V$ as the dilaton in complete analogy to the type IIA
compactifications. 

Before starting the derivation of the kinetic terms associated to the metric
deformations discussed above we mention that the metric variations can be
dealt with more easily in terms of the variations of either of the two $G_2$-structures 
which can be defined on seven-dimensional manifolds with $SU(3)$-structure \eqref{phiOJV}.
The expression (\ref{su3metvar}) can be written as 
\be
\label{g2metricvar}
  \delta g_{ab} = \frac12 {\vp^\pm_{(a}}^{mn} \delta \vp^{\pm}_{b)mn} 
- \frac13 \left( \vp^{\pm} \lrcorner \delta \vp^{\pm} \right) g_{ab} \;.
\ee
This is interesting since it shows that the complete metric variations can be parameterised in terms of either of the $G_2$-structures and the full 
$SU(3)$-structure is not required.
For each of the $G_2$-structures the formula coincides with the
metric variations on a manifold with $G_2$-structure \cite{House:2004pm}. 

We now proceed with the compactification of the Ricci scalar.
Initially, we do not decompose $\Omega$ and $J$ into
their four-dimensional scalar components but with the vector $V$ we write
\be
V(x,y) \equiv e^{\hat{\phi}(x)} z(y) \label{Vdef},
\ee
where $z$ is the single vector we have on the internal manifold from the
$SU(3)$-structure requirements. Note that it is still $V$ and not $z$
that features in the $SU(3)$ relations (\ref{su3rel}). The difference
between $V$ and $z$ can be understood as $V$ is the $SU(3)$ vector
which also encodes the possible deformations of the manifold, while
$z$ is only a basis vector in which we expand $V$. Therefore, the
factor $e^{\hat \phi}$ encodes information about the deformations
associated to the vector $V$. This is completely analogous to the
compactification of eleven-dimensional supergravity on a circle to
type IIA theory and in order to continue this analogy we call
the modulus in equation \eqref{Vdef} the dilaton.
We further define a quantity which, in the case where the
compactification manifold becomes a direct product of a
six-dimensional manifold and a circle, plays
the role of the volume of the six-dimensional space
\begin{equation}
  \label{defv6}
  \Vol_6 \equiv e^{-\hat{\phi}} \Vol \; ,
\end{equation}
where $\Vol$ is the volume of the full seven-dimensional space
\begin{equation}
  \label{defv7}
  \Vol \equiv \int \sqrt{g_7} d^7y = \frac16 \int J \wedge J \wedge J \wedge V \; .
\end{equation}
With this quantity we can construct the same structure for the complex structure moduli as in chapter \ref{cha:su3iia}.
We define the true holomorphic three-form $\Omn^{cs}$ as
\begin{equation}
  \label{eqn:omegaCS}
    e^{\half K^{cs}}\Omega^{cs} \equiv \frac{1}{\sqrt{8}}\Omega
    (\Vol_6)^{-\half} \; , \\ 
\end{equation}
where we have also introduced the K\"ahler potential for the complex structure
deformations, $K^{cs}$ 
\begin{equation}
  \label{mdefocs}  
  K^{cs} \equiv - \mathrm{ln} \left( || \Omega^{cs} ||^2 \Vol_6
  \right) =-\mathrm{ln}\;i<\Omega^{cs}|\bar\Omega^{cs}> \equiv -\mathrm{ln} \;i\int \Omega^{cs}
  \wedge \bar \Omega^{cs} \wedge z \; .
\end{equation}
The analogy with the IIA case of chapter \ref{cha:su3iia} holds for the spurious variations of $\Omn^{cs}$ so that the physical variations 
of $\Omn^{cs}$ correspond to the K\"ahler covariant derivative.

Using the expression for the metric fluctuations (\ref{su3metvar}) we derive the variation of the
eleven-dimensional Ricci scalar. The calculation
is presented in the appendix and here we recall the final result
\begin{eqnarray}
  \label{R11su3}
  \int_{\M_{11}}{\sqrt{-\hat{g}} d^{11} X \; \half \hat{R}} = \int_{\cal S}{\sqrt{-g_4}
  d^4x \Big[ \half R}  &-& \partial_{\mu} \phi \partial^{\mu} \phi \\ 
  &-& e^{-\hat{\phi}} e^{K_{cs}} \int_{\M_7}\sqrt{g_7}\; d^7y \; 
  D_{\mu} \Omega^{cs} \lrcorner D^{\mu} \bar{\Omega}^{cs} \nn \\
  &-& \quarter \Vol_6^{-1} 
  e^{-\hat{\phi}} \int_{\M_7} \sqrt{g_7} \; d^7y \; \partial_{\mu}J \lrcorner
  \partial^{\mu}J \Big] \; , \nn 
\end{eqnarray}
where we have dropped the $R_7$ term.
We define the four-dimensional dilaton as
\begin{equation}
  \label{eqn:4Ddilaton}
  \phi \equiv \hat{\phi} - \half \mathrm{ln} \Vol_6 \; .
\end{equation}
Finally we note that in order to arrive at the four-dimensional Einstein frame we performed the Weyl rescalings
\begin{equation}
  \label{weylrescaling}    
  \begin{aligned}
    g_{\mu\nu} \ra & ~\Vol^{-1} g_{\mu\nu} \; ,\\
    g_{mn} \ra & ~e^{-\frac{2}{3}\hat{\phi}} g_{mn}  \; .
  \end{aligned}
\end{equation}
The important thing to notice in this result is that the metric fluctuations
have naturally split into the dilaton, the $J$ and $\Omega^{cs}$ variations
with separate kinetic terms. Moreover, due to the dependence of $\sqrt{g_7}$
on the dilaton, it can be seen that the all the dilaton factors drop out from
the kinetic terms of the K\"ahler and complex structure moduli. 

%
\subsection{The form fields}
\label{sec:11daxions}
%

As we have seen in the previous subsection, the compactification of the
gravitational sector of M-theory on seven-dimensional manifolds with $SU(3)$-structure 
closely resembles the corresponding compactifications of type IIA
theory. Therefore we find it useful to continue
this analogy at the level of the matter fields and so we decompose
the three-form $\hat C_3$ along the vector direction which is featured in the
seven-dimensional manifolds with $SU(3)$-structure under consideration.
Consequently we write 
\be
\label{CB}
\hat{C_3} = C_3 + B_2 \w z \; ,
\ee
where $C_3$ is assumed to have no component along $z$, i.e. $C_3 \lrcorner z =
0$. As expected, in the type IIA picture $C_3$ corresponds to the RR
three-form, while $B_2$ represents the NS-NS two-form field.
Then compactifying the eleven-dimensional kinetic term, taking care to
perform the appropriate Weyl rescalings \eqref{weylrescaling}, we arrive at
\ba
\label{C11su3}
& & \hspace{-1cm}\int_{\M_{11}}{\sqrt{-\hat{g}} d^{11} X \; \left[ -\quarter \hat{F}
    \lrcorner \hat{F} \right]} \\ 
& & = \int_{\cal S}{\sqrt{-g_4} d^4x \left[ -\frac{1}{4} e^{2\phi} e^{-\hat{\phi}}
    \int_{\M_7} \sqrt{g_7}d^7y \;\partial_{\mu} C_3 \lrcorner \partial^{\mu} C_3 -
    \quarter \Vol_6^{-1} e^{-\hat{\phi}} \int_{\M_7} \sqrt{g_7}d^7y \;
    \partial_{\mu} B_2 \lrcorner \partial^{\mu} B_2 \right]} \; . \nn 
\ea
One immediately notices that the kinetic term for fluctuations of the
$B_2$-field along the internal manifold is the same as the kinetic term for the
fluctuations of the fundamental form $J$. Therefore we see
that these fluctuations pair up into the complex field
\be
T \equiv B_2 - iJ \; .\label{bigt}
\ee
%

%
\subsection{Flux}
\label{sec:11dflux}
%

The only background fluxes that can be turned on
in M-theory compactifications and which are compatible with
four-dimensional Lorentz invariance are given by 
\be
\ev{\hat{F}_4} = f\eta_4 + {\cal G} + d\ev{\hat{c}_3}  \mathrm{\;,}
\label{F4flux}
\ee
where
\be
{\cal G} = d\bg{C}_3 \;.
\ee
We have decomposed the internal parts of the fluxes into an exact and a non-exact part. The non-exact part is the usual flux, as discussed in chapter \ref{cha:su3iia}, 
that arises from the form $\bg{C}_3$ which is not globally well defined. This flux arises from the membrane and is 
quantised. There is also an exact part of the flux which arises from the vacuum expectation value of the field coming 
from $\hat{C}_3$ which we denote $\hat{c}_3$ to distinguish it from the part that is not globally well defined. See 
(\ref{hatc3}) for the explicit definition. 

The external part of the flux $f$ is called the Freund-Rubin flux.
As observed in the literature \cite{Lambert:2005sh,House:2004pm,Beasley:2002db}, the Freund-Rubin
flux is not the true constant parameter describing this degree of
freedom. Rather one has to consider the flux of the dual seven-form
field strength $\hat F_7$ 
\begin{equation}
  \label{F7}
  \hat F_7 = d \hat C_6 + \frac12 \hat C_3 \wedge \hat F_4 \; ,
\end{equation}
which should now be the true dual of the Freund-Rubin flux. As can
be seen the $\hat F_7$ flux also receives a contribution from the
ordinary $\hat F_4$ flux. Therefore, in general, the Freund-Rubin flux
parameter is given by
\begin{equation}
  \label{ftolambda} 
  f = \frac{1}{\Vol} \left(\lambda + \half \int_{\M_7}  \hat{c}_3 \wedge {\cal G} + \half \int_{\M_7} \hat{c}_3 \wedge d\hat{c}_3 
  \right) \; ,
\end{equation}
where $\lambda$ is a constant which parameterises the seven-form flux.

%
\subsection{The gravitini mass matrix}
\label{sec:11dmass}
%

Recall from section \ref{sec:constsusy} that on a seven-dimensional
manifold with $SU(3)$-structure one can define two independent (Majorana)
spinors which we have denoted $\epsilon_{1,2}$. We therefore consider the ansatz
\be
\label{4dgravitini}
\hat{\Psi}_{\mu} =  \Vol^{-\quarter} \left( \psi^{1}_{\mu} \otimes \epsilon_1
  + \psi^{2}_{\mu} \otimes \epsilon_2 \right) \; ,
\ee
where $\psi^{1,2}$ are the four-dimensional gravitini which are Majorana
spinors and the overall normalisation factor is chosen in order to reach
canonical kinetic terms in four-dimensions. It is more customary to work with
gravitini which are Weyl spinors in four dimensions and therefore we decompose
$\psi^{1,2}$ above as
\be
\psi_{\mu}^{\alpha} = \half \left( \psi_{+\mu}^{\alpha} + \psi_{-\mu}^{\alpha}
\right) \; ,
\ee
where $\alpha, \beta = 1,2$ and the chiral components of four-dimensional
gravitini satisfy 
\be
\gamma_5\psi^\alpha_{\pm\mu} = \pm\psi^\alpha_{\pm\mu} \; .
\ee
Then compactifying the eleven-dimensional gravitino terms in
(\ref{mtheoryaction}) and performing the appropriate Weyl rescalings
(\ref{weylrescaling}) we arrive at the four-dimensional action (\ref{n=2gravitini}).
The main steps in deriving the mass matrix are presented in appendix
\ref{cha:massiia}. The result reads
\begin{eqnarray}
  \label{massmatrix}
  S_{11} &=& \frac{ie^{\frac{7}{2}\hat{\phi}}}{8\Vol^{\frac{3}{2}}} \left\{
    \int_{{\cal M}_7} \left[ dU^+ \w U^+ + 2 {\cal G} \w U^+ \right] +
    2\lambda \right\} \; ,\nonumber \\ 
  S_{22} &=& \frac{ie^{\frac{7}{2}\hat{\phi}}}{8\Vol^{\frac{3}{2}}} \left\{
    \int_{{\cal M}_7} \left[ dU^- \w U^- + 2 {\cal G} \w U^- \right] +
    2\lambda \right\} \; ,  \\ 
  S_{12} &=& S_{21} = \frac{ie^{\frac{5}{2}\hat{\phi}}}{8\Vol^{\frac{3}{2}}}
    \int_{{\cal M}_7} \left[ 2i{\cal G}\w \Omega^+ + 2i d \hat{c}_3 \w \Omega^+ -
    2 dJ \w \Omega^+ \w z \right ] \; . \nn
\end{eqnarray}
Here $\mathcal{G}$ denotes the internal non-exact part of the background flux which was defined in equation
\eqref{F4flux}, $\lambda$ is the Freund-Rubin constant and we have further introduced 
\begin{equation}
  \label{eqn:N2field}
  U^{\pm}  \equiv  \hat c_3 + ie^{-\hat{\phi}}\vp^{\pm} = \hat c_3 \pm
  ie^{-\hat{\phi}}\Omega^- - iJ \w z \;.
\end{equation}

The eigenvalues of the gravitini mass matrix determine the amount of supersymmetry in the vacuum. As we did in chapter \ref{cha:su3iia} we use them 
to study partial supersymmetry breaking in section \ref{sec:pren=1}. An important concept, that is not present in the six-dimensional case of chapter \ref{cha:su3iia}, is the 
notion of an effective $G$-structure. Consider a manifold where the gravitini $\psi^{\alpha}_\mu$ in (\ref{4dgravitini}) are mass eigenstates. This corresponds to the 
case where the off-diagonal terms in the mass matrix vanish. Then a mass gap between the gravitini also implies a mass gap between 
the two internal spinors $\epsilon_{\alpha}$. This, in turn, through the definitions of the two $G_2$-structures (\ref{phi+-}) means that one 
$G_2$ form becomes heavier than 
the other. In that sense performing the truncation of the higher mass gravitini and its associated matter spectrum corresponds to a truncation of one of the 
$G_2$-structures. The effective $\N=1$ theory after the truncation is equivalent to a compactification on a manifold with $G_2$-structure where the $G_2$-structure is 
the lowest mass one, even though the full manifold has $SU(3)$-structure. The manifold can be said to have an effective $G_2$-structure. We study an explicit example 
of this phenomenon in section \ref{sec:11dsusyminima}. 
It is important to note however that not all truncations of the $SU(3)$-structure $\N=2$ theory to an effective $\N=1$ theory 
correspond to an effective $G_2$ compactification. It is only the case when the gravitino mass eigenstates correspond to the internal spinors from which the two 
$G_2$-structures are constructed. The notion of an effective $G$-structure is particularly important in M-theory compactifications since all manifolds with $G_2$-structure 
are known to have $SU(2)$-structure \cite{Behrndt:2005im} and so lead to a $\N=4$ four-dimensional theory. An $SU(2)$-structure contains two $SU(3)$-structures and so in this chapter we 
actually study compactifications on an effective $SU(3)$-structure where we assume the other $SU(3)$-structure is massive and can be truncated. This is not as strong 
an assumption as it appears. In fact in this chapter we give two explicit manifolds where such a mass hierarchy appears. These are the cosets of section \ref{sec:n=2coset} and 
section \ref{sec:su3u1u1u1}. Both of the cosets are such that only an $SU(3)$-structure is compatible with the coset symmetries. On coset manifolds the forms that satisfy the 
symmetries are precisely the lowest mass forms and so we recover an effective $SU(3)$-structure.

%
\subsection{The Kaluza-Klein basis}
\label{sec:11dkkbasis}
%

As was the case for six-dimensional manifolds with $SU(3)$-structure, specifying the basis of lowest mass forms on the internal manifold is a difficult task.
Given the absence of such a classification in the literature we propose a set of forms motivated through supersymmetry and explicit examples. 
We first specify the forms and then motivate the choice.
The basis includes a single one-form $z$. 
There are also a set of two-forms $\left\{ \omega_i \right\}$, and four-forms $\left\{ \tilde\omega^i \right\}$
\footnote{Note that these are not the Hodge duals of the two-forms since those are five-forms. Rather the correspondence between the four-forms and the two-forms is of the 
nature $\left( \tilde{\omega}^i\right)_{mnpq} = \left( \star \omega_i \right)_{mnpqr}z_{s}\delta^{rs}$.}
, and set of three-forms $\left\{\alpha_A, \beta^A\right\}$ which satisfy
\ba
 \label{7ddefot}
  \int_{\M_7}{\om_i \wg \tilde{\om}^j \wg z} &=& \delta^{j}_{i}  \;, \\
  \int_{\M_7}{\alpha_{A} \wg \beta^{B} \wg z} &=& \delta^{B}_{A} \; ,\\ 
  \int_{\M_7}{\alpha_{A} \wg \alpha_{B} \wg z} &=&  \int_{\M_7}{\beta^{A} \wg \beta^{B} \wg z}  = 0 \;.
\ea
The forms satisfy the following relations 
\begin{equation}
  \begin{aligned}
    \om_{i} \w \alpha_{A} = \om_{i} \w \beta^{A} = & ~0 \; ,\\
    z \lrcorner \om_{i} = 
    z \lrcorner \alpha_{A} = z \lrcorner \beta^{A} = &~ 0\; . 
  \end{aligned}
\label{7dformrel}
\end{equation}
These forms can in general depend on all seven internal coordinates and not be closed. We do not specify the differential relations in general but calculate them 
explicitly for the examples considered.

The strongest motivation for this choice of basis is the analogy with type IIA compactifications. We saw in chapter \ref{cha:su3iia} how the basis of forms 
lead to an $\N=2$ supergravity and since we should recover an $\N=2$ supergravity also in this case the basis should follow a similar structure. The major 
difference is the addition of a single one-form $z$ which is obviously motivated through the extra vector $V$ in the seven-dimensional $SU(3)$-structure. 
Since there is only one independent singlet vector we expect only one one-form. The other possible one-forms are triplets and are therefore ruled out by the same 
considerations as the six-dimensional case.
The relations between the forms (\ref{7dformrel}) are motivated by the anticipation that the $SU(3)$-structure forms should be decomposed in terms of the basis. 
The constraints then follow from the $SU(3)$-structure relations (\ref{su3rel}). As in the six-dimensional case the index range of the forms is not topological.

%
\subsection{The action as a gauged $\N=2$ supergravity}
\label{sec:11dn=2sugra}
%

The eleven-dimensional forms are expanded in the basis (\ref{7ddefot}) as
\ba
V(X) &=& e^{\hat{\phi}(x)} z(y) \label{vexpan}\,\\
J(X) &=& v^i(x)\om_i(y) \label{jexpan} \mathrm{\;,}\\
\Omn^{cs}(X) &=& X^A(x) \al_A(y) - F_A(x) \beta^A(y) \label{omexpan}\mathrm{\;,} \\
\hat{B}_2(X) &=& B(x) + \bg{B}(y) + b^i(x)\om_i(y) \label{bexpan} \mathrm{\;,} \\
\hat{C}_3(X) &=& C(x) + \bg{C}(y) + \xi^A(x)\al_A(y) - \widetilde{\xi}_A(x)\beta^A(y) + A^i(x) \wg \om_i(y) \label{cexpan} \mathrm{\;.}
\ea
Note that the $SU(3)$-structure relations (\ref{su3rel}) rule out any component along $z$ in the expansions of $J$ and $\Omn$. Similarly 
$B_2$ can not be expanded along the $z$ direction as it already comes
from a three-form with one leg along $z$, while $C_3$ was assumed not to have
any component along $z$ (\ref{CB}). As was the case in IIA the two-form field $B$ can be massive and so is not dualised to a scalar. 
The field content in four-dimensions is then identical to the IIA case as given in table \ref{iiaN=2multiplets}.

We also find it useful to introduce at this level one more notation. As
we are mostly interested in the scalar fields in the theory we denote all
the fluctuations of $\hat C_3$ which give rise to scalar fields in four
dimensions by $\hat c_3$. Just from its definition we can see that this is a
three-form on the internal manifold. In terms of the expansions above it takes
the form
\begin{equation}
  \label{hatc3}
  \hat c_3(x,y) = b^i(x) (\omega_i \wedge z)(y) + \xi^A(x) \alpha_A(y) - \tilde \xi_A(x) \beta^A(y)
  \; .
\end{equation}

The geometry of the scalar manifold follows in the same was as the IIA case from the K\"ahler potentials
\ba
K^{cs} &=& -\mathrm{ln} \;i\int_{\M_7}{ \Omega^{cs} \wedge \bar \Omega^{cs} \wedge z }\;, \\
K^{km} &=& -\mathrm{ln} \;\frac43 \int_{\M_7}{ J \wg J \wg J \wg z  } \;.
\ea
Since the two-form $B$ is massive the supergravity is electrically and magnetically gauged. 
To derive the prepotentials we work with special co-ordinates in the vector multiplet fields $t^i = b^i - iv^i$, 
and decompose $U^{\pm}$ into 
\be
U^{\pm} = t^i \omega_i \wg z + \tilde{U}^{\pm} \;,
\ee
where now $\tilde{U}^{\pm}$ is a function of only the hypermultiplet fields.
The prepotentials read
\ba
P^1_0 &=& -\sqrt{2}e^{2\phi} \int_{\M_7}{\left[ \left(\re{(d\tilde{U}^+)} + {\cal G}\right) \wg \im{(\tilde{U}^+)} \right]} \nn \;, \\
P^2_0 &=& -\frac{e^{2\phi}}{\sqrt{2}} \int_{\M_7}{\left[ \left(\re{(d\tilde{U}^+)}  
+ 2{\cal G}\right) \wg \re{(\tilde{U}^+)} - \im{(d\tilde{U}^+)} \wg \im{(\tilde{U}^+)} + 2\lambda \right]} \nn \;, \\
P^3_0 &=& \sqrt{2}e^{2\phi} \int_{\M_7}{ \left[\left(\re{(d\tilde{U}^+)}  + {\cal G}\right) \wg 
\left( e^{-\hat{\phi}} \Omn^+  \right)\right] } \nn \;, \\
P^1_i &=& -\sqrt{2}e^{2\phi} \int_{\M_7}{\left[  \im{(d\tilde{U}^+)} \wg \omega_i \wg z  \right]} \nn \;, \\
P^2_i &=& -\sqrt{2}e^{2\phi} \int_{\M_7}{\left[ \left(\re{(d\tilde{U}^+)}  + {\cal G}\right) 
\wg \omega_i \wg z  \right]} \nn \;, \\
P^3_i &=& -\sqrt{2}e^{2\phi} \int_{\M_7}{\left[ d\omega_i \wg \left( e^{-\hat{\phi}} \Omn^+ \right) 
\wg z    \right]} \nn \;, \\
Q^{2i} &=& -\sqrt{2}e^{2\phi} H^k \;,
\ea
where the parameters $H^k$ are defined as 
\be
dz = H^k \omega_k \;. \label{dzhk}
\ee
When $z$ is closed the magnetic gaugings vanish and the two-form field $B$ becomes massless. To 
completely determine the supergravity we need to specify the Killing vectors which are used in gauging the 
vector multiplet scalars. These are not fixed by the gravitini mass matrix and so we calculate them by 
explicit dimensional reduction of the relevant terms. The $\N=2$ action (\ref{n=2act}), after the 
gauging (\ref{gaugez}), contains the term
\be
{\cal L}_{k} = -2g_{i\bar{j}} \partial_{\mu} b^i \left( V^l \right)^{\mu}  k_l^{\bar{j}} \;,
\ee
where we have taken the Killing vectors $k_l^{\bar{j}}$ to be real anticipating the result. Note that since the 
Killing vectors should be holomorphic this is only possible if they are constant with respect to the vector scalars. Dimensionally reducing the third term in (\ref{mtheoryaction}) we find the term in the four-dimensional action 
\be
{\cal L}_{k} = -2 \partial_{\mu} b^i \left( V^l \right)^{\mu} 
\left[ \frac{1}{4} e^{2\phi} \int_{\M_7}{\omega_i \wg z \wg \star d\omega_l } \right] \;,
\ee
from which we determine the Killing vectors
\be
k^{\bar{l}}_{j} = \quarter e^{2\phi} g^{\bar{l}i} \int_{\M_7}{\omega_i \wg z \wg \star d\omega_j }  \;.
\ee
This completely determines the four-dimensional $\N=2$ gauged supergravity that results from compactifications 
of M-theory on manifolds with $SU(3)$-structure. The theory is quite rich with charged vector multiplets and hypermultiplets as well as massive tensor fields. Having specified the theory we move on to considering the vacuum 
structure.

%
\section{Preserving $\N=2$ supersymmetry}
\label{sec:pren=2}
%

In this section we consider the case where the $SU(3)$-structure manifolds are such that the vacuum preserves the full $\N=2$ supersymmetry. 
The case where the vacuum only preserves $\N=1$ supersymmetry is studied in section \ref{sec:pren=1}. We begin by finding the most general solution of 
eleven-dimensional supergravity on manifolds with $SU(3)$-structure that preserves $\N=2$ supersymmetry. We use the results to check the form of the 
gravitini mass matrix (\ref{massmatrix}). We then study an explicit example of a manifold that satisfies the solution and show how the torsion classes 
can be combined with the eleven-dimensional solution to determine the value of the moduli in the vacuum.

\subsection{$\N=2$ solution}
\label{sec:n=2solution}

In this section we classify the most general manifolds with
$SU(3)$-structure that are solutions to M-theory that preserve $\N=2$
supersymmetry with four-dimensional space-time being Einstein and admitting two Killing
spinors. In order to study such solutions in full generality we allow for a
warped product metric
\begin{equation}
  \label{warpedmetric}
  ds^2_{11} = e^{2A(y)}g_{\mu \nu}(x) dx^\mu dx^\nu + g_{mn}(x,y) dy^m dy^n \; ,
\end{equation}
but it turns out that the warp factor, $A(y)$, actually vanishes.  This
class of solutions has also been recently discussed in
\cite{Behrndt:2005im}. We look for solutions to the eleven-dimensional
Killing spinor equation
\begin{equation}
  \nabla_M \eta + \frac{1}{288} \left[ \Gamma_M^{\;\;NPQR} - 8 \delta_M^{[N}
  \Gamma^{PQR]} \right] \hat F_{NPQR} \;\eta = 0 \; .
\end{equation}
For the background field strength $\hat F_{MNPQ}$ above we consider
the most 
general ansatz compatible with four-dimensional Lorentz invariance. Therefore,
the only non-vanishing components of $\hat F$ are $\hat F_{mnpq}$
and $F_{\mu \nu \rho \sigma} = f \epsilon_{\mu \nu \rho \sigma}$.

Given that the internal manifold has $SU(3)$-structure we know that there exist at
least two globally defined Majorana spinors and so we take a Killing spinor
ansatz
\begin{equation}
  \eta = \theta_1(x) \otimes \epsilon_1(y) + \theta_2(x) \otimes \epsilon_2(y)
  \; .
\end{equation}
Since we are looking for a $\N=2$ solution we treat $\theta_1$ and $\theta_2$ as
independent. This leads to more stringent constraints than the $\N=1$
case, where they may be related, which makes finding the most general
solution straightforward. As we are looking for four-dimensional maximally
symmetric spaces, the Killing spinors $\theta_{1,2}$ satisfy
\begin{equation}
  \label{4dks}
  \nabla_{\mu} \theta_{i} = - \frac{i}{2} \Lambda^i_1 \gamma_{\mu} \gamma_5 
 \theta_{i} + \half \Lambda^i_2 \gamma_{\mu} \theta_{i}
  \;\;\;\;\;\mathrm{(no\ sum \ over \ } i) \; ,
\end{equation}
where the index $i=1,2$ labels the two spinors. The integrability condition
reads
\begin{equation}
  \label{einsteinsol} 
  R_{\mu\nu} = -3\left[ \left( \Lambda^i_1 \right)^2 + \left( \Lambda_2^i
  \right)^2 \right]g_{\mu\nu} \; , \quad i=1,2 \; ,
\end{equation}
and so one immediately sees that not all $\Lambda^i_{1,2}$ are independent,
but have to satisfy
\begin{equation}
  \left( \Lambda^1_1 \right)^2 + \left( \Lambda_2^1 \right)^2 = 
  \left(\Lambda^2_2 \right)^2 + \left( \Lambda_2^2 \right)^2 \; .
\end{equation}
Decomposing the Killing spinor equation into its external and internal
parts we arrive at the following equations
\begin{eqnarray}
  \label{internalderiv}
  \nabla_m \epsilon_{1,2} &=& \left( \frac{i}{12}e^{-4A}f\gamma_m
  \right)\epsilon_{1,2} \; ,\\  
  0 &=& \left( \gamma_m^{\;\;npqr}{\hat F}_{npqr} - 8 \gamma^{pqr}
    {\hat F}_{mpqr} \right)\epsilon_{1,2} \; , \\ 
  \left( \frac{i}{2} \Lambda^{1,2}_1 \right)\epsilon_{1,2} &=& 
  \left( \half e^{A} \gamma^n \partial_n A + \frac{i}{6}e^{3A}f
  \right)\epsilon_{1,2} \; , \\ 
  \left( \frac{1}{2} \Lambda^{1,2}_2 \right)\epsilon_{1,2} &=& \left(
  -\frac{1}{288} e^{A} \gamma^{npqr}{\hat F}_{npqr}\right)\epsilon_{1,2} \; .
\end{eqnarray}
In order to classify this solution from the point of view of the $SU(3)$-structure 
we find the corresponding non-vanishing torsion classes by
computing the exterior derivatives of the structure forms. Using their
definition in terms of the spinors \eqref{OJVdef} and applying the results
above we find
\begin{eqnarray}
  \label{nowarping}
  d V &=& \frac{1}{3} f J \; ,\nonumber \\ 
  d J &=& 0 \; ,\\
  d\Omega &=& -\frac{2i}{3} f \Omega \wedge V \; , \nonumber \\
  dA &=& 0 \; . \nn
\end{eqnarray}
The first thing to note is that the warp factor $A$ is constant in this
vacuum and therefore can be set to zero by a constant rescaling of the
metric.
The second thing to observe, comparing with equation \eqref{su3torsion}, is
that only the singlet classes $R$ and $c_2$ are non-vanishing. Moreover, they
are not independent, but proportional to each other as they can both be
expressed in terms of the Freund-Rubin parameter $f$. 

From equations \eqref{internalderiv} we can also determine the
parameters $\Lambda^i_{1,2}$, which give the value of the
cosmological constant, and are given by
\begin{equation}
  \label{fsol}
  \begin{aligned}
    \Lambda^1_1 = & ~ \Lambda^2_1 = \frac{f}{3} \; , \\
    \Lambda^1_2 = &~  \Lambda^2_2 = 0 \; .
  \end{aligned}
\end{equation}

The Killing spinor equations \eqref{internalderiv} also give constraints on
the internal flux that imply it should vanish. However an easier way to see
this is to consider the integral of the external part of the
eleven-dimensional Einstein equation which reads 
\begin{equation}
  \Vol R_{(4)} + \frac{4}{3}\Vol f^2 + \frac{1}{72}\int_{\M_7}{\sqrt{g_7}d^7y \; {\hat F}_{mnpq} 
    {\hat F}^{mnpq}} = 0 \; .
\end{equation}
We see that using (\ref{fsol}) and (\ref{einsteinsol}) we indeed recover ${\hat F}_{mnpq}=0$. Since $f$ is a constant, this means that the Bianchi identity $dF=0$ is indeed 
satisfied and so follow all the equations of motion.

Finally we note that in terms of the two $G_2$-structures $\vp^{\pm}$,
equations \eqref{nowarping} can be recast into a simple form
\begin{equation}
  d \vp^{\pm} = \frac{2}{3}f \star \vp^{\pm} \; ,
\end{equation}
which shows that both $G_2$-structures are in fact weak-$G_2$.

\subsubsection{The mass of the gravitini}
\label{sec:n=2massless}

We can now use this solution to illustrate the discussion on the relation between
the gravitini masses and supersymmetry and to check our form of the mass
matrix. Inserting the solution just derived into the mass matrix we should
find that the masses of the two gravitini degenerate and that they are both
physically massless. Taking the solution \eqref{nowarping} from the previous
section the mass matrix \eqref{massmatrix} reads 
\begin{equation}
  \label{SN2}
  \begin{aligned}
    S_{12} =& ~ 0 \; ,\\
    S_{11} =& ~ S_{22} = \frac{-ife^{\frac{7}{2}\hat{\phi}}}{3\Vol^{\half}}
    \; ,
  \end{aligned}
\end{equation}
which indeed shows that the masses of the two gravitini are the same. 
To show that the two gravitini are physically massless we recall that in AdS
space the physical mass of the gravitino is given by (\ref{mphys32}).
In order to obtain the AdS radius, $l$ in (\ref{mphys32}), correctly normalised we recall that the mass
matrix \eqref{SN2} was obtained in the Einstein frame which differs from the
frame used in the previous section by the Weyl rescaling
\eqref{weylrescaling}. Inserting this into \eqref{einsteinsol} we obtain the
properly normalised AdS inverse radius 
\begin{equation}
  \label{lN2}
  l = \frac{fe^{\frac{7}{2}\hat{\phi}}}{3\Vol^{\half}} \; .
\end{equation}
Note that here, as well as in equation \eqref{SN2}, the fields $\hat \phi$ and
$\mathcal{V}$ should be replaced with their particular values which they have
for this solution. 
Equation \eqref{lN2}, together with \eqref{SN2}, shows that the physical mass
of the gravitini \eqref{mphys32} vanishes confirming our expectations that
the vacuum determined in the previous section does indeed preserve $\N=2$
supersymmetry. 

\subsection{The coset $SO(5)/SO(3)$}
\label{sec:n=2coset}

In order to see the above considerations at work we go through an
explicit example of a manifold that satisfies the $N=2$ solution discussed in
the previous sections. The manifold we consider is the coset space
$SO(5)/SO(3)_{A+B}$. Cosets are particularly useful as examples of structure
manifolds because the spectrum of forms that respect the coset symmetries is
highly constrained. There are more details about cosets in general and about
this particular coset in the appendix. 
In this section we summarise the results and construct a
basis of forms with which we can perform the compactification.
 
We begin by finding the most general symmetric two-tensor that respects the
coset symmetries, this is the metric on the coset and is given by

\be
g = \left( \begin{array}{ccccccc} a  & 0  & 0  & 0 & d  & 0  & 0 \\
    0  & a  & 0  & 0 & 0 & d  & 0 \\
    0  & 0 & a  & 0 & 0 & 0 & d \\
    0  & 0  & 0  & b & 0 & 0  & 0 \\
    d  & 0  & 0  & 0  & c & 0  & 0 \\
    0  & d  & 0  & 0 & 0  & c & 0\\
    0  & 0  & d  & 0 & 0 & 0 & c \\
    \end{array}  \right) \; ,
\ee
where all the parameters are real. The parameters of the metric are the
geometrical moduli and we see that we have four real moduli on this coset.
Note that there is a positivity domain $ac > d^2 $. Having
established the metric on the coset we can move on to find the structure
forms. The strategy here is to find the most general one, two and three-forms
and then impose the $SU(3)$-structure relations on them. It is at this stage
that we really see what the $G$-structure of the coset is.  This analysis is
performed in the appendix and we find that the structure forms are given by
\ba
V &=& e^{\hat{\phi}} z \; , \nn \\
J &=& v \; \omega \; \label{so5struct} , \\
\Omega &=&  \cs_3 \alpha_0 + \cs_4 \alpha_1 + \cs_6 \beta^1 + \cs_7 \beta^0 \; , \nn
\ea 
where 
\ba
e^{\hat{\phi}} &=& \sqrt{b} \;, \nn \\
v &=& \sqrt{ac - d^2} \;,
\ea
and the relations between the $\cs$s and the metric moduli are given in the appendix.
The basis forms satisfy the algebraic relations (\ref{7ddefot}) and the differential relations 
\begin{eqnarray}
  \label{so5basis} 
  dz &=& - \omega \; , \nn \\ 
  d\omega &=& 0 \; , \nn \\ 
  d\alpha_0 &=& z \wedge \alpha_1 \; , \nn \\ 
  d\beta^0 &=& - z \wedge \beta^1 \; , \\ 
  d\alpha_1 &=& 2 z \wedge \beta^1 - 3 z \wedge \alpha_0 \; , \nn \\
  d\beta^1 &=& -2 z \wedge \alpha_1 + 3 z \wedge \beta^0 \; . \nn
\end{eqnarray}
The structure forms (\ref{so5struct}) show that indeed the
coset has $SU(3)$-structure. In terms of the moduli classification we
have been using it has a dilaton, one K\"ahler modulus and one complex
structure modulus\footnote{As is expected form $\N=2$ supergravity the
  parameters $\cs_3$,$\cs_4$,$\cs_6$ and $\cs_7$ describe only two real degrees of freedom.}
thus making up the four degrees of freedom in the metric.
We also show in appendix \ref{cha:cosets} that scalar functions are in general
not compatible with coset symmetries and therefore we conclude that for such
compactifications no warp factor can appear.

\subsubsection{Finding $\N=2$ minima}
\label{sec:coset2structurevacuum}

In this section we consider if the potential which arises from the
compactification on the coset above has a minimum where the moduli
are stabilised. In particular we wish to look for minima that preserve $\N=2$
supersymmetry and correspond to the solution discussed in section
\ref{sec:n=2solution}. As usual, in a bosonic background, the condition for
supersymmetry is the vanishing of the supersymmetry variations of the
fermions. This is precisely what we used in the previous section and thus
a supersymmetric solution should satisfy all the conditions derived there,
and in particular \eqref{nowarping}. Using (\ref{so5basis}) we see that the forms
\eqref{so5struct} obey 
\ba
\label{cosetderivs}
dV &=& -\frac{e^{\hat{\phi}}}{v} J \; , \nn \\
dJ &=& 0\;, \\ 
d \Omega  &=& z \wedge \left[ 
\left( -3 \cs_4 \right) \alpha_0 + 
\left( \cs_3 - 2 \cs_6 \right) \alpha_1 +
\left( 2 \cs_4 - \cs_7 \right) \beta^1 +
\left( 3 \cs_6 \right) \beta^0 \right].  \nn
\ea
Therefore these forms in general do not satisfy the solution constraints
(\ref{nowarping}). Requiring them to match the solution gives a set of
equations for the moduli that exactly determine the value of the moduli
in the vacuum. For the coset at hand the solution is given by
\begin{eqnarray}
  \label{N2sol}
  e^{\hat \phi} & = & \frac{6^{\frac13}\sqrt{42}}{14}
  \lambda^{\frac{1}{6}} \; , \nn \\  
  v & = & \frac{6^{\frac{2}{3}}}{7} \lambda^{\frac{1}{3}} ,\; \\
  \cs_3 &=& -\cs_6 = -i\cs_4 = i\cs_7 = \frac{6}{49}\left(i-1\right) \sqrt{7\lambda}, \; \nn
\end{eqnarray}
where we have replaced the Freund-Rubin flux $f$ by the true flux parameter
$\lambda$ from equation \eqref{ftolambda}.  Note that $\cs$ are not
the true complex structure moduli, but are related to them through \eqref{eqn:omegaCS}. 
However, the complex structure moduli
defined in \eqref{omexpan}, which can be most easily read off in special
coordinates, do not depend on the rescalings of $\Omega$ and therefore, in
our case the value of the single modulus is given by
\be
z^1 = \frac{X^1}{X^0} = \frac{\cs_4}{\cs_3} = i \;.
\ee

It can also be shown that the axionic scalar fields, which
come from the expansion of the three-form $\hat{C}_3$ in the forms
\eqref{so5basis} are also stabilised. A simple argument to support this
statement is that non-vanishing values of the axions leads to a
non-zero internal ${\hat F}_4$ flux at this vacuum solution due to the
non-trivial derivative 
algebra that the basis forms satisfy \eqref{so5basis}, which in turn is ruled out
by the supersymmetry conditions found in section \ref{sec:n=2solution}. Hence,
these scalar fields are forced by supersymmetry to stay at zero vacuum
expectation value and therefore are fixed.

It is also worth observing one more thing regarding this solution. If we think
in terms of the type IIA quantities we see that the K\"ahler modulus $v$ and
the dilaton $e^{\hat \phi}$ are not independent and choosing to stay in the
supergravity approximation on type IIA side, ie take $v \gg 1$, drives
the theory to the strong coupling regime which explains why such solutions
can not be seen in the perturbative type IIA approach.

Finally we note that as the solution above is supersymmetric, the
four-dimensional space-time is AdS with the AdS curvature which scales with
$\lambda$ as
\begin{equation}
  l \sim \frac{1}{\lambda^{\frac16}} \; .
\end{equation}
Thus, in the large volume limit, given by $\lambda \gg 1$, the four-dimensional space-time approaches flat space.

%
\section{The $\N=1$ theory}
\label{sec:pren=1}
%

In this section we analyse the case where the vacuum only preserves $\N=1$
supersymmetry. We show that this occurs due to spontaneous
partial supersymmetry breaking and that it is possible to write an effective
$\N=1$ theory about this vacuum. We derive the K\"ahler potential and
superpotential for this theory and go through an explicit example of a
manifold that leads to this phenomenon.

The partial supersymmetry breaking follows in exactly the same way as the IIA case. In order to derive the 
low energy superfield spectrum we need to restrict to the case where there are no complex structure moduli. In the case of $SU(3)$-structure 
in six dimensions this corresponded to the subset of half-flat manifolds specified by (\ref{hftc}). The equivalent restriction for the seven-dimensional 
case in terms of the torsion classes (\ref{su3torsion}) reads
\begin{equation}
  \label{nocomplex} 
  \begin{aligned}
    \mathrm{Re}(c_1) = &~ V_2 = S_1 = c_2 = W_2 = A_2 = 0 \; ,\\
    \mathrm{Im}(c_1) \neq& ~0 \;.
  \end{aligned}
\end{equation}
Under these conditions the three-form $\Omega^{+}$ is exact and so the absence of complex structure moduli follows from the same arguments as in chapter \ref{cha:su3iia}. 
The condition (\ref{nocomplex}) appears to be quite strong and we have
already come across an example where this is violated in section
\ref{sec:n=2coset}. On the other hand it was shown in \cite{Behrndt:2005im} that an $\N=1$ anti-deSitter vacuum, which
is required for all the moduli to be stabilised, necessarily means
that $J$ is not closed. Hence we always expect at least one of the
torsion classes in (\ref{nocomplex}) to be non-vanishing. Other than
this we must take the condition as a limitation of this work.

We further have to determine the gravitino mass matrix for this situation.
Using (\ref{massmatrix}), (\ref{eqn:4Ddilaton}), (\ref{defv6}) we find that in
the particular case considered above, \eqref{nocomplex}, the gravitino mass
matrix diagonalises. 
\ba
\label{eqn:newSmatrix}
S_{11}&=&\frac{i}{8}\frac{e^{2\phi}}{\sqrt{\Vol_6}}\int _{{\cal M}_7}
[dU^+\wedge U^+ + 2 \mathcal{G} \wedge U^+ + 2\lambda] \; , \nn \\ 
S_{22}&=&\frac{i}{8}\frac{e^{2\phi}}{\sqrt{\Vol_6}}\int _{{\cal M}_7}
[dU^-\wedge U^- + 2 \mathcal{G}\wedge U^- + 2\lambda] \; , \\
S_{12}&=&S_{21}=0 \;, \nn
\ea
where we used the fact that the internal flux $\mathcal {G}$
must be closed to satisfy the Bianchi identity.

The low energy superfields follow in the same way as chapter \ref{cha:su3iia} and we recover a single chiral superfield
\begin{equation}
  \label{msuperu}
  U^{0\pm} \equiv \xi^{0} \pm ie^{-\phi}\left( \frac{-4iZ^0}{F_0}
  \right)^{\half} \; ,
\end{equation}
where the sign $\pm$ is determined by which of the gravitini is massless and
we drop the index unless required for clarity. Recall that the quantity $-4iX^0/F_0$ is a positive real number.

To check that this is indeed the correct superfield we should make sure we recover the moduli space metric from the K\"ahler potential in 
the gravitino mass. The appropriate kinetic terms in (\ref{C11su3}) read 
\ba
S_{kin}^{U} &=&   \int_{\cal S} \sqrt{-g} d^{4}x \bigg[ - \left( \frac{F_0}{-4iZ^0} \right)e^{2\phi} \times \\ \nn 
&~& \;\;\;\;\;\;\;\;\;\;\;\;\;\;\;\;\;\;\;\;  \left. \partial_\mu
\left(\xi^0 + i e^{-\phi} \left( \frac{-4iZ^0}{F_0}\right)^\half \right)
  \partial^\mu \left( \xi^0 - i e^{-\phi} \left( \frac{-4iZ^0}{F_0}\right)^\half \right)  \right].
\ea
We can use (\ref{eqn:newSmatrix}) to read off the K\"ahler potential
\be
\label{11dn=1kahler}
e^{K/2} = \frac{e^{2\phi}}{\sqrt{8\Vol_6}}\;.
\ee
It is then easily shown that indeed the superfield and K\"ahler potential
satisfy  
\be
\partial_{U^0}\partial_{\bar{U^0}} \ln\left[ \frac{e^{4\phi}}{8\Vol_6} \right]
= -\left( \frac{F_0}{-4iZ^0} \right)e^{2\phi}.
\ee
Hence we have identified the correct superfield in the truncated spectrum. The superfields arising from the $\N=2$ vector 
multiplets are the usual 
\be
t^i \equiv b^i - iv^i \label{msupert}\;,
\ee
where the index $i$ now runs over the lower mass fields.

The superpotential for the $\N=1$ theory can be read off from the gravitino
mass to be  
\be
W = \frac{i}{\sqrt{8}} \left\{ \int_{{\cal M}_7} \left[ dU \w U
   + \mathcal {G} \wedge U \right] + 2\lambda  \right\} \label{n=1super} \;.
\ee
From this
expression for the superpotential we can see that we should generically expect
a constant term $\lambda$, linear terms in $U$, quadratic terms $~t^2$, $~U^2$
as well as mixed terms $~tU$. These type of potentials will, in general,
stabilise all the moduli and we go through an example in the next section.

It is instructive to note that finding a supersymmetric solution for this
superpotential automatically solves the equations which are required for a
solution of the full $\N=2$ theory to preserve some supersymmetry.
Therefore, for such a solution, it would be enough to show, using the mass
matrix \eqref{eqn:newSmatrix}, that a mass gap between the two gravitini forms
in order to prove that partial supersymmetry breaking does indeed occur.

%
\section{The coset $SU(3)\times U(1)/U(1)\times U(1)$}
\label{sec:su3u1u1u1}
%

In this section we study an explicit example of a manifold that
preserves $\N=1$ supersymmetry in the vacuum.  The manifold we 
consider is the coset $SU(3) \times U(1) / U(1) \times U(1)$.  
Details of
the structure of the coset can be found in the appendix and in this section we
summarise the relevant parts.  The coset is specified by three integers
$p$,$q$, and $r$ that determine the embeddings of the $U(1) \times U(1)$ in
$SU(3) \times U(1)$, where the integers satisfy
\be
0 \le 3p \le q \; ,
\ee
with all other choices corresponding to different parameterisations of the
$SU(3)$.  As with the previous coset example we can use the coset symmetries
to derive the invariant $SU(3)$-structure forms and the metric.  The metric is
given by
\begin{equation}
  g = \left( \begin{array}{ccccccc} 
      a & 0 & 0  & 0 & 0 & 0 & 0 \\ 
      0 & a & 0  & 0 & 0 & 0 & 0 \\  
      0 & 0 & b  & 0 & 0 & 0 & 0 \\  
      0 & 0 & 0  & b & 0 & 0 & 0 \\  
      0 & 0 & 0  & 0 & c & 0 & 0 \\  
      0 & 0 & 0  & 0 & 0 & c & 0 \\  
      0 & 0 & 0  & 0 & 0 & 0 & d \\  
    \end{array}  \right)\; ,
\end{equation}
where the parameters $a,b,c,d$ are all real. We can write the invariant forms
as  
\ba
\label{coset2su3} 
V &=& \sqrt{d} z \; ,\nonumber \\
J &=& a \omega_1 + b \omega_2 + c \omega_3 \; ,\\
\Omega &=& \sqrt{abc}\left( i \alpha_0 - 4 \beta^0 \right). \nn 
\ea
This basis can be shown to satisfy the following differential relations
\begin{eqnarray}
  \label{coset2diffrel}
  dz & = & m^i \omega_i \; , \nn \\
  d \omega_i & = & e_i \beta^0 \; , \quad d \tilde\omega^i = 0 \; , \\
  d \alpha_0 & = & e_i \tilde \omega^i \; , \quad d \beta^0 =0  \; , \nn
\end{eqnarray}
where we have introduced two vectors $e_i=(2,2,2)$, and $m^i=(\alpha, -\beta,
\gamma), ~ i = 1,2,3$ which encode the information about the metric
fluxes. The quantities $\alpha, ~\beta$ and $\gamma$ are not independent, but
satisfy $\alpha -\beta + \gamma=0$ and in terms of the integers $p$ and $q$ take the form
\begin{eqnarray}
  \alpha &\equiv& \frac{q}{\sqrt{3p^2 + q^2}} \; , \nonumber \\
  \beta &\equiv& \frac{3p+q}{2\sqrt{3p^2 + q^2}}\; ,  \\
  \gamma &\equiv& \frac{3p-q}{2\sqrt{3p^2 + q^2}} \; .\nn
\end{eqnarray}
This ends our summary of the relevant features of the coset. We see that this
manifold indeed has the required torsion classes (\ref{nocomplex}) and, as
expected, has no complex structure moduli and three K\"ahler moduli.

%
\subsection{$\N=1$ supersymmetric minima}
\label{sec:11dsusyminima}
%

As explained in \cite{Castellani:1983tc}, M-theory compactifications on the coset manifold
presented above are expected to preserve $\N=1$ supersymmetry in the
vacuum. Therefore we can use the machinery developed at the beginning of this
section and derive the $\N=1$ theory in the vacuum. For simplicity we turn off the
four-form flux $\mathcal {G}$ and so,
using equations (\ref{11dn=1kahler}) and (\ref{n=1super}) we find the
superpotential and K\"ahler potential to be
\ba
W &=& \frac{1}{\sqrt{8}} \left[ 4 U^0 \left( t^1 + t^2 + t^3 \right) + 2 \alpha t^2 t^3 - 2 \beta t^1 t^3 + 2 \gamma t^1 t^2 + 
2 \lambda \right] ,\\
K &=& - 4 \mathrm{ln} \left[ -i\left( U^0 - \bar{U}^0 \right) \right] - \mathrm{ln} \left[ -i \left( t^1 - \bar{t}^1 \right)
 \left( t^2 - \bar{t}^2 \right)\left( t^3 - \bar{t}^3 \right)\right] \;,
\ea
where the superfields $t^i$ were defined in \eqref{msupert} while for $U^0$ we
have
\begin{equation}
  \label{Udef}
  U^{0} = \xi^0 \pm i e^{-\phi} \; ,
\end{equation}
as (\ref{coset2su3}) gives $-4iZ^0/F_0=1$.
We can look for supersymmetric vacua to this action by solving the F-term
equations. For convenience we restrict to the family of cosets with $p=0$
though the results can be reproduced for more general choices of embeddings.
We find the solution to the F-term equations
\begin{equation}
  \label{n=1sol}
  \frac{t^1}{2} = t^2 = t^3 = U^0 = -i \sqrt{\frac{\lambda}{3}} \; .
\end{equation}

At this point we can go back to check which of the gravitini is more massive.
Inserting the solution (\ref{n=1sol}) into the expression of the mass matrix
(\ref{eqn:newSmatrix}) we obtain
\be
\left|S_{11}\right| > \left|S_{22}\right| \label{eabove}\; , 
\ee
which means $\psi^{2}$ is the lighter gravitino and the one that should be
kept in the truncated theory. This gravitino is physically massless as expected.
This also fixes the $\pm$ sign ambiguity in the
superfield and superpotential so that we have $U^0 \equiv U^{0-}$. Finally we
note that as this solution is a supersymmetric solution of the
truncated $\N=1$ theory and that according to (\ref{eabove}) the
gravitino masses are not degenerate we have indeed encountered the phenomenon
of partial supersymmetry breaking.

We have therefore obtained an explicit model where all the four-dimensional fields are stabilised. It forms an example of an $SU(3)$-structure compactification that 
stabilises all the moduli. We expect the moduli stabilisation property to be generic to $SU(3)$-structure compactifications since all the fields appear 
non-trivially in the mass matrix (\ref{massmatrix}) and we have already come across a 
different type of manifold where the moduli are stabilised in section \ref{sec:n=2coset}.

\subsubsection{The structure in the vacuum}
\label{sec:coset2weakg2}

It is informative to look at the form of the G-structure of the coset in the
vacuum in terms of the $G_2$-structures.  The two $G_2$ forms (\ref{phiOJV})
satisfy the vacuum differential and algebraic relations
\ba
d \vp^{\pm} &=& \sqrt{2} \left( \frac{\lambda}{3} \right)^{\frac{3}{4}} \left[ -8 \beta^0 \w z \pm 2 \om_1 \w \om_2 
 + (\pm 2 + 1 ) \om_2 \w \om_3 \pm 2 \om_1 \w \om_3 \right] , \nonumber \\
\frac{2}{3}f\star \vp^{\pm} &=& \sqrt{2} \left( \frac{\lambda}{3} \right)^{\frac{3}{4}} \left[ \pm 8 \beta^0 \w z - 2 \om_1 \w \om_2 
 - \om_2 \w \om_3 - 2 \om_1 \w \om_3 \right].
\ea
It is clear to see that only $\vp^-$ is weak-$G_2$, and this is indeed the
$G_2$-structure that features in the superpotential and is associated with the
lower mass gravitino. This shows an explicit mass gap appearing between the
two $G_2$-structures which is the same mass gap that corresponds to the
partial supersymmetry breaking which we have used to write an effective $\N=1$
theory. Hence we have shown an example of the idea of an effective G structure
where we could have arrived at this truncated $\N=1$ theory through a
$G_2$-structure compactification even though the manifold actually has
$SU(3)$-structure.  Finally we should note that we could have used the
condition that the manifold should be weak-$G_2$ in the vacuum to solve for
the values of the moduli as we did in section
\ref{sec:coset2structurevacuum} instead of solving the F-term equations.

%
\section{Summary}
\label{sec:11dsummary}
%

In this chapter we studied compactifications of eleven-dimensional supergravity on seven-dimensional manifolds with $SU(3)$-structure. We argued that these 
theories can be naturally cast into a form very similar to type IIA compactifications on six-dimensional manifolds with $SU(3)$-structure as studied in chapter 
\ref{cha:su3iia}. We then derived the resulting four dimensional $\N=2$ supergravity. We studied stable vacua of the supergravity that preserve the full $\N=2$ 
supersymmetry by finding the most general solution to M-theory on manifolds with $SU(3)$-structure that preserves $\N=2$ supersymmetry which, when combined with 
an explicit manifold, lead to a successful moduli stabilisation mechanism. We then used the gravitini mass matrix to study partial supersymmetry breaking and derived 
the resulting effective $\N=1$ theory for a certain class of manifolds. By studying an explicit example of such a manifold the $\N=1$ theory was shown to have 
stable supersymmetric vacua.

\appendix

%
\chapter{Mathematical Tools}
\label{cha:mathstools}
%

In this appendix we review some of the mathematics used in this thesis which also serves to fix the notation 
and conventions. It is not meant as an introduction to the mathematics involved but rather as a collection of 
the relevant ideas and definitions. For a more thorough treatment of the subjects covered we refer the reader to \cite{nakahara,VanProeyen:1999ni}. 

Throughout this work we use a bracket notation for symmetrising and anti-symmetrising indices. The anti-symmetric bracket notation $[...]$ is defined as
\be
M^{\left[\mu_{1}...\mu_{q}\right]} = \frac{1}{q!} \sum_{\sigma \in S_q}{ \mathrm{sgn}(\sigma) M^{ \mu_{\sigma(1)}...\mu_{\sigma(q)}} }  \mathrm{\;,}
\ee
where the sum is performed over the permutations group $S_q$ with $\sigma$ being a particular permutation and 
sgn$(\sigma)$ being its corresponding sign. The symmetric bracket notation $(...)$ is
defined in the same way but without the sgn$(\sigma)$ factor. The notation $(...|...|...)$ is used to omit 
an index range from the symmetry properties. Products of $\Gamma$ matrices or basis forms are taken to be anti-symmetrised
\be
\Gamma_{\mu_1\mu_2...\mu_d} \equiv \Gamma_{[\mu_1}\Gamma_{\mu_2}...\Gamma_{\mu_d]} \mathrm{\;.}
\ee

Since we are interested in both Euclidean and Minkowski signature manifolds a useful quantity to define is $1_{\pm}$ 
which takes the value of $+1$ for a Euclidean signature metric and $-1$ for a Minkowski signature.

The index notation in this work is such that in Minkoswki spaces indices range from $0,...$ while for Euclidean spaces they start from $1$. In the case 
where an index is split into two ranges $M=\left\{\mu,m\right\}$ we generally restart the second index from $1$ again unless otherwise specified.

The convention adopted regarding flat $\hat{\mu}$ and curved $\mu$ indices related by the vielbeins $e^{\hat{\mu}}_{\nu}$ defined so that 
\be
g_{\mu\nu} = e^{\hat{\rho}}_{\mu}e^{\hat{\sigma}}_{\nu} \eta_{\hat{\rho}\hat{\sigma}} \mathrm{\;,}
\ee
is that they are both denoted by $\mu$ to save clattered index expressions. The appropriate indices to use should be apparent from the particular expression 
but a general rule is that they are always curved with the two exceptions being the $\Gamma$ matrices in section \ref{sec:spinors} and the last two indices on the spin 
connection (\ref{spincov}).

%
\section{Riemannian geometry}
\label{sec:reigeo}
%

A Riemannian manifold is a differentiable manifold endowed with a metric that is a symmetric two-tensor $g_{\mu\nu}$ 
where the indices run over the number of dimensions. In this thesis the metric signature is taken to be of the 
Minkowski form $(-,+,+,+,...)$ in four, ten and eleven dimensions and of the Euclidean form $(+,+,...)$ in six and seven dimensions. 
We define connection components on the manifold $\Gamma^{\rho}_{\mu\nu}$ that describe how the basis vectors change throughout the manifold so that a connection $\nabla$ acting on a general vector $V^{\rho}$ and one-form $\omega_{\rho}$ (see section \ref{sec:fibreforms}) is given by
\ba
\nabla_{\mu} V^{\rho} &=& \partial_{\mu} V^{\rho} + \Gamma^{\rho}_{\mu\nu} V^{\nu} \mathrm{\;,} \nn \\
\nabla_{\mu} \omega_{\nu} &=& \partial_{\mu} \omega_{\nu} - \Gamma^{\rho}_{\mu\nu} \omega_{\rho} \mathrm{\;.}
\ea
The torsion $T^{\mu}_{\;\;\nu\rho}$ on the manifold is defined as
\be
T^{\mu}_{\;\;\nu\rho} = \Gamma^{\mu}_{\nu\rho} - \Gamma^{\mu}_{\rho\nu} \mathrm{\;.}
\ee
We can also construct measures of the curvature of the manifold that are the Riemann tensor, the Ricci tensor and the 
Ricci scalar which are related by contraction of indices and given respectively by
\ba
R^{\mu}_{\;\;\nu\rho\sigma} & \equiv & \partial_{\nu}\Gamma^{\mu}_{\rho\sigma} - \partial_{\rho}\Gamma^{\mu}_{\nu\sigma} + 
\Gamma^{\lambda}_{\rho\sigma}\Gamma^{\mu}_{\nu\lambda} - \Gamma^{\lambda}_{\nu\sigma}\Gamma^{\mu}_{\rho\lambda} \mathrm{\;,} \nn \\
R_{\mu\nu} & \equiv & R^{\sigma}_{\mu\sigma\nu} \mathrm{\;,} \nn \\
R &\equiv & g^{\mu\nu}R_{\mu\nu} \mathrm{\;.}
\ea
A connection $\nabla$ is said to be metric compatible if
\be
\nabla_{\mu} g_{\nu\rho} = \partial_{\mu}g_{\nu\rho} - \Gamma_{\mu\nu}^{\lambda}g_{\lambda\rho} - \Gamma_{\mu\rho}^{\lambda}g_{\lambda\nu} = 0 \mathrm{\;.}
\ee
For a metric compatible connection we can decompose the connection coefficients as
\be
\Gamma^{\rho}_{\mu\nu} = \Gamma^{\rho}_{(c)\mu\nu} + \kappa^{\rho}_{\mu\nu} \mathrm{\;,}
\ee
where $\Gamma_{(c)}$ are called Christoffel symbols and are given by
\be
\Gamma^{\rho}_{(c)\mu\nu} = \half g^{\rho\sigma} \left( \partial_{\mu} g_{\sigma\nu} + \partial_{\nu}g_{\sigma\mu} - \partial_{\sigma}g_{\mu\nu} \right) \mathrm{\;.}
\ee
$\kappa$ is called the contorsion and is given in terms of the torsion by
\be
\kappa^{\rho}_{\mu\nu} = \half \left(  T^{\rho}_{\;\;\mu\nu} + T_{\nu\;\;\mu}^{\;\;\rho} + T_{\mu\;\;\nu}^{\;\;\rho} \right) \mathrm{\;.}
\ee
The connection where the contorsion vanishes is called the Levi-Civita connection.
The determinant of the metric $g$ is defined as
\be
g \equiv \mathrm{det} (g_{\mu\nu}) \mathrm{\;.}
\ee
A useful identity involving the metric determinant is
\be
\partial_{\mu} \left( \sqrt{1_{\pm}g} V^{\mu} \right) = \sqrt{1_{\pm}g} \nabla_{\mu} V^{\mu} \mathrm{\;.} 
\label{derivdetide}
\ee
It is often important to consider small variations of the metric of the form $g_{\mu\nu} = g^{0}_{\mu\nu} + \delta g_{\mu\nu}$. Under such variations the relevant quantities vary as 
\ba
\delta g^{\mu\nu} &=& -g^{0\;\mu\rho}g^{0\;\nu\sigma}\delta g_{\rho\sigma} \mathrm{\;,} \\
\delta \sqrt{1_{\pm}g} &=& \half \sqrt{1_{\pm}g} g^{0\;\mu\nu} \delta g_{\mu\nu} \mathrm{\;,} \\
\delta \Gamma^{\rho}_{\mu\nu} &=& \half g^{0\;\rho\sigma}\left( \nabla_{\mu} \delta g_{\nu\rho} 
+ \nabla_{\nu} \delta g_{\mu\rho} - \nabla_{\rho} \delta g_{\mu\nu} \right) \mathrm{\;,} \\
\delta R^{\rho}_{\mu\nu\sigma} &=& \nabla_{\mu} \delta \Gamma^{\rho}_{\nu\sigma} - \nabla_{\nu} \delta \Gamma^{\rho}_{\mu\sigma}  \mathrm{\;.} 
\ea
Rescalings of the metric by a conformal factor are called Weyl rescalings. Under a Weyl rescaling 
\be
g^{\mu\nu} = \Omega^2 \hat{g}^{\mu\nu} \mathrm{\;,}
\ee
the following quantities transform as
\ba
\sqrt{1_{\pm}g} &=& \Omega^{-d}\sqrt{1_{\pm}\hat{g}} \mathrm{\;,} \\
\int{d^dx\;\sqrt{1_{\pm}g}\; \Omega^{d-2} R } &=& \int{d^dx \sqrt{1_{\pm}\hat{g}}\left[\hat{R} 
+ (d-1)(d-2)\left( \frac{\partial \Omega}{\Omn} \right)^2 \right] } \mathrm{\;,}
\ea
where $d$ denotes the number of dimensions. 
Finally a useful quantity is the Levi-Civita tensor (density) $\hat{\epsilon}$ defined in $d$-dimensions as purely anti-symmetric
\be
\hat{\epsilon}_{\mu_1...\mu_d} = \hat{\epsilon}_{\left[\mu_1...\mu_d\right]}  \mathrm{\;,}
\ee
where the indices are raised and lowered by the flat metric $\eta_{\mu\nu}$ and in our conventions
\be
\hat{\epsilon}_{01...d} = +1 \mathrm{\;.}
\ee
The Levi-Civita symbol satisfies the following useful identity 
\be
\hat{\epsilon}^{\mu_1...\mu_p...\mu_d} \hat{\epsilon}_{\nu_1...\nu_p\mu_{p+1}...\mu_d} = 1_{\pm} (d-p)!p! 
\delta^{\left[\mu_1\right.}_{\nu_1}...\delta^{\left. \mu_p \right]}_{\nu_p} \mathrm{\;.}
\ee
It is sometimes useful to define the quantity
\ba
\epsilon_{\mu_1...\mu_d} &=& \sqrt{1_{\pm}g} \;\hat{\epsilon}_{\mu_1...\mu_d} \mathrm{\;,} \nn \\
\epsilon^{\mu_1...\mu_d} &=& \frac{1}{\sqrt{1_{\pm}g}} \;\hat{\epsilon}^{\mu_1...\mu_d} \mathrm{\;,}
\ea
which is now raised and lowered by the full metric and in terms of which the volume form (see section \ref{sec:fibreforms}) $\eta$, which measures the volume of a space, 
is defined as 
\be
\eta \equiv \frac{1}{d!} \epsilon_{\mu_1...\mu_d} dx^{\mu_1...\mu_d} = \sqrt{1_{\pm}g} d^{d}x \mathrm{\;.}
\ee

%
\section{Fibre bundles and differential forms}
\label{sec:fibreforms}
%

A fibre bundle is a manifold $E$ that specified in terms of a base manifold $M$ and 
a \textit{typical fibre} manifold $F$ through a projection $\pi : E \ra M$ such that for each point 
$p \in M$ there exists a neighbourhood $U_i$ and a diffeomorphism $\phi_i$ 
\be
\phi_i : U_i \times F \ra \pi^{-1}(U_i) \mathrm{\;,}
\ee
which is the inverse of the projection 
\be
\pi \circ \phi_{i}(p,f) = p \mathrm{\;,}
\ee
where $f \in F$. The map $\phi_i$ is called the local trivialisation since it maps locally 
the fibre bundle to the direct product of the base and fibre. The structure of the bundle comes from 
the relation between maps on different patches. Consider two overlapping sets on $M$, $U_i \cap U_j \neq \emptyset$, then we require that the transition functions $t_{ij}$ defined as
\be
t_{ij} \equiv \phi^{-1}_{i}(p) \circ \phi_{j}(p) : F \ra F \mathrm{\;,}
\ee 
are smooth and form a group $G$ which is called the structure group of the bundle. The example where the 
transition functions are all unity is the trivial bundle which is just a direct product of the base and the fibre.
A \textit{section} of the fibre bundle $s : M \ra E$ is a smooth map which satisfies
\be
\pi \circ s = \mathrm{id}_M \mathrm{\;,}
\ee
where $id_M$ denotes the identity element on $M$. 

Consider a fibre bundle $E \underrightarrow{\pi} \Reals^k$ where we choose local sections $e_{\alpha}(p)$, $\alpha=1,...,k$ that are 
linearly independent for all $p \in U_i$. Such sections are said to define a frame. Consider another frame $\tilde{e}_{\alpha}$ defined 
on the patch $U_j$ where $p \in U_i \cap U_j$. Then then two frames are related by the transition functions as
\be
\tilde{e}_{\alpha}(p) = \left( t_{ij} \right)^{\;\;\beta}_{\alpha} e_{\beta}(p) \mathrm{\;.}
\ee
Therefore the structure group is the group that transforms between the possible frames.

A vector bundle is a fibre bundle where the typical fibre is a vector space. The tangent bundle $TM$ over a manifold $M$ is a vector bundle 
with fibre $T_p M$ for $p \in M$ with the typical fibre $\Reals^{d}$ (or ${\mathbb{C}}^d$) where $d = \mathrm{dim}(M)$.  For a co-ordinate basis $\{ x^{\alpha} \}$ on a patch 
$U_i \subset M$ we have the frame given by $\{ e_{\alpha} \} = \{ \partial_{\alpha} \}$. The co-tangent bundle $T^{*}M$ is the bundle dual to the tangent bundle which is 
spanned by co-tangent vectors or one-forms $\{ e^{\alpha} \}$ such that $e^{\alpha} e_{\beta} = \delta^{\alpha}_{\beta}$. In terms of co-ordinates the one-forms are given by $\{ dx^{\alpha} \}$. We continue to use this coordinate basis for clarity when dealing with components but the relations derived henceforth hold for general forms in a basis independent way. The wedge product $\wedge$ is defined as 
\be
dx^{\mu_1} \wedge ... \wedge dx^{\mu_q} \equiv  q! \;dx^{\left[\mu_{1}\right.} \otimes ... \otimes dx^{\left.\mu_{q}\right]} \mathrm{\;.}
\ee
Using the wedge product we can construct higher order forms with a $p$-form 
$\omega_p$ defined as
\be
\omega_p \equiv \frac{1}{p!} \omega_{\mu_1...\mu_p}dx^{\mu_1...\mu_p} \mathrm{\;,}
\ee
where we have adopted the notation
\be
dx^{\mu_1...\mu_p} = dx^{\mu_1} \wedge ... \wedge dx^{\mu_p} \mathrm{\;.}
\ee
Note that through its definition a form as anti-symmetric components only.
The space of $p$-forms on a manifold $M$ is denoted $\Lambda^p(M)$. 
The Hodge star $\star$ is a map $\Lambda^p{M} \ra \Lambda^{(d-p)}(M)$ which takes the form
\be
\star \omega_p \equiv \frac{1}{p!(d-p)!}\omega_{\mu_1...\mu_p}\epsilon^{\mu_1...\mu_p}_{
\;\;\;\;\;\;\;\;\;\;\;\mu_{p+1}...\mu_m} dx^{\mu_{p+1}...\mu_m} \mathrm{\;.}
\ee
This gives the relations
\ba
\pi_{p} \wedge \star \omega_{p} &=& \frac{1_{\pm}}{p!} \pi^{\mu_1...\mu_p}\omega_{\mu_1...\mu_p} \eta \mathrm{\;,}\\
\star \star  \omega_p &=& 1_{\pm} (-1)^{p(d-p)} \omega_p \mathrm{\;,} \label{starstar} \\
\star 1 &=& \eta \mathrm{\;.}
\ea
The inner product of two forms $\lrcorner$ is defined as
\be
\omega_p \lrcorner \pi_q \equiv \frac{1}{p!} \omega^{\mu_1...\mu_p} \pi_{\mu_1...\mu_p\mu_{p+1}...\mu_q} dx^{\mu_{p+1}...\mu_{q}} \mathrm{\;.}
\ee
The external derivative $d$ is a map $\Lambda^p(M) \ra \Lambda^{p+1}(M)$ such that
\be
d\omega_p \equiv \frac{1}{p!} \partial_{\nu}\omega_{\mu_1...\mu_p} dx^{\nu\mu_1...\mu_p} \mathrm{\;.}
\ee
It is nilpotent
\be
d d \omega_p = 0 \mathrm{\;,}
\ee
and operates on products as
\be
d\left( \omega_p \pi_q \right) = d\omega_p \pi_q + (-1)^{p}\omega_p d\pi_q  \mathrm{\;.}
\ee
The conjugate of the exterior derivative $d^{\dagger}$ is a map $\Lambda^p(M) \ra \Lambda^{p-1}(M)$ defined as
\be
d^{\dagger} \equiv - 1_{\pm} (-1)^{d(p+1)} \star d \star \mathrm{\;,}
\ee
which acts on forms as
\be
d^{\dagger}\omega_p = \frac{1}{(p-1)!}(-1)^{(p-1)(d-p)} \nabla^{\nu} \omega_{\nu\mu_1...\mu_{p-1}}dx^{\mu_1...\mu_{p-1}} \mathrm{\;.}
\ee
The Laplacian $\Delta$ takes $\Lambda^p(M) \ra \Lambda^{p}(M)$ and is defined as
\be
\Delta \equiv d d^{\dagger} + d^{\dagger}d \mathrm{\;,}
\label{lapdef}
\ee
which operates on forms as
\be
\Delta \omega_p = \frac{1}{(p!)^2} \left[ \nabla^{\nu}\nabla_{\nu} \omega_{\mu_1...\mu_p}
 + \sum_{i}{\left[ \nabla^{\nu}, \nabla_{\mu_i} \right]} \omega_{\mu_1...\nu_{(i)}...\mu_p}     \right]dx^{\mu_1...\mu_p} \mathrm{\;.}
\ee
Forms that satisfy $d\omega=0$ and $d^{\dagger}\omega=0$ are called closed and co-closed respectively.
Forms that satisfy $\Delta\omega = 0$ and are called harmonic. Harmonic forms are closed and co-closed and 
vice-versa. A form that can be written as an exterior derivative of another form $\omega = d\pi$ is called exact. Hodge's decomposition theorem states that every form can be written as
\be
\omega_p = d\alpha_{p-1} + d^{\dagger} \beta_{p+1} + \gamma_p \mathrm{\;,}
\label{hodgetheorem}
\ee
where $\alpha$, $\beta$ and $\gamma$ are forms that are uniquely specified for every $\omega$ and $\gamma$ is harmonic.

The Lie derivative ${\cal L}_X t$ of a general tensor $t^{(p,q)}\in T^p_q$, where $p$ and $q$ denote the number of components in the tangent and co-tangent bundles respectively, is done with respect to a vector $X=X^{\mu}\partial_{\mu}$ and is a map $T^p_q \ra T^p_q$ that takes the following forms. For the case where the tensor is a function $f$ ($p=0,q=0$) we have
\be
{\cal L}_X f = X(f) \mathrm{\;,}
\ee
where $X(f)=X^{\mu}\partial_{\mu}f$.
For a vector $Y$ ($p=1,q=0$) it reads
\be
{\cal L}_X Y = \left[ X,Y \right] \equiv \left[ X^{\mu} \partial_{\mu} Y^{\nu} - Y^{\mu} \partial_{\mu} X^{\nu}\right] \partial_{\nu} \mathrm{\;,}
\ee
where the second term defines the Lie bracket of two vectors.
For a general $p$-form
\be
{\cal L}_X \omega_p = \left( d i_X + i_X d \right) \omega_p \mathrm{\;,}
\ee
where $i_{X}\omega_p$ is the interior product between a vector and a form defined as 
\be
i_X \omega_p \equiv \frac{1}{(p-1)!} X^{\nu} \omega_{\nu\mu_1...\mu_{p-1}} dx^{\mu_1...\mu_{p-1}} \mathrm{\;.}
\ee
Note that sometimes the notation $X \lrcorner \omega_p = i_X \omega_p$ is used in analogy with the inner product of forms.
A Killing vector is a vector whose Lie derivative of the metric vanishes 
\be
{\cal L}_X g_{\mu\nu} = \nabla_{(\mu} X_{\nu)}= 0  \mathrm{\;.}
\ee

%
\section{Homology and co-homology}
\label{sec:homo}
%

A $p$-chain $C_p(M)$ on a manifold $M$ is a formal sum
\be
C_p(M) = \sum_{i}{a_i N_i(M)} \mathrm{\;,}
\ee
where $N_i$ are p-dimensional submanifolds of $M$ and $a_i \in \Reals$. The boundary operator $\partial$ maps 
a $p$-chain to a $p-1$ chain that is its geometrical boundary.  The sets of cycles $Z_p(M)$ and 
boundaries $B_p(M)$ on a manifold $M$ are defined as 
\ba
Z_p(M) &\equiv & \left\{ C_p | \partial C_p = \emptyset \right\} \mathrm{\;,} \\
B_p(M) &\equiv &\left\{ C_p | C_p = \partial d_{p+1} \right\} \mathrm{\;,}
\ea
where $d_{p+1}$ is some $p+1$ chain. The homology group $H_p(M)$ is defined as the set of $p$-cycles that do not differ by a boundary
\be
H_p(M) \equiv \frac{Z_p(M)}{B_p(M)} \mathrm{\;,}
\ee
where the modding out is done through the equivalence relation between two cycles $C_p(M) \sim C_p'(M)$ if $C_p(M) - C_p'(M) \subset B_p(M)$. The dimension of $H_p(M)$ is called the Betti number $b_p(M)$. The 
Euler number $\chi(M)$ is defined as
\be
\chi(M) \equiv \sum_{i=0}^{d}{(-1)^i b_i(M)} \mathrm{\;,}
\ee
and can be thought of as a measure of the number of holes in the manifold.

A relation between a $p$-form $\omega_p$ and a $p$-chain $C_p$ can be induced through the inner product 
\be
\left( \omega_p, C_p \right) \equiv \int_{C_p}{\omega_p} \in \Reals \mathrm{\;.}
\ee
Stokes's theorem states that
\be
\left( C_p,d\omega_p \right) = \left( \partial C_p,\omega_p \right) \mathrm{\;,}
\ee
and can be used to define a correspondence between the homology of the manifold and its co-homology $H^p(M)$ defined as 
\ba
Z^p(M) &\equiv & \left\{ \omega_p | d \omega_p = 0 \right\} \mathrm{\;,} \\
B^p(M) &\equiv &\left\{ \omega_p | \omega_p = d \pi_{p-1} \right\} \mathrm{\;,} \\
H^p(M) &\equiv &\frac{Z^p(M)}{B^p(M)} \mathrm{\;,}
\ea
where forms are split into equivalence classes through $\omega_p \sim \omega_p'$ if $\omega_p - \omega_p' \in B^p(M)$.
Hodge's theorem (\ref{hodgetheorem}) states that each equivalence class in $H^p(M)$ has a unique harmonic representative and so 
$b^p(M)$ measures the number of harmonic forms. We can then identify representative cycles in the homology for the 
harmonic forms through
\be
\int_{C^i}{\omega_j} = \delta^i_j \mathrm{\;.}
\ee
This is analogous to the basis forms and cycles used in CY compactifications.

%
\section{Complex and almost complex manifolds}
\label{sec:compman}
%

Any even-dimensional real manifold $M$ admits locally, for a point $p$, a two-tensor $J^{\;\;m}_{n}$ of type $(1,1)$, i.e. $J \in T_p^*M \otimes T_pM$ , which satisfies
\be
\label{almostcom}
J^{\;\;n}_p J^{\;\;m}_n  = -\delta^m_p \mathrm{\;.}
\ee
If the tensor $J$ is also globally well defined it is called an almost complex structure and the manifold $M$ is called an almost complex manifold. In that case it is possible to define projection operators
\be
\left(P_{\pm}\right)^{\;\;m}_n \equiv \half \left( \delta^m_n \mp i J^{\;\;m}_n \right) \mathrm{\;,}
\ee
that split the tangent and cotangent spaces into $T_pM \ra T_pM^{\pm}$, $T_p^*M \ra  T_p^*M^{\pm}$. Tensors that are in $T_pM^{+}$ and $T_p^*M^{+}$ are called holomorphic and tensors that are in $T_pM^{-}$ and $T_p^*M^{-}$ are called anti-holomorphic. 

The relation between a complex manifold and an almost complex manifold arises from global properties. Consider a vector field ${\cal X}$ that is a set of smoothly connected 
vectors assigned to each point on $M$. If at some point $p$ this vector field is holomorphic $\left.{\cal X}\right|_{p} \in T_pM^{+}$ it need not be so in some other point $p'$ since the 
transition functions may not be holomorphic themselves. The condition that the transition functions are holomorphic is the condition for an almost complex manifold to 
be a complex manifold and is equivalent to the requirement that the Lie bracket of two holomorphic vector fields remains holomorphic. In that case the almost 
complex structure is said to be integrable and the Nijenhuis tensor $N$, defined as
\be
N_{nm}^{\;\;\;\;\;k} = J_m^{\;\;l}\left(\nabla_l J_n^{\;\;k} - \nabla_n J_l^{\;\;k}  \right) - J_n^{\;\;l}\left(\nabla_l J_m^{\;\;k} - \nabla_m J_l^{\;\;k}  \right) \mathrm{\;,} 
\ee
vanishes. Correspondingly if $N=0$ the manifold is said to be complex. This condition can be thought of as a restriction on the torsion on the manifold. For example using the 
decomposition of torsion for six-dimensional manifolds with $SU(3)$-structure (\ref{torsionclasses}), a complex manifold satisfies $\tc{1}=\tc{2}=0$. 

For a complex manifold it is possible to define complex co-ordinates $s^{\alpha},\bar{s}^{\bar{\alpha}}$ where $\alpha$ runs over half the real dimension of the manifold. In 
terms of these co-ordinates the complex structure takes the form
\be
J_{\alpha}^{\;\;\beta} = i\delta_{\alpha}^{\;\;\beta} \;\;,\;\; J_{\bar{\alpha}}^{\;\;\bar{\beta}} = -i\delta_{\bar{\alpha}}^{\;\;\bar{\beta}} \;\;,\;\; 
J_{\alpha}^{\;\;\bar{\beta}} = J_{\bar{\alpha}}^{\;\;\beta} = 0 \mathrm{\;.}
\ee
It is always possible to find an Hermitian metric $g_{\alpha\bar{\beta}}$ on the manifold satisfying 
\be
g_{\bar{\alpha}\bar{\beta}} = 0 \;\;,\;\; g_{\alpha\bar{\beta}} = \overline{g_{\bar{\alpha}\beta}} \mathrm{\;.}
\ee
Forms and vectors can be decomposed in terms of the bases $\left\{ds^{\alpha},d\bar{s}^{\bar{\alpha}}\right\}$ and 
$\left\{\partial_{\alpha},\partial_{\bar{\alpha}}\right\}$ respectively. A form with $p$ holomorphic and $q$ anti-holomorphic components is classified as 
a $(p,q)$ form. 
The construction of homology and cohomology can be extended to the case of complex manifold where now the number of harmonic forms (or cycles) of type $(p,q)$ is 
given by the Hodge number $h^{(p,q)}$ which is the dimension of $H^{(p,q)}$.

An important $(1,1)$ form is the K\"ahler form $J$ defined as
\be
J \equiv i g_{\alpha\bar{\beta}} ds^{\alpha} \wedge d\bar{s}^{\bar{\beta}} \mathrm{\;.}
\ee
If the K\"ahler form is closed, $dJ=0$, the manifold is called a K\"ahler manifold and such manifolds have the property that the metric is given in terms of a 
real function $K(s,\bar{s})$, termed the K\"ahler potential, as
\be
g_{\alpha\bar{\beta}} = \partial_{\alpha}\partial_{\bar{\beta}} K(s,\bar{s}) \mathrm{\;.}
\ee
The only non-vanishing Levi-Civita connections on a K\"ahler manifold are 
\be
\Gamma^{\alpha}_{\beta\gamma} = g^{\alpha\bar{\delta}}\partial_{\beta}g_{\gamma\bar{\delta}} \mathrm{\;,}
\ee
and its conjugate.

%
\section{Spinors}
\label{sec:spinors}
%

In this section we consider spin representations of $SO(1,d-1)$ and $SO(0,d)$ corresponding to rotations in $d$-dimensional 
Minkowski and Euclidean spaces respectively. For simplicity of notation we denote both Minkowski and Euclidean metrics as 
\be
\eta_{\mu\nu} \equiv \mathrm{diag}(1_{\pm},+1,...,+1) \mathrm{\;.} 
\ee
The Clifford algebra is generated by $d$, $2^{k+1} \times 2^{k+1}$ matrices $\Gamma_{\mu}$ satisfying
\be
\left\{ \Gamma_{\mu}, \Gamma_{\nu} \right\} = 2\eta_{\mu\nu} \mathrm{\;,} \label{cliffalg}
\ee
which gives
\be
\left( \Gamma_{\mu} \right)^2 = \eta_{\mu\mu} \;\;\;\;\;\mathrm{(no\;contraction)} \mathrm{\;.}
\ee
$k$ is the rank of the Cartan sub-algebra which is given by 
\be
k = \left\{ \begin{array}{c} \frac{d-2}{2} \mathrm{\;\;for\;d\;even} \\ \frac{d-3}{2} \mathrm{\;\;for\;d\;odd} \end{array} \right. \mathrm{\;.}
\ee
The matrices defined as 
\be
\sigma_{\mu\nu} \equiv \frac{i}{4}\left[ \Gamma_{\mu}, \Gamma_{\nu} \right] \mathrm{\;,}\label{sigmn}
\ee
span a representation of the Lie algebra of the $SO(d)$ groups.
A set of matrices satisfying (\ref{cliffalg}) can always be constructed as follows. We consider first the Euclidean case and define the Pauli $\sigma$ matrices as
\be
\sigma_1 = \left( \begin{array}{cc} 0&1 \\ 1&0 \end{array} \right) \;\;,\;\; \sigma_2 = \left( \begin{array}{cc} 0&-i \\ i&0 \end{array} \right) \;\;,\;\; 
\sigma_3 = \left( \begin{array}{cc} 1&0 \\ 0&-1 \end{array} \right) \mathrm{\;.}
\label{pauli}
\ee
It is possible to construct $\Gamma$ matrices as
\ba
\Gamma_1 &=& \sigma_1 \otimes \one \otimes \one \otimes ... \mathrm{\;,} \nn \\
\Gamma_2 &=& \sigma_2 \otimes \one \otimes \one \otimes ... \mathrm{\;,}\nn \\
\Gamma_3 &=& \sigma_3 \otimes \sigma_1 \otimes \one \otimes ... \mathrm{\;,}\nn \\
\Gamma_4 &=& \sigma_3 \otimes \sigma_2 \otimes \one \otimes ... \mathrm{\;,}\nn \\
\Gamma_5 &=& \sigma_3 \otimes \sigma_3 \otimes \sigma_1 \otimes ... \mathrm{\;,}\\
&...& \mathrm{\;.}
\ea
To generalise this construction to the Minkowski case we set $\Gamma_0 \ra i\Gamma_0$. 
At this point the reader is reminded of the index conventions as set out in the beginning of this appendix under which for Minkowski space 
we would begin the index ranges from $0$.
In even dimensions a complete set of $\Gamma$ matrices $\left\{ \Gamma_{n}\right\}$ 
, where $n=1,...,d+1$, is provided by
\be
\Gamma_{n} \equiv \Gamma_{[\mu_1}\Gamma_{\mu_2}...\Gamma_{\mu_n]} = \Gamma_{\mu_1...\mu_n} \mathrm{\;,} \label{gammadef}
\ee
where the last matrix we write as\footnote{In Minkowski space this reads $\Gamma_{*} = i\left( -i \right)^{\frac{d}{2}} \Gamma_{0}...\Gamma_{d-1}$.}
\be
\Gamma_{*} = \left( -i \right)^{\frac{d}{2}} \Gamma_{1}...\Gamma_{d} \;\;,\;\; \left( \Gamma_{*} \right)^2=1\mathrm{\;.}
\label{gammastar}
\ee
$\Gamma_{*}$ anti-commutes with all the other $\Gamma$s and so can be used as $\Gamma_{d+1}$ in the next odd dimension. 
The matrices defined in (\ref{gammadef}) satisfy the Clifford algebra of (\ref{cliffalg}) and also 
\be
\Gamma_{\mu}^{\dagger} = 1_{\pm} A \Gamma_{\mu} A^{-1} \mathrm{\;,}
\label{gammacom}
\ee
where
\be
A = \left\{ \begin{array}{c} \Gamma_0 \mathrm{\;\;for\;Minkoswki} \\ \one \mathrm{\;\;for\;Euclidean} \end{array} \right. \mathrm{\;.}
\ee
Products of $\Gamma$ matrices are often encountered and the following formulae are useful for manipulating them. The identities can be derived by writing down all the 
terms with the correct symmetry properties (numerical coefficients are determined by the symmetries) and determining the signs by substituting in explicit values for the 
indices. The expressions are valid for all dimensions compatible with the index ranges 
\ba
\Gamma_{\mu_1...\mu_p}\Gamma^{\mu_1...\mu_p} &=& \frac{d!}{(d-p)!} \mathrm{\;,} \nn \\
\Gamma_{\mu}\Gamma_{\nu} &=& \Gamma_{\mu\nu} + \delta_{\mu\nu} \mathrm{\;,} \nn \\
\Gamma_{\mu\nu}\Gamma_{\rho} &=& \Gamma_{\mu\nu\rho} + 2\Gamma_{[\mu}\delta_{\nu]\rho} \mathrm{\;,} \nn \\
\Gamma_{\mu\nu\rho}\Gamma_{\sigma} &=& \Gamma_{\mu\nu\rho\sigma} + 3\Gamma_{[\mu\nu}\delta_{\rho]\sigma} \mathrm{\;,} \nn \\
\Gamma_{\mu\nu}\Gamma^{\rho\sigma} &=& \Gamma_{\mu\nu}^{\;\;\;\;\;\rho\sigma} - 4\delta_{[\mu}^{[\rho}\Gamma_{\nu}^{\;\;\sigma} -2\delta_{\mu\nu}^{\rho\sigma} \mathrm{\;,} \nn \\
\Gamma_{\mu\nu\rho}\Gamma^{\sigma\delta} &=& \Gamma_{\mu\nu\rho}^{\;\;\;\;\;\;\;\sigma\delta} - 6\Gamma_{[\mu\nu}^{\;\;\;\;\;[\sigma}\delta_{\rho]}^{\delta]} 
-6 \Gamma_{[\mu}\delta_{\nu\rho]}^{\sigma\delta} \mathrm{\;.}
\ea

Dirac spinors $\lambda_{\alpha}$ have $2^{k+2}$ real components and transform under $SO(d)$ as
\be
\delta \lambda = i \hat{\epsilon}^{\mu\nu} \sigma_{\mu\nu} \lambda \mathrm{\;,}
\ee
and so form a representation of $SO(d)$. 
The covariant derivative of a spinor is taken with respect to the spin connection $\om_{\mu}^{\;\;\nu\rho}$ so that
\be
\nabla_{\mu}\lambda = \partial_{\mu} \lambda - \quarter \om_{\mu}^{\;\;\nu\rho}\Gamma_{\nu\rho} \lambda \mathrm{\;.}
\label{spincov}
\ee
The most convenient way to discuss different types spinors is for each particular case of dimensions and metric signature. This is because the different types 
of spinors can only exist in particular dimensions and take particular forms for certain signatures. In general a charge conjugation matrix $C$ can 
always be introduced so that 
\be
C^{T} = \pm C \;\;,\;\; \Gamma^T_\mu = \pm C \Gamma_{\mu} C^{-1} \mathrm{\;,}
\ee
where the $\pm$ signs are fixed for particular dimensions and signatures. 
If the number of dimensions also allows a consistent\footnote{The consistency here is that for a spinor $\lambda$ we can define a complex conjugation operator $*$ such that $\lambda^{**}=\lambda$.}
introduction of a complex conjugation $*$ for (Dirac) spinors $\lambda$ then we can define
the Dirac conjugate $\bar{\lambda}$ as
\be
\bar{\lambda} = \lambda^{\dagger} A \mathrm{\;.}
\ee 
We can also define the Majorana conjugate $\lambda^c$ as
\be
\lambda^c = \lambda^T C \mathrm{\;.} 
\ee 
A Majorana spinor is then one whose Majorana conjugate is equal to its Dirac conjugate $\lambda^c = \bar{\lambda}$.
In practice this translates to a reality condition and halves the degrees of freedom in the spinor. 

If the number of dimensions is even then a $\Gamma_*$ exists and we can define Weyl (chiral) spinors by
\be
\lambda_L = \half \left( \one - \Gamma_{*}\right) \lambda \;\;,\;\; \lambda_R = \half \left( \one + \Gamma_{*}\right) \lambda \mathrm{\;,}
\ee
which again have half the number of degrees of freedom. Left handed Weyl spinors have negative chirality and right handed spinors have positive chirality
\be
\Gamma_{*} \lambda_L = -\lambda_L \;\;,\;\; \Gamma_{*} \lambda_R = +\lambda_R \mathrm{\;.}
\ee

%
\subsubsection{Spinors in four Minkowski dimensions}
\label{sec:spin4}
%

The $\Gamma$ matrices are denoted by $\Gamma=\left\{ \gamma_{\mu},\gamma_5 \right\}$ with $\mu=0,...,3$. The general relations (\ref{gammastar}) and (\ref{gammacom}) read 
\ba
\left( \gamma_{\mu} \right)^{\dagger} &=& \gamma_0 \gamma_\mu \gamma_0 \mathrm{\;,} \\
\gamma_5 &=& \frac{i}{4!}\hat{\epsilon}_{\mu\nu\rho\sigma} \gamma^{\mu\nu\rho\sigma} = -i \gamma_0\gamma_1\gamma_2\gamma_3 \mathrm{\;.}
\label{gamma5}
\ea
The definition of $\gamma_5$ can be used to derive useful relations of the type
\be
\gamma_{\mu\rho\sigma} = -i \hat{\epsilon}_{\mu\rho\sigma\delta}\gamma^{\delta}\gamma_5 \mathrm{\;.}
\ee
The form of these relations is obvious up to an overall complex factor in front that can be deduced by putting in particular values for the indices and multiplying 
both sides by the appropriate $\gamma$ matrices.
 
We can choose a representation where the $\gamma_{\mu}$ matrices are all real ($\gamma_5$ imaginary). 
In this representation $C=\gamma_0$ and the Majorana condition becomes a reality condition. Both Majorana and Weyl spinors in four dimensions have four real components.
Weyl spinors in four-dimensions satisfy the useful relations
\ba
\bar{\lambda}_L \gamma^{(n)} \lambda_{L} &=& 0 \mathrm{\;\;for\;n\;even} \;,\nn \\
\bar{\lambda}_R \gamma^{(n)} \lambda_{L} &=& 0 \mathrm{\;\;for\;n\;odd} \;,
\ea
where $\gamma^{(n)}$ denotes $n$ $\gamma$ matrices.

The number of supersymmetry generators (charges) in four dimensions is given by $4\times\N$ so that $\N=1$ supersymmetry in four dimensions is parameterised in terms of one Weyl spinor, $\N=2$ using two Weyl spinors and so on.

%
\subsubsection{Spinors in six Euclidean dimensions}
\label{sec:spin6}
%

The $\Gamma$ matrices are denoted by $\Gamma=\left\{ \gamma_{m}, \gamma_7 \right\}$ with $m=1,..,6$. They satisfy
\ba
\left( \gamma_{m} \right)^{\dagger} &=& \gamma_m \mathrm{\;,} \\
\gamma_7 &=& \frac{i}{6!}\hat{\epsilon}_{mnpqrs} \gamma^{mnpqrs} = i \gamma_1\gamma_2\gamma_3\gamma_4\gamma_5\gamma_6 \mathrm{\;.}
\label{gamma7}
\ea
We choose a representation where all the $\gamma_m$ are imaginary which means Majorana spinors are real and have eight components. 
Weyl spinors in six-dimensions satisfy the relations
\ba
\bar{\lambda}_L \gamma^{(n)} \lambda_{L} &=& 0 \mathrm{\;\;for\;n\;odd} \nn \;,\\
\bar{\lambda}_R \gamma^{(n)} \lambda_{L} &=& 0 \mathrm{\;\;for\;n\;even} \mathrm{\;.} 
\ea

%
\subsubsection{Spinors in seven Euclidean dimensions}
\label{sec:spin7}
%

The $\Gamma$ matrices are denoted by $\Gamma=\left\{ \gamma_{m} \right\}$ with $m=1,..,7$. They satisfy
\ba
\left( \gamma_{m} \right)^{\dagger} &=& \gamma_m \mathrm{\;,} \\
\gamma_1\gamma_2\gamma_3\gamma_4\gamma_5\gamma_6\gamma_7 &=& -i \one \mathrm{\;.}
\label{gamma8}
\ea
We choose a representation where all the $\gamma_m$ are imaginary which means Majorana spinors are real and have eight components. 
Note that we can not define Weyl spinors in seven dimensions.

%
\subsubsection{Spinors in ten Minkowski dimensions}
\label{sec:spin10}
%

The $\Gamma$ matrices are denoted by $\Gamma=\left\{ \Gamma_{M},\Gamma_{11} \right\}$ with $M=0,..,9$. They satisfy
\ba
\left( \Gamma_{M} \right)^{\dagger} &=& \Gamma_0 \Gamma_M \Gamma_0 \mathrm{\;,} \\
\Gamma_{11} &=& \frac{-1}{10!}\hat{\epsilon}_{MNPQRSTUVW} \Gamma^{MNPQRSTUVW} = \Gamma_0...\Gamma_9 \mathrm{\;.}
\label{gamma11}
\ea
We choose a representation where all the $\Gamma_M$ are real which means Majorana spinors are real and have 32 components. In ten dimensions it also possible to define Majorana-Weyl spinors which are real Weyl spinors with 16 components. 

The number of supercharges is given by $16\times\N$ which means $\N$-supersymmetry is parameterised in terms of $\N$ Majorana-Weyl spinors. 

In a compactification the full ten-dimensional manifold ${\cal M}_{10}$ is decomposed into the product of four-dimensional space-time, with co-ordinates $\mu$, and 
some internal manifold $\M_6$ with co-ordinates $m$. Under this decomposition the ten-dimensional $\Gamma$ matrices decompose as
\ba
\Gamma_{\mu} &=& \gamma_{\mu} \otimes \one \mathrm{\;,} \nn \\
\Gamma_{m} &=& \gamma_5 \otimes \gamma_{m} \mathrm{\;,} \nn \\
\Gamma_{11} &=& \gamma_5 \otimes \gamma_7 \mathrm{\;.}
\label{10dgammadec}
\ea
Note that this decomposition is consistent with the reality conditions on the matrices in various dimensions.

%
\subsubsection{Spinors is eleven Minkowski dimensions}
\label{sec:spin11}
%

We denote the $\Gamma$ matrices $\Gamma=\left\{ \Gamma_{M} \right\}$ with $M=0,..,10$. We have
\ba
\left( \Gamma_{M} \right)^{\dagger} &=& \Gamma_0 \Gamma_M \Gamma_0 \mathrm{\;,} \\
\Gamma_0...\Gamma_{10} &=& \one \mathrm{\;.}
\label{gamma12}
\ea
We choose a representation where all the $\Gamma_M$ are real which means Majorana spinors are real and have 32 components. 

There can only be 32 supercharges in eleven dimensions which means supersymmetry is parameterised in terms of a single Majorana spinor.

The decomposition under compactification is
\ba
\Gamma_{\mu} &=& \gamma_{\mu} \otimes \one \mathrm{\;,} \nn \\
\Gamma_{m} &=& \gamma_5 \otimes \gamma_{m} \mathrm{\;.} 
\label{11dgammadec}
\ea

%
\section{Useful identities}
\label{sec:iden}
%

%
\subsection{Six-dimensional $SU(3)$-structure identities}
\label{sec:6dsu3iden}
%

The procedure for calculating the identities presented in this section is to use the fact that we can always go to an orthonormal real frame $\left\{ e^m \right\}$ where the 
$SU(3)$-structure forms take the explicit form
\ba
J &=& e^{12} + e^{34} + e^{56} \mathrm{\;,} \\
\Omn &=& \left( e^{135} - e^{146} - e^{236} - e^{245} \right) + i\left( e^{136} + e^{145} + e^{235} - e^{246} \right) \mathrm{\;.} 
\ea
We can calculate relations explicitly using these forms and since they are tensor relations they will hold in all frames. A more formal way to derive 
the following is by using Fierz identities on the spinor bilinears used to define the forms.
Some of the relations are most neatly written using projectors
\be
\left(P_{\pm}\right)^{\;\;m}_n \equiv \half \left( \delta^m_n \mp i J^{\;\;m}_n \right) \mathrm{\;,}
\ee
in terms of which $\Omn$ is holomorphic
\ba
\left( P_+ \right)_m^{\;\;\;n} \Omn_{npq} &=& \Omn_{mpq} \mathrm{\;,}\\
\left( P_- \right)_m^{\;\;\;n} \Omn_{npq} &=& 0 \mathrm{\;.}
\ea
Some useful $SU(3)$-structure identities read
\ba
\star \Omn &=& -i \Omn \mathrm{\;,} \label{omnsd}\\
\Omn_{[mnp}\Omb_{qrs]} &=& \frac{2i}{5} \epsilon_{mnpqrs} \mathrm{\;,} \\
\Omn_{mnp}\Omb^{qrs} &=& 48 \left( P_+ \right)_{[m}^{\;\;\;[q}  \left( P_+ \right)_{n}^{\;\;r}  \left( P_+ \right)_{p]}^{\;\;\;s]} \mathrm{\;,} \\
\Omn_{mnp}\Omb^{qrp} &=& 16  \left( P_+ \right)_{[m}^{\;\;\;[q}  \left( P_+ \right)_{r]}^{\;\;\;n]} \mathrm{\;,}\\
\Omn_{mnp}\Omb^{qnp} &=& 16  \left( P_+ \right)_{m}^{\;\;\;q} \mathrm{\;,}\\
\Omn_{mnp}\Omb^{mnp} &=& ||\Omn||^2 = 48 \mathrm{\;,}\\
\Omn_{mnp}\Omn^{qrp} &=& 0 \mathrm{\;,} \\
\epsilon_{mnpqrs}J^{pq}J^{rs} &=& 8 J_{mn} \mathrm{\;,}\\
\epsilon_{mnpqrs}J^{rs} &=& 6 J_{[mn}J_{pq]} \mathrm{\;,}\\
\epsilon_{mnpqrs} &=& 15 J_{[mn}J_{pq}J_{rs]} \mathrm{\;,}\\
2\left( P_+ \right)_{(mn)} &=& g_{mn} \mathrm{\;.}
\ea

It is informative to go through an example derivation of the expressions that relate derivatives of the forms to the torsion (\ref{su3rel}).  
Acting with the Levi-Civita connection $\nabla$ on the spinor bi-linear for $J$ is evaluated as
\ba
\nabla_m J_{np} &=& -\frac{i}{4} \kappa_{m}^{\;\;\;rs} \left[ \left( \gamma_{rs}\eta_+ \right)^{\dagger} \gamma_{np} \eta_+ 
+ \eta^{\dagger}_+ \gamma_{np}\gamma_{rs} \eta_+ \right] \nn \\
&=& -\frac{i}{4}\kappa_{m}^{\;\;\;rs} \eta^{\dagger}_+ \left( \gamma_{np}\gamma_{rs} - \gamma_{rs}\gamma_{np} \right) \eta_+ \nn \\
&=& -\frac{i}{2}\kappa_{m}^{\;\;\;rs} \eta^{\dagger}_+ \left( \delta_{rn}\gamma_{sp} - \delta_{sn}\gamma_{rp} + \delta_{sp}\gamma_{rn} - \delta_{rp}\gamma_{sn} \right) \eta_+\nn \\
&=& -2i\kappa_{m[n}^{\;\;\;\;\;\;r}\eta^{\dagger}_+ \gamma_{r|p]}\eta_+ \nn \\
&=& 2\kappa_{m[n}^{\;\;\;\;\;\;r}J_{r|p]} \mathrm{\;.}
\ea
Now using the fact that 
\be
\nabla_{[m}J_{np]} = \partial_{[m}J_{np]} = \frac{1}{3}\left( dJ \right)_{mnp} \mathrm{\;,}
\ee
we recover 
\be
\left( dJ \right)_{mnp} = 6\kappa_{[mn}^{\;\;\;\;\;\;r}J_{r|p]} \mathrm{\;.}
\ee
The same method can be applied to derive all such relations.

%
\subsection{Seven-dimensional $SU(3)$-structure identities}
\label{sec:7dsu3iden}
%

In seven dimensions the $SU(3)$-structure takes the form
\ba
J &=& e^{12} + e^{34} + e^{56} \mathrm{\;,} \\
\Omn &=& \left( e^{135} - e^{146} - e^{236} - e^{245} \right) + i\left( e^{136} + e^{145} + e^{235} - e^{246} \right) \mathrm{\;,} \\
V &=& e^7 \mathrm{\;.}
\ea
We can define projectors that can be thought of as projections along and perpendicular to the direction of the vector
\ba
\left(P_{\pm}\right)^{\;\;m}_n &\equiv & \half \left( \delta^m_n \mp i J^{\;\;m}_n - V_mV^n\right) \mathrm{\;,} \\
\left( P_7 \right)^m_n &\equiv & V_mV^n  \mathrm{\;.}
\ea
In terms of $P_{\pm}$, $\Omn$ is holomorphic. The set of objects $\left\{ \Omn,J,P_{\pm} \right\}$ satisfy the same subset of relations as in six dimensions. 
Some useful relations involving also $V$ read
\ba
\star \Omega^{\pm} &=& \pm \Omega^{\mp} \wedge V  \mathrm{\;,} \\
\star \left( J\wedge V \right) &=& \half J \wedge J \mathrm{\;,} \\
V^t \epsilon_{mnpqrst} &=& 15 J_{[mn}J_{pq}J_{rs]} \mathrm{\;,} \\
J^{rs} \epsilon_{mnpqrst} &=& 30 J_{[mn}J_{pq}V_{t]} \mathrm{\;,} \\
J^{rs} V^t \epsilon_{mnpqrst} &=& 6 J_{[mn}J_{pq]} \mathrm{\;.} 
\ea

%
\subsection{$G2$-structure identities}
\label{sec:g2iden}
%

$G2$-structure in seven dimensions can be mapped to the form
\be
\vp = e^{136} + e^{235} + e^{145} - e^{246} - e^{127} - e^{347} - e^{567} \mathrm{\;,}
\ee
from which it is possible to derive the following identities
\ba
\epsilon_{mnpqrst} &=& 5 \vp_{mnp}\left( \star \vp \right)_{qrst} \mathrm{\;,}\\
\vp_{m}^{\;\;\;pq}\vp^{m}_{\;\;\;ab} &=& \left( \star \vp \right)^{pq}_{\;\;\;\;ab} + 2 \delta^{[pq]}_{ab}  \mathrm{\;,} \label{g2ide1} \\
\vp_{mpq}\vp^{npq} &=& 6\delta_m^n \mathrm{\;,}\\
\vp_{mnp}\vp^{mnp} &=& 42 \mathrm{\;,} \\
9\left( \star \vp \right)^{[pq}_{\;\;\;[ab}\delta^{m]}_{n]} &=& \left( \star \vp \right)^{pqmt}\left( \star \vp \right)_{abnt}
 + \vp^{pqm}\vp_{abn} - 6 \delta^{pqm}_{abn} \mathrm{\;.} \label{g2ide2}
\ea

%
\subsection{'t Hooft symbols identities}
\label{sec:hooftiden}
%

The 't Hooft symbols \cite{'tHooft:1976fv} $\eta^x_{uv}$ have index ranges $x=1,2,3$ and $u=1,2,3,4$. They are defined as
\ba
\eta^x_{uv}&=&\bar\eta^x_{uv}=\epsilon^x_{\;uv} \;,\; \mathrm{if\;} u,v = 1,2,3 \mathrm{\;\;,}\\
\eta^x_{u4}&=&\bar\eta^x_{4u}=\delta^x_u \mathrm{\;.}
\ea
Some useful relations that can be derived from their definition read
\ba
\eta^x_{\mu\nu} &=& \half \hat{\epsilon}_{\mu\nu\rho\sigma} \eta^{x,\rho\sigma} \mathrm{\;,}\\
\bar{\eta}^x_{\mu\nu} &=& -\half \hat{\epsilon}_{\mu\nu\rho\sigma} \bar{\eta}^{x,\rho\sigma} \mathrm{\;,}\\
\eta^x_{\mu\nu}\bar{\eta}^y_{\mu\rho} &=& \eta^{x}_{\mu\rho} \bar{\eta}^y_{\mu\nu} \mathrm{\;,}\\
\eta^x_{\mu\nu}\eta^x_{\rho\sigma} &=& \delta_{\mu\rho}\delta_{\nu\sigma} - \delta_{\mu\sigma}\delta_{\nu\rho} + \hat{\epsilon}_{\mu\nu\rho\sigma} \mathrm{\;,}\\
\hat{\epsilon}^{xyz}\eta^y_{\mu\nu}\eta^z_{\rho\sigma} &=& \delta_{\mu\rho}\eta^{x}_{\nu\sigma} - \delta_{\mu\sigma}\eta^{x}_{\nu\rho} 
- \delta_{\nu\rho}\eta^{x}_{\mu\sigma} + \delta_{\nu\sigma}\eta^{x}_{\mu\rho} \mathrm{\;.}
\ea

%
\chapter{Reduction of the Ricci Scalar in String and M-theory}
\label{cha:riccireduction}
%

In this appendix we derive the kinetic terms for the geometric moduli in both type IIA and eleven-dimensional supergravity by reducing the Ricci scalar on a manifold
of the product type. We begin by deriving the kinetic terms for metric variations of seven-dimensional manifolds with $SU(3)$-structure in eleven-dimensional supergravity.
The result are cast into a form from which it is very easy to extract the kinetic terms for type IIA theory on six-dimensional manifolds with $SU(3)$-structure by integrating 
out the extra dimension. During the derivation we only assume the $SU(3)$-structure relations (\ref{6dsu3alg}) and (\ref{su3rel}) which, 
since the set of forms $\left\{ J,\Omn \right\}$ 
follow the same algebraic relations in both cases, allows us to simply integrate out $V$ from the eleven-dimensional expression to reach the IIA case. 

We consider the eleven-dimensional manifold $\M_{11}$ to take an unwarped product form $\M_{11}=\S \otimes \M_7$.
The eleven-dimensional metric, including the fluctuations, takes the following form
\begin{eqnarray}
  \label{metfluct}
  \hat{g}_{MN}dX^MdX^N & = & {\bar g}_{\mu \nu}(x) dx^\mu dx^\nu + {\bar g}_{mn}(x,y) dy^m
  dy^n \\
  & = & {\bar g}_{\mu \nu}(x) dx^\mu dx^\nu + [\bar g_{mn}^0(y) + {\bar
  h}_{mn}(x,y)] dy^m dy^n \; .\nn
\end{eqnarray}
Direct computation of the eleven-dimensional Ricci scalar gives
\begin{eqnarray}
 & & \int_{\M_{11}}{\sqrt{-g_{11}} d^{11} X \; \half \hat{R}} \\
 &=& \int_{\M_{11}}{\sqrt{-g_{11}} d^{11} X \half \left[{\bar R}_4 + {\bar R}_7 - {\bar g}^{mn} {\bar \Box}_4 {\bar g}_{mn}
  +\left(\frac34 {\bar g}^{mp} {\bar g}^{nq} - \frac14 {\bar g}^{mn} {\bar  g}^{pq} \right) \left( \partial {\bar g}_{mn}
  \right) \left(\partial {\bar g}_{pq} \right) \right]} \nn \\
  & = & \int_{\S}{ \sqrt{-{\bar g}_4} d^4 x} \int_{\M_{7}}{ \sqrt{\bar{g}_7} d^7 y \half \left[ {\bar R}_4 + {\bar R}_7 
  - \frac14 \left({\bar g}^{mp} {\bar g}^{nq} - {\bar g}^{mn} {\bar g}^{pq} \right) \left( \partial {\bar g}_{mn} \right) 
  \left(\partial {\bar g}_{pq} \right) \right]} \; ,\nonumber
\end{eqnarray}
where in the last equation we have performed a partial integration with respect to the four-dimensional integral using (\ref{derivdetide}).
At this point we replace the metric variations with variations of the structure forms. Although eventually
we wish to parameterise the variations in terms of the $SU(3)$-structure forms
at this point it is easier to work with the $G_2$-forms. Using equation (\ref{g2metricvar}) we arrive at
\ba
  \label{R11}
  \int_{\M_{11}}{\sqrt{-g_{11}} d^{11} X \; \hat{R}} = \int_{\S}{\sqrt{-{\bar g}_4} d^4 x} \int_{\M_{7}}{
  \sqrt{\bar g_7}d^7y \; \bigg[} {\bar R}_4 + {\bar R}_7 &-& \frac{1}{12} (\partial
  \bar \varphi)_{mnp} (\partial \bar \varphi)^{mnp}  \nn \\
&+& \left. \frac32 \; \frac{(\partial \bar{\Vol})^2}{\bar{\Vol}^2} \right] \;.
\ea
To reach this expression we used the $G_2$-identities (\ref{g2ide1}) and (\ref{g2ide2}) as well as the expression for the variations
\be
\vp \lrcorner \delta \vp = 3 \Vol^{-1} \delta \Vol \;.
\ee
We also used the fact that only the symmetric part of $\vp_{m}^{\;\;pq} \delta
\vp_{npq}$ contributes to the gauge independent metric variations.  Here
$\bar{\Vol}$ is the volume of the internal manifold as measured with the
metric $\bar g_{mn}$ which thus contains the metric fluctuations.  Note that
because we only consider the lowest KK states, ${\bar R}_4$ is independent of
the internal coordinates and thus its integration produces a factor of the
seven-dimensional volume $\bar{\Vol}$.  In order to put the four-dimensional
action in the standard form we further need to rescale the four dimensional
metric as
\be 
{\bar g}_{\mu \nu} = \frac1{\bar{\Vol}} g_{\mu \nu} \; .
\label{mtheoryweyl}
\ee
Apart from normalising the Einstein-Hilbert term correctly this rescaling
also produces a term which precisely cancels the last term of \eqref{R11}. 
The compactified eleven-dimensional Ricci scalar takes the form
\begin{equation}
  \label{R11fin}
  \int_{\M_{11}}{\sqrt{-g_{11}} d^{11} X \; \hat{R}} = \int_{\S}{\sqrt{-g_4} d^4
  x \Big[ R_4 + \int
  \sqrt{\bar g_7}d^7y \; \big( {\bar R}_7 - \frac{1}{12} (\partial
  \bar \varphi)_{mnp} (\partial \bar \varphi)^{mnp} \big) \Big]} \; .
\end{equation}
At this stage we move back to using the $SU(3)$-structure forms using the translation equation (\ref{phiOJV}). We also move to the string 
frame by rescaling the internal metric
\begin{equation}
  \label{gintstring}
  \bar g_{mn} =  e^{-\frac23 \hat \phi} g_{mn} \; ,
\end{equation}
where the dilaton is defined as in equation (\ref{Vdef}).
Defining the $SU(3)$-structure forms with respect to the metric $g_{mn}$ the
decomposition \eqref{phiOJV} becomes
\begin{equation}
  \bar \vp^\pm = e^{-\hat \phi} (\pm \Omega^- - J\wedge V) \; .
\end{equation}
Before identifying the correct degrees of freedom in four dimensions, as
discussed in section \ref{sec:iiaricciscalar}, we need to take out the K\"ahler
moduli dependence from $\Omn$ and we do this by defining a 'six-dimensional'
volume $\Vol_6$ and the true 'holomorphic' three-form $\Omega^{cs}$ as in
equations (\ref{defv6}) and (\ref{defocs}). With these definitions we have
\begin{equation}
  \label{delphi}
  \partial \bar \vp^\pm = e^{-\phi} \big( \pm \left(\partial \phi\right) e^{\half K_{cs}}\Omega_{cs}^- \pm
  \partial \left( e^{\half K_{cs}}\Omega_{cs}^- \right)-
  \frac{1}{\sqrt{\Vol_6}} \partial J \wedge V)  \;,
\end{equation}
where we have introduced the four-dimensional dilaton 
\be
\label{4Ddilaton}
  \phi \equiv \hat{\phi} - \half \mathrm{ln} \Vol_6 \; .
\ee
It can be checked that the following condition holds
\begin{equation}
  \label{delO}
  \left(\partial \left(e^{\half K_{cs}}\Omega^-_{cs}\right) \right)_{mnp}
  \left(e^{\half K_{cs}}\Omega^-_{cs}\right)^{mnp} = 0 \;,
\end{equation}
and so when we square the expression (\ref{delphi}) there is no mixing between
the various terms.
Substituting (\ref{delphi}) into (\ref{R11fin}) and using the fact that
\be
\left(\partial \Omega^{cs}\right) \lrcorner \left( \Omn^{cs}\right) = 0 \;\;,\;\;\; \left(\partial \Omega^{cs}\right) \lrcorner \left( \partial \Omn^{cs}\right) = 0\;,
\ee
we find
\ba
  \label{mR11su3}
  \int_{\M_{11}}{\sqrt{-g_{11}} d^{11} X \; \half \hat {R}} = \int_{\S}{ \sqrt{-g_4}
  d^4x \Big[} \half {\cal R}_4  &-& \partial_{\mu} \phi \partial^{\mu} \phi \\ &+&
  \half e^{2\phi} \Vol^{-1} \int_{\M_7}{\sqrt{g_7}d^7y \; {\cal R}_7 } \nn \\ 
  &-& e^{-\hat{\phi}} e^{K_{cs}} \int_{\M_{7}}{ \sqrt{g_7}\; d^7y} \; 
  D_{\mu} \Omega^{cs} \lrcorner D^{\mu} \bar{\Omega}^{cs} \nn \\
  &-& \frac{1}{4\Vol_6} 
  e^{-\hat{\phi}} \int_{\M_7} \sqrt{g_7} \; d^7y \; \partial_{\mu}J \lrcorner
  \partial^{\mu}J \Big] \;. \nn
\ea
We have replaced the variations of $\Omega$ with the K\"ahler covariant derivative as discussed in section \ref{sec:iiaricciscalar}.

To reach the IIA expression we simply take the forms $J$ and $\Omn$ to only have six-dimensional dependence and components and use $V=e^{\hat{\phi}}z$ to integrate out the 
unit vector $z$. This directly gives 
\ba
  \label{R11su36d}
  \int_{\M_{11}}{\sqrt{-g_{11}} d^{11} X \; \half \hat {R}} = \int_{\S}{ \sqrt{-g_4}
  d^4x \Big[} \half {\cal R}_4  &-& \partial_{\mu} \phi \partial^{\mu} \phi \\ 
  &+& \half e^{2\phi} \Vol^{-1} \int_{\M_6}{\sqrt{g_6} d^6y \;{\cal R}_6 } \nn \\ 
  &-& e^{K_{cs}} \int_{\M_{6}}{ \sqrt{g_6}\; d^6y} \; D_{\mu} \Omega^{cs} \lrcorner D^{\mu} \bar{\Omega}^{cs} \nn \\
  &-&\frac{1}{4\Vol} \int_{\M_6}{\sqrt{g_6} \; d^6y \; \partial_{\mu}J \lrcorner
  \partial^{\mu}J \Big]} \;. \nn 
\ea
It can be easily checked that the Weyl rescalings (\ref{mtheoryweyl}) and (\ref{gintstring}) are equivalent to the Weyl rescaling to reach the ten-dimensional 
Ricci scalar first, and then performing the Weyl rescalings (\ref{iiaweyl}).

%
\chapter{The IIA Gravitini Mass Matrix}
\label{cha:massiia}
%

In this appendix we derive the four-dimensional gravitini mass matrix through dimensional reduction of the appropriate 
terms in the ten-dimensional action of massive type IIA supergravity (\ref{iia10daction}). 
We consider the ten-dimensional gravitino decomposition (\ref{iiagravitinoansatz})
\be
\hat{\Psi}_\mu = \frac{1}{2\sqrt2} \Vol^{-1/4} \left[ \left( \psi^1_{+\mu} + \psi_{-\mu}^1 \right)\otimes \left(\eta_+ + \eta_- \right)
 -i \left( \psi^2_{+\mu} + \psi_{-\mu}^2 \right) \otimes \left(\eta_+ - \eta_- \right) \right] \;.
\ee 
We proceed to go through each term in (\ref{givegravitinomasses}). 
\newline
\newline
\large{\textbf{The kinetic term}}
\newline
\newline
The ten-dimensional kinetic term for the gravitino induces a four-dimensional mass for the particular index ranges
\be
{\cal L}_1 = - \hat{\bar{\Psi_{\mu}}}\Gamma^{\mu n \nu}D_{n}\hat{\Psi}_{\nu} \; . \label{iiatorterm}
\ee
This term is only non-vanishing when the internal spinors are not covariantly constant and so corresponds to the potential induced 
by the torsion on the manifold. To dimensionally reduce this term we substitute the gravitino ansatz and 
decompose the $\Gamma$ matrices as in (\ref{10dgammadec}). This gives 
\ba
{\cal L}_1 &=& -\frac{1}{8\Vol^{\half}} \left\{ \bar{\psi}^1_{+\mu} \gamma^{\mu\nu} \psi^{1}_{-\nu} \left[ \frac{i}{2} \kappa_{[mnp]}\Omega^{-mnp}\right] \right. \nonumber \\
& &\;\;\;\;\; + \; i\bar{\psi}^1_{+\mu} \gamma^{\mu\nu} \psi^{2}_{-\nu} \left[ -\frac{1}{2} \kappa_{[mnp]}\Omega^{+mnp}\right] \\
& &\;\;\;\;\; + \;  i\bar{\psi}^2_{+\mu} \gamma^{\mu\nu} \psi^{1}_{-\nu} \left[ -\frac{1}{2} \kappa_{[mnp]}\Omega^{+mnp}\right] \nonumber \\
& &\;\;\;\;\; -  \left. \bar{\psi}^{2}_{+\mu} \gamma^{\mu\nu} \psi^{2}_{-\nu} \left[ \frac{i}{2} \kappa_{[mnp]}\Omega^{-mnp}\right] 
+ \mathrm{c.c.}
\;\right\} \;. \nn \label{l1tor} 
\ea
We have acted on the spinors with the derivative and replaced the resulting spinor bi-linears with the corresponding structure forms. We can now use the 
relations (\ref{djtor}) to eliminate the contorsion for differential relations of the structure forms which gives
\ba
{\cal L}_1 &=& -\frac{1}{8\Vol^{\half}} \left\{ 
\bar{\psi}^1_{+\mu} \gamma^{\mu\nu} \psi^{1}_{-\nu} \left[ -\frac{i}{12} \left( dJ \right)_{mnp} \left( \Omega^+ \right)^{mnp} \right] \right. \nonumber \\
& &\;\;\;\;\; + \; \bar{\psi}^2_{+\mu} \gamma^{\mu\nu} \psi^{2}_{-\nu}  \left[ \frac{i}{12} \left( dJ \right)_{mnp} \left( \Omega^+ \right)^{mnp} \right] \nonumber \\
& &\;\;\;\;\; + \;  \bar{\psi}^1_{+\mu} \gamma^{\mu\nu} \psi^{2}_{-\nu} \left[ -\frac{i}{12} \left( dJ \right)_{mnp} \left( \Omega^- \right)^{mnp}\right] \nonumber \\
& &\;\;\;\;\; +  \left.\bar{\psi}^2_{+\mu} \gamma^{\mu\nu} \psi^{1}_{-\nu} \left[ - \frac{i}{12} \left( dJ \right)_{mnp} \left( \Omega^- \right)^{mnp}
  \right] \;  + \mathrm{c.c.} \; \right\} \; . \label{l1diff}
\ea
This concludes the reduction of the kinetic term and we now move on to the flux terms.
\newline
\newline
\large{\textbf{The flux terms}}
\newline
\newline
We begin by reducing the term
\be
{\cal L}_2 = -\half m e^{\frac{5}{4}\hat{\phi}}\hat{\overline{\Psi}}_\mu \Ga^{\mu\nu} \hat{\Psi}_\nu \mathrm{\;.}
\ee
Simple substitution of the gravitino ansatz yields
\be
{\cal L}_2 =- \frac{1}{8\Vol^{\half}} me^{\frac{5}{4}\hat{\phi}} \left[ \bar{\psi}^1_{+\mu} \gamma^{\mu\nu} \psi^{1}_{-\nu} 
+ \bar{\psi}^2_{+\mu} \gamma^{\mu\nu} \psi^{2}_{-\nu} +  \mathrm{c.c.} \right] \mathrm{\;.}
\ee
The rest of the flux terms follow again by simple substitution and $\gamma$ matrix algebra. They read
\ba
{\cal L}_3 &=& \frac{1}{24} e^{-\frac{1}{2}\hat{\phi}} (\hat{F}_3)_{prs} 
 \hat{\overline{\Psi}}^\mu \Ga_{[\mu}\Ga^{prs}\Ga_{\nu]}\Ga_{11}\hat{\Psi}^\nu  \\ 
&=& -\frac{1}{8\Vol^{\half}} e^{-\frac{1}{2}\hat{\phi}} \left\{ 
\bar{\psi}^1_{+\mu} \gamma^{\mu\nu} \psi^{1}_{-\nu} \left[ \half \hat{F}_3 \lrcorner \Omn^+ \right] \right.
+ \bar{\psi}^2_{+\mu} \gamma^{\mu\nu} \psi^{2}_{-\nu}  \left[ -\half  \hat{F}_3 \lrcorner \Omn^+ \right] \nonumber \\
& &\;\;\;\;\; + \;  \bar{\psi}^1_{+\mu} \gamma^{\mu\nu} \psi^{2}_{-\nu} \left[ \half \hat{F}_3 \lrcorner \Omn^- \right] 
+  \left.\bar{\psi}^2_{+\mu} \gamma^{\mu\nu} \psi^{1}_{-\nu} \left[ \half  \hat{F}_3 \lrcorner \Omn^-\right] \; 
+ \mathrm{c.c.} \; \right\} \;, \nn \\
{\cal L}_4 &=&  \quarter m e^{\frac{3}{4}\hat{\phi}} \hat{B}_{pr} 
 \hat{\overline{\Psi}}^\mu \Ga_{[\mu}\Ga^{pr}\Ga_{\nu]}\Ga_{11}\hat{\Psi}^\nu  \\ \nn 
&=& -\frac{1}{8\Vol^{\half}} e^{\frac{3}{4}\hat{\phi}}\left\{ 
\bar{\psi}^1_{+\mu} \gamma^{\mu\nu} \psi^{1}_{-\nu} \left[ im \hat{B} \lrcorner J   \right] 
+ \bar{\psi}^2_{+\mu} \gamma^{\mu\nu} \psi^{2}_{-\nu}  \left[ im \hat{B} \lrcorner J  \right] \; 
+ \mathrm{c.c.} \; \right\},\\
{\cal L}_5 &=&  -\frac{1}{96} e^{\frac{1}{4}\hat{\phi}} (\hat{F}_4)_{prst}  
 \hat{\overline{\Psi}}^\mu \Ga_{[\mu}\Ga^{prst}\Ga_{\nu]}\hat{\Psi}^\nu  \\ 
&=& -\frac{1}{8\Vol^{\half}} e^{\frac{1}{4}\hat{\phi}}\left\{ 
\bar{\psi}^1_{+\mu} \gamma^{\mu\nu} \psi^{1}_{-\nu} \left[ \frac{1}{16}\left( \hat{F}_4 \right)_{mnpq} J^{[mn}J^{pq]} \right] \right. \nn \\
& & \;\;\;\;\;\;\;\;\;\;\;\;\;\;\;\;\;\;
\left. +\; \bar{\psi}^2_{+\mu} \gamma^{\mu\nu} \psi^{2}_{-\nu}  \left[ \frac{1}{16}\left( \hat{F}_4 \right)_{mnpq} J^{[mn}J^{pq]} \right] \; 
+ \mathrm{c.c.} \; \right\} \;,\nn \\
{\cal L}_6 &=&  -\frac{1}{96} e^{\frac{1}{4}\hat{\phi}} (\hat{F}_4)_{\rho\sigma\delta\epsilon}  
 \hat{\overline{\Psi}}^\mu \Ga_{[\mu}\Ga^{\rho\sigma\delta\epsilon}\Ga_{\nu]}\hat{\Psi}^\nu \\ \nn 
&=& -\frac{1}{8\Vol^{\half}} e^{\frac{1}{4}\hat{\phi}}\left\{ 
\bar{\psi}^1_{+\mu} \gamma^{\mu\nu} \psi^{1}_{-\nu} \left[ -\half if \right] 
+ \bar{\psi}^2_{+\mu} \gamma^{\mu\nu} \psi^{2}_{-\nu}  \left[ -\half if \right] \; 
+ \mathrm{c.c.} \; \right\} \;.
\ea
After performing the Weyl rescalings (\ref{iiaweyl}), 
under which $J \ra e^{-\half \hat{\phi}}J$ and $\Omn \ra e^{-\frac34 \hat{\phi}}\Omn$,
the contributions computed above yield the mass matrix (\ref{m1m2d}).

%
\chapter{The M-theory Gravitini Mass Matrix}
\label{cha:massmtheory}
%

In this appendix we derive the four-dimensional gravitini mass matrix
through dimensional reduction of the appropriate terms in the
eleven-dimensional action. We work in terms of the $SU(3)$-structure
quantities as defined in section \ref{sec:constsusy}. We begin by writing the
eleven-dimensional gravitino ansatz (\ref{4dgravitini}) in terms of the
four-dimensional chiral gravitini (\ref{4dgravitini}) and the complex internal
spinors (\ref{xipm}) 
\be
\hat{\Psi}_{\mu} =  \Vol^{-\quarter} \left[ \left( \psi^{1}_{+\mu} + \psi^{1}_{-\mu} \right) \otimes \left( \eta_+ + \eta_- \right)  
 -i \left( \psi^{2}_{+\mu} + \psi^{2}_{-\mu} \right) \otimes \left( \eta_+ - \eta_- \right) \right] \; . \label{su3ans}
\ee
We now go through each term in (\ref{mtheoryaction}) that contributes to the four-dimensional mass matrix. 
\newline
\newline
\large{\textbf{The kinetic term}}
\newline
\newline
We begin with the eleven-dimensional kinetic term which produces a mass term in four dimensions for the particular index range choices
\be
{\cal L}_1 = -\half\bar\Psi_{\mu}\hat\Gamma^{\mu n \nu}\hat D_{n}\Psi_{\nu} \; . \label{torterm}
\ee
To calculate this we use the relation for the covariant derivative acting on the spinors
\be
D_m \eta_{\pm} = \quarter \kappa_{mnp}\gamma^{np}\eta_{\pm} \;, \label{spinorderiv}
\ee 
where $\kappa_{mnp}$ is the contorsion on the internal manifold which is anti-symmetric in its last two indices. 
Inserting (\ref{su3ans}) into (\ref{torterm}) and using (\ref{spinorderiv}) to evaluate the derivative on the spinors as well 
as (\ref{OJVdef}) to replace the spinor bi-linears with the $SU(3)$ forms we arrive at 
\ba
{\cal L}_1 &=& -\frac{1}{2\Vol^{\half}} \left\{ \bar{\psi^{1}}_{+\mu} \gamma^{\mu\nu} \psi^{1}_{-\nu} \left[  \frac{i}{2}\kappa_{[mnp]} 
\left(J \wedge V \right)^{mnp} -\frac{i}{2} \kappa_{[mnp]} \Omega^{-mnp} \right] \right. \nonumber \\
& &\;\;\;\;\; + \; \bar{\psi^{2}}_{+\mu} \gamma^{\mu\nu} \psi^{2}_{-\nu} \left[ \frac{i}{2}\kappa_{[mnp]} 
\left(J \wedge V \right)^{mnp} + \frac{i}{2} \kappa_{[mnp]} \Omega^{-mnp} \right] \\
& &\;\;\;\;\; + \; \bar{\psi^{1}}_{+\mu} \gamma^{\mu\nu} \psi^{2}_{-\nu} \left[ -i \kappa_{m[np]}V^{[n}\delta^{p]m} 
-\frac{i}{2} \kappa_{[mnp]} \Omega^{+mnp} \right] \nonumber \\
& &\;\;\;\;\; +  \left. \bar{\psi^{2}}_{+\mu} \gamma^{\mu\nu} \psi^{1}_{-\nu} \left[ i \kappa_{m[np]}V^{[n}\delta^{p]m} 
-\frac{i}{2} \kappa_{[mnp]} \Omega^{+mnp} \right] \; + \mathrm{c.c.}
\;\right\} \; . \nn \label{ml1tor} 
\ea
Now using the identity
\be
\bar{\psi^{2}}_{+\mu} \gamma^{\mu\nu} \psi^{1}_{-\nu}  = \bar{\psi^{1}}_{+\mu}
\gamma^{\mu\nu} \psi^{2}_{-\nu} \; ,
\ee
we can see that actually the first terms in the third and fourth lines cancel. This can be reasoned from the fact that the mass matrix 
should be symmetric. 
Using (\ref{torel}) we eliminate the contorsion from (\ref{ml1tor}) in favour of differential relations of the structure forms and thus obtain
\ba
{\cal L}_1 &=& -\frac{1}{2\Vol^{\half}} \left\{ \bar{\psi^{1}}_{+\mu} \gamma^{\mu\nu} \psi^{1}_{-\nu} 
\left[ \frac{i}{4} \left(dV\right)_{mn}J^{mn} + \frac{i}{96} \left( d\Omega^- \right)_{mnpq} \left( \star \Omega^- \right)^{mnpq} \right. \right.\nn \\
& & \;\;\;\;\;\;\;\;\;\;\;\;\;\;\;\;\;\;\;\;\;\;\;\;\;\;\;\;\;\;\;\;\;\;\;\;\;\;\;\;\;\;\;\;\;\;\;\;\;\;\;\;\;\;\;\; 
+ \frac{i}{12} \left( dJ \right)_{mnp} \left( \Omega^+ \right)^{mnp} \bigg] \nonumber \\
& &\;\;\;\;\; + \; \bar{\psi^{2}}_{+\mu} \gamma^{\mu\nu} \psi^{2}_{-\nu} \left[ \frac{i}{4} \left(dV\right)_{mn}J^{mn} + \frac{i}{96} \left( d\Omega^- \right)_{mnpq} \left( \star \Omega^- \right)^{mnpq} \right. \nn \\
& & \;\;\;\;\;\;\;\;\;\;\;\;\;\;\;\;\;\;\;\;\;\;\;\;\;\;\;\;\;\;\;\;\;\;\;\;\;\;\;\;\;\;\;\;\;\;\;\;\;\;\;\;\;\;\;\; 
 - \frac{i}{12} \left( dJ \right)_{mnp} \left( \Omega^+ \right)^{mnp} \bigg] \nonumber \\
& &\;\;\;\;\; + \; \bar{\psi^{1}}_{+\mu} \gamma^{\mu\nu} \psi^{2}_{-\nu} \left[ 
- \frac{i}{12} \left( dJ \right)_{mnp} \left( \Omega^- \right)^{mnp}\right] \nonumber \\
& &\;\;\;\;\; +  \left. \bar{\psi^{2}}_{+\mu} \gamma^{\mu\nu} \psi^{1}_{-\nu}
  \left[ - \frac{i}{12} \left( dJ \right)_{mnp} \left( \Omega^- \right)^{mnp}
  \right] \;  + \mathrm{c.c.} \; \right\} \; . \label{ml1diff}
\ea
This concludes the reduction of the kinetic term and we now move on to the flux terms.
\newline
\newline
\large{\textbf{The flux terms}}
\newline
\newline
We begin be reducing the term
\be
{\cal L}_2 = -\frac{1}{16}\bar\Psi^{\mu}\hat\Gamma^{\rho \sigma } \Psi^{\nu} F_{\mu\rho\sigma\nu}\; . \label{flux1term}
\ee
This term arises from the purely external Freud-Rubin flux which we write as in (\ref{ftolambda}). 
Substituting (\ref{su3ans}) into (\ref{flux1term}) and after some gamma matrix algebra we arrive at
\be
\label{l2diff}
{\cal L}_2 = \left[ i\bar{\psi^{1}}_{+\mu} \gamma^{\mu\nu} \psi^{1}_{-\nu} + 
i\bar{\psi^{2}}_{+\mu} \gamma^{\mu\nu} \psi^{2}_{-\nu} + \mathrm{c.c.} \right] \left[ \frac{1}{4\Vol^{\frac{3}{2}}} 
\left( \lambda + \half \int  \hat{c}_3 \wedge F \right)  \right] \; .
\ee
The second flux term reads 
\be
{\cal L}_3 = -\frac{3}{4(12)^2} \bar\Psi_{\mu}\hat\Gamma^{\mu\nu
  lmnp}\Psi_{\nu}F_{lmnp} \; . 
\ee
This is the term from the purely internal flux. Again the reduction is simple
and gives  
\ba
{\cal L}_3 &=&\frac{1}{4(12)^2\Vol^{\half}}  \left\{ \bar{\psi^{1}}_{+\mu} \gamma^{\mu\nu} \psi^{1}_{-\nu} 
\left[ F^{lmnp} \left( J \wedge V - \Omega^- \right)^{rst} \hat{\epsilon}_{lmnprst} \right] \right. \nonumber \\
& &\;\;\;\;\;\;\;\;\;\; + \; \bar{\psi^{2}}_{+\mu} \gamma^{\mu\nu} \psi^{2}_{-\nu} \left[ F^{lmnp} \left( J \wedge V 
+ \Omega^- \right)^{rst} \hat{\epsilon}_{lmnprst} \right] \\
& &\;\;\;\;\;\;\;\;\;\; + \; \bar{\psi^{1}}_{+\mu} \gamma^{\mu\nu} \psi^{2}_{-\nu} \left[ - F^{lmnp} \left( \Omega^+ \right)^{rst} 
\hat{\epsilon}_{lmnprst}\right] \nonumber \\
& &\;\;\;\;\;\;\;\;\;\; +  \left. \bar{\psi^{2}}_{+\mu} \gamma^{\mu\nu} \psi^{1}_{-\nu} \left[ - F^{lmnp} \left( \Omega^+ \right)^{rst} 
\hat{\epsilon}_{lmnprst} \right] \;  + \mathrm{c.c.} \; \right\} \; . \nn
\label{l3diff}  
\ea
Finally we recall that the purely internal flux has a contribution from the
the background flux $\mathcal{G}$, and one which is due to the torsion of the
internal manifold $d \hat{c}_3$, which combine into
\be
F_{lmnp} = {\cal G}_{lmnp} + \left( d\hat{c}_3 \right)_{lmnp} \; .
\ee
After performing the Weyl rescalings \eqref{weylrescaling},
under which $\Omn \ra e^{-\hat{\phi}} \Omn$, $J \ra e^{-\frac23 \hat{\phi}}J$ and $V \ra e^{-\frac13 \hat{\phi}}V$, the contributions computed above yield the following mass terms for the gravitini in four dimensions 
\begin{equation}
  \tilde{S}_{\mathrm{mass}} = \int_{{\cal M}_{11}} \sqrt{-\hat{g}} \left[ {\cal L}_1 +
    {\cal L}_2 + {\cal L}_3 \right] = \int_{{\cal M}_4} \sqrt{-g} \left[
    S_{\alpha\beta} \bar{\psi}^{\alpha}_{+\mu } \gamma^{\mu\nu}
    \psi^{\beta}_{-\nu} + \mathrm{c.c.} \right] \; ,  
\end{equation}
where
\ba
\label{S}
S_{11} &=& -\frac{ie^{\frac{7}{2}\hat{\phi}}}{8\Vol^{\frac{3}{2}}} 
 \left\{ \int_{{\cal M}_7} \left[ e^{-2\hat{\phi}}  d\Omega^- \wedge \Omega^- + e^{-2\hat{\phi}} dV \wedge V \wedge J \wedge J  \right.\right. \nn \\
 & & \hspace{30mm} + 2e^{-2\hat{\phi}}dJ \wedge \Omega^- \wedge V  \nonumber \\
 & & \hspace{30mm} - 2 {\cal G} \wedge \left( \hat{c}_3 + i e^{-\hat{\phi}}\left( \Omega^- - J \wedge V\right) \right)
 - d\hat{c}_3 \wedge \hat{c}_3 \nonumber \\ 
  & & \left. \hspace{30mm} - 2ie^{-\hat{\phi}} d\hat{c}_3 \wedge \left( \Omega^- - J \wedge V\right) \right] 
- 2 \lambda \bigg\} \; , \nonumber \\
S_{22} &=& -\frac{ie^{\frac{7}{2}\hat{\phi}}}{8\Vol^{\frac{3}{2}}} 
 \left\{ \int_{{\cal M}_7} \left[ e^{-2\hat{\phi}}d\Omega^- \wedge \Omega^- + e^{-2\hat{\phi}}dV \wedge V \wedge J \wedge J  \right.\right.\nn \\
 & & \hspace{30mm} - 2e^{-2\hat{\phi}}dJ \wedge \Omega^- \wedge V  \nonumber \\
 & & \hspace{30mm} - 2 {\cal G} \wedge \left( \hat{c}_3 + ie^{-\hat{\phi}} \left( - \Omega^- - J \wedge V\right) \right)
 - d\hat{c}_3 \wedge \hat{c}_3 \nonumber \\ 
  & & \left. \hspace{30mm} - 2ie^{-\hat{\phi}} d\hat{c}_3 \wedge \left( -\Omega^- - J \wedge V\right) \right] 
- 2 \lambda \bigg\}  \; , \nonumber \\
S_{12} = S_{21} &=& -\frac{ie^{\frac{7}{2}\hat{\phi}}}{8\Vol^{\frac{3}{2}}} \int_{{\cal M}_7} \left[ 
 2e^{-2\hat{\phi}}dJ \wedge \Omega^+ \wedge V 
  - 2i e^{-\hat\phi}{\cal G} \wedge \Omega^+
\right. \nn \\ & & \;\;\;\;\;\;\;\;\;\;\;\;\;\;\;\;\;\;\;\;\;\;\;\;\;\;\;\;\;\;\;\;\;\;\;\;\;\;\;\;\;\;\;\;\;\;\;\; \left.  
-\; 2i e^{-\hat\phi}d\hat{c}_3\wedge \Omega^+  \right]. 
\ea
This action can be written in the form (\ref{massmatrix}) using (\ref{eqn:N2field}).

%
\chapter{Coset Manifolds}
\label{cha:cosets}
%

In this appendix we briefly describe the procedure through which we can derive explicit information on the coset such as 
the metric, the $G$-structure forms and the basis forms and their differential relations.

Consider a compact group $G$ with some subgroup $H$ then we can decompose the Lie algebra as $g = h \oplus k$. So the Lie manifold 
${\cal M}_G$ is a fibration of the Lie manifold ${\cal M}_H$ over the base ${\cal M}_K$. The base manifold ${\cal M}_K$ is the coset 
manifold $\frac{G}{H}$. We now follow the discussion in \cite{Mueller-Hoissen:1987cq} and construct a set of Lie valued 
one-forms from elements on the fibre 
$L_y$ at a point $y$ on the coset manifold, which we then expand in terms of the generators of the groups $H$ and $K$
\be
\Theta \equiv L_y^{-1} d L_y \equiv \sigma^a H_a + e^i K_i \; ,
\ee
where the indices run over the number of generators of the subgroup. The forms $e^i$ form the basis forms on the coset manifold and 
we take them to be orthonormal so that 
\be
\int_{\M_K}{e^1\wg...\wg e^d} = 1\;,
\ee 
where $d$ is the dimension of the coset.
The expression  
\be
d\Theta = dL_y^{-1} \w dL_y  = -\Theta \w \Theta \; ,
\ee
gives that the basis forms satisfy the differential relations 
\begin{equation}
  \label{cosetdiff}
  \begin{aligned}
    d\sigma^a = & -\half f^{a}_{\;\;bc}\sigma^b \w \sigma^c -\half
    f^{a}_{\;\;ij}e^i \w e^j \; ,\\ 
    de^i =& -\half f^{i}_{\;\;jk}e^j \w e^k - f^{i}_{\;\;aj}\sigma^a \w e^j \;
    , 
  \end{aligned}
\end{equation}
where $f$ are the structure constants of the group $G$.
These expressions allow us to calculate the differential relations on the coset. The useful property of the coset is that requiring 
$G$-invariance
\be
g L_y = L_{y'} h \; ,
\ee 
where $g \in G$ and $h \in H$, we recover the transformation rules for a basis forms on the coset
\be
e^i(y') K_i = e^i(y) h K_i h^{-1} \;.
\ee
Now $hK_ih^{-1}$ is the (inverse) adjoint action of $h$ on $K_i$, so we can consider the adjoint representation of $H$ and write
\be
hK_ih^{-1} = D_{i}^{\;\;j}\left(h^{-1}\right) K_j \;.
\ee
Then a general $n$-tensor on the coset transforms as
\be
g = g_{i_1...i_n}  e^{i_1} \otimes ... \otimes e^{i_n} \ra g_{j_1...j_n}  D_{i_1}^{\;\;j_1}...D_{i_n}^{\;\;j_n} e^{i_1} \otimes ... \otimes e^{i_n}\;.
\ee
Expanding the elements $D$ in the generators, which in the adjoint are the structure constants
\be
D_{i}^{\;j} = \delta_i^j + \omega^a f^i_{\;aj} \;,
\ee
we find that for the tensor to remain invariant under the group action it must satisfy the relation
\be
f^{j}_{a i_1} g_{j i_2 ... i_n} + ... + f^{j}_{a i_n} g_{i_1 ... j} = 0\; ,
\;\;\; \forall a \; ,\label{cosetsym}
\ee
and should have constant co-efficients $g_{i_1...i_n}$.
This is the expression that restricts the possible forms that respect the
coset symmetries which we can use to solve for the most general one, two or
three-forms on the coset and also the metric. Having quickly derived the
relevant expressions (\ref{cosetdiff}) and (\ref{cosetsym}) we can move on to
consider the particular examples used in this paper. One immediate conclusion
we can draw is that scalar functions must be constant. This is the general result that cosets can
not support warping.

%
\section{$SU(3)/U(1)\times U(1)$}
%

This coset was first studied in \cite{Mueller-Hoissen:1987cq}. The group $SU(3)$ is represented in terms of the 
Gell-Mann matrices $\lambda^A$ with $A=1,..,8$ which satisfy
\be
\left[\lambda_A,\lambda_B \right] = f_{AB}^{\;\;\;C}\lambda_C \;.
\ee
It has two $U(1)$ sub-groups generated by $\lambda_3$ and $\lambda_8$. In terms of the quantities of the previous section $H$ is spanned by 
$\left\{ \lambda_3,\lambda_8\right\}$ and $K$ is spanned by $\left\{ \lambda_1,\lambda_2,\lambda_4,\lambda_5,\lambda_6,\lambda_7 \right\}$.
Applying the constraint (\ref{cosetsym}) we find that the most general symmetric two-tensor, which we interpret as the metric, on the coset must take the form
\be
g = a (e^1 \otimes e^1 +  e^2 \otimes e^2 ) + b (e^3 \otimes e^3 +  e^4
\otimes e^4 ) + c (e^5 \otimes e^5 +  e^6 \otimes e^6 ) \;,
\ee
where $a$, $b$ and $c$ are real parameters which are the metric degrees of freedom or the geometrical moduli. Similarly the most general two and three-forms read 
\ba
\Psi_2 &=& \cs_1 e^{12} + \cs_2 e^{34} + \cs_3 e^{56} \; , \\
\Psi_3 &=& \cs_4 \left( e^{135} + e^{146} - e^{236} + e^{245} \right) + 
+ \cs_5 \left( e^{136} - e^{145} + e^{235} + e^{246} \right) \;,
\ea
where all the parameters are complex. There are no consistent one-forms present. 
Imposing the six-dimensional $SU(3)$-structure relations (\ref{6dsu3alg}), we arrive at the expressions (\ref{jandomegadecompose}) where the basis forms explicitly read
\ba
\omega_1 &\equiv& -e^{12} \;,\;\; \omega_2 \equiv e^{34} \;,\;\; \omega_3 = -
e^{56} \;, \\ 
\tilde{\omega}^1 &\equiv& -e^{3456} \;,\;\; \tilde{\omega}^2 \equiv e^{1256}
\;,\;\; \tilde{\omega}^3 = - e^{1234} \; , \nn \\ 
\alpha_0 &\equiv& \left( -e^{136} +e^{145} -e^{235} -e^{246} \right) \;,\;\;
\beta^0 \equiv -\quarter \left( e^{135} + e^{146} - e^{236} + e^{245}\right)
\;.
\ea 
The differential relations on these basis forms are derived from 
\ba
de^1 &=& - \half e^{36} + \half e^{45}\; , \nn \\
de^2 &=& - \half e^{35} - \half e^{46} \; , \nn \\ 
de^3 &=& \half e^{25} + \half e^{16} \; , \nn \\ 
de^4 &=& - \half e^{15} + \half e^{26}  \; , \\ 
de^5 &=& \half e^{14} - \half e^{24}  \; , \nn\\
de^6 &=& - \half e^{13} - \half e^{24}  \;. \nn
\ea
These then give the differential relations (\ref{coset2diffrel}).

%
\section{$SO(5)/SO(3)$}
%

This coset was first studied in \cite{Castellani:1983yg}.
The group $SO(5)$ has two commuting $SO(3)$ subgroups. Hence there are a number of ways to mod out the $SO(3)$ 
and we consider the case where the subgroup $H$ is taken to be a linear combination of the 
two $SO(3)$s\footnote{For more details on this process see \cite{Karthauser:2006wb} where the case we study is denoted $SO(5)/SO(3)_{A+B}$.}. 
Then by calculating the structure 
constants and imposing (\ref{cosetsym}) we find that the most general symmetric two tensor on the coset must take the form
\ba
g & = & a (e^1 \otimes e^1 +  e^2 \otimes e^2 + e^3 \otimes e^3) + b e^4
\otimes e^4 + c (e^5 \otimes e^5 + e^6 \otimes e^6 + e^7 \otimes e^7) \nn \\ 
&& + 2 d(e^{(1} \otimes e^{5)} + e^{(2} \otimes e^{6)} + e^{(3} \otimes
e^{7)}) \; ,
\ea
where all the parameters are real. Similarly, the most general one, two and
three-forms are  
\ba
\label{cosetgenforms}
\Psi_1 &=& \cs_1 e^4 \;, \nn \\
\Psi_2 &=& \cs_2 \left( e^{15} + e^{26} + e^{37} \right) \; , \\
\Psi_3 &=& \cs_3 e^{123} + \cs_4 \left( e^{127} - e^{136} + e^{235} \right) +
 \cs_5 \left( e^{145} + e^{246} + e^{347} \right) \nn \\
  && \hspace{-.27cm} + \cs_7 e^{567} + \cs_6 \left( e^{167} - e^{257} + e^{356} \right) \; , \nn
\ea
where all the parameters can be complex. The structure forms $V$, $J$ and $\Omega$ must fall within the restrictions 
of (\ref{cosetgenforms}) and they can be uniquely determined by imposing the algebraic $SU(3)$-structure relations on the forms 
in (\ref{su3rel}). This leads to equations relating the complex parameters to the real metric moduli,
if we identify $\Psi_1$ with $V$, $\Psi_2$ with $J$, $\Psi_3$ with $\Omega$, we have
\ba
\cs_1 &=& \sqrt{b} \label{metricmodulisolution} \; , \nn\\ 
\cs_2 &=& \left( ac - d^2 \right)^{\half} \; , \nn\\
\cs_3 &=& \frac{\cs_6}{a^2} \left( d + i\left( ac - d^2 \right)^{\half}
\right)^2  \; , \nn  \\
\cs_4 &=& \frac{\cs_6 a}{\left( d + i\left( ac - d^2 \right)^{\half} \right)}
\; ,  \\
\cs_5 &=& 0 \; , \nn \\ 
\cs_6 &=& \frac{2\left( ac - d^2 \right)^{\half} a \sqrt{c} }{a + ic}  \; ,
\nn \\ 
\cs_7 &=& \frac{\cs_6 c}{\left( d - i\left( ac - d^2 \right)^{\half} \right)}
\; . \nn
\ea
Equations (\ref{metricmodulisolution}) give the form of $V$, $J$ and $\Omega$
and we see that the natural basis of forms on the manifold is 
\ba
z &\equiv& e^4 \; , \nn\\
\omega &\equiv& \left( e^{15} + e^{26} + e^{37} \right) \;, \nn\\ 
\alpha_0 &\equiv& e^{123} \;\;\; \beta^0 \equiv e^{567} \;, \nn\\ 
\alpha_1 &\equiv& \left( e^{127} - e^{136} + e^{235} \right) \;, \nn \\ 
\beta^1 &\equiv& \left( e^{167} - e^{257} + e^{356} \right)  \;, 
\ea 
in terms of which we can write the forms as given in equation (\ref{so5struct}). The differential relations on the coset basis forms 
can be calculated using (\ref{cosetdiff}) and are given by 
\ba
d\sigma^1 &=& - \sigma^{23} - e^{23} - e^{67} \; , \nn \\ 
d\sigma^2 &=& \sigma^{13} + e^{13} + e^{57} \; , \nn \\ 
d\sigma^3 &=& - \sigma^{12} - e^{12} - e^{56} \; , \nn \\ 
de^1 &=& -\sigma^2 e^3 + \sigma^3 e^2 + e^{45} \; , \nn\\ 
de^2 &=& \sigma^1 e^3 - \sigma^3 e^1 + e^{46} \; ,\\ 
de^3 &=& -\sigma^1 e^2 + \sigma^2 e^1 + e^{47} \; , \nn\\
de^4 &=& -e^{15} - e^{26} - e^{37} \; , \nn\\
de^5 &=& -\sigma^2 e^7 + \sigma^3 e^6 + e^{14} \; , \nn \\
de^6 &=& \sigma^1 e^7 - \sigma^3 e^5 + e^{24} \; , \nn \\ 
de^7 &=& -\sigma^1 e^6 + \sigma^2 e^5 + e^{34} \; , \nn 
\ea
From these expressions it is easy to calculate the basis form differential relations (\ref{so5basis}).

%
\section{$SU(3)\times U(1)/U(1)\times U(1)$}
%

This coset was first studied in \cite{Castellani:1983tc}.
In this case we have $G=SU(3)\times U(1)$. Now $U(1) \times U(1) \subset SU(3)$ so once we modded out by the $U(1) \times U(1)$ we will 
be left with a single $U(1)$ that is in general a linear combination of the three $U(1)$s in $G$ which we parameterise by 
three integers $p$,$q$ and $r$ \footnote{The case where $p=q=0$ is the trivial fibration case where the coset becomes 
$\left[ SU(3)/U(1) \times U(1) \right]\times U(1)$. In that case this is the same as compactifying type IIA supergravity on the manifold
$SU(3)/U(1) \times U(1)$.}. We can repeat the analysis in the previous section and we find 
\ba
\label{cosetgenform}
g &=& a (e^1 \otimes e^1 +  e^2 \otimes e^2 ) + b (e^3 \otimes e^3 +  e^4
\otimes e^4 ) + c (e^5 \otimes e^5 +  e^6 \otimes e^6 ) + d e^7 \otimes e^7 \;
, \nonumber \\ 
\Psi_1 &=& \cs_1 e^7 \; , \nn \\
\Psi_2 &=& \cs_2 e^{12} + \cs_3 e^{34} + \cs_4 e^{56} \; , \\
\Psi_3 &=& \cs_5 \left( e^{135} + e^{146} - e^{236} + e^{245} \right) + 
\cs_6 \left( e^{136} - e^{145} + e^{235} + e^{246} \right) \;. \nn
\ea
Imposing the $SU(3)$ relations we arrive at equation (\ref{coset2su3}) where the basis forms explicitly read
\ba
z &\equiv& e^7 \; , \nn \\
\omega_1 &\equiv& -e^{12} \;,\;\; \omega_2 \equiv e^{34} \;,\;\; \omega_3 = -
e^{56} \; , \\ 
\tilde{\omega}^1 &\equiv& -e^{3456} \;,\;\; \tilde{\omega}^2 \equiv e^{1256}
\;,\;\; \tilde{\omega}^3 = - e^{1234} \; , \nn \\ 
\alpha_0 &\equiv& \left( -e^{136} +e^{145} -e^{235} -e^{246} \right) \;\;\;
\beta^0 \equiv -\quarter \left( e^{135} + e^{146} - e^{236} + e^{245}\right)
\;. \nn
\ea 
The differential relations on these basis forms are derived from 
\ba
de^1 &=& \alpha e^{72} - \half e^{36} + \half e^{45}\; , \nn \\
de^2 &=& \alpha e^{17} - \half e^{35} - \half e^{46} \; , \nn \\ 
de^3 &=& \beta e^{74} + \half e^{25} + \half e^{16} \; , \nn \\ 
de^4 &=& \beta e^{37} - \half e^{15} + \half e^{26}  \; , \\ 
de^5 &=& -\gamma e^{67} + \half e^{14} - \half e^{24}  \; , \nn\\
de^6 &=& \gamma e^{57} - \half e^{13} - \half e^{24}  \; , \nn\\ 
de^7 &=& -\alpha e^{12} - \beta e^{34} - \gamma e^{56} \; . \nn
\ea
These then give the differential relations (\ref{coset2diffrel}) where we have defined the structure constants
\ba
\alpha &\equiv& f^7_{\;\;12} = \frac{q}{\sqrt{3p^2 + q^2}} \; , \nonumber \\
\beta &\equiv& f^7_{\;\;34} = \frac{3p+q}{2\sqrt{3p^2 + q^2}} \; , \\
\gamma &\equiv& f^7_{\;\;56} = \frac{3p-q}{2\sqrt{3p^2 + q^2}} \; . \nn
\ea

\bibliographystyle{style.bst}
\bibliography{thesis}

\begin{thebibliography}{100}

\bibitem{Eidelman:2004wy}
S.~Eidelman et~al.
\newblock Review of particle physics.
\newblock {\em Phys. Lett.}, B592:1, {\tt 2004}.

\bibitem{colman-mandula}
S.~Coleman and J.~Mandula.
\newblock All Possible Symmetries of the {S} Matrix.
\newblock {\em Phys. Rev.}, 159:5, {\tt 1967}.

\bibitem{wess-bagger}
J.~Wess and J.~Bagger.
\newblock {\em Supersymmetry and {S}upergravity}.
\newblock Princeton University Press, second edition, 1992.

\bibitem{polchinski}
J.~Polchinski.
\newblock {\em String {T}heory. 2 volumes.}
\newblock Cambridge University Press, 2005.

\bibitem{Polchinski:1995mt}
Joseph Polchinski.
\newblock Dirichlet-{B}ranes and {R}amond-{R}amond Charges.
\newblock {\em Phys. Rev. Lett.}, 75:4724--4727, 1995, {\tt hep-th/9510017}.

\bibitem{Witten:1995ex}
Edward Witten.
\newblock String theory dynamics in various dimensions.
\newblock {\em Nucl. Phys.}, B443:85--126, 1995, {\tt hep-th/9503124}.

\bibitem{Duff:1986hr}
M.~J. Duff, B.~E.~W. Nilsson, and C.~N. Pope.
\newblock KALUZA-{K}LEIN SUPERGRAVITY.
\newblock {\em Phys. Rept.}, 130:1--142, {\tt 1986}.

\bibitem{kaluza}
T.~Kaluza.
\newblock On the problem of unity in physics.
\newblock {\em Math. Phys.}, K1, {\tt 1921}.

\bibitem{klein}
O.~Klein.
\newblock Quantum theory and five-dimensional theory of relativity.
\newblock {\em Z. Phys.}, 37, {\tt 1926}.

\bibitem{Gauntlett:2002sc}
Jerome~P. Gauntlett, Dario Martelli, Stathis Pakis, and Daniel Waldram.
\newblock {G}-structures and wrapped {NS}5-branes.
\newblock {\em Commun. Math. Phys.}, 247:421--445, 2004, {\tt hep-th/0205050}.

\bibitem{Gauntlett:2003cy}
Jerome~P. Gauntlett, Dario Martelli, and Daniel Waldram.
\newblock Superstrings with intrinsic torsion.
\newblock {\em Phys. Rev.}, D69:086002, 2004, {\tt hep-th/0302158}.

\bibitem{Dall'Agata:2003ir}
Gianguido Dall'Agata and Nikolaos Prezas.
\newblock {N} = 1 geometries for {M}-theory and type {IIA} strings with fluxes.
\newblock {\em Phys. Rev.}, D69:066004, 2004, {\tt hep-th/0311146}.

\bibitem{Behrndt:2005im}
Klaus Behrndt, Mirjam Cvetic, and Tao Liu.
\newblock Classification of supersymmetric flux vacua in {M} theory.
\newblock 2005, {\tt hep-th/0512032}.

\bibitem{Gurrieri:2003st}
Sebastien Gurrieri.
\newblock {N} = 2 and {N} = 4 supergravities as compactifications from string
  theories in 10 dimensions.
\newblock 2003, {\tt hep-th/0408044}.

\bibitem{joyce}
D.~Joyce.
\newblock {\em Compact Manifolds with Special Holonomy}.
\newblock Oxford University Press, 2000.

\bibitem{Friedrich:1995dp}
Thomas Friedrich, Ines Kath, Andrei Moroianu, and Uwe Semmelmann.
\newblock {\tt On nearly parallel {G}(2) structures}.
\newblock SFB-288-162.

\bibitem{deWit:1984px}
B.~de~Wit, P.~G. Lauwers, and Antoine Van~Proeyen.
\newblock LAGRANGIANS OF {N}=2 SUPERGRAVITY - MATTER SYSTEMS.
\newblock {\em Nucl. Phys.}, B255:569, {\tt 1985}.

\bibitem{D'Auria:1990fj}
Riccardo D'Auria, Sergio Ferrara, and Pietro Fre.
\newblock Special and quaternionic isometries: General couplings in {N}=2
  supergravity and the scalar potential.
\newblock {\em Nucl. Phys.}, B359:705--740, {\tt 1991}.

\bibitem{Andrianopoli:1996cm}
L.~Andrianopoli et~al.
\newblock {N} = 2 supergravity and {N} = 2 super Yang-Mills theory on general
  scalar manifolds: Symplectic covariance, gaugings and the momentum map.
\newblock {\em J. Geom. Phys.}, 23:111--189, 1997, {\tt hep-th/9605032}.

\bibitem{Dall'Agata:2003yr}
Gianguido Dall'Agata, Riccardo D'Auria, Luca Sommovigo, and Silvia Vaula.
\newblock {D} = 4, {N} = 2 gauged supergravity in the presence of tensor
  multiplets.
\newblock {\em Nucl. Phys.}, B682:243--264, 2004, {\tt hep-th/0312210}.

\bibitem{Sommovigo:2004vj}
Luca Sommovigo and Silvia Vaula.
\newblock {D} = 4, {N} = 2 supergravity with Abelian electric and magnetic
  charge.
\newblock {\em Phys. Lett.}, B602:130--136, 2004, {\tt hep-th/0407205}.

\bibitem{D'Auria:2004yi}
Riccardo D'Auria, Luca Sommovigo, and Silvia Vaula.
\newblock {N} = 2 supergravity Lagrangian coupled to tensor multiplets with
  electric and magnetic fluxes.
\newblock {\em JHEP}, 11:028, 2004, {\tt hep-th/0409097}.

\bibitem{Cecotti:1984rk}
S.~Cecotti, L.~Girardello, and M.~Porrati.
\newblock TWO INTO ONE WON'T GO.
\newblock {\em Phys. Lett.}, B145:61, {\tt 1984}.

\bibitem{Cecotti:1985sf}
S.~Cecotti, L.~Girardello, and M.~Porrati.
\newblock AN EXCEPTIONAL {N}=2 SUPERGRAVITY WITH FLAT POTENTIAL AND PARTIAL
  SUPERHIGGS.
\newblock {\em Phys. Lett.}, B168:83, {\tt 1986}.

\bibitem{Ferrara:1995xi}
Sergio Ferrara, Luciano Girardello, and Massimo Porrati.
\newblock Spontaneous Breaking of {N}=2 to {N}=1 in Rigid and Local
  Supersymmetric Theories.
\newblock {\em Phys. Lett.}, B376:275--281, 1996, {\tt hep-th/9512180}.

\bibitem{Taylor:1999ii}
Tomasz~R. Taylor and Cumrun Vafa.
\newblock {RR} flux on Calabi-Yau and partial supersymmetry breaking.
\newblock {\em Phys. Lett.}, B474:130--137, 2000, {\tt hep-th/9912152}.

\bibitem{Andrianopoli:2001gm}
Laura Andrianopoli, Riccardo D'Auria, and Sergio Ferrara.
\newblock Consistent reduction of {N} = 2 $\to$ {N} = 1 four dimensional
  supergravity coupled to matter.
\newblock {\em Nucl. Phys.}, B628:387--403, 2002, {\tt hep-th/0112192}.

\bibitem{Louis:2002vy}
Jan Louis.
\newblock Aspects of spontaneous {N} = 2 $\to$ {N} = 1 breaking in
  supergravity.
\newblock 2002, {\tt hep-th/0203138}.

\bibitem{Gunara:2003td}
B.~E. Gunara.
\newblock {\tt Spontaneous {N}=2 $\to$ {N}=1 supersymmetry breaking and the
  super-Higgs effect in supergravity}.
\newblock Goettingen, Germany: Cuvillier (2003) 64 p.

\bibitem{Bodner:1989cg}
M.~Bodner and A.~C. Cadavid.
\newblock DIMENSIONAL REDUCTION OF TYPE {IIB} SUPERGRAVITY AND EXCEPTIONAL
  QUATERNIONIC MANIFOLDS.
\newblock {\em Class. Quant. Grav.}, 7:829, {\tt 1990}.

\bibitem{Bohm:1999uk}
Robert Bohm, Holger Gunther, Carl Herrmann, and Jan Louis.
\newblock Compactification of type {IIB} string theory on {C}alabi-{Y}au
  threefolds.
\newblock {\em Nucl. Phys.}, B569:229--246, 2000, {\tt hep-th/9908007}.

\bibitem{Candelas:1989bb}
Philip Candelas and Xenia de~la Ossa.
\newblock MODULI SPACE OF {C}ALABI-{Y}AU MANIFOLDS.
\newblock {\em Nucl. Phys.}, B355:455--481, {\tt 1991}.

\bibitem{tian}
G.~Tian.
\newblock {\em Mathematical Aspects of String Theory}.
\newblock World Scientific, 1987.

\bibitem{Ibanez:2004iv}
Luis~E. Ibanez.
\newblock The fluxed {MSSM}.
\newblock {\em Phys. Rev.}, D71:055005, 2005, {\tt hep-ph/0408064}.

\bibitem{Aldazabal:2000dg}
G.~Aldazabal, S.~Franco, Luis~E. Ibanez, R.~Rabadan, and A.~M. Uranga.
\newblock D = 4 chiral string compactifications from intersecting branes.
\newblock {\em J. Math. Phys.}, 42:3103--3126, 2001, {\tt hep-th/0011073}.

\bibitem{Will:2001mx}
Clifford~M. Will.
\newblock The confrontation between general relativity and experiment.
\newblock {\em Living Rev. Rel.}, 4:4, 2001, {\tt gr-qc/0103036}.

\bibitem{Candelas:1988di}
Philip Candelas, Paul~S. Green, and Tristan Hubsch.
\newblock FINITE DISTANCES BETWEEN DISTINCT {C}ALABI-{Y}AU VACUA: (OTHER WORLDS
  ARE JUST AROUND THE CORNER).
\newblock {\em Phys. Rev. Lett.}, 62:1956, {\tt 1989}.

\bibitem{Candelas:1989ug}
Philip Candelas, Paul~S. Green, and Tristan Hubsch.
\newblock ROLLING AMONG {C}ALABI-{Y}AU VACUA.
\newblock {\em Nucl. Phys.}, B330:49, {\tt 1990}.

\bibitem{Greene:1996cy}
Brian~R. Greene.
\newblock String theory on {C}alabi-{Y}au manifolds.
\newblock 1996, {\tt hep-th/9702155}.

\bibitem{Lukas:2004du}
Andre Lukas, Eran Palti, and P.~M. Saffin.
\newblock Type {IIB} conifold transitions in cosmology.
\newblock {\em Phys. Rev.}, D71:066001, 2005, {\tt hep-th/0411033}.

\bibitem{Palti:2005kv}
Eran Palti, Paul Saffin, and Jon Urrestilla.
\newblock The effects of inhomogeneities on the cosmology of type {IIB}
  conifold transitions.
\newblock {\em JHEP}, 03:029, 2006, {\tt hep-th/0510269}.

\bibitem{Gaida:1998km}
Ingo Gaida, Swapna Mahapatra, Thomas Mohaupt, and Wafic~A. Sabra.
\newblock Black holes and flop transitions in {M}-theory on {C}alabi-{Y}au
  threefolds.
\newblock {\em Class. Quant. Grav.}, 16:419--433, 1999, {\tt hep-th/9807014}.

\bibitem{Brandle:2002fa}
Matthias Brandle and Andre Lukas.
\newblock Flop transitions in {M}-theory cosmology.
\newblock {\em Phys. Rev.}, D68:024030, 2003, {\tt hep-th/0212263}.

\bibitem{Jarv:2003qx}
Laur Jarv, Thomas Mohaupt, and Frank Saueressig.
\newblock Effective supergravity actions for flop transitions.
\newblock {\em JHEP}, 12:047, 2003, {\tt hep-th/0310173}.

\bibitem{Jarv:2003qy}
Laur Jarv, Thomas Mohaupt, and Frank Saueressig.
\newblock {M}-theory cosmologies from singular {C}alabi-{Y}au
  compactifications.
\newblock {\em JCAP}, 0402:012, 2004, {\tt hep-th/0310174}.

\bibitem{Greene:1995hu}
Brian~R. Greene, David~R. Morrison, and Andrew Strominger.
\newblock Black hole condensation and the unification of string vacua.
\newblock {\em Nucl. Phys.}, B451:109--120, 1995, {\tt hep-th/9504145}.

\bibitem{Strominger:1995cz}
Andrew Strominger.
\newblock Massless black holes and conifolds in string theory.
\newblock {\em Nucl. Phys.}, B451:96--108, 1995, {\tt hep-th/9504090}.

\bibitem{Candelas:1989js}
Philip Candelas and Xenia~C. de~la Ossa.
\newblock COMMENTS ON CONIFOLDS.
\newblock {\em Nucl. Phys.}, B342:246--268, {\tt 1990}.

\bibitem{Hubsch:1992nu}
T.~Hubsch.
\newblock {\em {C}alabi-{Y}au manifolds: {A} bestiary for physicists}.
\newblock World Scientific, 1992.

\bibitem{Green:1988bp}
Paul~S. Green and Tristan Hubsch.
\newblock CONNECTING MODULI SPACES OF {C}ALABI-{Y}AU THREEFOLDS.
\newblock {\em Commun. Math. Phys.}, 119:431--441, {\tt 1988}.

\bibitem{Green:1988wa}
Paul~S. Green and Tristan Hubsch.
\newblock PHASE TRANSITIONS AMONG (MANY OF) {C}ALABI-{Y}AU COMPACTIFICATIONS.
\newblock {\em Phys. Rev. Lett.}, 61:1163, {\tt 1988}.

\bibitem{Candelas:1990rm}
Philip Candelas, Xenia~C. De~La~Ossa, Paul~S. Green, and Linda Parkes.
\newblock A pair of {C}alabi-{Y}au manifolds as an exactly soluble
  superconformal theory.
\newblock {\em Nucl. Phys.}, B359:21--74, {\tt 1991}.

\bibitem{Greene:1996dh}
Brian~R. Greene, David~R. Morrison, and Cumrun Vafa.
\newblock A geometric realization of confinement.
\newblock {\em Nucl. Phys.}, B481:513--538, 1996, {\tt hep-th/9608039}.

\bibitem{lefschetz}
S.~Lefschetz.
\newblock L'Analysis Situs et la Geometrie Algebrique, {G}authier-{V}illars,
  Paris.
\newblock {\tt 1924}.

\bibitem{Klemm:1997gg}
A.~Klemm.
\newblock On the geometry behind {N} = 2 supersymmetric effective actions in
  four dimensions.
\newblock 1997, {\tt hep-th/9705131}.

\bibitem{Kofman:2004yc}
Lev Kofman et~al.
\newblock Beauty is attractive: {M}oduli trapping at enhanced symmetry points.
\newblock {\em JHEP}, 05:030, 2004, {\tt hep-th/0403001}.

\bibitem{Watson:2004aq}
Scott Watson.
\newblock Moduli stabilization with the string Higgs effect.
\newblock {\em Phys. Rev.}, D70:066005, 2004, {\tt hep-th/0404177}.

\bibitem{Joyce:2001nm}
Dominic Joyce.
\newblock Lectures on special {L}agrangian geometry.
\newblock 2001, {\tt math.dg/0111111}.

\bibitem{Becker:1995kb}
Katrin Becker, Melanie Becker, and Andrew Strominger.
\newblock Five-branes, membranes and nonperturbative string theory.
\newblock {\em Nucl. Phys.}, B456:130--152, 1995, {\tt hep-th/9507158}.

\bibitem{Vafa:1995ta}
Cumrun Vafa.
\newblock A Stringy test of the fate of the conifold.
\newblock {\em Nucl. Phys.}, B447:252--260, 1995, {\tt hep-th/9505023}.

\bibitem{Ceresole:1995ca}
Anna Ceresole, R.~D'Auria, and S.~Ferrara.
\newblock The Symplectic Structure of {N}=2 Supergravity and its Central
  Extension.
\newblock {\em Nucl. Phys. Proc. Suppl.}, 46:67--74, 1996, {\tt
  hep-th/9509160}.

\bibitem{Mohaupt:2004pr}
Thomas Mohaupt and Frank Saueressig.
\newblock Dynamical conifold transitions and moduli trapping in {M}- theory
  cosmology.
\newblock {\em JCAP}, 0501:006, 2005, {\tt hep-th/0410273}.

\bibitem{'tHooft:1976fv}
Gerard 't~Hooft.
\newblock Computation of the quantum effects due to a four- dimensional
  pseudoparticle.
\newblock {\em Phys. Rev.}, D14:3432--3450, {\tt 1976}.

\bibitem{Alexander:2000xv}
S.~Alexander, Robert~H. Brandenberger, and D.~Easson.
\newblock Brane gases in the early universe.
\newblock {\em Phys. Rev.}, D62:103509, 2000, {\tt hep-th/0005212}.

\bibitem{Battefeld:2005av}
Thorsten Battefeld and Scott Watson.
\newblock String gas cosmology.
\newblock {\em Rev. Mod. Phys.}, 78:435--454, 2006, {\tt hep-th/0510022}.

\bibitem{Achucarro:1998er}
A.~Achucarro, M.~de~Roo, and L.~Huiszoon.
\newblock Unstable vortices do not confine.
\newblock {\em Phys. Lett.}, B424:288--292, 1998, {\tt hep-th/9801082}.

\bibitem{Moriarty:1988fx}
K.~J.~M. Moriarty, Eric Myers, and Claudio Rebbi.
\newblock DYNAMICAL INTERACTIONS OF FLUX VORTICES IN SUPERCONDUCTORS.
\newblock {\em Phys. Lett.}, B207:411, {\tt 1988}.

\bibitem{Achucarro:1997cx}
Ana Achucarro, Julian Borrill, and Andrew~R Liddle.
\newblock Semilocal string formation in two dimensions.
\newblock {\em Phys. Rev.}, D57:3742--3748, 1998, {\tt hep-ph/9702368}.

\bibitem{Urrestilla:2001dd}
Jon Urrestilla, Ana Achucarro, Julian Borrill, and Andrew~R. Liddle.
\newblock The evolution and persistence of dumbbells in electroweak theory.
\newblock {\em JHEP}, 08:033, 2002, {\tt hep-ph/0106282}.

\bibitem{House:2005yc}
Thomas House and Eran Palti.
\newblock Effective action of (massive) {IIA} on manifolds with {SU}(3)
  structure.
\newblock {\em Phys. Rev.}, D72:026004, 2005, {\tt hep-th/0505177}.

\bibitem{Derendinger:2004jn}
Jean-Pierre Derendinger, Costas Kounnas, P.~Marios Petropoulos, and Fabio
  Zwirner.
\newblock Superpotentials in {IIA} compactifications with general fluxes.
\newblock {\em Nucl. Phys.}, B715:211--233, 2005, {\tt hep-th/0411276}.

\bibitem{Villadoro:2005cu}
Giovanni Villadoro and Fabio Zwirner.
\newblock {N} = 1 effective potential from dual type-{IIA} {D}6/{O}6
  orientifolds with general fluxes.
\newblock {\em JHEP}, 06:047, 2005, {\tt hep-th/0503169}.

\bibitem{Camara:2005dc}
P.~G. Camara, A.~Font, and L.~E. Ibanez.
\newblock Fluxes, moduli fixing and {MSSM}-like vacua in a simple {IIA}
  orientifold.
\newblock {\em JHEP}, 09:013, 2005, {\tt hep-th/0506066}.

\bibitem{Grana:2005ny}
Mariana Grana, Jan Louis, and Daniel Waldram.
\newblock Hitchin functionals in {N} = 2 supergravity.
\newblock {\em JHEP}, 01:008, 2006, {\tt hep-th/0505264}.

\bibitem{Acharya:2006ne}
Bobby~S. Acharya, Francesco Benini, and Roberto Valandro.
\newblock Fixing Moduli in Exact Type IIA Flux Vacua.
\newblock 2006, {\tt hep-th/0607223}.

\bibitem{Curio:2000dw}
Gottfried Curio and Axel Krause.
\newblock Four-flux and warped heterotic {M}-theory compactifications.
\newblock {\em Nucl. Phys.}, B602:172--200, 2001, {\tt hep-th/0012152}.

\bibitem{Becker:2002sx}
Katrin Becker and Keshav Dasgupta.
\newblock Heterotic strings with torsion.
\newblock {\em JHEP}, 11:006, 2002, {\tt hep-th/0209077}.

\bibitem{Curio:2003ur}
Gottfried Curio and Axel Krause.
\newblock Enlarging the parameter space of heterotic {M}-theory flux
  compactifications to phenomenological viability.
\newblock {\em Nucl. Phys.}, B693:195--222, 2004, {\tt hep-th/0308202}.

\bibitem{Gurrieri:2004dt}
Sebastien Gurrieri, Andre Lukas, and Andrei Micu.
\newblock Heterotic on half-flat.
\newblock {\em Phys. Rev.}, D70:126009, 2004, {\tt hep-th/0408121}.

\bibitem{Micu:2004tz}
Andrei Micu.
\newblock Heterotic compactifications and nearly-{K}aehler manifolds.
\newblock {\em Phys. Rev.}, D70:126002, 2004, {\tt hep-th/0409008}.

\bibitem{deCarlos:2005kh}
Beatriz de~Carlos, Sebastien Gurrieri, Andre Lukas, and Andrei Micu.
\newblock Moduli stabilisation in heterotic string compactifications.
\newblock {\em JHEP}, 03:005, 2006, {\tt hep-th/0507173}.

\bibitem{Manousselis:2005xa}
Pantelis Manousselis, Nikolaos Prezas, and George Zoupanos.
\newblock Supersymmetric compactifications of heterotic strings with fluxes and
  condensates.
\newblock {\em Nucl. Phys.}, B739:85--105, 2006, {\tt hep-th/0511122}.

\bibitem{Anguelova:2006qf}
Lilia Anguelova and Konstantinos Zoubos.
\newblock Flux superpotential in heterotic {M}-theory.
\newblock 2006, {\tt hep-th/0602039}.

\bibitem{Gurrieri:2002iw}
Sebastien Gurrieri and Andrei Micu.
\newblock Type {IIB} theory on half-flat manifolds.
\newblock {\em Class. Quant. Grav.}, 20:2181--2192, 2003, {\tt hep-th/0212278}.

\bibitem{Behrndt:2005bv}
Klaus Behrndt, Mirjam Cvetic, and Peng Gao.
\newblock General type {IIB} fluxes with {SU(3)} structures.
\newblock {\em Nucl. Phys.}, B721:287--308, 2005, {\tt hep-th/0502154}.

\bibitem{Louis:2002ny}
Jan Louis and Andrei Micu.
\newblock Type {II} theories compactified on {C}alabi-{Y}au threefolds in the
  presence of background fluxes.
\newblock {\em Nucl. Phys.}, B635:395--431, 2002, {\tt hep-th/0202168}.

\bibitem{Giddings:2001yu}
Steven~B. Giddings, Shamit Kachru, and Joseph Polchinski.
\newblock Hierarchies from fluxes in string compactifications.
\newblock {\em Phys. Rev.}, D66:106006, 2002, {\tt hep-th/0105097}.

\bibitem{Grimm:2004uq}
Thomas~W. Grimm and Jan Louis.
\newblock The effective action of {N} = 1 {C}alabi-{Y}au orientifolds.
\newblock {\em Nucl. Phys.}, B699:387--426, 2004, {\tt hep-th/0403067}.

\bibitem{kklt}
Shamit Kachru, Renata Kallosh, Andrei Linde, and Sandip~P. Trivedi.
\newblock De Sitter vacua in string theory.
\newblock {\em Phys. Rev.}, D68:046005, 2003, {\tt hep-th/0301240}.

\bibitem{Balasubramanian:2005zx}
Vijay Balasubramanian, Per Berglund, Joseph~P. Conlon, and Fernando Quevedo.
\newblock Systematics of moduli stabilisation in {C}alabi-{Y}au flux
  compactifications.
\newblock {\em JHEP}, 03:007, 2005, {\tt hep-th/0502058}.

\bibitem{Kachru:2004jr}
Shamit Kachru and Amir-Kian Kashani-Poor.
\newblock Moduli potentials in type {IIA} compactifications with {RR} and {NS}
  flux.
\newblock {\em JHEP}, 03:066, 2005, {\tt hep-th/0411279}.

\bibitem{Behrndt:2004km}
Klaus Behrndt and Mirjam Cvetic.
\newblock General {N} = 1 supersymmetric flux vacua of (massive) type {IIA}
  string theory.
\newblock {\em Phys. Rev. Lett.}, 95:021601, 2005, {\tt hep-th/0403049}.

\bibitem{Behrndt:2004mj}
Klaus Behrndt and Mirjam Cvetic.
\newblock General {N} = 1 supersymmetric fluxes in massive type {IIA} string
  theory.
\newblock {\em Nucl. Phys.}, B708:45--71, 2005, {\tt hep-th/0407263}.

\bibitem{Lust:2004ig}
Dieter Lust and Dimitrios Tsimpis.
\newblock Supersymmetric {AdS}(4) compactifications of {IIA} supergravity.
\newblock {\em JHEP}, 02:027, 2005, {\tt hep-th/0412250}.

\bibitem{Romans:1985tz}
L.~J. Romans.
\newblock MASSIVE {N}=2a SUPERGRAVITY IN TEN-DIMENSIONS.
\newblock {\em Phys. Lett.}, B169:374, {\tt 1986}.

\bibitem{Behrndt:2003ih}
Klaus Behrndt and Mirjam Cvetic.
\newblock Supersymmetric intersecting {D}6-branes and fluxes in massive type
  {IIA} string theory.
\newblock {\em Nucl. Phys.}, B676:149--171, 2004, {\tt hep-th/0308045}.

\bibitem{DeWolfe:2005uu}
Oliver DeWolfe, Alexander Giryavets, Shamit Kachru, and Washington Taylor.
\newblock Type {IIA} moduli stabilization.
\newblock {\em JHEP}, 07:066, 2005, {\tt hep-th/0505160}.

\bibitem{Breitenlohner:1982jf}
Peter Breitenlohner and Daniel~Z. Freedman.
\newblock STABILITY IN GAUGED EXTENDED SUPERGRAVITY.
\newblock {\em Ann. Phys.}, 144:249, {\tt 1982}.

\bibitem{Breitenlohner:1982bm}
Peter Breitenlohner and Daniel~Z. Freedman.
\newblock POSITIVE ENERGY IN ANTI-DE {S}ITTER BACKGROUNDS AND GAUGED EXTENDED
  SUPERGRAVITY.
\newblock {\em Phys. Lett.}, B115:197, {\tt 1982}.

\bibitem{Tomasiello:2005bp}
Alessandro Tomasiello.
\newblock Topological mirror symmetry with fluxes.
\newblock {\em JHEP}, 06:067, 2005, {\tt hep-th/0502148}.

\bibitem{Gurrieri:2002wz}
Sebastien Gurrieri, Jan Louis, Andrei Micu, and Daniel Waldram.
\newblock Mirror symmetry in generalized {C}alabi-{Y}au compactifications.
\newblock {\em Nucl. Phys.}, B654:61--113, 2003, {\tt hep-th/0211102}.

\bibitem{Gutowski:1999tu}
J.~Gutowski, G.~Papadopoulos, and P.~K. Townsend.
\newblock Supersymmetry and generalized calibrations.
\newblock {\em Phys. Rev.}, D60:106006, 1999, {\tt hep-th/9905156}.

\bibitem{Mueller-Hoissen:1987cq}
Folkert Mueller-Hoissen and Richard Stuckl.
\newblock COSET SPACES AND TEN-DIMENSIONAL UNIFIED THEORIES.
\newblock {\em Class. Quant. Grav.}, 5:27, {\tt 1988}.

\bibitem{Micu:2006ey}
Andrei Micu, Eran Palti, and P.~M. Saffin.
\newblock M-theory on seven-dimensional manifolds with {SU(3)} structure.
\newblock 2006, {\tt hep-th/0602163}.

\bibitem{newcvetic}
Mirjam Cvetic and Tao Liu.
\newblock Moduli Stabilization in {M}-theory with {SU}(3) structure (in
  preparation).
\newblock {\tt 2006}.

\bibitem{Lambert:2005sh}
Neil Lambert.
\newblock Flux and {F}reund-{R}ubin superpotentials in {M}-theory.
\newblock {\em Phys. Rev.}, D71:126001, 2005, {\tt hep-th/0502200}.

\bibitem{House:2004pm}
Thomas House and Andrei Micu.
\newblock M-theory compactifications on manifolds with {G}(2) structure.
\newblock {\em Class. Quant. Grav.}, 22:1709--1738, 2005, {\tt hep-th/0412006}.

\bibitem{Dall'Agata:2005fm}
Gianguido Dall'Agata and Nikolaos Prezas.
\newblock Scherk-{S}chwarz reduction of {M}-theory on {G}2-manifolds with
  fluxes.
\newblock {\em JHEP}, 10:103, 2005, {\tt hep-th/0509052}.

\bibitem{D'Auria:2005rv}
Riccardo D'Auria, Sergio Ferrara, and M.~Trigiante.
\newblock Supersymmetric completion of {M}-theory 4{D}-gauge algebra from
  twisted tori and fluxes.
\newblock {\em JHEP}, 01:081, 2006, {\tt hep-th/0511158}.

\bibitem{Kaste:2003zd}
Peter Kaste, Ruben Minasian, and Alessandro Tomasiello.
\newblock Supersymmetric {M}-theory compactifications with fluxes on
  seven-manifolds and {G}-structures.
\newblock {\em JHEP}, 07:004, 2003, {\tt hep-th/0303127}.

\bibitem{Behrndt:2004mx}
Klaus Behrndt and Claus Jeschek.
\newblock Superpotentials from flux compactifications of {M}-theory.
\newblock {\em Class. Quant. Grav.}, 21:S1533--1538, 2004, {\tt
  hep-th/0401019}.

\bibitem{Lukas:2004ip}
Andre Lukas and P.~M. Saffin.
\newblock {M}-theory compactification, fluxes and {AdS}(4).
\newblock {\em Phys. Rev.}, D71:046005, 2005, {\tt hep-th/0403235}.

\bibitem{Behrndt:2004bh}
Klaus Behrndt and Claus Jeschek.
\newblock Fluxes in {M}-theory on 7-manifolds: {G2}, {SU}(3) and {SU}(2)
  structures.
\newblock 2004, {\tt hep-th/0406138}.

\bibitem{Gauntlett:2004hs}
Jerome~P. Gauntlett, Dario Martelli, James Sparks, and Daniel Waldram.
\newblock Supersymmetric {AdS} backgrounds in string and {M}-theory.
\newblock 2004, {\tt hep-th/0411194}.

\bibitem{Franzen:2005ve}
Anne Franzen, Payal Kaura, Aalok Misra, and Rajyavardhan Ray.
\newblock Uplifting the {I}wasawa.
\newblock {\em Fortsch. Phys.}, 54:207--224, 2006, {\tt hep-th/0506224}.

\bibitem{Beasley:2002db}
Chris Beasley and Edward Witten.
\newblock A note on fluxes and superpotentials in {M}-theory compactifications
  on manifolds of {G}(2) holonomy.
\newblock {\em JHEP}, 07:046, 2002, {\tt hep-th/0203061}.

\bibitem{Castellani:1983tc}
L.~Castellani and L.~J. Romans.
\newblock {N}=3 AND {N}=1 SUPERSYMMETRY IN A NEW CLASS OF SOLUTIONS FOR d = 11
  SUPERGRAVITY.
\newblock {\em Nucl. Phys.}, B238:683, {\tt 1984}.

\bibitem{nakahara}
M.~Nakahara.
\newblock {\em Geometry, Topology and Physics}.
\newblock Institute of Physics, 2003.

\bibitem{VanProeyen:1999ni}
Antoine Van~Proeyen.
\newblock Tools for supersymmetry.
\newblock 1999, {\tt hep-th/9910030}.

\bibitem{Castellani:1983yg}
L.~Castellani, L.~J. Romans, and N.~P. Warner.
\newblock A CLASSIFICATION OF COMPACTIFYING SOLUTIONS FOR d = 11 SUPERGRAVITY.
\newblock {\em Nucl. Phys.}, B241:429, {\tt 1984}.

\bibitem{Karthauser:2006wb}
Josef L.~P. Karthauser and P.~M. Saffin.
\newblock The dynamics of coset dimensional reduction.
\newblock {\em Phys. Rev.}, D73:084027, 2006, {\tt hep-th/0601230}.

\end{thebibliography}

\end{document}